\documentclass[longauth]{aa} 

\usepackage{graphicx}
\usepackage{multicol}
\usepackage{txfonts}
\usepackage{array}
\newcommand{\nustar}{\textsl{N\MakeLowercase{u}STAR}\xspace}
\begin{document}

   \title{Hard X-ray Properties of \textit{NuSTAR}  Blazars}

   \author{Gopal Bhatta
          \inst{1},
          Maksym Mohorian\inst{2},
              \and
          Illya Bilinsky\inst{2}
          }

   \institute{Astronomical Observatory, Jagiellonian University, ul. Orla 171, 30-244 Krak\'ow, Poland\\
              \email{gopal@oa.uj.edu.pl}
         \and
             Taras Shevchenko National University of Kyiv, Akademika Hlushkova Ave, 4b, 02000 Kyiv, Ukraine\\
                     }

   \date{Received xxxxxxx ; accepted xxxxxxx}

 
  \abstract
   {  Investigation of the hard X-ray emission properties of blazars is key to the understanding of the central engine of the sources and associated jet process. In particular, simultaneous spectral and timing analyses of the intra-day hard X-ray observations provide us a means to peer into the compact innermost blazar regions, not accessible to our current instruments. }
   {The primary objective of the work is to associate the observed hard X-ray variability properties in the blazars to their flux and spectral states, thereby, based on the correlation among them, extract the details about the emission regions and the processes occurring near the central engine.}
   { We carried out timing, spectral and cross-correlation analysis of 31 \textit{NuSTAR} observations of 13 blazars. We investigated the spectral shapes of the sources using single power-law, broken power-law and log-parabola models.  We also studies the co-relation between the soft and the hard emission using  z-transformed discrete correlation function. In addition, we attempted to constrain smallest emission regions using minimum variability timescales derived from the light curves.}
   {We found that for most of the sources the hard X-ray emission can be well represented by log-parabola model; and that the spectral slopes for different blazar sub-classes are consistent with so called ``blazar sequence''.  We also report a steepest spectra ($\Gamma \sim3$ ) in the BL Lacertae PKS 2155--304 and a hardest spectra ($\Gamma \sim1.4$) in the flat-spectrum radio quasar PKS 2149--306. In addition, we noted a close connection between the flux and spectral slope within the source sub-class in the sense that high flux and/or flux states tend to be harder in spectra.   In BL Lacertae objects, assuming particle acceleration by diffusive shocks and synchrotron cooling as the dominant processes governing the observed flux variability, we constrain the magnetic field of the emission region to be a few gauss; whereas in flat-spectrum radio quasars, using external Compton models, we estimate the energy of the lower end of the injected electrons to be a few Lorentz factors.}
   {}


   \maketitle
%

\section{Introduction}

Blazars, a subclass of active galactic nuclei (AGN), are radio-loud sources with their relativistic  jets closely aligned to the line of sight. The Doppler boosted non-thermal emission is  highly variable over a wide range of spatial and temporal frequencies. The broadband spectral energy distribution (SED) of blazars features two distinct spectral peaks:  The lower peak, usually observed between the radio and the X-ray, is widely accepted to be result of the synchrotron emission by the energetic particles; however, the origin of high energy component,  mostly peaking between UV to $\gamma$-ray, is still debated. There are two widely discussed models based on the origin of seed photons: according to synchrotron self-Compton (SSC) model (e.g. \citealt{Maraschi1992,Mastichiadis2002})  the same population of the electrons emitting synchrotron radiation up-scatters the softer photon to high energy; whereas in external Compton (EC)  model the seed photons for the Compton up-scattering are provided by the various components of an AGN e.g.,  accretion disk (AD; \citealt{Dermer1993}), broad-line region (BLR; \citealt{Sikora1994}) and dusty torus (DT; \citealt{Blazejowski2000}).

Blazars consists of  further  two sub-classes: flat-spectrum radio quasars (FSRQ) and  BL Lacertae (BL Lac) sources. FSRQs are more luminous sources which show emission lines over the continuum; and they have the synchrotron peak in the lower frequency.  As the sources are found to have abundant seed photons due to accretion disk, BLR and dusty torus, the high energy emission is most likely due to  EC process, as opposed to SSC  \citep{Ghisellini2011}.  BL Lac objects constitute  less powerful sub-class having weak or no emission lines over the continuum, and the synchrotron peak in them lies in the UV to X-rays bands. BL Lacs represent an extreme class of sources with an excess of high energy emission (hard X-rays to TeV emission) resulting from the synchrotron and inverse-Compton processes. However, their apparent low luminosity could be due to lack of strong circum-nuclear photon fields and relatively low accretion rates. Blazar sources can have further sub-division  based  on the frequency of the synchrotron peak ($\nu_s$):  high synchrotron peaked blazars (HSP; $\nu_s > 10^{15}$ Hz), intermediate synchrotron  peaked blazars (ISP; $10^{14}<\nu_s <10^{15} $ Hz) and  low synchrotron peaked blazars (LSP; $\nu_s < 10^{14}$ Hz) \citep[see][]{Abdo2010}. In the unifying scheme known as ``blazar sequence''  as we move from FSRQ to HSP,  bolometric luminosity decreases but  $\gamma$-ray emission increases  \citep{Fossati1998,Ghisellini2017}.  This means, while FSRQs are $\gamma$-ray dominated, in HSP sources synchrotron and $\gamma$-ray emission become comparable. In other words, with the increase in their bolometric luminosities, blazars become ``redder''  and ``Compton dominant'' as the ratio of the luminosities at the Compton peak to the synchrotron peak frequency increases.  

\begin{table*}
	\caption{General information about the studied blazar sources}
	\centering
	\label{info}
	\begin{tabular}{l|l|l|l|c}
		\hline
		Source name & Source class &R.A. (J2000) & Dec. (J2000) & Redshift (z) \\
		\hline
		S5 0014+81 &FSRQ & $00^h17^m08.4748^s$ & $+81^d35^m08.136^s$ & 3.366\\
		B0222+185 &FSRQ & $02^h25^m04.6688^s$ & $+18^d46^m48.766^s$ & 2.690 \\
		HB 0836+710 &FSRQ & $08^h41^m24.3652^s$ & $+70^d53^m42.173^s$ & 2.172 \\
		3C 273 &FSRQ & $12^h29^m06.6997^s$ & $+02^d03^m08.598^s$ &0.158 \\
		3C 279 &FSRQ, TeV & $12^h56^m11.1665^s$ & $-05^d47^m21.523^s$ & 0.536 \\
		PKS 1441+25 &FSRQ, TeV &$14^h43^m56.9^s$ & $+25^d01^m44^s$ & 0.939 \\
		PKS 2149--306 &FSRQ & $21^h51^m55.5239^s$ & $-30^d27^m53.697^s$ & 2.345 \\
		1ES 0229+200 &BL Lac, HSP, TeV & $02^h32^m48.616^s$ & $+20^d17^m17.45^s$ & 0.140 \\
		S5 0716+714 &BL Lac, ISP, TeV &$07^h21^m53.4^s$ & $+71^d20^m36^s$ & 0.300\\
		Mrk 501 & BL Lac, HSP, TeV & $16^h53^m52.2167^s$ & $+39^d45^m36.609^s$ & 0.0334 \\
		1ES 1959+650 &BL Lac, HSP, TeV & $19^h59^m59.8521^s$ & $+65^d08^m54.652^s$ & 0.048\\
		PKS 2155--304 &BL Lac, HSP, TeV & $21^h58^m52.0651^s$ & $-30^d13^m32.118^s$ & 0.116 \\
		BL Lac &BL Lac, ISP, TeV &$22^h02^m43.3^s$ & $+42^d16^m40^s$ & 0.068 \\

		\hline
	\end{tabular}
\end{table*}

Blazar continuum emission is characterized by broadband emission  which is variable on diverse timescales. The variability timescales can be  long-term (years to decades), short-term (weeks to months) and intra-day/night (minutes to hours). Long-term variability most likely arises due to variable accretion rates; short-term flaring episodes lasting a few weeks could be due to the shock waves propagating down the jets; and the low-amplitude rapid variability known as intra-day variability might arise due to the turbulent flow of the plasma in the innermost regions of the jets  \citep[e.g.][]{bhatta13,Cawthorne2006,Lister2005,Hughes98,Marscher1996}.  In general, the variability shown by AGNs appears predominantly aperiodic in nature, although quasi-periodic oscillations on various timescales have been detected for a number of sources \citep[see][]{Bhatta2017,bhatta16c,Zola2016}

  Blazar variability in the X-ray bands has been extensively studied using numerous instruments over  the past several decades as following.  In a study including  a large sample of BL Lac sources observed with Einstein Observatory Imaging Proportional Counter (IPC), the source spectra were well described by single power-law model  with spectral slope indexes ($\alpha_{X}$) in the range of 0.1--0.5 \citep{Worrall1990}. The soft X-ray study of a sample of radio selected BL Lacs (RBL; \citealt{Urry1996}) and  X-ray selected BL Lacertae objects (XBL; \citealt{Perlman1996}) using ROSAT Position Sensitive Proportional Counter (PSPC) showed  that the 0.2--2.0 keV spectra of the sources could be well described mostly by single power-law with $\alpha_{X}$ between 0.5 --2.3.   The single power-law  and the broken power-law models were successfully used to describe the X-ray spectra from various instruments such as Advanced Satellite for Cosmology and Astrophysics (ASCA)  \citep[e.g.][]{Kubo1998}, BeppoSAX  \citep[][]{Wolter1998,Padovani2002}, European X-ray Observatory Satellite (EXOSAT) \citep[e.g.][]{Sambruna1994} and the ROentgen SATellite (ROSAT) observations  \citep[e.g.][] {Perlman1996,Urry1996}.   In the ASCA spectra of 4 FSRQs,  \citet{Sambruna2000} found steep ($\Gamma_{X}\sim2-2.5)$ soft X-ray (0.2--2.4  keV)  photon indexes similar to those observed in synchrotron-dominated BL Lac objects;  and the spectra were found to be consistent with power-law models. However, the ASCA spectra were observed to be flatter than their ROSAT spectra. Similarly, in some cases  continuously curved, log-parabola model provided better representation for the X-ray spectral distribution of some  sources  \citep[][]{Donato2005}.  Also, \citet{Massaro2004a, Massaro2004b}  found the log-parabola as the best model for the characterization of X-ray spectra of Mrk 421 and Mrk 501 in their multiple flux states.  Spectral curvature have also been detected in  the  XMM-Newton spectra of  a number of X-ray bright BL Lac objects from the Einstein Slew Survey  \citep[see][]{Perlman2005} and  several  BeppoSAX blazars  \citep[see][]{Donato2005}.  Using Swift/XRT  spectra  of a sample of TeV  blazars,  \citet{Wierzcholska2016b} decomposed the synchrotron and the inverse Compton components. Furthermore, in a few sources a linear relation between the flux and the hardness ratio, also called  “harder-when-brighter” trend, has been reported  by \citet{Zhang2005,Zhang2006}. Similarly, soft and hard lags were observed during the correlation study between the emission in different X-ray bands   \citep[e.g.][]{Fossati2000a,Zhang2006}. In addition, hysteresis loops in the  spectral index and flux intensity plane have been reported  \citep[e.g.][]{Ravasio2004,Falcone2004,Brinkmann2005}. To sum up, these studies over the decades suggest  that the sources  exhibit high amplitude rapid variability on diverse timescales ranging from a few hours to a few months, and   that the nature of the X-ray blazar spectra in various energy bands behaves in a variable and complex fashion. 
  
  \begin{table*}
	\centering
	\caption{Observational data and variability properties of the \textit{NuSTAR} blazar sources.}
	\begin{tabular}{|c|l|c|c|c|c|c|c|}
		\hline
		\# & Source & Obs. date & Obs. ID & Obs. time (ks) & $F_{var}$ (per cent) & VA & $\tau_{var}$ (ks)\\
		\hline
		1 & S5 0014+81 & 2014-12-21 & 60001098002 & 46.80 & 30.02$\pm$1.38 & 3.15$\pm$1.23 & 0.91$\pm$0.83\\
		2 && 2015-01-23 & 60001098004 & 39.60 & 14.29$\pm$1.73 & 2.02$\pm$1.07 & 1.77$\pm$0.74\\
		
		3 & B0222+185 & 2014-12-24 & 60001101002 & 61.00 & 6.92$\pm$1.37 & 0.97$\pm$0.40 & 4.48$\pm$2.96\\
		4 && 2015-01-18 & 60001101004 & 70.00 & 8.90$\pm$1.40 & 1.12$\pm$0.60 & 3.58$\pm$1.67\\
		
		5 & HB 0836+710 & 2013-12-15 & 60002045002 & 47.00 & 12.92$\pm$0.87 & 1.43$\pm$0.35 & 2.53$\pm$0.91\\
		6 && 2014-01-18 & 60002045004 & 67.00 & 8.85$\pm$0.52 & 1.11$\pm$0.35 & 4.99$\pm$1.94\\
		
		7 & 3C 273 & 2016-06-26 & 10202020002 & 74.70 & 10.05$\pm$6.01 & 1.48$\pm$2.16 & 8.81$\pm$3.34\\
		8 && 2017-06-26 & 10302020002 & 72.00 & 14.86$\pm$4.79 & 4.06$\pm$6.09 & 1.24$\pm$1.70\\
		
		9 & 3C 279 & 2013-12-16 & 60002020002 & 78.00 & 16.59$\pm$0.77 & 2.28$\pm$0.52 & 2.31$\pm$1.26\\
		10 && 2013-12-31 & 60002020004 & 78.00 & 17.26$\pm$0.28 & 1.50$\pm$0.17 & 5.61$\pm$3.99\\
		
		11 & PKS 1441+25 & 2015-04-25 & 90101004002 & 72.00 & 26.01$\pm$3.82 & 2.82$\pm$1.34 & 1.24$\pm$0.62\\
		
		12 & PKS 2149--306 & 2013-12-17 & 60001099002 & 71.10 & 9.30$\pm$0.65 & 1.21$\pm$0.21 & 3.31$\pm$2.27\\
		13 && 2014-04-18 & 60001099004 & 90.00 & 10.60$\pm$0.88 & 1.64$\pm$0.80 & 2.24$\pm$1.00\\
		
		14 & 1ES 0229+200 & 2013-10-05 & 60002047004 & 38.00 & 13.33$\pm$0.85 & 1.61$\pm$0.47 & 2.35$\pm$1.23\\
		
		15 & S5 0716+714 & 2015-01-24 & 90002003002 & 32.00 & 14.93$\pm$1.45 & 1.49$\pm$0.58 & 2.79$\pm$1.43\\
		
		16 & Mrk 501 & 2013-04-13 & 60002024002 & 35.00 & 5.24$\pm$0.66 & 0.75$\pm$0.14 & 6.30$\pm$2.21\\
		17 && 2013-05-08 & 60002024004 & 55.00 & 17.76$\pm$0.42 & 1.52$\pm$0.14 & 4.89$\pm$1.56\\
		18 && 2013-07-12 & 60002024006 & 20.00 & 5.23$\pm$0.43 & 0.59$\pm$0.17 & 18.79$\pm$10.01\\
		19 && 2013-07-13 & 60002024008 & 20.40 & 9.79$\pm$0.30 & 1.05$\pm$0.11 & 2.25$\pm$0.89\\
		
		20 & 1ES 1959+650 & 2014-09-17 & 60002055002 & 32.00 & 33.60$\pm$0.58 & 2.48$\pm$0.35 & 3.76$\pm$1.14\\
		21 && 2014-09-22 & 60002055004 & 32.00 & 13.93$\pm$0.66 & 0.68$\pm$0.14 & 8.31$\pm$5.59\\
				
		22 & PKS 2155--304 & 2012-07-08 & 10002010001 & 71.00 & 19.66$\pm$0.75 & 3.44$\pm$2.65 & 0.95$\pm$0.64\\
		23 && 2013-04-23 & 60002022002 & 90.00 & 25.17$\pm$1.07 & 2.21$\pm$0.80 & 1.86$\pm$0.84\\
		24 && 2013-07-16 & 60002022004 & 26.10 & 27.78$\pm$0.98 & 5.65$\pm$3.40 & 0.79$\pm$0.40\\
		25 && 2013-08-02 & 60002022006 & 29.70 & 22.03$\pm$1.92 & 2.96$\pm$3.33 & 0.30$\pm$0.12\\
		26 && 2013-08-08 & 60002022008 & 36.00 & 18.67$\pm$6.63 & 2.10$\pm$1.59 & 1.93$\pm$1.09\\
		27 && 2013-08-14 & 60002022010 & 31.50 & 37.69$\pm$6.61 & 3.76$\pm$2.89 & 1.59$\pm$0.86\\
		28 && 2013-08-26 & 60002022012 & 24.30 & 19.74$\pm$1.33 & 1.70$\pm$0.26 & 3.13$\pm$1.98\\
		29 && 2013-09-04 & 60002022014 & 29.70 & 18.90$\pm$1.98 & 1.52$\pm$0.24 & 3.41$\pm$1.41\\
		30 && 2013-09-28 & 60002022016 & 25.20 & 31.34$\pm$2.42 & 7.28$\pm$7.74 & 0.77$\pm$0.66\\
		
		31 & BL Lac & 2012-12-11 & 60001001002 & 42.30 & 25.03$\pm$4.12 & 3.55$\pm$2.86 & 1.88$\pm$0.96\\
		\hline
		\end{tabular}
	\label{table:obs}
\end{table*}

Recently, several sources have been observed in the hard X-ray regime by Nuclear Spectroscopic Telescope Array (\textit{NuSTAR} ), mostly to complement the contemporaneous multi-frequency observing campaigns:  \citet{Madsen2015}  described the \textit{NuSTAR} spectra of the blazar 3C 273 by an exponentially cutoff power-law with a weak reflection component from cold, dense material; the spectra revealed an evidence of a weak  neutral iron line as well.  
   In the \textit{NuSTAR}  observations of the  FSRQ 3C 279, \citet{Hayashida2015} observed a  spectral softening by $\Delta \Gamma_{X} \approx$ 0.4 at $\sim$4 keV between the two observation epochs.  Blazar S5 0836+71 was found to be highly variable in hard X-ray during the  broadband study by \citet{Paliya2015}.  Similarly, \citet{Furniss2015} found that the combined Swift and \textit{NuSTAR} of the blazar Mrk 501, during both a low and high flux state,  could be well fitted  by a log-parabolic spectrum.  In the combined   \textit{NuSTAR} and Swift/XRT  spectra of S5 0716+714,  \citet{Wierzcholska2016a} reported a break energy at $\sim$8 keV revealing both low and high energy components.  \citet{Sbarrato2016} in their study of the two high red-shifted blazars,   S5 0014+81 and B0222+185,  concluded that the two sources harbored  the most luminous accretion disk and the most powerful jet, respectively, placing  them at the extreme end of the disk-jet relation for $\gamma$-ray blazars. \citet{Rani2017}  observed  rapid hard X-ray variability on hour timescales in a few blazar sources. Similarly, \citet{Pandey2017}  reported the instances of intraday variability in the \textit{NuSTAR}  light curves of  a number of TeV blazars, and also noticed a general “harder-when-brighter” trend.

 In this paper, we conduct a thorough analysis of all the blazar sources from the \textit{NuSTAR} data archive by carrying out timing, spectral and cross-correlation analyses to study  the nature of the  variability properties of blazars in the hard X-ray regime. Our work is mainly motivated to understand the physical process in the blazars by exploring the possible relation of variability properties, particularly variability and the minimum variability timescale, with the mean flux  and the spectral state of the sample sources, and thereby shed light into the innermost regions of blazars, hidden from our direct view.
 We organize our presentations in the following way: In section 2, the observation and the data processing of 31 \textit{NuSTAR} observations of 13 blazar sources are discussed. We present our timing, spectral and cross-correlation study on the light curves and the spectra in section 3. In Section 4, we report  several  interesting observational features such as  rapid flux and spectra variability, a connection between higher flux and harder spectra, and hard and soft lags; and discuss the observed features in the light of current blazar models. Finally we summarize our conclusions in Section 5.

\section{Observations and Data Reduction \label{sec:obs2}}

\subsection{Source Sample }

We selected the sample sources from the \textit{NuSTAR}  archive that were classified as blazar sources. Moreover, only the observations with observation period greater than 10 kilo-seconds (ks) and carrying the issue flag 0 were included in the study. The name,  class, position and redshift for the sources are listed in Table \ref{info}. The source sample consists of 7 FSRQs, 2 ISPs and 4 HSPs\footnote{We did not include Mrk 421 in the sample because it is being exclusively studied by our research group.} which are also TeV blazars. The redshift of the sources has a diverse range from the nearest one (z=0.0334; Mrk 501) to the farthest one (z=3.366; S5 0014+81).

\subsection{\textit{NuSTAR}  Observations }
Nuclear Spectroscopic Telescope Array (\textit{NuSTAR} ) is a sensitive hard X-ray (3 -- 79 keV) instrument with two  focal plane modules: FPMA and FPMB. The observatory operates within the bandpass with spectral resolution of $\sim$1 keV. The field of view of each telescope is $\sim13'$, and the half-power diameter of an image of a point source is $\sim$1' \citep[see][]{Harrison2013}.  The raw data products were processed using \textit{NuSTAR} Data Analysis Software (NuSTARDAS) package version 1.3.1. We reduced and analyzed the observations using HEASOFT\footnote{https://heasarc.nasa.gov/lheasoft/} version 6.21 and CALDB version 2017-06-14. By using the  standard \textsl{nupipeline} script, calibrated and cleaned event files were produced. Source flux and spectra were extracted from a region of 30'' radius centered around  the source location, and the background was extracted from a 70'' radius region relatively close to the source but also far enough to be free from contamination by the source. The light curves were generated using a time bin of 15 minutes. Similarly, in order to have at least 30 counts  per channel, the  spectra were re-binned using the task \emph{grppha}.
\begin{figure}
\begin{center}
\resizebox{.4\textwidth}{!}{\includegraphics[angle=0]{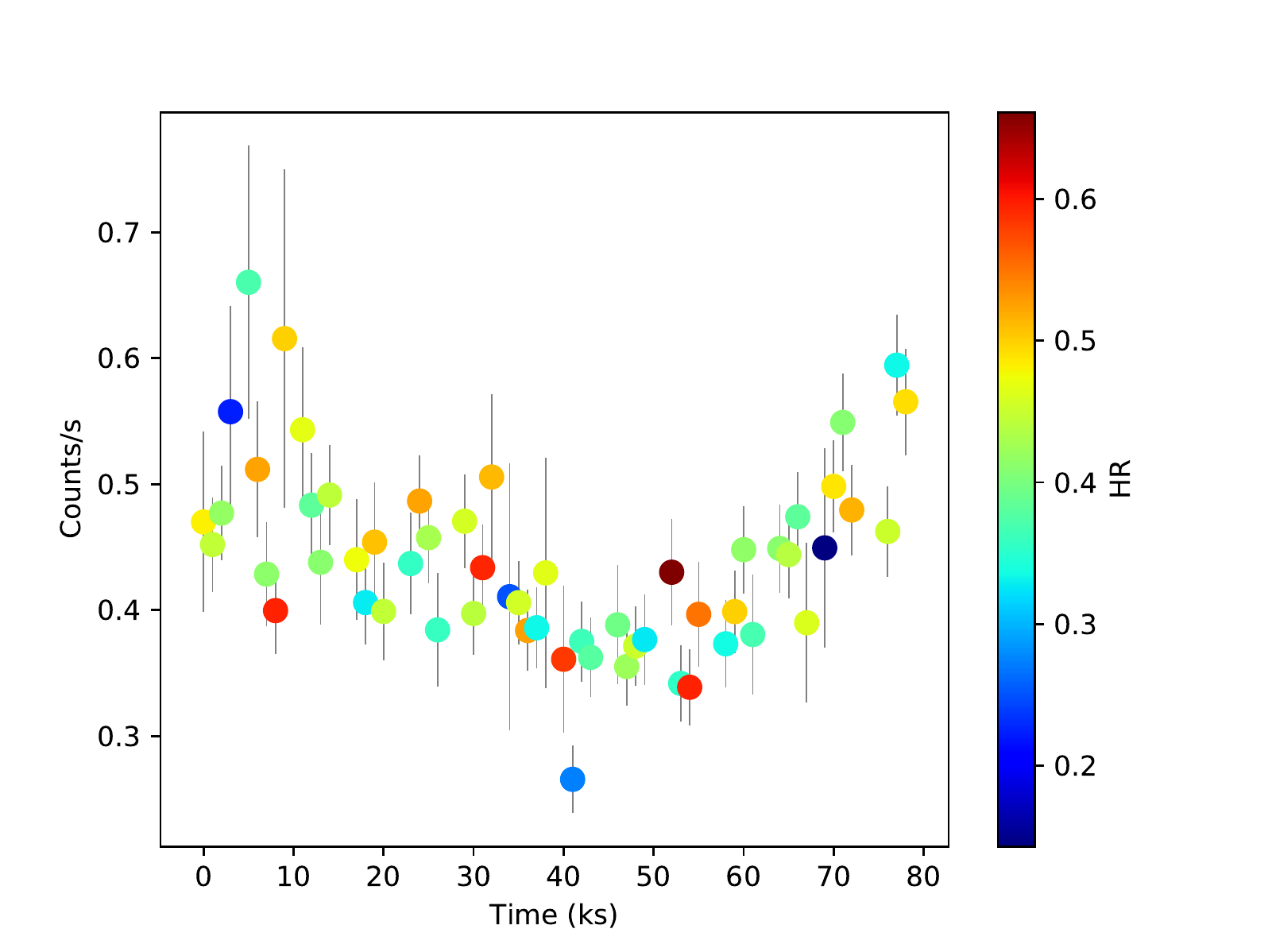}} \\
3C 279, 60002020002\\
\resizebox{.4\textwidth}{!}{\includegraphics[angle=0]{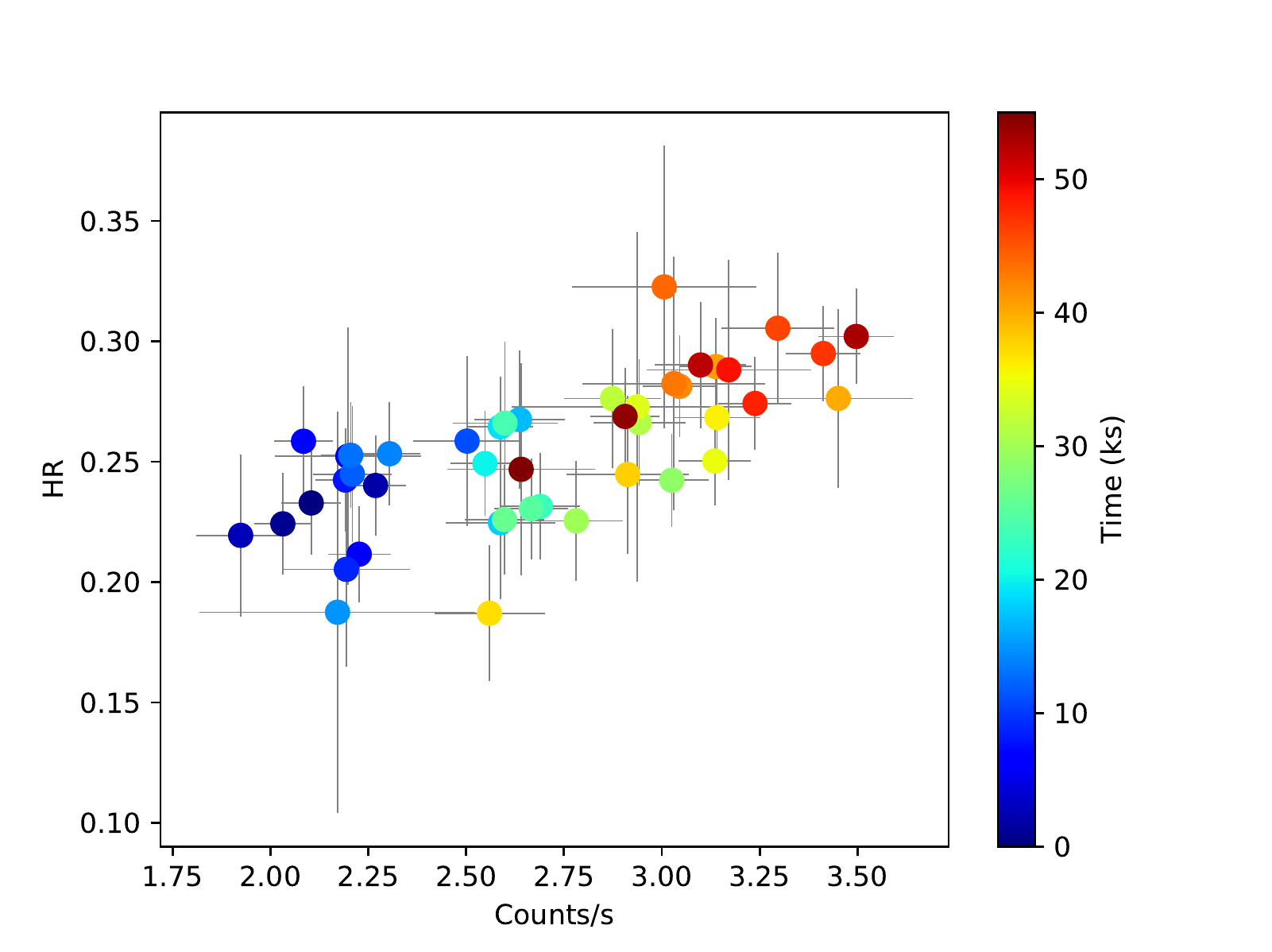}} \\
Mrk 501, 60002024004\\
\resizebox{.35\textwidth}{!}{\includegraphics[angle=-90]{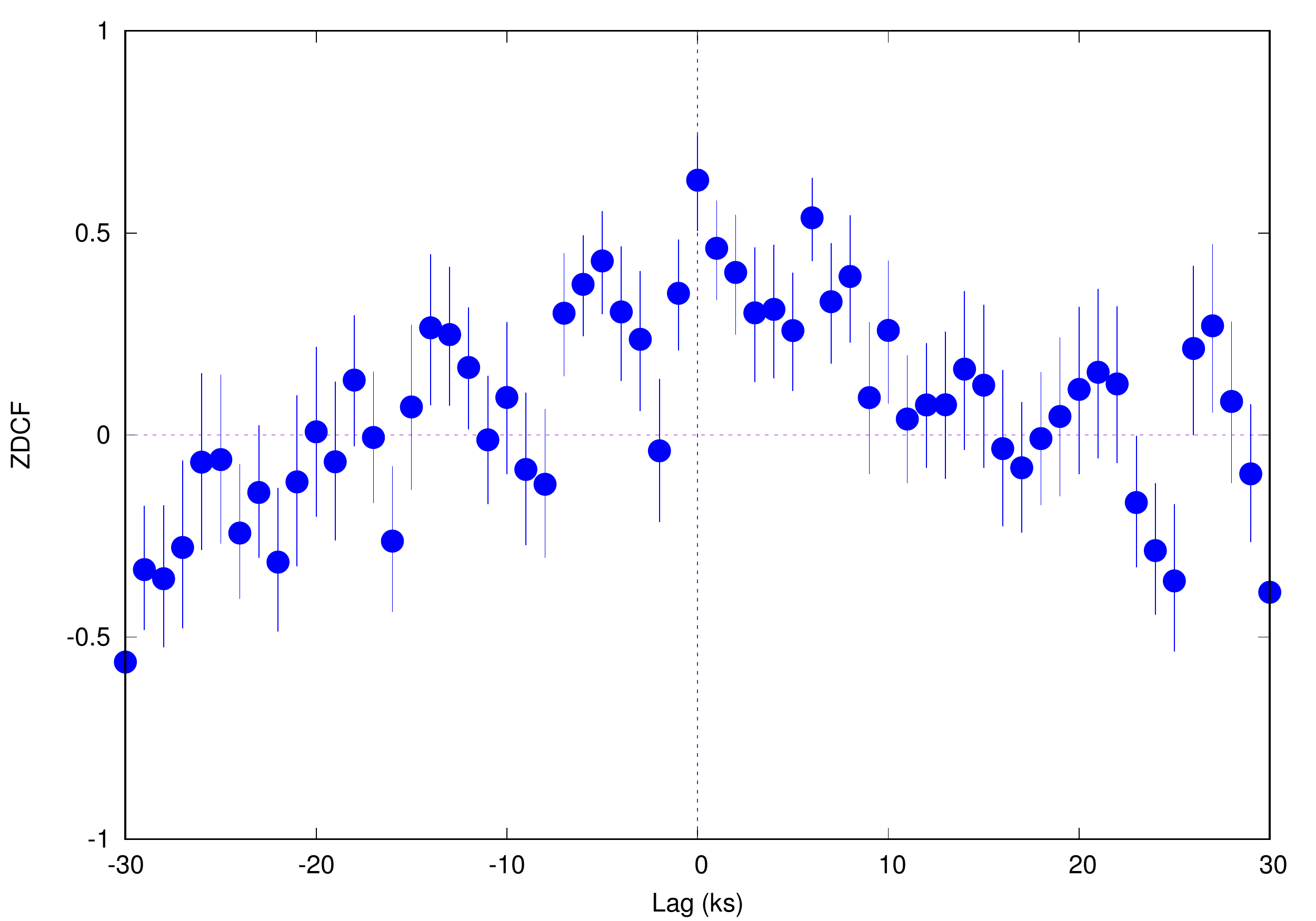}} \\
3C 279, 60002020002
\resizebox{.4\textwidth}{!}{\includegraphics[angle=-90]{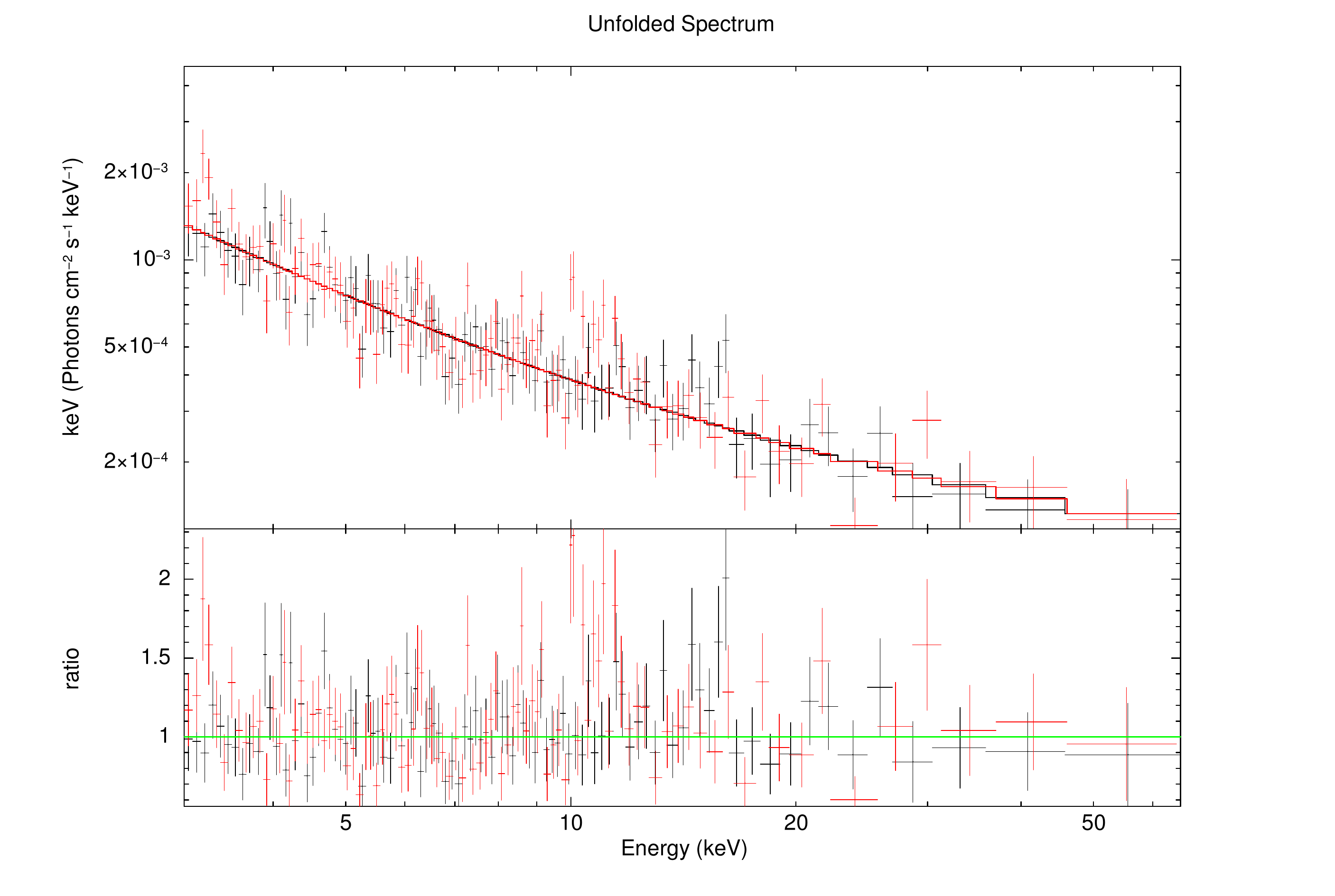}} \\
S5 0716+714, 90002003002

\caption{Hard X-ray  NuSTAR  observations of blazar sources showing light curve  and flux-HR relation, ZDCF  and  spectral fit, from the top to the bottom respectively, for the sources 3C 279,  Mrk 501, 3C 279, and S5 0716+714, respectively. The color bars in light curve and the flux-HR plots represent the HR and time respectively. Similar plots for other sources are shown in the online material section.}
\label{fig:20}
\end{center}
\end{figure}

  \section{Analysis \label{sec:analysis}}
 The \textit{NuSTAR} observations  of the  blazar sources  discussed in this paper along with their observation ID and observation dates are listed in Table \ref{table:obs}. The light curve of the source 3C 279 (obs. ID: 60002020002), displaying modulations in the hard X-ray emission, is presented on the top panel of Figure \ref{fig:20}. To see the spectral states of the individual flux points, the plot symbols are color-coded according to the hardness ratio (defined below). The light curves for the other observations are presented similarly in the on-line material. In order to examine the hard X-ray variability properties of the sample sources, we performed timing, spectral and cross-correlation analyses which are discussed below. 
 
\subsection{Flux variability}
Most of the observations  for the sample sources were found to be rapidly variable within the observation period. The observed variability is quantified by defining two measures: Variability amplitude (VA) measuring the peak-to-peak flux oscillations is given as

 \begin{equation}
  VA=\frac{F_{max}-F_{min}}{F_{min}},
  \label{VA}
  \end{equation}
where $F_{max}$ and $F_{min}$ are the maximum and minimum flux in counts/sec. This kind of variability measure, derived only from the extreme fluxes, may not represent the overall variability. In such case, fractional variability \citep[FV; see][]{vau03,Bhatta2018}, which considers all the fluxes in the light curve, may be more suitable measure to represent the observed variability. Following \citet[][]{Burbidge1974}, the minimum timescale of such variability is determined  using the expression 

 \begin{equation}
   \tau_{var}=  \left | \frac{\Delta t}{ \Delta lnF} \right |,
  \end{equation}
where $\Delta t$ is the time interval between flux measurements  \citep[see also][]{Hagen-Thorn2008}. To compute the uncertainty in $\tau_{var}$, we followed the general error propagation rule i.e. for a general function $y=f(x_{1}, x_{2}, . . x_{n})$ with the corresponding uncertainties $\Delta x_{1}, \Delta x_{2},  . . \Delta x_{n}$ in $x_{1}, x_{2}, . . x_{n}$, respectively, uncertainty in y can be expressed as \citep[similar to Equation 3.14 in][]{Bevington2003}

\begin{equation}
\Delta y\simeq \sqrt{\left ( \frac{\partial y }{\partial x_{1}} \Delta x_{1} \right )^{2} +\left ( \frac{\partial y }{\partial x_{2}} \Delta x_{2} \right )^{2}+...+\left ( \frac{\partial y }{\partial x_{n}} \Delta x_{n} \right )^{2}
}
\label{error_prop}
\end{equation}

 Thus using Equation  \ref{error_prop},  uncertainty in $\tau_{var}$ are estimated as
\begin{equation}
\Delta \tau _{var}
\simeq \sqrt{\frac{F_{1}^{2} \Delta F_{2}^{2} +F_{2}^{2} \Delta F_{1}^{2}}{F_{1}^{2}F_{2}^{2}\left (  ln \left [ F_{1}/F_{2} \right ]\right )^{4}}}\ \Delta t,
\end{equation}

\noindent where $F_{1}$ and $F_{2}$ are the count rates  used to estimate the minimum variability timescales, and $\Delta F_{1}$ and $ \Delta F_{2}$ their corresponding uncertainties.

 All these quantities characterizing flux variability in the sources i.e. fractional variability, variability amplitude and minimum variability timescales for the source sample are listed in the 6, 7, and 8th column, respectively, of Table 2.

\begin{figure}
\includegraphics[width=\columnwidth]{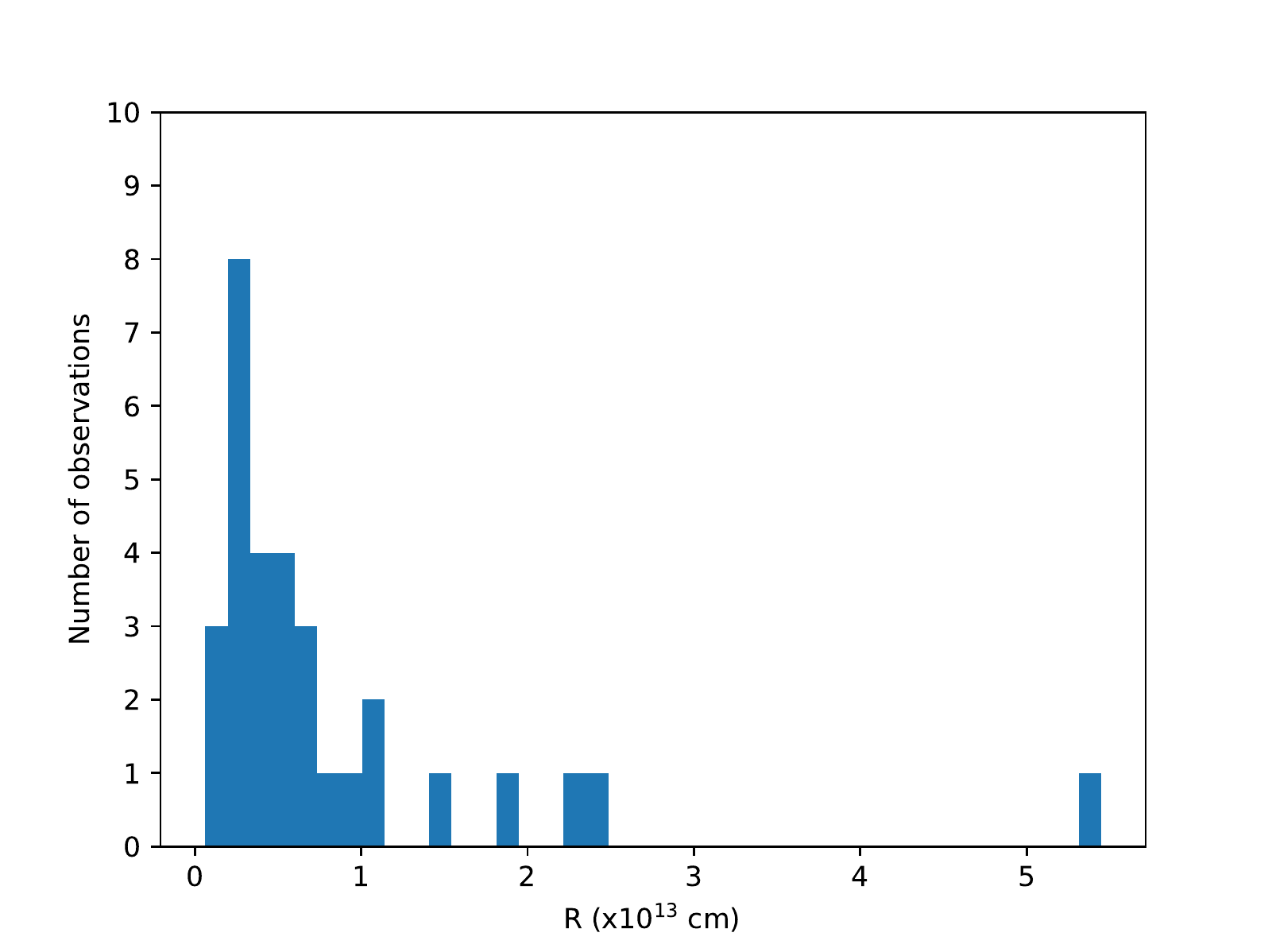}
\caption{Distribution of the emission region sizes in the \textit{NuSTAR}  blazars derived from their minimum variability timescales. \label{Fig2}}
\end{figure}

Now, using the causality argument, the minimum variability timescale $\tau_{var}$ can be used to estimate the upper limit for the minimum size of the emission region ($R$) as given by

\begin{equation}
R\geq\frac{\delta }{\left ( 1+z \right )}c\tau_{var},
\label{size}
\end{equation}
where  $\delta$, Doppler factor, is defined as $\delta =(\Gamma \left ( 1-\beta cos\theta \right ))^{-1}$; and for the velocity $\beta=v/c$ the bulk Lorentz factor can be written as $\Gamma=1/\sqrt{1-\beta^{2}}$. Here it is assumed that the emission originates from the innermost regions of the blazar jets which move with high speeds along the path that makes an angle, $\theta$, with the line of sight. For a moderate value of $\delta=10$, the distribution of the emission region sizes are shown in Figure \ref{Fig2}. 

\subsection{ Spectral Analysis: Hardness Ratio and Spectral Fitting}
To study the spectral variability of the X-ray emission from the sources, the source light curves are produced in two energy bands: a soft band between 3--10 keV and a hard band between 10--79 keV. Then we define hardness ratio (HR) as
\begin{equation}
HR=\frac{F_{hard} }{F_{soft}},
\label{HR}
\end{equation}
where $F_{hard}$ and $F_{soft}$ are the flux in count rates in the hard (10--79 keV) and soft (3--10 keV) bands, respectively. The hardness ratio is a commonly used  model-independent method to study spectral variations over time and flux states. In this work, we particularly examine the relation between flux and HRs over the observation period to constrain the underlying physics. The middle panel of Figure \ref{fig:20} shows the flux hardness ratio plot for the source Mrk 501 (obs. ID: 6000202400), with clearly visible  harder-when-brighter trend. To look for  possible hysteresis loops in the flux-HR plane, the symbols were color-coded according to the time.

Spectral analysis of the \textit{NuSTAR} blazars were carried out by the spectral fitting the source spectra using \emph{xspec} \citep{Arnaud1996} models and using the $\chi ^{2}$ minimization statistics. The spectra from the instruments FPMA and  FPMB were simultaneously fitted in \emph{xspec}. To account for any possible subtle differences between the instruments, an inter-calibration constant was included in the spectral models. The values of the constant, ranging  from 0.97 to 1.04, indicated that there were no major differences between the observations obtained by the two instruments. To ascertain the best representation of the spectral behavior, each spectrum was  fitted using three spectral models: power-law (PL), log-parabola (LP) and broken power-law (BPL). The power-law model can be given as
\begin{equation}
\frac{dN}{dE}=NE^{-\Gamma }
\end{equation}
\noindent where N, E and  $\Gamma$ are normalization, photon energy and photon index, respectively.
Similarly, the log-parabola model having a continuous break is given by

\begin{equation}
\frac{dN}{dE}=N_{0}\left ( \frac{E}{E_{0}} \right )^{-\left ({\Gamma +\beta log\left (E/E_{0}  \right )}  \right )}
\end{equation}

\noindent where $N_{0}$ and $E_{0}$ are normalization, and the reference energy fixed to 10 keV, respectively; and $\Gamma$ and $\beta$ are the photon index and the curvature parameter, respectively \citep[see][]{Massaro2004a}. Finally, the broken-power law is expressed as
\begin{equation}
   \frac{dN}{dE}= N_{0}
\begin{cases}
    E^{-\Gamma_{1}},& \text{if } E\geq E_{b}\\
    E^{-\Gamma_{2}},              & \text{otherwise}
\end{cases}
\end{equation}

\noindent where $\Gamma_{1}$ and $\Gamma_{2}$  represent the high- and low-energy photon indexes; and $N_{0}$ and $E_{b}$ are normalization and the break energy, respectively. To account for the galactic absorption \emph{tbabs} (Tuebingen-Boulder ISM absorption model; \citealt{Wilms2000}) was multiplied with these models, while the hydrogen column density were taken from \citet{Kalberla2005}. 

 Of the three models, we chose the best-fit spectral model after performing F-test\footnote{The F-test tool used in this work is available in xpsec}. In particular the significance of LP and BPL was estimated against PL (null hypothesis), and the model was accepted as better-fit if the probability under the null hypothesis was equal or smaller than 0.1 - equivalently, significance equal or greater than 90\%. If not, PL was considered to be the best representation. Further, between two models, i.e., LP and BPL, the model with higher significance (or lower probability value) was chosen to be the best one. Based on such criteria, out of 31 observation spectra,  7, 17  and 7 spectra were found to be best represented by PL, LP and BPL spectral models, respectively. The fitting parameters for all the observations are listed in Table 3. Spectral fitting for the source S5 0716+714 is presented in the bottom panel of Figure \ref{fig:20}, and the similar figures for the rest of the observations are presented in the on-line material. The distribution of the photon indexes, resulting from the best-fit models, over the mean flux in count rates is shown in Figure \ref{fig:F_Gamma}.

\subsection{Discrete Correlation Function}
Cross-correlation study between emission in different energy bands offers insights that can shed light into the on-going processes at the emission sites e. g. the dominant radiative processes involved and distribution of the emitting particles \citep[see e.g.][]{Zhang2002}.   We analyzed the correlation between the \textit{NuSTAR} blazar light curves in the soft energy (3-10 keV) and hard energy (10-79 keV) using z-transformed discrete correlation function (ZDCF) along with the likelihood of the ZDCF peaks and the associated uncertainties as described in  \citealt[][]{Alexander2013}\footnote{The software is publicly available at \url{ http://www.weizmann.ac.il/particle/tal/research-activities/software}}  \citep[see also][]{ Bhatta2018}. The ZDCFs between the lower energy (LE) and higher energy (HE) light curves for the source 3C 279 (Obs. ID 60002020002) are shown in the bottom panel of Figure\,\ref{fig:20}, and the similar plots for the rest of the observations discussed in the paper are presented in the online material section; besides the results are also tabulated in Table \ref{DCF}.
\clearpage
\onecolumn
\begin{longtable}[H]{|p{.12\textwidth}|c|c|c|c|c|c|c|}

		\caption{Spectral fitting of the \nustar blazars. Col. 1: source name, Col. 2: Obs. ID, Col. 3: spectral models, power-law (PL), log-parabola (LP), broken power-law (BPL), Col. 4 photon index (PL and LP),  high-energy photon index (BPL), Col. 5: break energy (keV), Col. 6: Curvature Parameter (LP), low-energy photon index (BPL), Col. 7: reduced $\chi^{2}$/degrees of freedom, and  Col. 8: F-test and probability value. The best-fit spectral models are given in bold font. \label{table:3}} \\  \hline
		(1) & (2) & (3) & (4) & (5) & (6) & (7) & (8)\\ 
		Source & Obs. ID & Model & $ \Gamma, \Gamma_{1} $ & $E_{b} (keV) $ & $\beta,\ \Gamma_2 $ & $\chi^2_{red}$/dof & F-value (prob.)\\ \hline
		\endfirsthead
		\caption{ Continued... }\\
		\hline
		Source & Obs. ID & Model & $\Gamma $ & $E_{b} (keV) $ & $\beta,\ \Gamma_2 $ & $\chi^2_{red}$/dof & F-value (prob.)\\ \hline
		\endhead
		\hline
		\endfoot
		\hline
		\endlastfoot
		S5 0014+81 & 60001098002 & PL & 1.82$\pm$0.03 & -- & -- & 1.1851/153 &  \\
		&& LP & 1.84$\pm$0.04 & -- & 0.36$\pm$0.12 & 1.1226/152 & 9.52 (2.42$\times10^{-3}$)\\
		&& \textbf{BPL} & 1.73$\pm$0.04 & 21.79$\pm$2.56 & 3.40$\pm$0.73 & 1.0706/151 & 9.18 (1.73$\times10^{-4}$)\\
		& 60001098004 & \textbf{PL} & 1.70$\pm$0.03 & -- & -- & 1.1128/164 &  \\
		&& LP & 1.70$\pm$0.03 & -- & 0.00$\pm$0.11 & 1.1197/163 & --\\ 
		&& BPL & 1.71$\pm$0.04 & 19.55$\pm$16.11 & 1.56$\pm$0.28 & 1.1231/162 & 0.25 (7.81$\times10^{-1}$)\\ \hline
		B0222+185 & 60001101002 & PL & 1.54$\pm$0.02 & -- & -- & 0.9783/479 &  \\
		&& \textbf{LP} & 1.54$\pm$0.02 & -- & 0.22$\pm$0.05 & 0.9380/478 & 21.58 (4.39$\times10^{-6}$)\\
		&& BPL & 1.47$\pm$0.03 & 14.04$\pm$2.09 & 1.75$\pm$0.07 & 0.9405/477 & 10.63 (3.06$\times10^{-5}$)\\
		& 60001101004 & PL & 1.64$\pm$0.02 & -- & -- & 0.9882/366 &  \\
		&& LP & 1.66$\pm$0.02 & -- & 0.29$\pm$0.06 & 0.9270/365 & 25.16 (8.24$\times10^{-7}$)\\
		&& \textbf{BPL} & 1.34$\pm$0.08 & 6.54$\pm$0.70 & 1.75$\pm$0.03 & 0.9149/364 & 15.66 (2.99$\times10^{-7}$)\\ \hline	
		HB 0836+710 & 60002045002 & PL & 1.69$\pm$0.02 & -- & -- & 0.9106/452 &  \\
		&& LP & 1.69$\pm$0.02 & -- & --0.08$\pm$0.05 & 0.9075/451 & 2.54 (1.11$\times10^{-1}$)\\
		&& \textbf{BPL} & 1.73$\pm$0.03 & 12.65$\pm$4.07 & 1.60$\pm$0.06 & 0.9045/450 & 2.52 (8.13$\times10^{-2}$)\\
		& 60002045004 & PL & 1.66$\pm$0.01 & -- & -- & 1.0267/664 &  \\
		&& \textbf{LP} & 1.66$\pm$0.01 & -- & 0.10$\pm$0.04 & 1.0146/663 & 8.92 (2.93$\times10^{-3}$)\\
		&& BPL & 1.59$\pm$0.03 & 7.98$\pm$1.83 & 1.70$\pm$0.03 & 1.0156/662 & 4.63 (1.01$\times10^{-2}$)\\ \hline
		3C 273 & 10202020002 & PL & 1.62$\pm$0.00 & -- & -- & 1.0871/1335 &  \\
		&& \textbf{LP} & 1.62$\pm$0.00 & -- & 0.11$\pm$0.01 & 1.0326/1334 & 71.46 (7.31$\times10^{-17}$)\\
		&& BPL & 1.57$\pm$0.01 & 13.43$\pm$1.05 & 1.72$\pm$0.02 & 1.0299/1333 & 38.07 (8.32$\times10^{-17}$)\\
		& 10302020002 & PL & 1.66$\pm$0.01 & -- & -- & 0.9334/1017 &  \\
		&& \textbf{LP} & 1.66$\pm$0.01 & -- & 0.08$\pm$0.02 & 0.9164/1016 & 19.87 (9.23$\times10^{-6}$)\\
		&& BPL & 1.64$\pm$0.01 & 19.35$\pm$3.23 & 1.78$\pm$0.05 & 0.9190/1015 & 8.97 (1.38$\times10^{-4}$)\\ \hline
		3C 279 & 60002020002 & PL & 1.73$\pm$0.02 & -- & -- & 0.9442/480 &  \\
		&& LP & 1.73$\pm$0.02 & -- & 0.08$\pm$0.05 & 0.9411/479 & 2.58 (1.09$\times10^{-1}$)\\
		&& \textbf{BPL} & 1.71$\pm$0.02 & 29.87$\pm$8.06 & 2.15$\pm$0.33 & 0.9386/478 & 2.43 (8.90$\times10^{-2}$)\\
		& 60002020004 & PL & 1.74$\pm$0.01 & -- & -- & 0.9031/691 &  \\
		&& \textbf{LP} & 1.74$\pm$0.01 & -- & 0.07$\pm$0.03 & 0.8982/690 & 4.77 (2.93$\times10^{-2}$)\\
		&& BPL & 1.69$\pm$0.03 & 8.66$\pm$2.40 & 1.78$\pm$0.03 & 0.8980/689 & 2.96 (5.24$\times10^{-2}$)\\ \hline
		PKS 1441+25 & 90101004002 & \textbf{PL} & 2.08$\pm$0.08 & -- & -- & 1.030/49 &  \\
		&& LP & 2.01$\pm$0.09 & -- & --0.32$\pm$0.28 & 1.027/48 & 1.14 (2.90$\times10^{-1}$)\\
		&& BPL & 2.09$\pm$0.09 & 23.56$\pm$23.32 & 1.51$\pm$1.38 & 1.070/47 & 0.08 (9.19$\times10^{-1}$)\\ \hline
		PKS 2149--306 & 60001099002 & PL & 1.37$\pm$0.01 & -- & -- & 0.9722/824 &  \\
		&& LP & 1.36$\pm$0.01 & -- & 0.05$\pm$0.03 & 0.9686/823 & 4.06 (4.42$\times10^{-2}$)\\
		&& \textbf{BPL} & 1.34$\pm$0.02 & 12.48$\pm$3.57 & 1.42$\pm$0.03 & 0.9668/822 & 3.30 (3.73$\times10^{-2}$)\\
		& 60001099004 & \textbf{PL} & 1.46$\pm$0.01 & -- & -- & 0.9730/744 &  \\
		&& LP & 1.46$\pm$0.01 & -- & 0.04$\pm$0.03 & 0.9716/743 & 2.07 (1.50$\times10^{-1}$)\\
		&& BPL & 1.42$\pm$0.03 & 8.86$\pm$2.94 & 1.49$\pm$0.02 & 0.9701/742 & 2.11 (1.22$\times10^{-1}$)\\ \hline	
		1ES 0229+200 & 60002047004 & PL & 2.03$\pm$0.02 & -- & -- & 1.0547/387 &  \\ 
		&& \textbf{LP} & 2.06$\pm$0.02 & -- & 0.23$\pm$0.07 & 1.0255/386 & 12.02 (5.86$\times10^{-4}$)\\
		&& BPL & 1.99$\pm$0.03 & 16.04$\pm$3.42 & 2.30$\pm$0.15 & 1.0390/385 & 3.92 (2.06$\times10^{-2}$)\\ \hline
		S5 0716+714 & 90002003002 & PL & 1.90$\pm$0.03 & -- & -- & 1.2050/194 &  \\
		&& \textbf{LP} & 1.87$\pm$0.03 & -- & --0.33$\pm$0.09 & 1.1428/193 & 11.56 (8.19$\times10^{-4}$)\\
		&& BPL & 1.94$\pm$0.04 & 19.60$\pm$5.08 & 1.50$\pm$0.23 & 1.1922/192 & 2.04 (1.33$\times10^{-1}$)\\ \hline
		Mrk 501 & 60002024002 & PL & 2.27$\pm$0.01 & -- & -- & 0.8889/562 &  \\
		&& \textbf{LP} & 2.30$\pm$0.02 & -- & 0.16$\pm$0.04 & 0.8649/561 & 16.59 (5.29$\times10^{-5}$)\\
		&& BPL & 2.26$\pm$0.01 & 19.77$\pm$5.55 & 2.48$\pm$0.16 & 0.8848/560 & 2.30 (1.01$\times10^{-1}$)\\
		& 60002024004 & PL & 2.24$\pm$0.01 & -- & -- & 1.0918/730 &  \\
		&& \textbf{LP} & 2.26$\pm$0.01 & -- & 0.13$\pm$0.03 & 1.0650/729 & 19.37 (1.24$\times10^{-5}$)\\
		&& BPL & 2.23$\pm$0.01 & 24.50$\pm$4.37 & 2.55$\pm$0.14 & 1.0786/728 & 5.47 (4.40$\times10^{-3}$)\\
		& 60002024006 & PL & 2.09$\pm$0.01 & -- & -- & 1.0474/765 &  \\
		&& \textbf{LP} & 2.12$\pm$0.01 & -- & 0.19$\pm$0.03 & 0.9817/764 & 52.20 (1.22$\times10^{-12}$)\\
		&& BPL & 2.00$\pm$0.02 & 8.45$\pm$0.70 & 2.20$\pm$0.02 & 0.9836/763 & 25.81 (1.42$\times10^{-11}$)\\
		& 60002024008 & PL & 2.13$\pm$0.01 & -- & -- & 1.0916/720 &  \\
		&& \textbf{LP} & 2.17$\pm$0.01 & -- & 0.29$\pm$0.03 & 0.9538/719 & 105.02 (4.27$\times10^{-23}$)\\
		&& BPL & 1.98$\pm$0.02 & 8.04$\pm$0.47 & 2.28$\pm$0.02 & 0.9548/718 & 52.58 (4.90$\times10^{-22}$)\\ \hline	
		1ES 1959+650 & 60002055002 & PL & 2.28$\pm$0.01 & -- & -- & 1.0531/561 &  \\
		&& \textbf{LP} & 2.30$\pm$0.02 & -- & 0.10$\pm$0.04 & 1.0444/560 & 5.67 (1.76$\times10^{-2}$)\\
		&& BPL & 2.27$\pm$0.01 & 20.25$\pm$8.80 & 2.41$\pm$0.15 & 1.0537/559 & 0.84 (4.32$\times10^{-1}$)\\
		& 60002055004 & PL & 2.54$\pm$0.01 & -- & -- & 1.1642/540 &  \\
		&& \textbf{LP} & 2.59$\pm$0.02 & -- & 0.21$\pm$0.05 & 1.1230/539 & 20.81 (6.28$\times10^{-6}$)\\
		&& BPL & 2.50$\pm$0.02 & 13.69$\pm$1.55 & 2.86$\pm$0.10 & 1.1192/538 & 11.86 (9.15$\times10^{-6}$)\\ \hline
		PKS 2155-304 & 10002010001 & PL & 3.00$\pm$0.02 & -- & -- & 1.1986/377 &  \\
		&& LP & 3.10$\pm$0.04 & -- & 0.26$\pm$0.09 & 1.1774/376 & 7.79 (5.53$\times10^{-3}$)\\
		&& \textbf{BPL} & 2.84$\pm$0.06 & 5.92$\pm$0.70 & 3.13$\pm$0.05 & 1.1612/375 & 7.07 (9.67$\times10^{-4}$)\\
		& 60002022002 & PL & 2.70$\pm$0.03 & -- & -- & 0.9128/307 &  \\
		&& \textbf{LP} & 2.63$\pm$0.04 & -- & --0.21$\pm$0.10 & 0.9023/306 & 4.57 (3.33$\times10^{-2}$)\\
		&& BPL & 2.72$\pm$0.03 & 15.40$\pm$3.32 & 2.25$\pm$0.26 & 0.9031/305 & 2.65 (7.24$\times10^{-2}$)\\
		& 60002022004 & PL & 2.55$\pm$0.04 & -- & -- & 0.9447/151 &  \\
		&& LP & 2.48$\pm$0.05 & -- & --0.23$\pm$0.14 & 0.9366/150 & 2.31 (1.31$\times10^{-1}$)\\
		&& \textbf{BPL} & 2.59$\pm$0.04 & 21.85$\pm$3.19 & 0.87$\pm$0.52 & 0.8691/149 & 7.57 (7.41$\times10^{-4}$)\\
		& 60002022006 & \textbf{PL} & 3.04$\pm$0.05 & -- & -- & 0.9465/120 &  \\
		&& LP & 3.05$\pm$0.08 & -- & 0.02$\pm$0.19 & 0.9543/119 & 0.02 (8.90$\times10^{-1}$)\\
		&& BPL & 3.04$\pm$0.05 & 17.42$\pm$131.65 & 3.13$\pm$4.34 & 0.9624/118 & 0.01 (9.91$\times10^{-1}$)\\
		& 60002022008 & PL & 2.88$\pm$0.05 & -- & -- & 0.9755/94 &  \\
		&& \textbf{LP} & 2.70$\pm$0.08 & -- & --0.51$\pm$0.20 & 0.9242/93 & 6.22 (1.44$\times10^{-2}$)\\
		&& BPL & 2.99$\pm$0.09 & 9.14$\pm$2.01 & 2.48$\pm$0.21 & 0.9381/92 & 2.87 (6.16$\times10^{-2}$)\\
		& 60002022010 & \textbf{PL} & 2.98$\pm$0.05 & -- & -- & 0.7921/106 &  \\
		&& LP & 3.03$\pm$0.09 & -- & 0.16$\pm$0.22 & 0.7939/105 & 0.76 (3.85$\times10^{-1}$)\\
		&& BPL & 2.94$\pm$0.06 & 13.74$\pm$4.22 & 3.80$\pm$1.07 & 0.7792/104 & 1.88 (1.58$\times10^{-1}$)\\
		& 60002022012 & PL & 2.66$\pm$0.03 & -- & -- & 1.0162/210 &  \\
		&& \textbf{LP} & 2.79$\pm$0.05 & -- & 0.48$\pm$0.13 & 0.9483/209 & 16.04 (8.63$\times10^{-5}$)\\
		&& BPL & 2.55$\pm$0.04 & 11.14$\pm$1.44 & 3.20$\pm$0.22 & 0.9413/208 & 9.35 (1.29$\times10^{-4}$)\\
		& 60002022014 & \textbf{PL} & 2.80$\pm$0.04 & -- & -- & 0.9787/182 &  \\
		&& LP & 2.79$\pm$0.06 & -- & --0.02$\pm$0.15 & 0.9840/181 & 0.02 (8.88$\times10^{-1}$)\\
		&& BPL & 2.80$\pm$0.04 & 39.70$\pm$48.09 & -2.50$\pm$26.91 & 0.9788/180 & 0.99 (3.73$\times10^{-1}$)\\
		& 60002022016 & PL & 2.61$\pm$0.06 & -- & -- & 1.024/78 &  \\
		&& \textbf{LP} & 2.52$\pm$0.07 & -- & --0.35$\pm$0.20 & 1.000/77 & 2.87 (9.42$\times10^{-2}$)\\
		&& BPL & 2.71$\pm$0.11 & 8.21$\pm$2.92 & 2.41$\pm$0.17 & 1.013/76 & 1.42 (2.47$\times10^{-1}$)\\ \hline
		BL Lac & 60001001002 & \textbf{PL} & 1.85$\pm$0.02 & -- & -- & 0.9482/409 &  \\
		&& LP & 1.85$\pm$0.02 & -- & 0.02$\pm$0.06 & 0.9503/408 & 0.10 (7.57$\times10^{-1}$)\\
		&& BPL & 1.84$\pm$0.03 & 13.76$\pm$14.96 & 1.89$\pm$0.09 & 0.9515/407 & 0.29 (7.48$\times10^{-1}$)
		\label{fitting}
\end{longtable}
\twocolumn

  In the figure, we see that most of the cases we do not find a strong correlation between low and high energy emission at the zero lag, and in a few cases hints of hard and soft lags can be seen. It should be pointed out that between two similar DCF values at the different lags, the one closer to zero lag would be statistically more significant as the number of observations that go into the calculation of DCF value decreases with the increase in the lead/lag.

\section{Results}
The results of all of the above analyses on the individual sources along with their brief introduction are presented below.

\subsection*{S5 0014+81}
  FSRQ S5 0014+81, detected by multiple X-ray instruments,  possesses the most luminous accretion disk among blazars \citep[see][and references therein]{Sbarrato2016}. Also, of the sources discussed in this paper, it is the most distant source at the redshift of 3.366. The high-redshift blazar reveals contributions due to thermal emission from the accretion disk in its optical continuum \citep{Ghisellini2010}. We looked into two \textit{NuSTAR} observations separated by one month. The first observation (Obs. ID: 60001098002) shows one of the largest variability with FV$\sim 30\%$  and rapid ($\tau_{var}=0.91\pm0.93$ ks) minimum variability timescale within 46 ks observation period; while the second observation shows a moderate variability (FV $\sim14\%$)  within 39 ks.  No obvious trend in flux-HR plane could be observed. While in the first observation, we do not see any significant correlation between the low and high energy emission, in the second observation we found a hint of soft lag of $\sim 0.9$ ks with z-transformed discrete correlation coefficient $(ZDCF)\sim 0.34$ and likelihood $(LH)=0.62$.  The spectra for the first observation is fitted with BPL with a break at $\sim20$ keV energy, whereas for the second one power-law model with $\Gamma_{X}\sim1.7$ is fitted well.

\subsection*{B0222+185}
 Blazar B0222+185 has been  widely studied by X-ray instruments, e.g. Swift/BAT \citep{Ajello2012,Baumgartner2013}. In the hard X-ray study, it was found to be one the most powerful blazars ever observed \citep{Sbarrato2016}; the optical flux showed an evidence for the thermal emission from the accretion disk  \citep{Ghisellini2010}. It is one of the most distant sources (z=2.69) discussed in this work. We studied two \textit{NuSTAR} observations spanning 61 and 70 ks. In the light curves, the flux points appear to be scattered showing no coherent variability pattern. Similarly, no clear trend in the flux-HR plane can be observed. The correlation between the soft and hard emission shows a sign of hard lag of $\sim 9.0$ ks and $\sim 2$ ks with ZDCF  $\sim0.48$ and $\sim0.53$. However the larger associated errors and small values of LH make them inconclusive. The first observation is fitted with LP with $\beta \sim0.2$ and the second observation is well fitted with BPL with $E_{b} \sim 6.5$ keV.

 \subsection*{HB 0836+710}
Source HB 0836+710 is a high-redshift blazar, extensively studied in multi-band emission \citep[see][and reference therein]{Akyuz2013}. The source is identified with a prominent  kpc-scale radio jet \citep{Hummel1992}. The optical-UV spectrum  is dominated by thermal emission from the accretion disk \citep{Ghisellini2010}, and  the $\gamma$-ray emission region is found to be located  $\sim35$ pc away from the central engine \citep{Jorstad2013}.  In the two \textit{NuSTAR} observations which we examined, the source showed rapid variability with the minimum variability timescales as small as $2.53\pm0.91$ ks. The second observation shows a systematic modulation of flux-HR plane. However, the ZDCF values $\sim$ 0.31 and $\sim$0.30 at the zero lag show that there is not much correlation between the low and high energy emission. For the first and second observations BPL and LP models, respectively, were fitted.

\subsection*{3C 273}
 3C 273 is the nearest bright quasar with a large-scale visible jet. Being highly variable across nearly all energies  the source has been the subject of numerous broadband observation campaigns \citep[e.g.][] {Soldi2008,Abdo2010}. In the optical-UV band there is a bright excess, commonly called \emph{blue bump}, possibly a signature of  thermal reprocessing from the accretion disk \citep{Paltani1998}. We examined two \textit{NuSTAR} observations (Obs. ID  10202020002 and 10302020002) exactly one year apart. In the first observation, we found moderate  (FV $\sim10\%$) but rapid variability ($\tau_{var} =8.81\pm3.34$ ks). We observe that the flux is stable and HR changes randomly; whereas in the second observation, the source became more variable with FV$\sim15\%$ and rapid  ($\tau_{var}=1.24\pm1.70$ ks) in flux and HR. In the first observation, we  find a good correlation (ZDCF=$0.45$ and  LH=$0.44$) between the high and the low energy emission at zero lag. In the second epoch, although not much significant (ZDCF=$0.46$ and  LH=$0.45$), we see a possible soft lag of $\sim$4.8 ks. The spectra for both of the observations were well fitted with LP with  $\beta \sim 0.1$.

\begin{table*}
\renewcommand{\arraystretch}{1.5}
		\centering
		\caption{Discrete cross-correlation function between the low- (3-10 keV) and high-energy (10-79 keV) emission of the \textit{NuSTAR} blazars. The +ve lag indicates hard lag.}
	\begin{tabular}{|l|c|r|c|c|}
	\hline
		Source &  Obs. ID & Lag (ks) & ZDCF&Likelihood\\
		\hline
		S5 0014+81  & 60001098002 &$+5.40^{+ 0.58}_ { -10.43}$ & $0.32^{+0.14}_{-0.13}$&0.22\\
		 & 60001098004 &$-0.90^{+ 0.40}_ { -0.72}$ & $0.34^{+0.14}_{-0.13}$&0.62\\
		 
		 B0222+185  & 60001101002 &$+9.00^{+ 0.44}_ { -10.36}$ & $0.48^{+0.17}_{-0.18}$&0.29\\
		& 60001101004 &$+2.00^{+ 0.75}_ { -3.89}$ & $0.53^{+0.15}_{-0.14}$&0.42\\
		
		HB 0836+710  & 60002045002 & $-6.70^{+ 9.13}_ { -7.72}$ & $0.31^{+0.12}_{-0.18}$&0.35\\
		 & 60002045004 & $+3.00^{+ 9.36}_ { -2.34}$ & $0.30^{+0.17}_{-0.18}$&0.33\\
		 
		3C 273  & 10202020002 & $0.00^{+ 0.91}_ { -0.58}$ & $0.45^{+0.15}_{-0.16}$&0.44\\
		 & 10302020002 & $-4.80^{+ 1.04}_ { -2.13}$ & $0.46^{+0.15}_{-0.16}$&0.45\\
		3C 279  & 60002020002 &  $0.00^{+ 5.53}_ { -0.68}$ & $0.63^{+0.12}_{-0.11}$&0.43\\
		 & 60002020004 &$+7.00^{+ 5.97}_ { -10.07}$ & $0.79^{+0.06}_{-0.05}$&0.45\\
				
		PKS 1441+25  & 90101004002 & $+4.00^{+ 4.29}_ { -10.21}$ & $0.15^{+0.23}_{-0.24}$&0.08\\
		PKS 2149--306  & 60001099002 & $-1.80^{+ 5.71}_ { -4.27}$ & $0.47^{+0.14}_{-0.13}$&0.72\\
		 & 60001099004 & $-3.60^{+ 6.46}_ { -4.34}$ & $0.23^{+0.14}_{-0.14}$&0.19\\
		 		 
		S5 0716+714 & 90002003002 & $+3.00^{+ 6.37}_ { -7.02}$& $0.50^{+0.20}_{-0.18}$&0.13\\

Mrk 501 & 60002024002 & $+3.00^{+ 0.58}_ { -6.45}$& $0.40^{+0.27}_{-0.24}$&0.13\\
 & 60002024004 & $+0.00^{+ 0.47}_ { -0.48}$& $0.92^{+0.37}_{-0.28}$&0.57\\
& 60002024006 & $-4.00^{+ 6.02}_ { -0.44}$& $0.75^{+0.16}_{-0.12}$&0.40\\
 & 60002024008 & $0.00^{+ 1.98}_ { -2.52}$& $0.78^{+0.98}_{-0.81}$&0.72\\

1ES 1959+650 & 60002055002 & $0.00^{+ 1.12}_ { -0.52}$& $0.95^{+0.31}_{-0.23}$&0.41\\

 & 60002055004 & $+0.00^{+ 0.62}_ { -0.57}$& $0.67^{+0.18}_{-0.15}$&0.54\\
		PKS 2155–304 & 10002010001 & $0.00^{+ 0.52}_ { -3.56}$& $0.49^{+0.15}_{-0.14}$&0.61\\
			     & 60002022002 & $-1.25^{+ 1.63}_ { -1.78}$& $0.49^{+0.13}_{-0.13}$&0.37\\
				& 60002022004 & $-1.50^{+ 1.55}_ { -0.94}$& $0.75^{+0.15}_{-0.12}$&0.38\\
				& 60002022006 & $-1.50^{+ 4.37}_ { -2.67}$& $0.29^{+0.24}_{-0.22}$&0.15\\
				& 60002022008 & $+0.00^{+ 3.69}_ { -0.42}$& $0.75^{+0.14}_{-0.11}$&0.53\\
				& 60002022010 & $+1.88^{+ 1.39}_ { -1.17}$& $0.49^{+0.27}_{-0.23}$&0.36\\
				& 60002022012 & $+2.02^{+ 1.55}_ { -1.42}$& $0.55^{+0.24}_{-0.19}$&0.37\\
				& 60002022014 & $-5.27^{+ 3.12}_ { -1.02}$& $0.32^{+0.27}_{-0.25}$&0.30\\
				& 60002022016 & $-2.70^{+ 1.76}_ { -0.65}$& $0.32^{+0.29}_{-0.28}$&0.20\\

		BL Lac  & 60001001002 &$-2.70^{+ 2.14}_ { -7.72}$ & $0.31^{+0.12}_{-0.18}$&0.35\\
		\hline
	\end{tabular}
	
 	\label{DCF}
 	\end{table*}

\subsection*{3C 279}

Blazar  3C 279 is a FSRQ source profusely emitting in hard X-ray and $\gamma$-rays. The source,  highly variable  across a wide range of spectral bands \citep[see][and the references therein]{Hayashida2015}, is one of a handful of sources  detected above 100 GeV  (MAGIC Collaboration et al. 2008). The source reveals a compact, milliarcsecond-scale radio core ejecting radio knots with a bulk Lorentz factor, $\Gamma= 15.5\pm 2.5$  along the direction making an angle, $\theta_{obs}=2.1\pm 1.1^{\circ}$, to the line of sight  \citep{Jorstad2005,Jorstad2004}. Our study on two \textit{NuSTAR} observations show that source displays moderate variability in hour-like timescales ($\tau_{var}=2.31 \pm1.26$ ks and $5.61\pm3.99$ ks), the correlation between soft and hard emission showed a hard lag by a few ks, particularly distinguished (ZDCF $\sim 0.79$ and LH=$0.45$) in the second observation (obs. ID: 60002020004).  We could not see any clear trend in flux-HR plane. Of the three spectral models, first observation was fitted with BPL model with $E_{b}\sim30$ keV and the second one was well represented by LP with  a small $\beta \sim0.07$.

 \subsection*{ PKS 1441+25}
 PKS 1441+25, a TeV blazar, has been detected in  very high energy (VHE) $\gamma$-rays  by VERITAS and MAGIC  \citep[see][]{Abeysekara2015}.  The source showed rapid variability when flux doubled within  a few hours; and  it also exhibited one of the most rapid ($\tau_{var} =1.24\pm0.62$ ks) and largest variability (FV $\sim 26\%$)  observed within the observation period of 72 ks. We did not see a simple correlation between the flux and the hardness ratio,  and there was no apparent correlation (ZDCF $\sim 0.00$) between the low and high energy emission  at the zero lag. The spectrum was fitted well with a PL model with photon index $\sim2$.
 
 \subsection*{PKS 2149--306}
PKS 2149-306 is a   X-ray bright FSRQ  often marked by dramatic flux and spectral variability as observed by most of the X-ray telescopes  \citep[see][and references therein]{DAmmando2016}. In both of the \textit{NuSTAR} observation we studied, the source showed significant variability (FV $\sim$10\%) in the timescale of a few hours. In the first observation, we see a hint of a soft lag near 1.8 ks with ZDCF$=0.47$ and LH$=0.72$ and harder-when-brighter trend whereas in the second there is no much correlation between the low and high energy emission, and a complex flux HR relation was observed. For both the observations, the source spectra were fitted with BPL, and PL  having flattest photon indexes of $\sim1.5$.

 \subsection*{1ES 0229+200}

BL Lac 1ES 0229+200  is one of the important TeV sources which has been used to study the properties of the extragalactic background light and the intergalactic magnetic field through its very high energy emission \citep[][and references therein]{Aliu2014}. We examined the 38 ks long \textit{NuSTAR} observation for its hard X-ray properties. The source was found to display a significant (FV$\sim13\%$) and rapid ($\tau_{var} = 2.35\pm1.23$ ks) variability. The flux did not appear to be correlated with the HR. The source spectra were best-fitted using LP model with  photon index, $\Gamma_{X}\sim2$.

 \subsection*{ S5 0716+714}
  S5~0716+714 is one of the best studied sources across broad bands. The TeV source is widely famous for its variability with almost 100\% duty cycle \citep[see][and the references therein]{bhatta16b}.  In the \textit{NuSTAR} observation we studied, the source showed rapid variability; the flux nearly doubled within the observation period of 32 ks. In addition, significant average flux variability (FV $\sim 15\%$) with $2.79\pm1.43$ ks minimum variability timescale was noticed. However, we could not detect any obvious HR-flux relation; however, the correlation between the high and low energy emission revealed a possible hard lag of $\sim3.00$ ks with ZDCF value $\sim 0.5$ however with a small LH, 0.13. The spectrum was fitted using LP model with $\Gamma_{X}\sim1.9$  and a negative curvature, $\beta\sim-0.33$.

\subsection*{Mrk 501}
  Mrk 501, shining bright in X-ray, is one of the most favored targets for multi-frequency observations \citep[see][and references therein]{Furniss2015}. We studied 4 \textit{NuSTAR} observations between April to July 2013. The light curve of the first observation (Obs. ID: 60002024002)  showed low variability (fractional variability $\sim$ 5\% ) and no clear trend in hardness ratio variability.  In the second observation  (Obs. ID: 60002024004), the overall flux followed a rising trend  for $\sim$47 ks and later declined during remaining  8 ks; the source displays significant variability with FV $\sim$17\%. The  harder-when-brighter behavior is clearly visible in the flux-HR plane as shown in Figure 1 (middle panel).  During the third observation (60002024006) the source is nearly twice brighter than in the other observations but with decreased variability (FV $\sim$ 5\% ). The last data set for Mrk 501 (60002024008) shows the source getting fainter with random flux-HR trend. Similarly, we found that of the 4 observations, the correlation between LE and HE light curves were significant for Obs ID 60002024004 and 60002024008, whereas for the other two observations we did not find any clear lead/lag. The spectra are fitted well with different power-law models for different observations, the $\Gamma_{X}$ ranging from $2.1-2.3$ (refer to Table 3). 
  
  \subsection*{1ES 1959+650}

BL Lac 1ES 1959+650,  a HSP  \citep{Giebels2002} and a TeV blazar \citep[see][]{Holder2003}, was first detected in X-rays by \citet{Elvis1992}. We analyzed data for two observations: 60002055002 and 60002055004. In the first observation, we  see the flux rising by the factor $\sim 2$, displaying the harder-when-brighter trend.  Although the fractional variability does not differ significantly from the first one, in the second epoch the light curve is relatively stable and does not display a well defined trend in the flux-HR plane. The ZDCF analysis showed that LE and HE light curves had relatively strong correlation around zero lag. For both of the observations, LP model with $\Gamma_{X} \sim$2.3 and 2.6 best describes the source spectra.

\subsection*{PKS 2155--304}

 PKS 2155--304, one of the brightest HSP blazars  and widely studied in X-ray bands  \citep[see][and references therein]{Madejski2016}. The source is known to frequently exhibit rapid variability in the X-ray bands on hourly timescales \citep[e.g.][]{Rani2017,Tanihata2001,Zhang1999}.  We analyzed 9 \textit{NuSTAR} observations between July 2012 to October 2013, and found that the source displayed several interesting features including large fractional variability ($\sim27\%$) and the most rapid variability with smallest minimum variability timescale $0.30\pm0.12$ ks. Besides, in three of the observations, the flux changes by twice within a few hours. However, the flux-HR relation does not show any obvious trend. The ZDCF analysis did not reveal any clear lead/lag between LE and HE light curves (refer to Table 4). The spectra for different observations were fitted with all three models i. e., PL, LP and BPL models, separately, while the photon indexes ranged between  $\sim$2.5 -- 3.0. We noted that with $\Gamma_{X}\sim3.0$ the source displayed one of the steepest spectra usually found in any BL Lac objects.

\subsection*{BL Lac}
BL Lac is a proto-type source of the class with the same name. The source has been observed by several multi-wavelength campaigns \citep[see][and the references therein]{Bhatta2018}. The 42 ks long \textit{NuSTAR} observation, we examined, showed large (FV$\sim 25\%$) and rapid ($\tau_{var} = 1.88\pm0.96$ ks) variability. However, HR did not appear to be correlated with the flux. We observed a relatively smaller correlation ($\sim 0.30$) between the high and low energy emission at the zero lag. The source spectra, best-fitted with PL model with $\Gamma_{X}\sim1.85$.

\section{Discussion  \label{discussion}}
In this section, we attempt to explain the results of the above analyses in the light of the existing blazar models.

\subsection{Rapid Hard  X-ray Variability}
  Hard X-ray observations offer a direct access to the heart of an AGN revealing important processes occurring at the innermost regions of the central engine. The variable hard X-ray emission in AGN is  considered to originate at the corona, a  compact region  above the accretion disc.  Hard X-ray emission from most of the AGNs mainly consists of three components: soft-access, neutral iron line and the Compton hump;  and in Seyfert I type galaxies these components are distinctly  observed in their spectra \citep[e.g. see][]{Walton2014}. 
 However, in case of blazars, as the Doppler boosted jet emission is dominant over the coronal emission, the spectra exhibits pure power-law shapes devoid of emission or absorption features.  Hard X-ray variability in blazars over various timescales could be resulted by the up-scattering of the soft photon fields located at various geometrical components of an AGN including accretion disk, jets, dusty torus, and BLR.  Consequently, any modulation in the photon field,  high-energy electron population and magnetic filed  \textsl{in situ} can produce the hard X-ray variability which can propagate along the jets.  Besides, the distribution of the emission region sizes, as presented in Figure \ref{Fig2}, points out that such variability originates in the compact ($\sim 10^{12}$ cm) volumes of the sources.

 We note that thus estimated sizes of the emission regions are smaller than the gravitational radius of an  AGN with a typical black hole mass of 10$^{9}\ M_{\odot}$ $\sim1.5\times10^{12}$ cm. This suggests the observed short timescale modulations could either be ascribed to the changes occurring at a fraction of the entire black hole regions or the fluctuations could reflect  small scale instabilities intrinsic to the jet \citep[see][]{Begelman2008}.  In the relativistic turbulence scenario by  \citet{Narayan2012},  magnetohydrodynamic turbulence in the jet can lead to compact sub-structures that move relativistically in random directions. Alternatively, very high  bulk Lorentz factors (e.g. $\Gamma$$\sim 100$) associated with the emitting regions can make them appear comparable to $r_{g}$. It is possible to achieve a such high  $\Gamma$s with the \emph{jets-in-a-jet} model in which magnetic reconnection \citep[e.g.][]{Giannios2009} or  turbulence \citep[e.g.][]{Narayan2012}  can produce relativistic outflows the bulk jet frame.
 
 The  rapid flux variations could  also be explained as the emission from the shocked regions in the blazar jets viewed close to line of sight  \citep[e.g.][]{Marscher1985,Spada2001,Joshi2011}. The non-thermal emission modulations can also be attributed to various instabilities in the jet e.g. turbulence behind the shocks \citep[see][]{bhatta13,Marscher2014}.

\begin{figure}
\begin{center}
\includegraphics[width=\columnwidth]{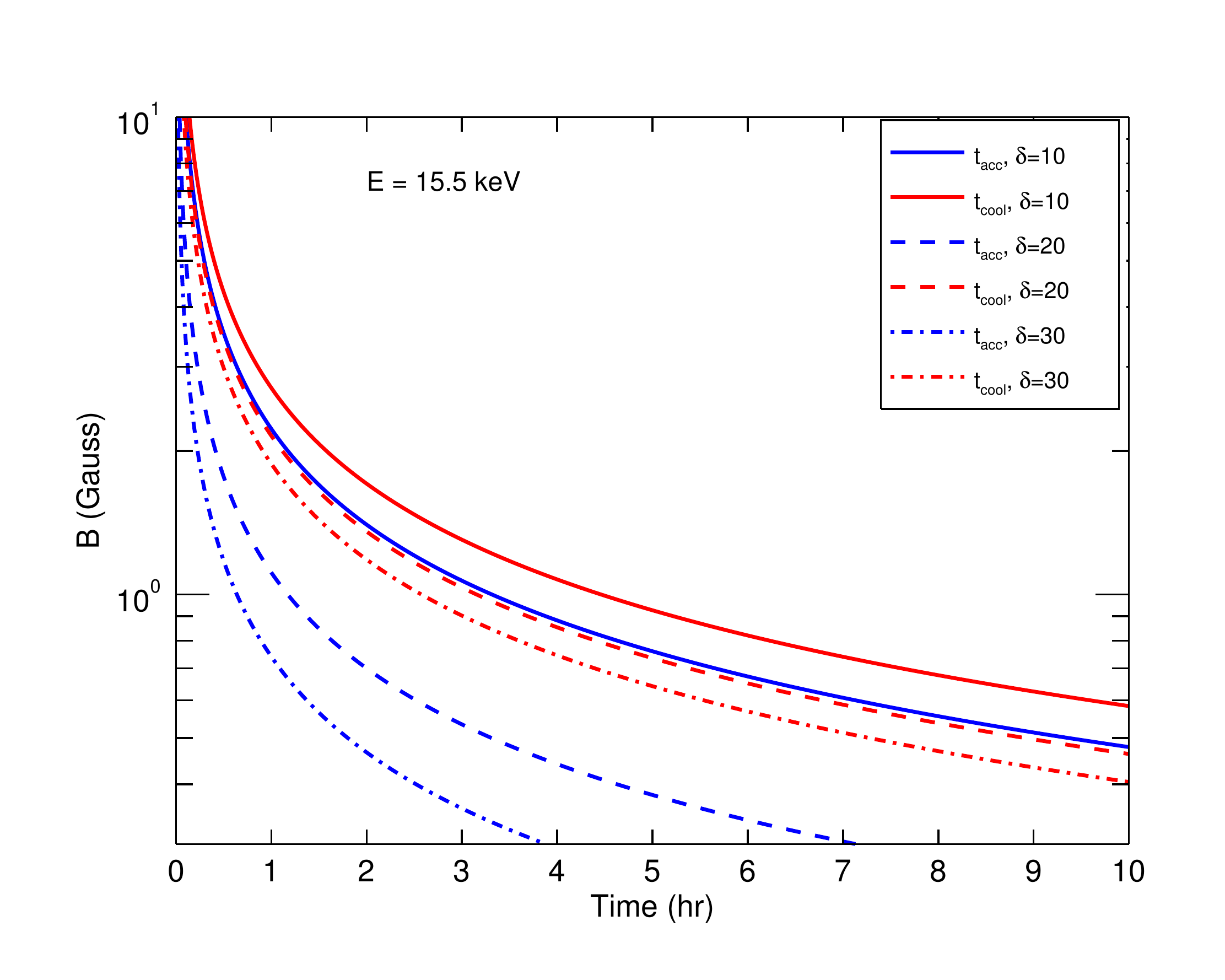}
\caption{The relation between the magnetic field and the particle acceleration and synchrotron cooling timescales in the observed frame for the moderate values of Doppler factors.  \label{Fig4}}
\end{center}
\end{figure}

In HSPs,  the hard X-ray emission is probably due to the high-energy tail of the synchrotron emission from the large scale jets.  The variable emission then can be related to the particle acceleration and synchrotron emission by the electrons of the highest energy. In such scenario, the variability timescales can be directly linked with the particle acceleration and cooling timescales. To estimate the synchrotron cooling timescale, i. e. cooling due to synchrotron emission, we define $t_{cool}=$(Energy of an electron)/(Synchrotron power loss) =$\gamma m_{e} c^{2}/P_{syn}$. This gives
 
   \begin{equation}
  t_{cool}=\frac{3}{4}\frac{m_{e}c}{\sigma_{T}U_{B}\gamma\beta ^{2}} \sim 7.74\times 10^{8}\gamma^{-1} B^{-2}  \ s
  \end{equation}
  \noindent  where we use  $\beta \sim 1$ considering ultra-relativistic electrons.  We note that such energy dependent cooling timescale can produce more rapid variability at hard X-ray energies than at soft X-ray energies.  If we assume that the cooling takes place mainly due to synchrotron process, and that the most of the synchrotron emission is emitted in the \textit{NuSTAR} energy band ($\sim15$ keV; logarithmic mean of the \textit{NuSTAR} range), then following \citet{Zhang2002},  magnetic field corresponding to the cooling timescales can be given as 

\begin{equation}
B=\frac{2.09 \times 10^{2}(1+z)^{1/3}}{t^{2/3}_{cool}\delta^{1/3}E^{1/3}}
\end{equation}

\noindent where E is the energy of the observed photons expressed in keV, and B and $\delta$ are the magnetic field and the Doppler factor of the emitting region, respectively. Similarly, assuming the particle acceleration due to  diffusive shock acceleration e. g. \citet{Blandford87}, magnetic field corresponding to the particle acceleration timescales can be given as
\begin{equation}
B=\frac{21.04 \times 10^{-2}(1+z)\xi^{2/3} E^{1/3}}{t^{2/3}_{acc}\delta},
\end{equation}
\noindent where $\xi$ is the acceleration parameter conveniently expressed in the fiducial scale of $10^{5}$ \citep[for details see][]{Zhang2002} indicating the acceleration rate of electrons. For moderate $\delta$ and z=0.1, the curves showing the relation between the magnetic field and the acceleration and synchrotron cooling timescales, within the \textit{NuSTAR} band, are presented in Figure \ref{Fig4}. It is interesting to note that for a reasonable $\xi=0.2 \times 10^{5}$, the cooling curves closely follow the acceleration curves. From these curves, it can be inferred that  for the given variability timescales of a few hours as seen in the source (refer to Table 2 last column), reasonable value of the magnetic field could be in the order of a few Gauss. Once we constrain the magnetic field, we can also estimate the energy of the high-tail synchrotron emitting electrons. Assuming most of the emission is concentrated near the maximum synchrotron frequency $\nu _{s}$ (in Hz), it can be expressed as
  \begin{equation}
  \nu _{s}=\frac{2e}{3\pi m_{e}c}\gamma_{max} ^{2}B \sim 3.73 \times 10^{6} \gamma_{max} ^{2}B. 
 \end{equation}

\noindent Using B=1 G, the Lorentz factor for the highest energy electrons can be estimated as  $\gamma_{max} \sim 9.8\times 10^{5}$; such a high  value of $\gamma_{max}$ is particularly consistent with the fact that most of the BL Lacs discussed in the paper are TeV blazars.

\begin{figure}
		\centering
		\includegraphics[width=0.98\linewidth]{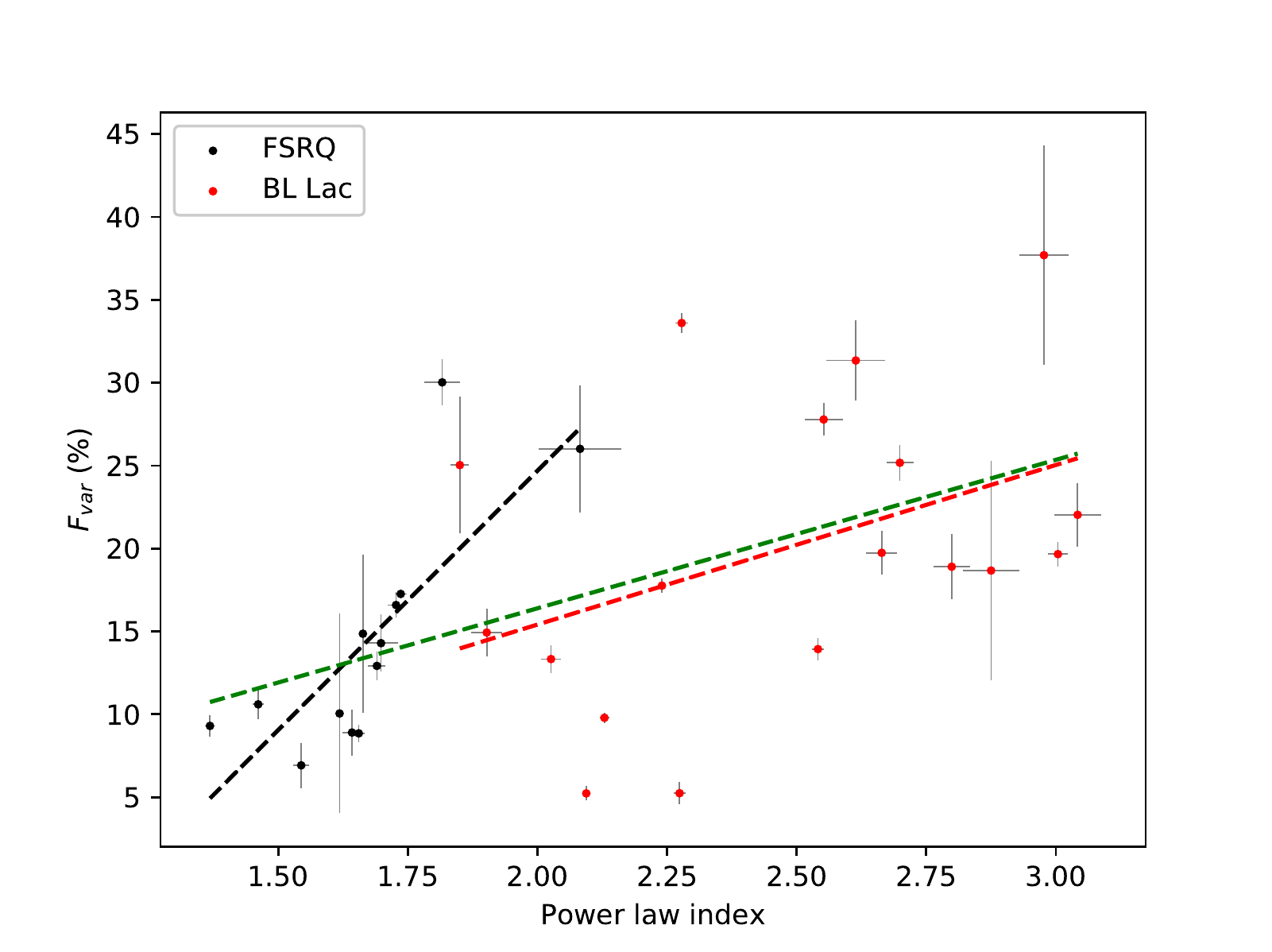}
		\caption{Fractional variability of \textit{NuSTAR} light curves plotted against the photon power-law index for the corresponding observations for  FSRQs (black) and BL Lacs (red). The green, black and the red dashed lines represent the linear fit to all the sources, only FSRQs and only BL Lacs, respectively.}
		\label{fig:Gamma_Fvar}
		\end{figure}

	\begin{figure}
		\centering
		\includegraphics[width=0.98\linewidth]{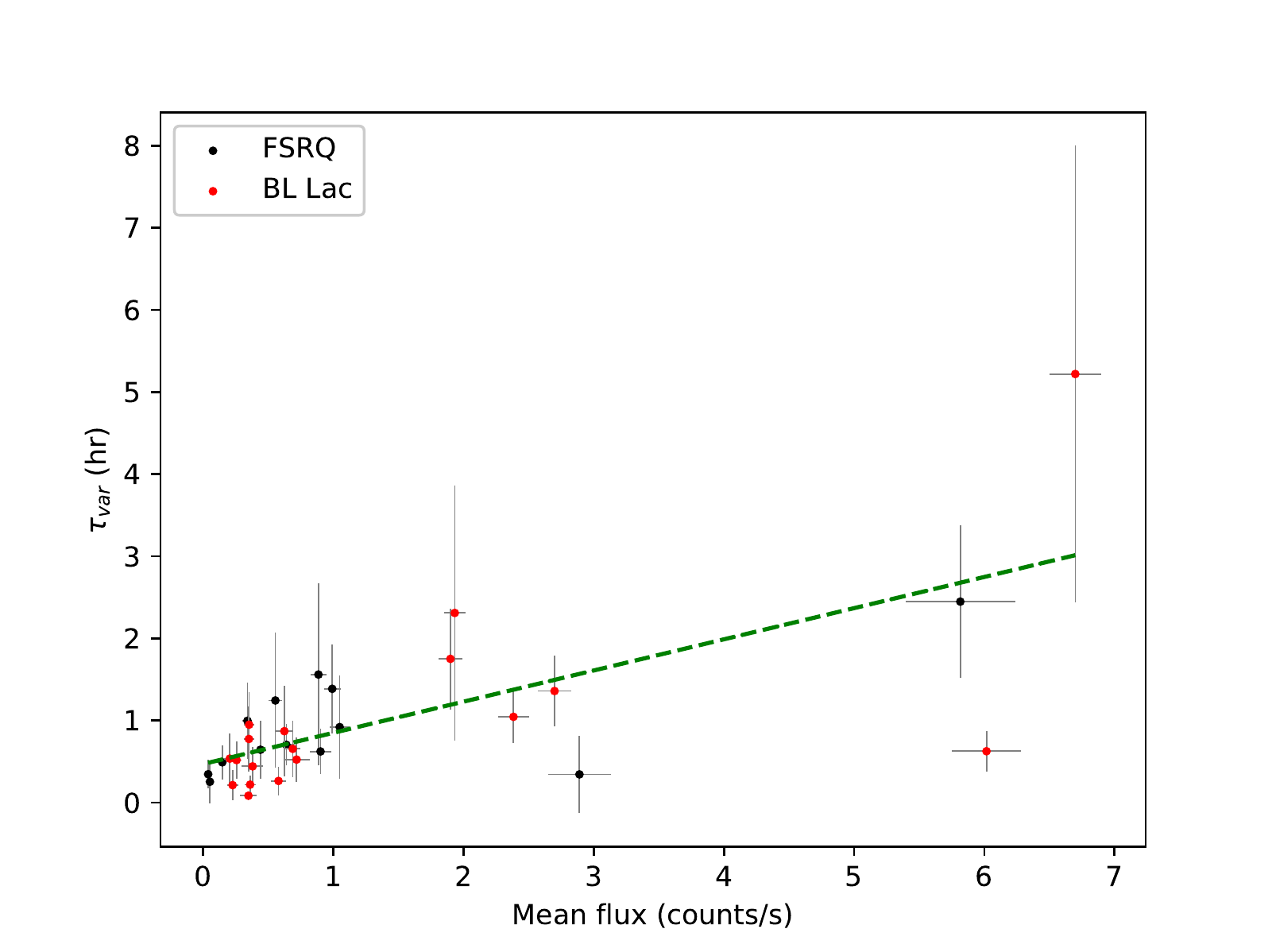}
		\caption{Correlation between minimum variability timescales of the \textit{NuSTAR} blazar sources and their mean fluxes. The green dashed line represents the linear best-fit to the data.}
		\label{fig:F_tvar}
		\end{figure}

In powerful FSRQs, the hard X-ray could be resulted from a number of processes such as synchrotron radiation of pair cascades powered by ultra-relativistic protons,  synchrotron radiation by ultra-relativistic protons and inverse-Compton scattering of the external soft photons \citep[see][and references therein]{Sikora2009}. In the more likely EC  scenario, the IC cooling timescale, depending on the energy density of the external photon field and the electron energy, can be written as
  
   \begin{equation}
  t_{IC} \sim \frac{3}{4}\frac{m_{e}c}{\sigma_{T}U_{ext}\gamma} 
  \end{equation}
  
 \noindent Now, the external photon field can be attributed to hot dusty torus (HDR), broad line region (BLR) or even the accretion disk. As a more realistic example, assuming that HDR with monotonic photon field energy $h\nu_{0} \sim 0.1$ eV (in infra-red range; see \citealt{Kataoka2008}) poses for the U$_{ext}$, and that the most of the IC emission lies within the \textit{NuSTAR} band,  the energy of the injected electrons in the source rest frame can be estimated using $ \nu \sim \gamma^{2}\nu_{0}$,  where $\nu_{0}$ and $\nu$ are the frequencies of the soft and up-scattered emission, respectively. Moreover, to account for the fact that the emission zone is moving with a Doppler factor $\delta$, the relation can be written as $ \nu \sim \gamma^{2}\delta^{2}\nu_{0}$. Now using $h\nu_{0}$=0.1 eV for HDR and 10 eV for BLR \citep[see][]{Nalewajko2012}, the energy of the lower-tail of the high energy particles ($\gamma_{min}$) turns out to be $\sim 40$ and 4 Lorentz factors, respectively. It is preferable to have lower $\gamma_{min}$ because the jet power is very sensitive to the minimum energy of the emitting electrons. A large $\gamma_{min}$ (typically, $\gtrsim$100)  would drastically reduce the kinetic jet power and can make it even smaller than the radiative power. All the kinetic power of the jet would then be consumed by the radiation making the jet weaker and eventually stop. In such case, we would not expect to see the Mpc scale radio jets, which is against the observations  \citep[for relevant discussion refer to] []{Ghisellini2010a}.

Figure \ref{fig:Gamma_Fvar}, presenting the distribution of the FV over the photon indexes $\Gamma_{X}$, suggests that the sources tend to be more variable in their steeper spectral states. The strength of the correlation between the quantities are measured by Spearman rank correlation coefficient ($\rho$). The correlation looks more pronounced in FSRQs (black symbols) as indicated by the higher value of $\rho$ = 0.84 with $p$-value = $3\times 10^{-4}$  compare to $\rho$ = 0.59 with $p$-value = 0.01 for HSPs (red  symbols).  When included  all the sources (green symbols) the correlation becomes moderate with $\rho$=0.60 and $p$-value = $2\times 10^{-4}$. The best linear fit for FSRQ only, HSP only  and all the sources are shown by black, red and green dashed lines, respectively.

\begin{figure}
		\centering
		\includegraphics[width=0.98\linewidth]{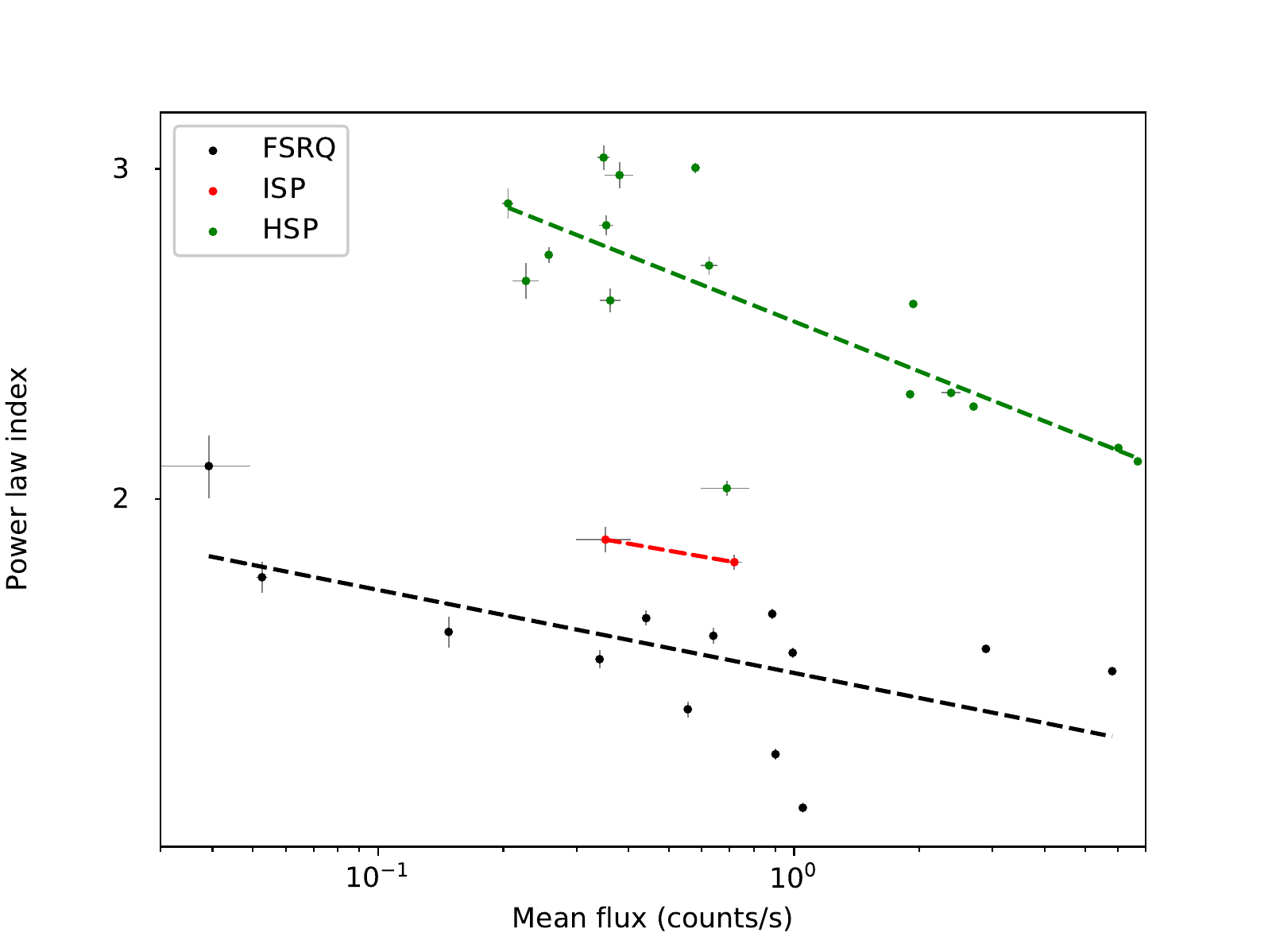}
		\caption{Distribution of the \textit{NuSTAR} power law photon indexes over the mean fluxes for FSRQ (black), ISP (red) and HSP (green). The dashed lines represent the corresponding linear best-fits.}
		\label{fig:F_Gamma}
		\end{figure}

In combination with the close relation between high flux and harder photon index seen in Figure \ref{fig:F_Gamma} (discussed more in Section \ref{spectral_shape}), this indicates that the observed overall variability could be dominantly contributed by the softer photons. The idea also seems to be reflected in Figure \ref{fig:F_tvar} showing correlation between mean flux and minimum variability timescale ($\rho=0.60$ with $p$-value = $1.88\times 10^{-7}$) which indicates more rapid minimum variability timescales for fainter flux states. Such a rapid variability associated with low-flux level could be linked to  small-scale sub-structures resulting from the turbulence at the innermost blazar regions \citep[e.g.][]{Narayan2012,bhatta13,Marscher2014} in contrast to the processes involving a large injection (e.g. due to shocks) or release (e.g. due to magnetic reconnection) of energy over a large volume that are capable of producing big flares in the light curves \citep[e.g.][]{Hayashida2015}. As also seen in Figure \ref{fig:Gamma_Fvar}, the relation between FV and $\Gamma_{X}$ in BL Lacs does not look as distinct as in the FSRQs, as suggested by the relatively poor linear-fit (dashed red line). It is possible that the relation might have been diluted in BL Lacs due to the rapid synchrotron cooling timescales for the particles at the high-energy end of the power-law distribution contributing to the photons at the high-energy end of the spectrum, making it harder; and hence the opposite trend i.e. indicative of energetic photons being more variable.

 Alternatively, the hard X-ray variability exhibited by the sources can be related to extrinsic effects e.g. rapid swing in the angle  of the emission regions about the line of sight.  A small deflection in viewing angle and/or bulk Lorentz factor leading to change in Doppler factor can also result in a large flux variations, in the order of the variability amplitudes displayed by the sources as listed in the 6th column of Table 2. \citep[for detailed discussion refer to] []{Ghisellini97, Bhatta2017}. 

\subsection{Flux Hardness Ratio Relation}

 We explored the relation between flux and the hardness ratio in the source by plotting one against another. However, we did not observe any obvious correlation between the flux and the hardness ratio that could be applied to all the observation. Only in one case (see Figure 1, second panel),  we could observe a clear evidence of harder-when-brighter within the observation period. We also looked for the sign of hysteresis loops in the flux-HR plane. However, no such loops could be found. 
 
 In blazars, the nature of the correlation between the flux and the spectral state, so far, is somewhat uncertain.  In the optical band,   bluer-when-brighter tendency is more associated with BL Lacs in the intra-day timescales, whereas  redder-when-brighter trend seems to be frequently observed in FSRQs .  In $\gamma$-ray regime also  blazars  were found to behave in the similar fashion i.e., in some cases the spectrum hardens with the source intensity and in other cases  the spectrum softens with the flux enhancements \citep[see][for the discussion]{Bhatta2017}. The bluer-when-brighter trend seen in the \textit{NuSTAR} observation of Mrk 501 (Figure 1, second panel) could be due to  local enhancements of the magnetic field in the jets leading to elevated synchrotron emission with an excess of hard photons.
 
 \subsection{Spectral shapes and photon index distribution \label{spectral_shape}}
Figure \ref{fig:F_Gamma} shows the distribution of the photon indexes of the best-fit models over the mean fluxes. The distribution clearly distinguishes the photon indexes for FSRQs and HSPs. It can be seen that the spectral indexes $\alpha_{X}$ ($\alpha_{X} = \Gamma_{X}-1)$ for the HSPs are steep ranging from $\sim2$--1 and the ones for the FSRQs are
$\sim1$--$0.3$. These results are consistent with the previous similar works \citep[e.g.][]{Donato2005,Tramacere2007}. Although, there are only two ISPs, their $\alpha_{X}$s in appropriate place in the figures  between the $\Gamma_{X}$s for HSPs and FSRQs.  The results are consistent with the standard blazar paradigm, so called blazar sequence, that in the high energy regime the FSRQs with large Compton dominance \citep{ Padovani1997} exhibit harder spectra in comparison to the BL Lacs (in present case TeV blazars). A similar distinction between FSRQs and BL Lacs, based their $\Gamma_{X}$ in the Swift X-ray range and $\Gamma_{\gamma}$ in the Fermi/LAT $\gamma$-ray range, was observed by \citet{Sambruna2010}. The BL Lacs in general are dim possibly due to  the sub-Eddington accretion rates; and they  are usually identified with flatter spectra in the hard X-ray/$\gamma$-region. Figure \ref{fig:F_Gamma}  also suggests a close connection between the flux and spectral slope within the source class in the sense that high flux and/or flux states tend to be of harder spectra $\rho$=-0.67, $p$-value = 0.019 (FSRQ) and    $\rho$ = -0.74, $p$-value  = 0.001 (HSP). This might indicate that high flux and/or flux states are most likely linked to small scale instantaneous changes in the mass accretion rate and disk efficiency which could be modulated by the disk instabilities \citep[e.g.][]{Mangalam1993} due to the formation of the hot spots; and thus hard X-ray flux modulations seen in the sources are possibly triggered at the innermost regions of the central engine.

\subsection{Correlation between low and high energy emission}

As we examined the correlation between the low and high energy emission by the sources, there does not seem to be a single behavior that can be generalized for  all the observations.  Instead, all kinds of relation are observed: In some cases there is a strong correlation between the emission in the two energy bands, whereas in some cases they appear  completely uncorrelated  as indicated by their low ZDCF values at all lags. Similarly, we also observed possible signatures of  hard and soft lags. However, due to the Poisson noise like behavior of the variability, it is hard to be conclusive. The apparent uncorrelated energy bands might be the result of emission from completely unrelated population of the particles, or reflection from the uncorrelated regions of varying sizes \citep[similar to][]{Tanihata2000}.  On the other hand, the hard and soft lags can be interpreted within the framework of the particle injection and synchrotron cooling at the emission sites \citep[see][]{Kirk98,Zhang2002}. In such a frame work,  depending upon whether the cooling  or the particle acceleration mechanism dominates the variability processes, the soft and hard lag , respectively, can be expected \citep[for details see][]{Zhang2002}.

\section{Conclusion \label{conclusion}}
We analyzed the 31 \textit{NuSTAR} observation for 13 blazars including 7 FSRQs, 4 HSPs, and 2 ISPs. The source displayed high amplitude rapid variability within a timescale of a few hours; the minimum variability timescales range from 0.3 to 18.8 ks, whereas the FV range from $\sim$5--38 \%.  In one occasion, the relation between the hardness ratio and the flux could be dubbed as harder-when-brighter trend,  but in general the relation between the flux and the HR seemed more complex. Similarly, we did not detect any trend in the correlation between the hard and soft energy emission that could be generalized for all the observations. We also found the hints of the presence of soft and hard lags by a few hours. However the low values of the associated likelihood render
the results inconclusive. For most of the observations, log-parabolic model revealing spectral curvature seems to be the best representation of the \textit{NuSTAR} blazar spectra, although some of the source spectra were better fitted with single power-law and broken power-law models. Moreover, the distribution of the spectral slopes appear consistent with the current blazar paradigm in which the HSPs possess the steepest and FSRQs have the flattest spectral slope. In addition, we detected close connection between the photon indexes and the mean flux states that could been seen within the blazar sub-classes. We also noted that the sources tend to be more variable in their steeper spectral states. However, the last feature should be explored further involving a larger sample of blazars.

\begin{acknowledgements}
We are thankful to the anonymous reviewer for his/her constructive comments that helped improve the quality of the paper significantly. GB acknowledges the financial support by the Polish National Science Centre through the grants UMO-2017/26/D/ST9/01178 and DEC- 2012/04/A/ST9/00083. We would also like to thank  Prof. Micha{\l} Ostrowski and {\L}ukasz Stawarz for their useful comments and suggestions during the work. This paper made use of data from the \textit{NuSTAR} mission, a
project led by the California Institute of Technology, managed by the Jet Propulsion Laboratory, funded by the National
Aeronautics and Space Administration. We thank the \textit{NuSTAR} Operations, Software and Calibration teams for support
with the execution and analysis of these observations.
This research made use of the \textit{NuSTAR} Data Analysis Software
(NuSTARDAS) jointly developed by the ASI Science
Data Center (ASDC, Italy), and the California Institute of
Technology (USA).
\end{acknowledgements}

%

\begin{thebibliography}{99}
\bibitem[Abeysekara et al.(2015)]{Abeysekara2015} Abeysekara, A.~U., Archambault, S., Archer, A., et al.\ 2015, \apjl, 815, L22

\bibitem[Abdo et al.(2010)]{Abdo2010} Abdo, A.~A., Ackermann, M., Ajello, M., et al.\ 2010, \apjs, 188, 405

\bibitem[Ajello et al.(2012)]{Ajello2012} Ajello, M., Alexander, D.~M., Greiner, J., et al.\ 2012, \apj, 749, 21

\bibitem[Akyuz et al.(2013)]{Akyuz2013} Akyuz, A., Thompson, D.~J., Donato, D., et al.\ 2013, \aap, 556, A71

\bibitem[Aliu et al.(2014)]{Aliu2014} Aliu, E., Archambault, S., Arlen, T., et al.\ 2014, \apj, 782, 13

\bibitem[Aharonian(2000)]{Aharonian2000} Aharonian, F.~A.\ 2000, New Astron., 5, 377

\bibitem[Aleksi{\'c} et al.(2015)]{Aleksic2015} Aleksi{\'c}, J., Ansoldi, S., Antonelli, L.~A., et al.\ 2015, A\&A, 576, A126

\bibitem[Alexander (2013)]{Alexander2013} Alexander T.\ ,  arXiv:1302.1508,  2013 

\bibitem[Arnaud (1996)]{Arnaud1996} Arnaud, K. A. 1996, in Astronomical Society of the Pacific Conference Series, Vol. 101, Astronomical Data Analysis Software and Systems V, ed. G. H. Jacoby \& J. Barnes, 17

\bibitem[Baumgartner et al.(2013)]{Baumgartner2013} Baumgartner, W.~H., Tueller, J., Markwardt, C.~B., et al.\ 2013, \apjs, 207, 19


\bibitem[Begelman et al.(2008)]{Begelman2008} Begelman, M.~C., Fabian, A.~C., \& Rees, M.~J.\ 2008, \mnras, 384, L1

\bibitem[Bevington \& Robinson(2003)]{Bevington2003}Bevington, P. R., \& Robinson, D. K. 2003, in Data Reduction and Error Analysis for the Physical Science

\bibitem[Bhatta \& Webb (2018)]{Bhatta2018} Bhatta, G. \& Webb J., \ 2018, Galaxies, 6, 2

\bibitem[Bhatta (2017)]{Bhatta2017}Bhatta, G. 2017, ApJ 487, 7

\bibitem[Bhatta et al.(2013)]{bhatta13}Bhatta, G., et. al. 2013, A\&A, 558A, 92B

\bibitem[Bhatta et al.(2016b)]{bhatta16b} Bhatta, G., Stawarz, {\L}., Ostrowski, M., et al.\ 2016b, ApJ, 831, 92

\bibitem[Bhatta et al.(2016c)]{bhatta16c} Bhatta, G., Zola S., Stawarz, {\L}., et al.\ 2016c, ApJ, 832, 47

\bibitem[B{\l}a{\.z}ejowski et al.(2000)]{Blazejowski2000} B{\l}a{\.z}ejowski, M., Sikora, M., Moderski, R., \& Madejski, G.~M.\ 2000, \apj, 545, 107

\bibitem[Blandford \& Eichler(1987)]{Blandford87} Blandford, R., \& Eichler, D.\ 1987, \physrep, 154, 1

\bibitem[B{\"o}ttcher \& Els(2016)]{Bttcher2016} B{\"o}ttcher, M., \& Els, P.\ 2016, \apj, 821, 102

\bibitem[Brinkmann et al.(2005)]{Brinkmann2005} Brinkmann, W., Papadakis, I.~E., Raeth, C., Mimica, P., \& Haberl, F.\ 2005, \aap, 443, 397

\bibitem[Burbidge et al.(1974)]{Burbidge1974} Burbidge, G.~R., Jones, T.~W., \& Odell, S.~L.\ 1974, \apj, 193, 43

\bibitem[Camenzind \& Krockenberger(1992)]{Camenzind92}Camenzind M., Krockenberger M., 1992, A\&A, 255, 59


\bibitem[Cawthorne(2006)]{Cawthorne2006} Cawthorne, T.~V.\ 2006, \mnras, 367, 851

\bibitem[D'Ammando \& Orienti(2016)]{DAmmando2016} D'Ammando, F., \& Orienti, M.\ 2016, \mnras, 455, 1881

\bibitem[Dermer \& Schlickeiser(1993)]{Dermer1993} Dermer, C.~D., \& Schlickeiser, R.\ 1993, \apj, 416, 458

\bibitem[Donato et al.(2005)]{Donato2005} Donato, D., Sambruna, R.~M., \& Gliozzi, M.\ 2005, \aap, 433, 1163

\bibitem[Edelson et al.(2014)]{Edelson2014}Edelson, R., Vaughan, S., Malkan, M., et al.\ 2014, \apj, 795, 2

\bibitem[Edelson \& Krolik(1988)]{EK88} Edelson, R. A., \& Krolik, J. H.\ 1988, \apj, 333, 646

\bibitem[Elvis et al.(1992)]{Elvis1992} Elvis, M., Plummer, D., Schachter, J., \& Fabbiano, G.\ 1992, \apjs, 80, 257

\bibitem[Falcone et al.(2004)]{Falcone2004} Falcone, A.~D., Cui, W., \& Finley, J.~P.\ 2004, \apj, 601, 165

\bibitem[Fossati et al.(2000b)]{Fossati2000b} Fossati, G., Celotti, A., Chiaberge, M., et al.\ 2000, \apj, 541, 166

\bibitem[Fossati et al.(2000a)]{Fossati2000a} Fossati, G., Celotti, A., Chiaberge, M., et al.\ 2000, \apj, 541, 153 

\bibitem[Fossati et al.(1998)]{Fossati1998} Fossati, G., Maraschi, L., Celotti, A., Comastri, A., \& Ghisellini, G.\ 1998, \mnras, 299, 433

\bibitem[Furniss et al.(2015)]{Furniss2015} Furniss, A., Noda, K., Boggs, S., et al.\ 2015, \apj, 812, 65

\bibitem[Giebels et al.(2002)]{Giebels2002} Giebels, B., Bloom, E.~D., Focke, W., et al.\ 2002, \apj, 571, 763

\bibitem[Ghisellini et al.(2017)]{Ghisellini2017} Ghisellini, G., Righi, C., Costamante, L., \& Tavecchio, F.\ 2017, \mnras, 469, 255

\bibitem[Ghisellini et al.(2011)]{Ghisellini2011} Ghisellini, G., Tavecchio, F., Foschini, L., \& Ghirlanda, G.\ 2011, \mnras, 414, 2674

\bibitem[Ghisellini et al.(2010a)]{Ghisellini2010} Ghisellini, G., Della Ceca, R., Volonteri, M., et al.\ 2010, \mnras, 405, 387


\bibitem[Ghisellini et al.(2010b)]{Ghisellini2010a} Ghisellini, G., Tavecchio, F., Foschini, L., et al.\ 2010, \mnras, 402, 497

\bibitem[Ghisellini et al.(1997)]{Ghisellini97}Ghisellini, G., Villata, M., Raiteri. et al., 1997, A\&A, 327, 61

\bibitem[Giannios et al.(2009)]{Giannios2009} Giannios, D., Uzdensky, D.~A., \& Begelman, M.~C.\ 2009, \mnras, 395, L29

\bibitem[Giommi et al.(1999)]{Giommi1999} Giommi, P., Massaro, E., Chiappetti, L., et al.\ 1999, \aap, 351, 59

\bibitem[Hagen-Thorn et al.(2008)]{Hagen-Thorn2008} Hagen-Thorn, V.~A., Larionov, V.~M., Jorstad, S.~G., et al.\ 2008, \apj, 672, 40-47

\bibitem[Harrison et al.(2013)]{Harrison2013}Harrison, F. A., et al. 2013, ApJ, 770, 103

\bibitem[Hayashida et al.(2015)]{Hayashida2015} Hayashida, M., Nalewajko, K., Madejski, G.~M., et al.\ 2015, \apj, 807, 79

\bibitem[Holder et al.(2003)]{Holder2003} Holder, J., Bond, I.~H., Boyle, P.~J., et al.\ 2003, \apjl, 583, L9

\bibitem[Homan et al.(2001)]{Homan2001} Homan, D.~C., Attridge, J.~M., \& Wardle, J.~F.~C.\ 2001, \apj, 556, 113

\bibitem[Hummel et al.(1992)]{Hummel1992} Hummel, C.~A., Muxlow, T.~W.~B., Krichbaum, T.~P., et al.\ 1992, \aap, 266, 93


\bibitem[Hughes et al.(1998)]{Hughes98}Hughes P. A., Aller H. D., Aller M. F., 1998, ApJ, 503, 662

\bibitem[Impey \& Neugebauer(1988)]{Impey1988} Impey, C.~D., \& Neugebauer, G.\ 1988, \aj, 95, 307


\bibitem[Jorstad et al.(2013)]{Jorstad2013} Jorstad, S., Marscher, A., Larionov, V., et al.\ 2013, European Physical Journal Web of Conferences, 61, 04003

\bibitem[Jorstad et al.(2004)]{Jorstad2004} Jorstad, S.~G., Marscher, A.~P., Lister, M.~L., et al.\ 2004, \aj, 127, 3115

\bibitem[Jorstad et al.(2005)]{Jorstad2005} Jorstad, S.~G., Marscher, A.~P., Lister, M.~L., et al.\ 2005, \aj, 130, 1418

\bibitem[Joshi \& B{\"o}ttcher(2011)]{Joshi2011} Joshi, M., \& B{\"o}ttcher, M.\ 2011, \apj, 727, 21

\bibitem[Kalberla et al.(2005)]{Kalberla2005} Kalberla, P.~M.~W., Burton, W.~B., Hartmann, D., et al.\ 2005, \aap, 440, 775

\bibitem[Kataoka et al.(2008)]{Kataoka2008} Kataoka, J., Madejski, G., Sikora, M., et al.\ 2008, \apj, 672, 787-799

\bibitem[Kirk et al.(1998)]{Kirk98}Kirk, J. Reiger, F.M., \& Mastichiadis, A., 1998, A\&A, 333, 452.

\bibitem[Kubo et al.(1998)]{Kubo1998} Kubo, H., Takahashi, T., Madejski, G., et al.\ 1998, \apj, 504, 693

\bibitem[Lister \& Homan(2005)]{Lister2005} Lister, M.~L., \& Homan, D.~C.\ 2005, \aj, 130, 1389

\bibitem[Madejski et al.(2016)]{Madejski2016} Madejski, G.~M., Nalewajko, K., Madsen, K.~K., et al.\ 2016, \apj, 831, 142

\bibitem[Madsen et al.(2015)]{Madsen2015} Madsen, K.~K., F{\"u}rst, F., Walton, D.~J., et al.\ 2015, \apj, 812, 14


\bibitem[MAGIC Collaboration et al.(2008)]{MAGIC2008} MAGIC Collaboration, Albert, J., Aliu, E., et al.\ 2008, Science, 320, 1752

\bibitem[Mangalam \& Wiita(1993)]{Mangalam1993} Mangalam, A.~V., \& Wiita, P.~J.\ 1993, \apj, 406, 420

\bibitem[Maraschi et al.(1992)]{Maraschi1992} Maraschi, L., Ghisellini, G., \& Celotti, A.\ 1992, \apjl, 397, L5

\bibitem[Marscher \& Travis(1996)]{Marscher1996} Marscher, A.~P., \& Travis, J.~P.\ 1996, \aaps, 120, 537

\bibitem[Marscher \& Gear(1985)]{Marscher1985} Marscher, A.~P., \& Gear, W.~K.\ 1985, \apj, 298, 114

\bibitem[Marscher(2014)]{Marscher2014} Marscher, A.~P.\ 2014, \apj, 780, 87 

\bibitem[Massaro et al.(2008)]{Massaro2008} Massaro, F., Giommi, P., Tosti, G., et al.\ 2008, \aap, 489, 1047

\bibitem[Massaro et al.(2006)]{Massaro2006} Massaro, E., Tramacere, A., Perri, M., Giommi, P., \& Tosti, G.\ 2006, \aap, 448, 861

\bibitem[Massaro et al.(2004b)]{Massaro2004b} Massaro, E., Perri, M., Giommi, P., Nesci, R., \& Verrecchia, F.\ 2004, \aap, 422, 103

\bibitem[Massaro et al.(2004a)]{Massaro2004a} Massaro, E., Perri, M., Giommi, P., \& Nesci, R.\ 2004, \aap, 413, 489

\bibitem[Mastichiadis \& Kirk(2002)]{Mastichiadis2002} Mastichiadis, A., \& Kirk, J.~G.\ 2002, PASA, 19, 138

\bibitem[Nalewajko et al.(2012)]{Nalewajko2012} Nalewajko, K., Begelman, M.~C., Cerutti, B., Uzdensky, D.~A., \& Sikora, M.\ 2012, \mnras, 425, 2519

\bibitem[Narayan \& Piran(2012)]{Narayan2012} Narayan, R., \& Piran, T.\ 2012, \mnras, 420, 604

\bibitem[Paliya(2015)]{Paliya2015} Paliya, V.~S.\ 2015, \apj, 804, 74

\bibitem[Paltani et al.(1998)]{Paltani1998} Paltani, S., Courvoisier, T.~J.-L., \& Walter, R.\ 1998, \aap, 340, 47

\bibitem[Pandey et al.(2017)]{Pandey2017} Pandey, A., Gupta, A.~C., \& Wiita, P.~J.\ 2017, \apj, 841, 123

\bibitem[Padovani et al.(2002)]{Padovani2002} Padovani, P., Costamante, L., Ghisellini, G., Giommi, P., \& Perlman, E.\ 2002, \apj, 581, 895

\bibitem[Padovani et al.(1997)]{Padovani1997} Padovani, P., Giommi, P., \& Fiore, F.\ 1997, \mnras, 284, 569

\bibitem[Perlman et al.(1996)]{Perlman1996} Perlman, E.~S., Stocke, J.~T., Wang, Q.~D., \& Morris, S.~L.\ 1996, \apj, 456, 451

\bibitem[Perlman et al.(2005)]{Perlman2005} Perlman, E.~S., Madejski, G., Georganopoulos, M., et al.\ 2005, \apj, 625, 727

\bibitem[Peterson(1997)]{Peterson1997} Peterson, B.~M.\ 1997, An introduction to active galactic nuclei, Publisher: Cambridge, New York Cambridge University Press, 1997 Physical description xvi, 238 p.~ISBN 0521473489


\bibitem[Rani et al.(2017)]{Rani2017} Rani, P., Stalin, C.~S., \& Rakshit, S.\ 2017, \mnras, 466, 3309

\bibitem[Ravasio et al.(2004)]{Ravasio2004} Ravasio, M., Tagliaferri, G., Ghisellini, G., \& Tavecchio, F.\ 2004, \aap, 424, 841

\bibitem[Ravasio et al.(2002)]{Ravasio2002} Ravasio, M., Tagliaferri, G., Ghisellini, G., et al.\ 2002, \aap, 383, 763

\bibitem[Raiteri et al.(2013)]{Raiteri2013} Raiteri, C.~M., Villata, M., D'Ammando, F., et al.\ 2013, \mnras, 436, 1530


\bibitem[Sambruna et al.(2010)]{Sambruna2010} Sambruna, R.~M., Donato, D., Ajello, M., et al.\ 2010, \apj, 710, 24

\bibitem[Sambruna et al.(2007)]{Sambruna2007} Sambruna, R.~M., Tavecchio, F., Ghisellini, G., et al.\ 2007, \apj, 669, 884

\bibitem[Sambruna et al.(2000)]{Sambruna2000} Sambruna, R.~M., Chou, L.~L., \& Urry, C.~M.\ 2000, \apj, 533, 650

\bibitem[Sambruna et al.(1994)]{Sambruna1994} Sambruna, R.~M., Barr, P., Giommi, P., et al.\ 1994, \apj, 434, 468

\bibitem[Sbarrato et al.(2016)]{Sbarrato2016} Sbarrato, T., Ghisellini, G., Tagliaferri, G., et al.\ 2016, \mnras, 462, 1542


\bibitem[Sikora \& Begelman(2013)]{Sikora13} Sikora, M., \& Begelman, M.~C.\ 2013, ApJL, 764, L24

\bibitem[Sikora et al.(2009)]{Sikora2009} Sikora, M., Stawarz, {\L}., Moderski, R., Nalewajko, K., \& Madejski, G.~M.\ 2009, \apj, 704, 38

\bibitem[Sikora(1994)]{Sikora1994} Sikora, M.\ 1994, \apjs, 90, 923

\bibitem[Soldi et al.(2008)]{Soldi2008} Soldi, S., T{\"u}rler, M., Paltani, S., et al.\ 2008, \aap, 486, 411

\bibitem[Spada et al.(2001)]{Spada2001} Spada, M., Ghisellini, G., Lazzati, D., \& Celotti, A.\ 2001, \mnras, 325, 1559

\bibitem[Tanihata et al.(2001)]{Tanihata2001} Tanihata, C., Urry, C.~M., Takahashi, T., et al.\ 2001, \apj, 563, 569

\bibitem[Tanihata et al.(2000)]{Tanihata2000} Tanihata, C., Takahashi, T., Kataoka, J., et al.\ 2000, \apj, 543, 124


\bibitem[Tramacere et al.(2007)]{Tramacere2007} Tramacere, A., Giommi, P., Massaro, E., et al.\ 2007, \aap, 467, 501

\bibitem[Urry et al.(1996)]{Urry1996} Urry, C.~M., Sambruna, R.~M., Worrall, D.~M., et al.\ 1996, \apj, 463, 424

\bibitem[Vaughan et al.(2003)]{vau03}Vaughan, S., Edelson, R., Warwick, R. S., \& Uttley, P., 2003, MNRAS, 345, 1271

\bibitem[Walton et al.(2014)]{Walton2014} Walton, D.~J., Risaliti, G., Harrison, F.~A., et al.\ 2014, \apj, 788, 76


\bibitem[Wagner \& Witzel(1995)]{Wagner95}Wagner S. J., Witzel A., 1995, ARA\&A, 33, 163

\bibitem[Welsh(1999)]{Welsh99}Welsh, W. F. 1999, PASP, 111, 1347

\bibitem[Wierzcholska \& Siejkowski(2016)]{Wierzcholska2016a} Wierzcholska, A., \& Siejkowski, H.\ 2016, \mnras, 458, 2350

\bibitem[Wierzcholska \& Wagner(2016)]{Wierzcholska2016b} Wierzcholska, A., \& Wagner, S.~J.\ 2016, \mnras, 458, 56


\bibitem[Wilms et al.(2000)]{Wilms2000} Wilms, J., Allen, A., \& McCray, R.\ 2000, \apj, 542, 914

\bibitem[Wolter et al.(1998)]{Wolter1998} Wolter, A., Comastri, A., Ghisellini, G., et al.\ 1998, \aap, 335, 899

\bibitem[Worrall \& Wilkes(1990)]{Worrall1990} Worrall, D.~M., \& Wilkes, B.~J.\ 1990, \apj, 360, 396

\bibitem[Zhang et al.(2006)]{Zhang2006} Zhang, Y.~H., Bai, J.~M., Zhang, S.~N., et al.\ 2006, \apj, 651, 782

\bibitem[Zhang et al.(2005)]{Zhang2005} Zhang, Y.~H., Treves, A., Celotti, A., Qin, Y.~P., \& Bai, J.~M.\ 2005, \apj, 629, 686

\bibitem[Zhang(2002)]{Zhang2002} Zhang, Y.~H.\ 2002, \mnras, 337, 609

\bibitem[Zhang et al.(1999)]{Zhang1999} Zhang, Y.~H., Celotti, A., Treves, A., et al.\ 1999, \apj, 527, 719

\bibitem[Zola et al.(2016)]{Zola2016} Zola, S., Valtonen, M., Bhatta, G., et al.\ 2016, Galaxies, 4, 41
\end{thebibliography}
%

\clearpage

\onecolumn

\section{On-line Material: Light curves, hardness ratio plots and spectral fits of \textit{NuSTAR} blazars \label{sec:online}}


\begin{figure*}
\begin{multicols}{4}
    \includegraphics[width=1.05\linewidth]{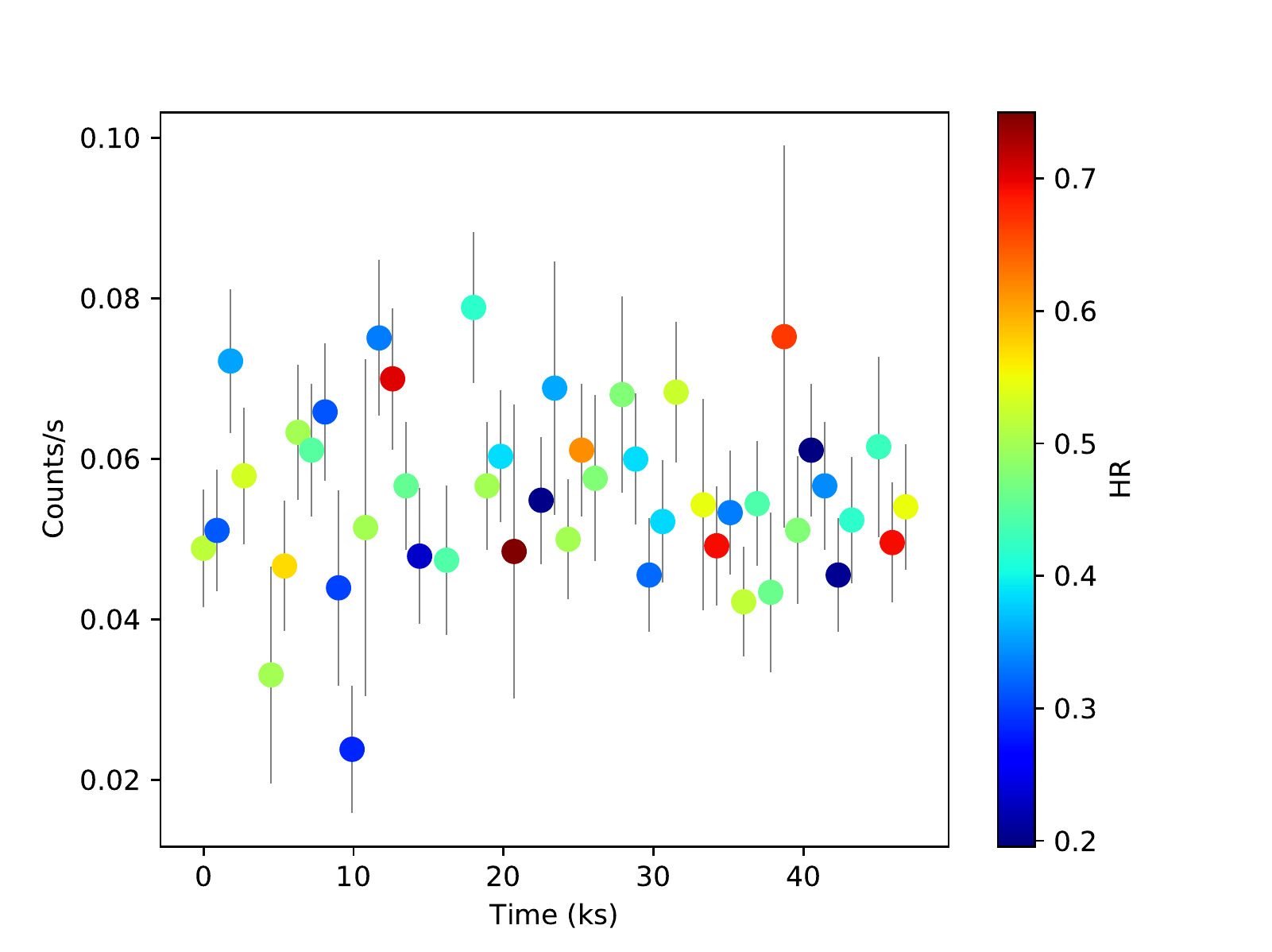}\par 
    \includegraphics[width=1.05\linewidth,angle=0]{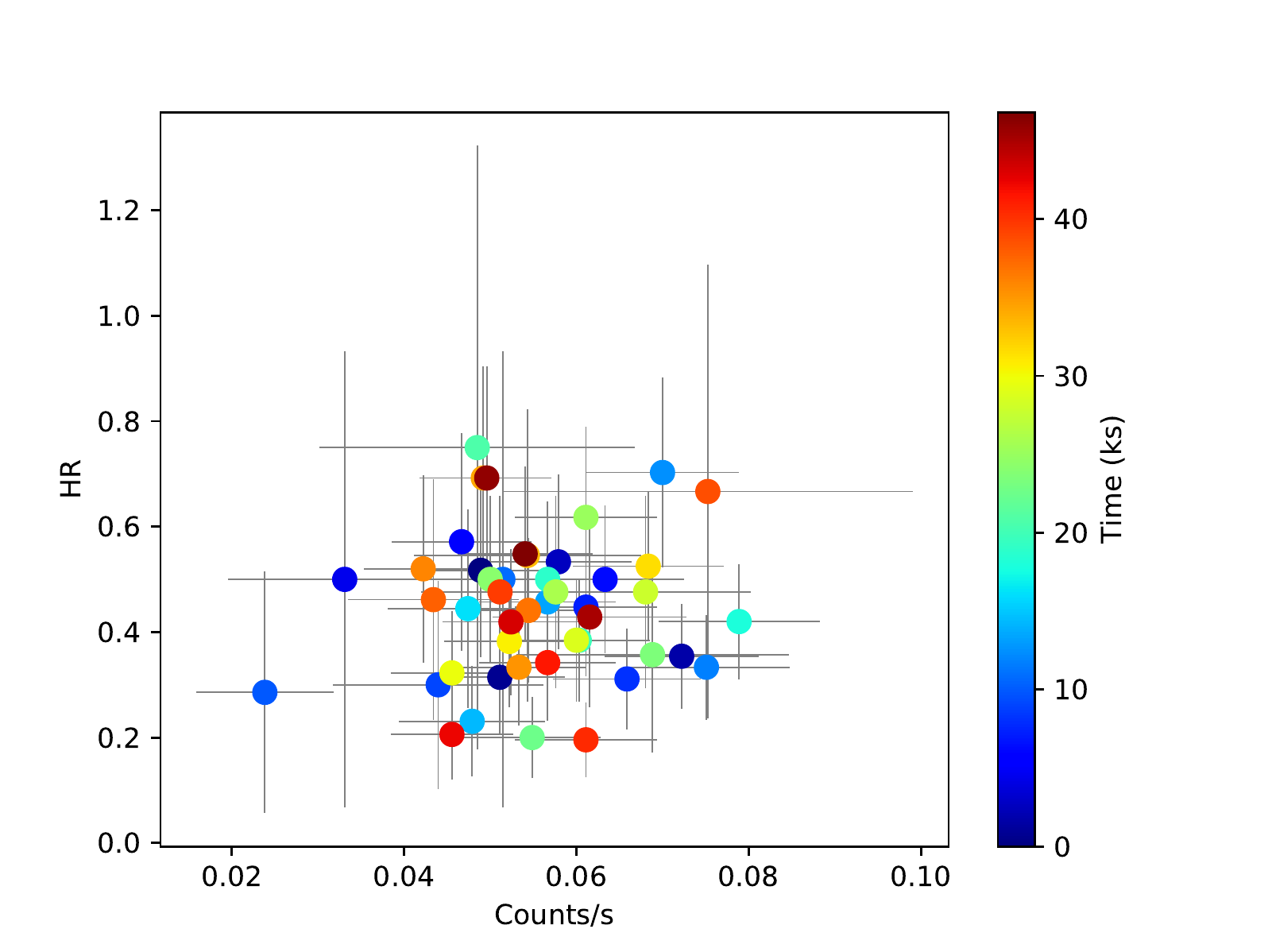}\par
    \includegraphics[width=0.72\linewidth,angle=-90]{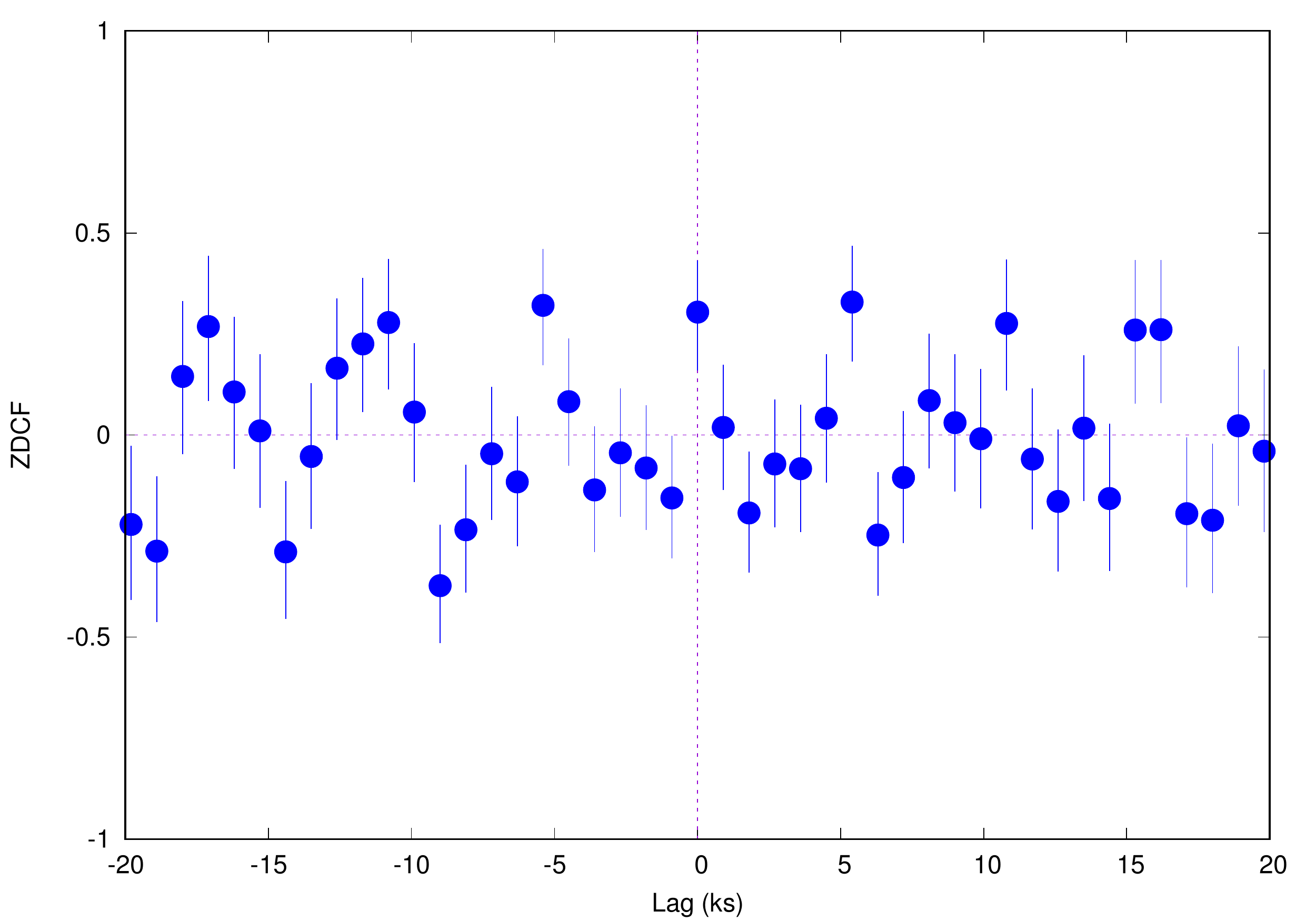}\par
    \includegraphics[width=0.73\linewidth, angle=-90]{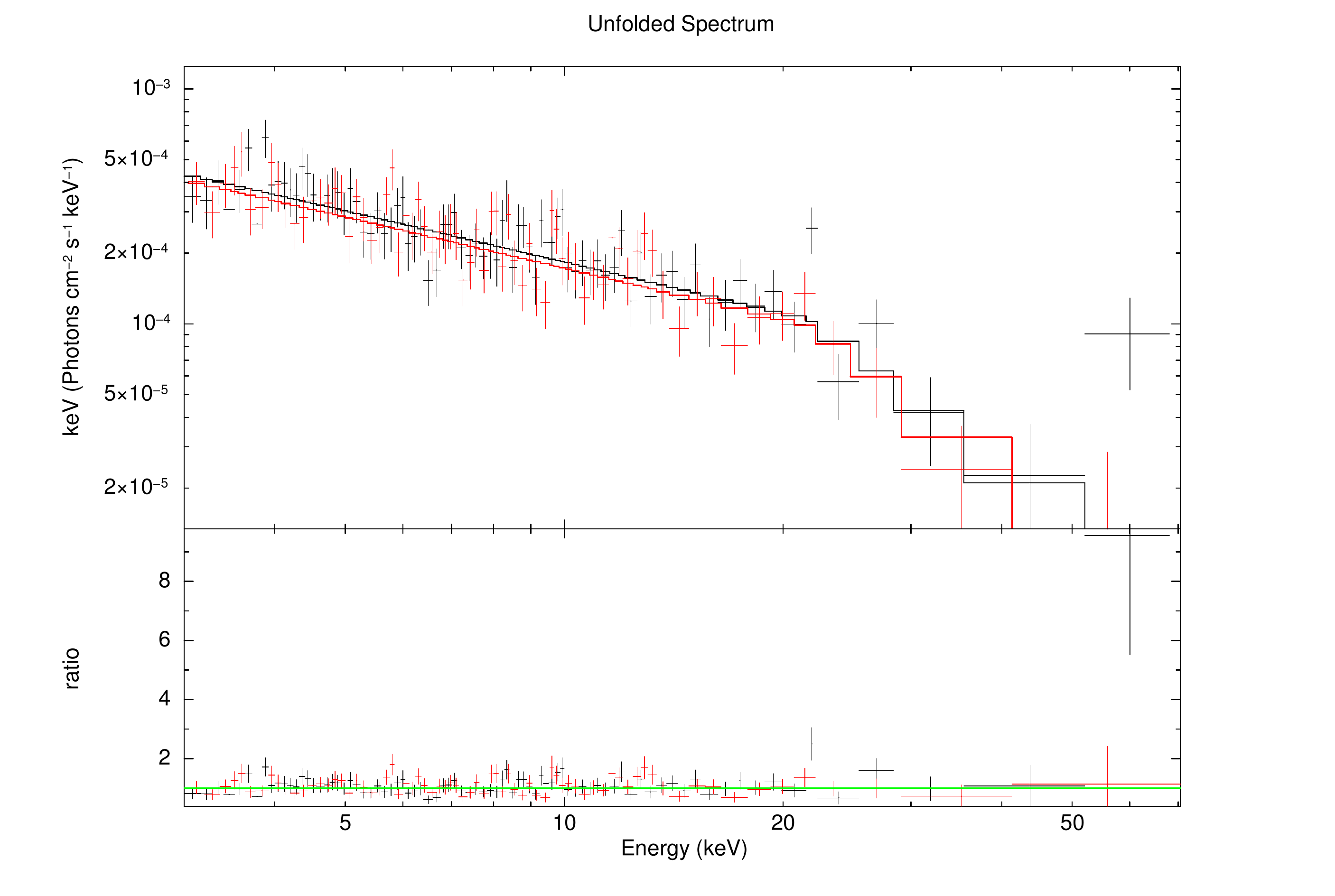}
    \end{multicols}
\center{S5 0014+81, 60001098002}

\begin{multicols}{4}
    \includegraphics[width=1.05\linewidth]{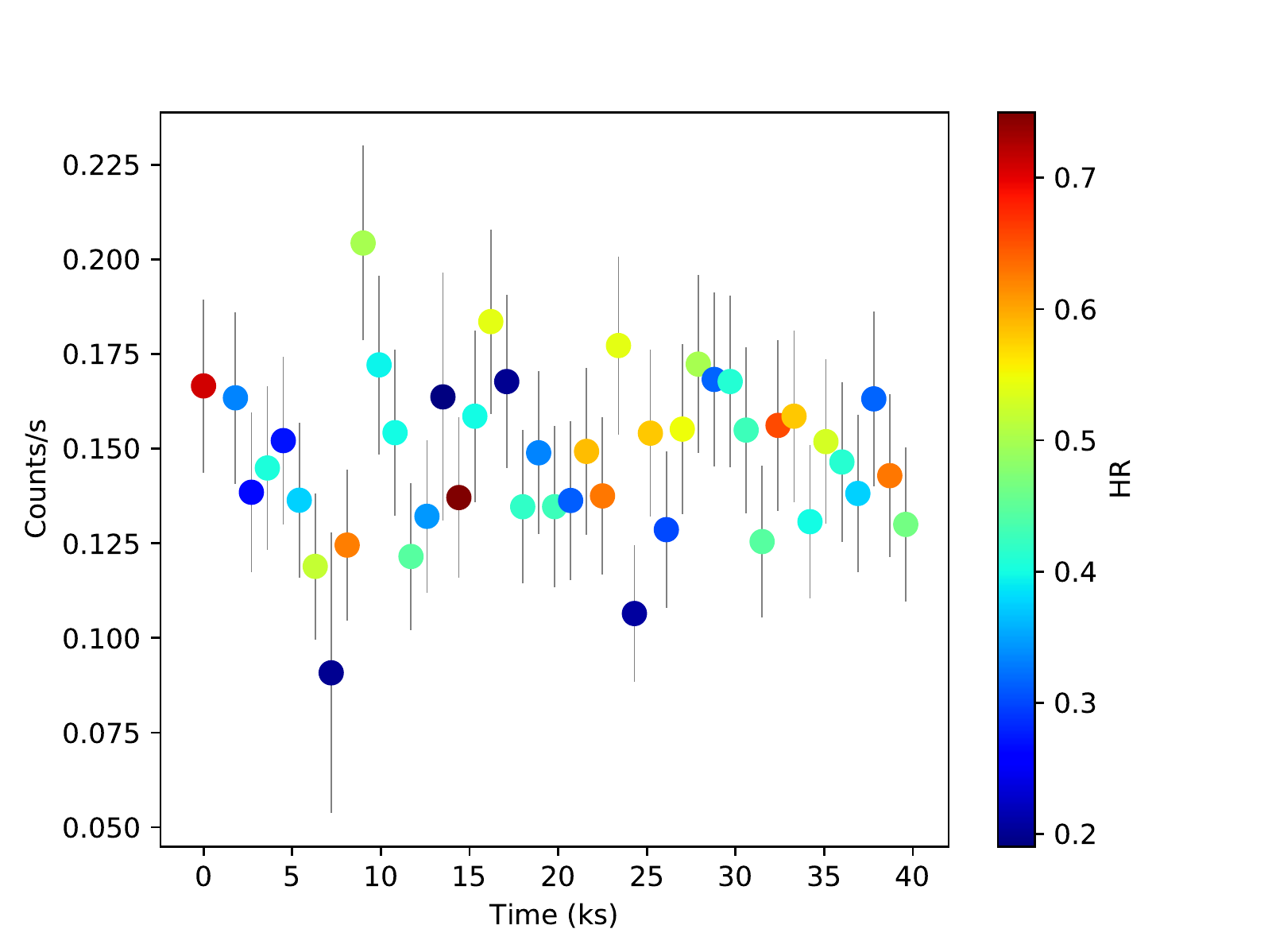}\par 
    \includegraphics[width=1.05\linewidth,angle=0]{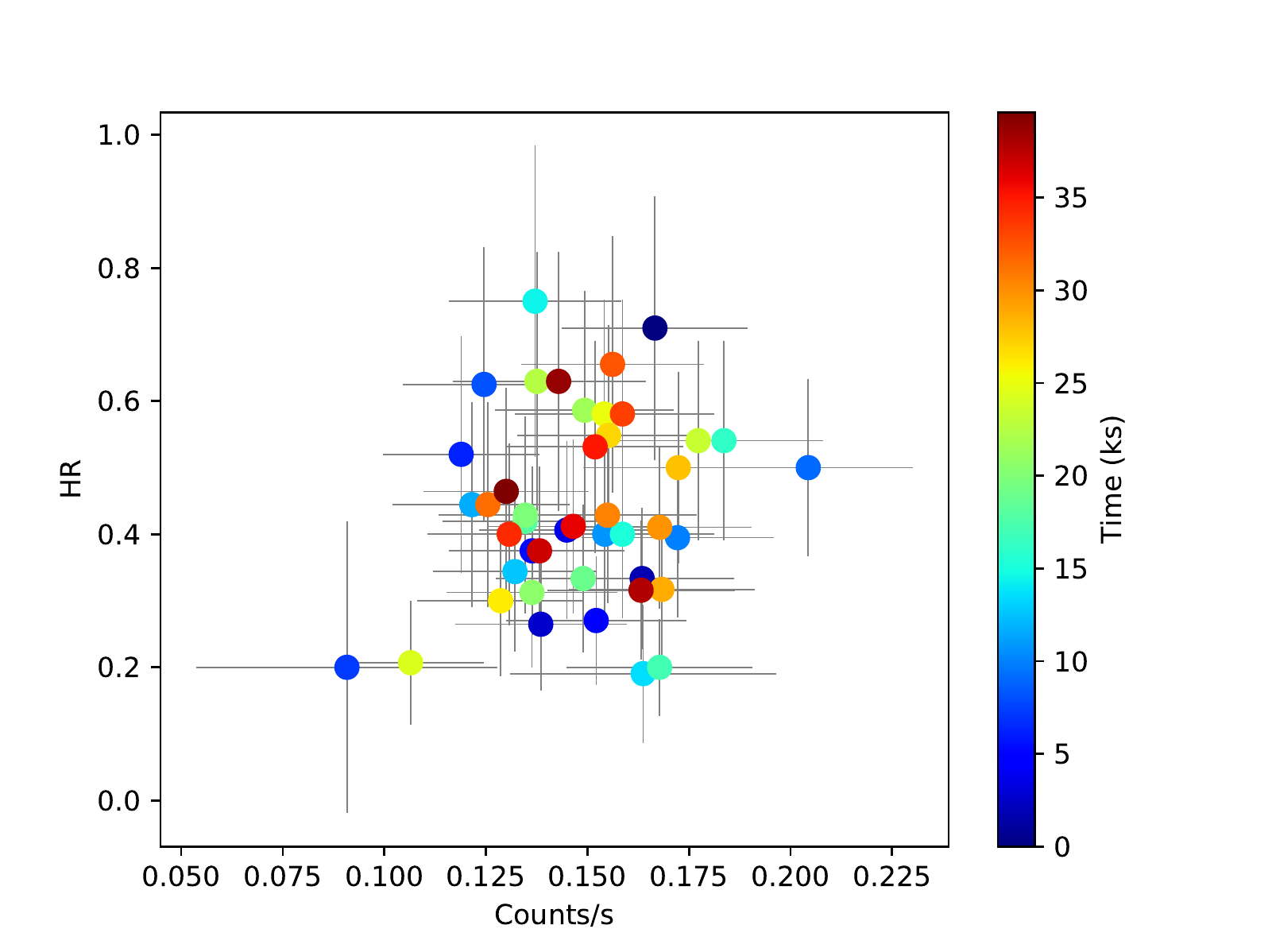}\par
    \includegraphics[width=0.72\linewidth,angle=-90]{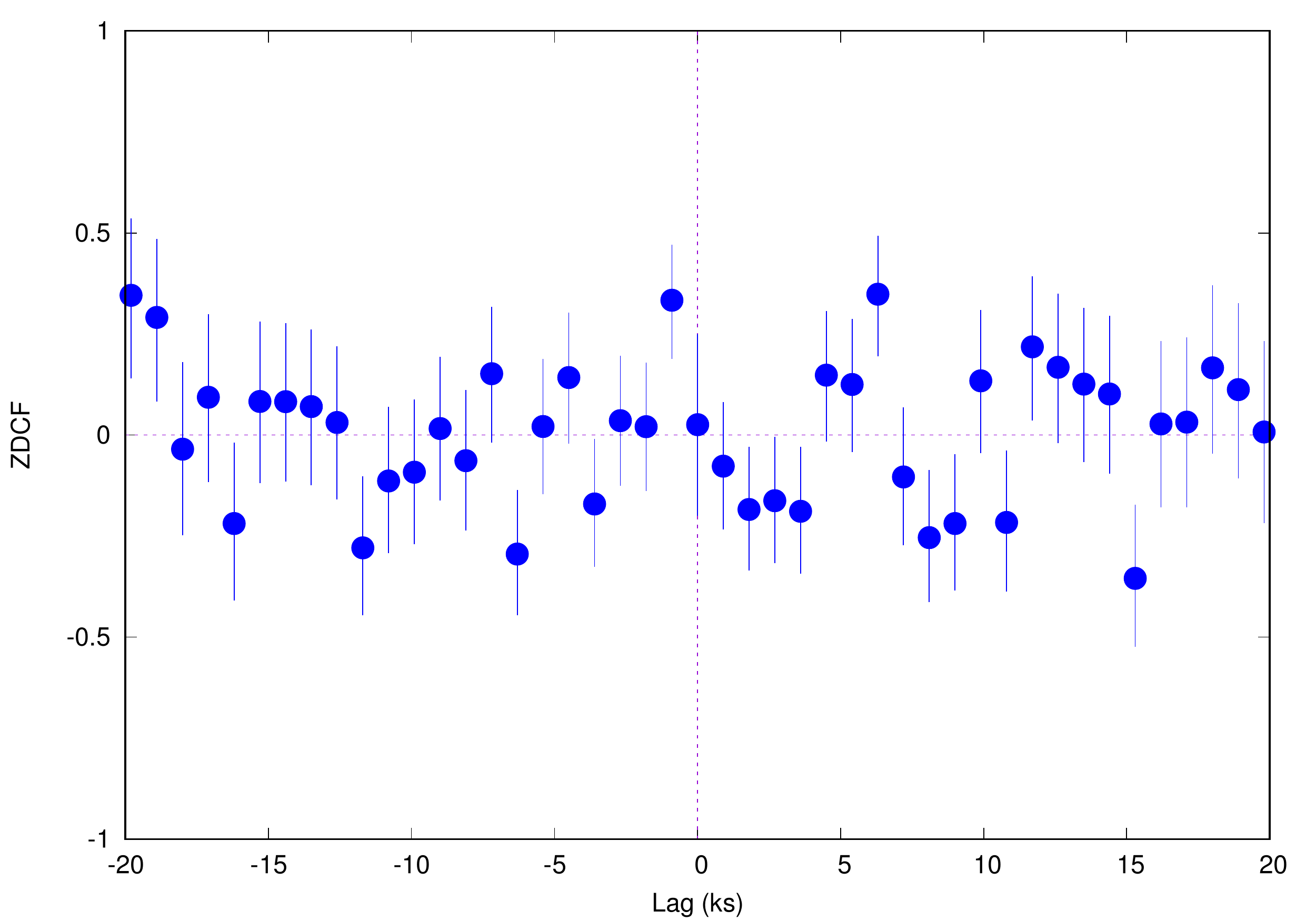}\par
    \includegraphics[width=0.73\linewidth, angle=-90]{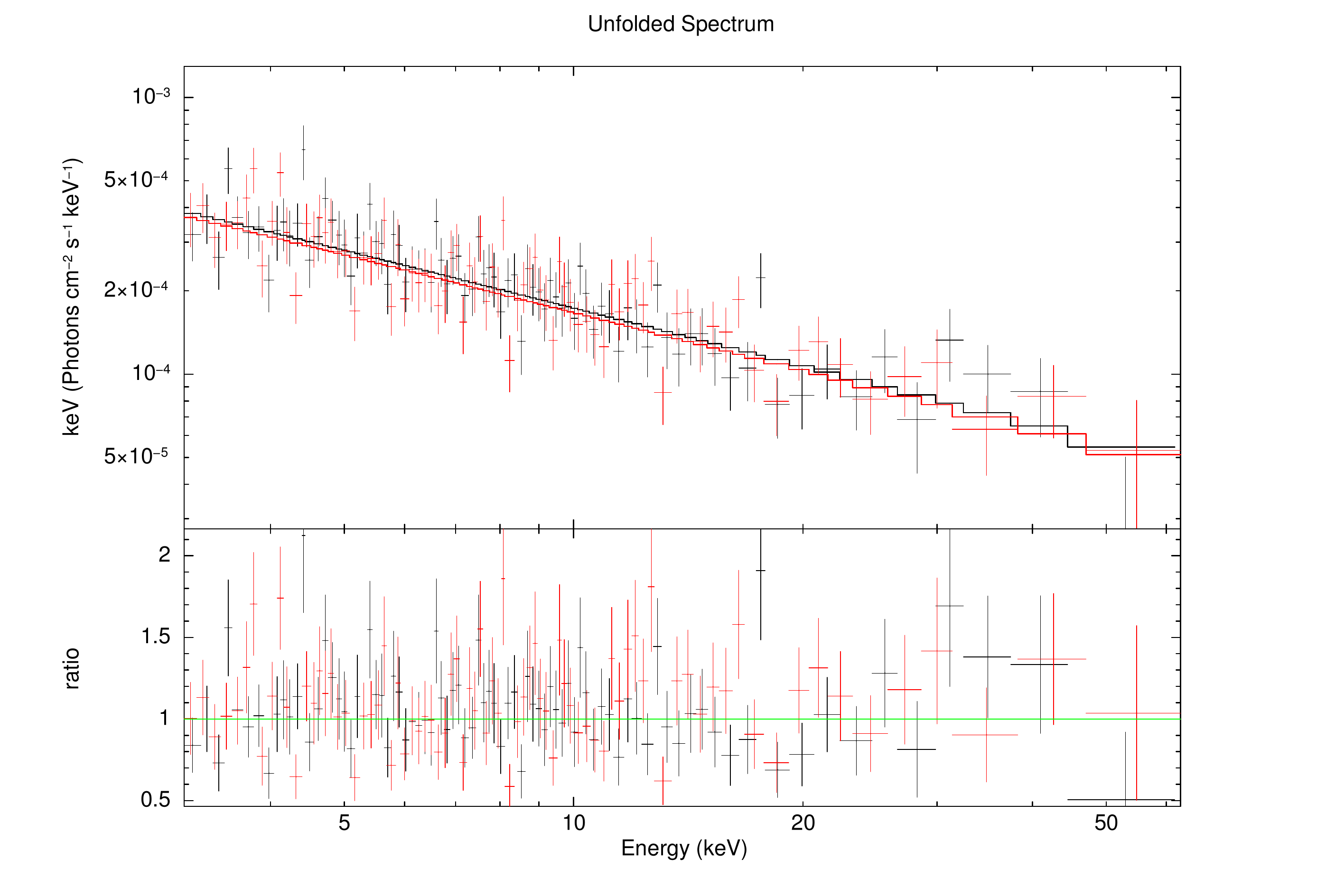}
    \end{multicols}
\center{S5 0014+81, 60001098004}

\begin{multicols}{4}
    \includegraphics[width=1.05\linewidth]{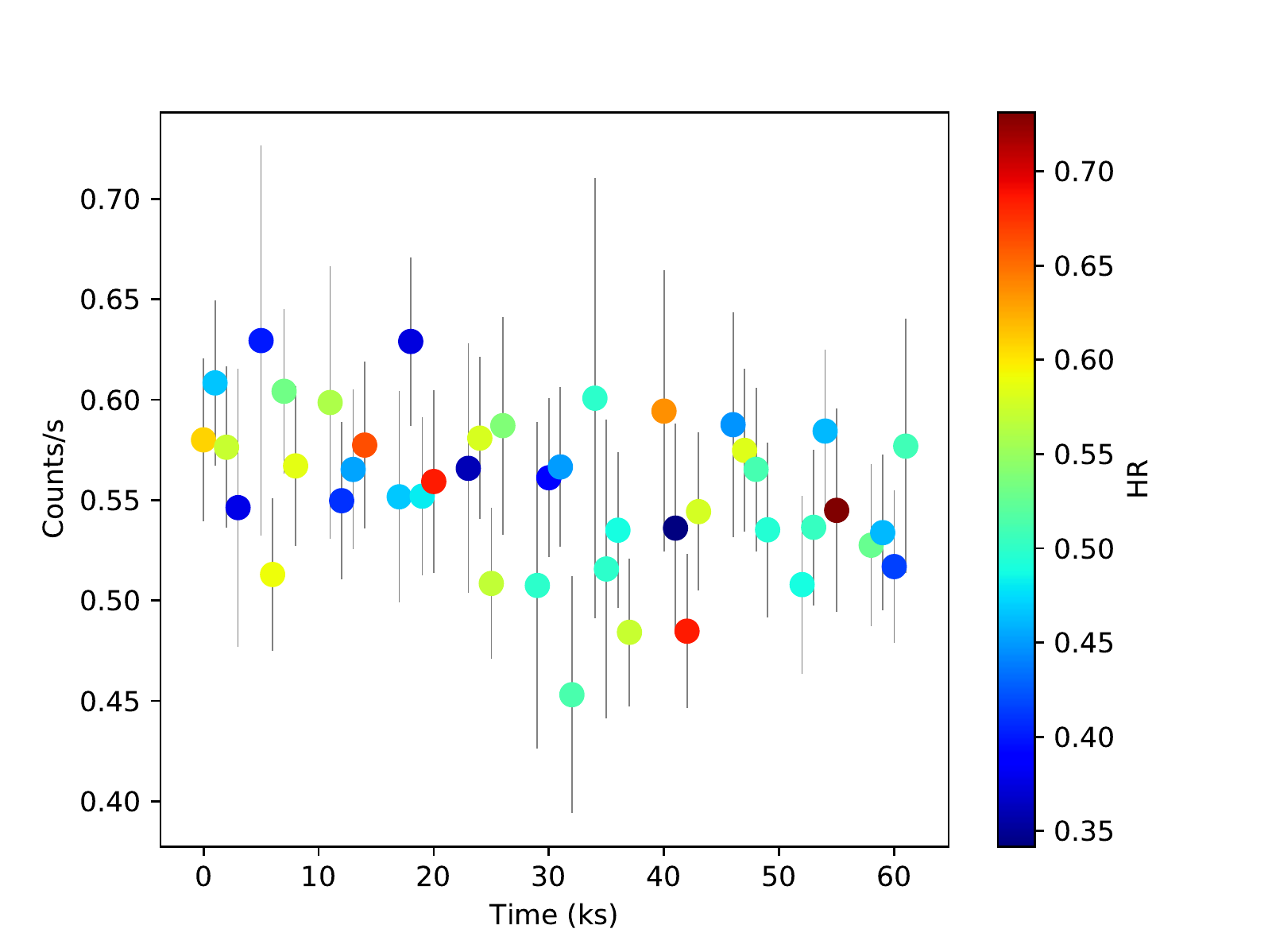}\par 
    \includegraphics[width=1.05\linewidth,angle=0]{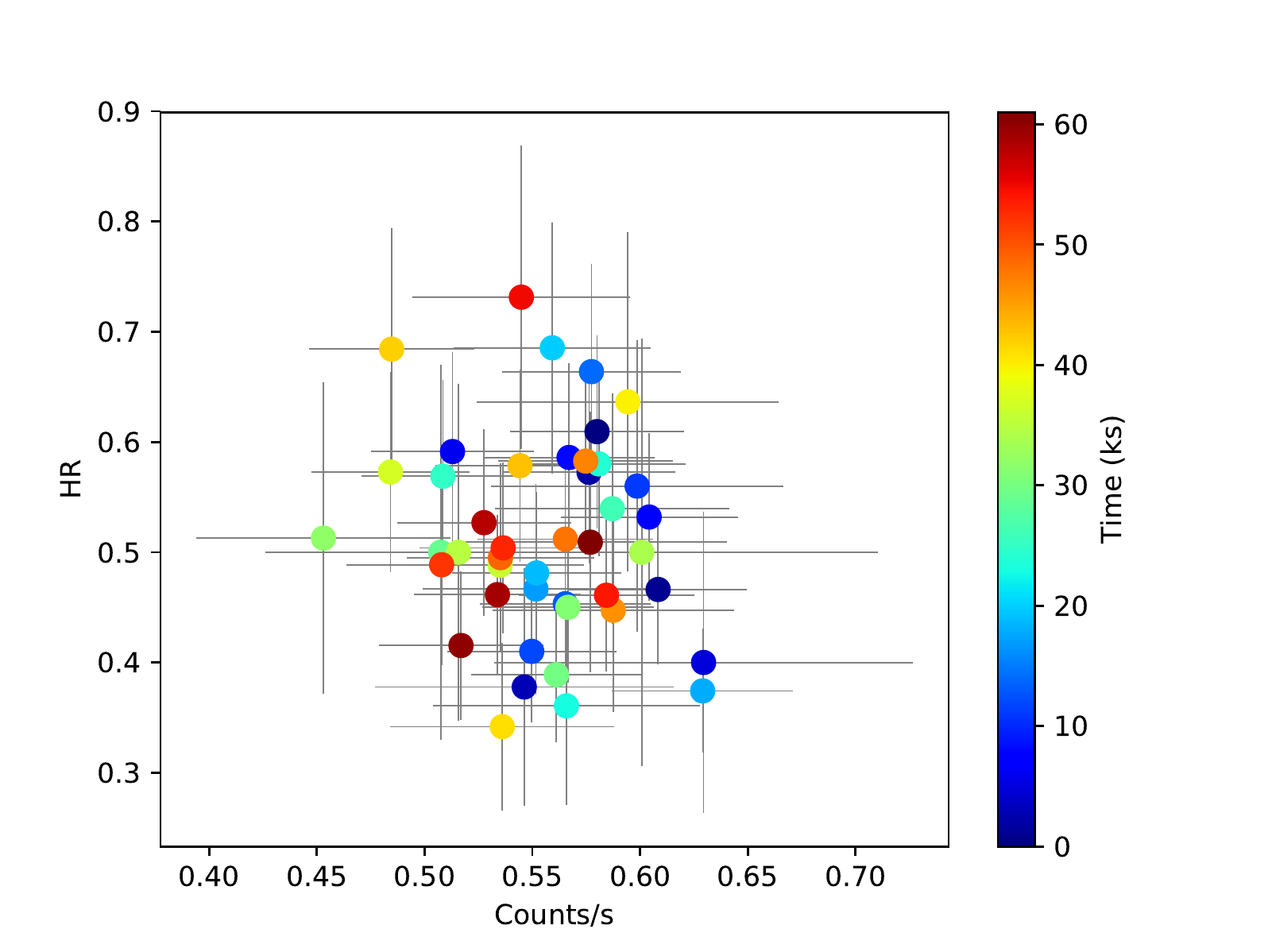}\par
    \includegraphics[width=0.72\linewidth,angle=-90]{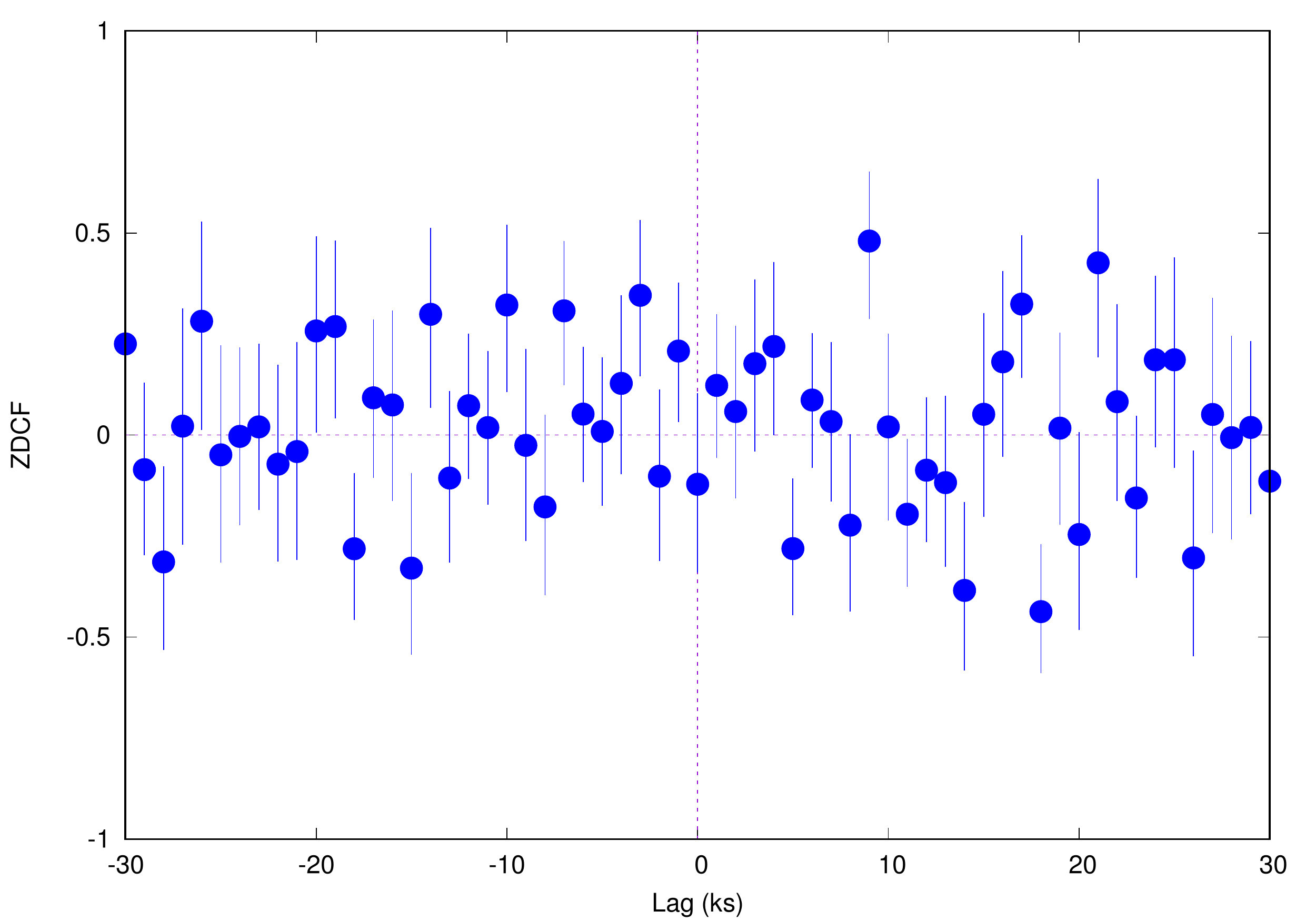}\par
    \includegraphics[width=0.73\linewidth, angle=-90]{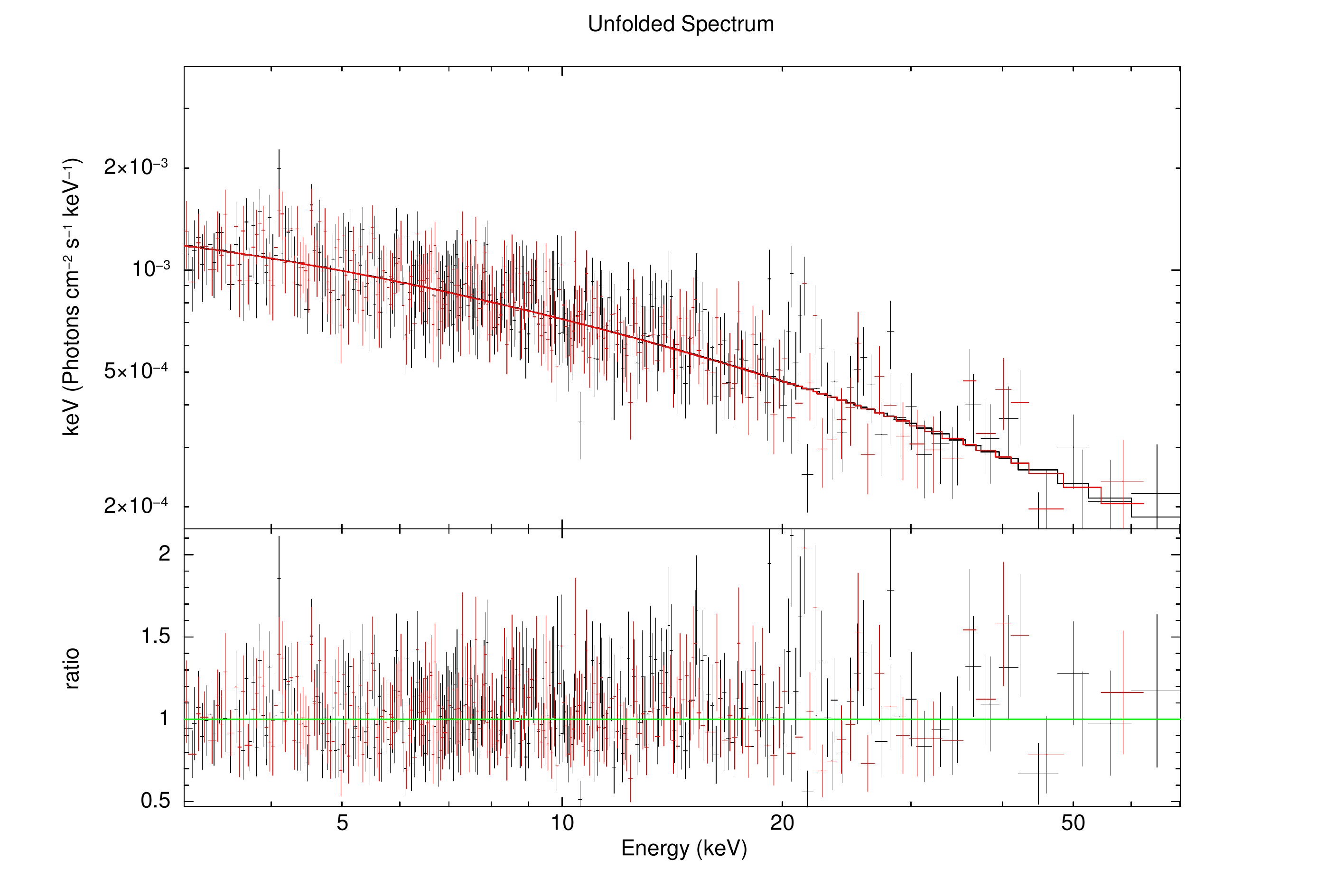}
    \end{multicols}
\center{B0222+185, 60001101002}

\begin{multicols}{4}
    \includegraphics[width=1.05\linewidth]{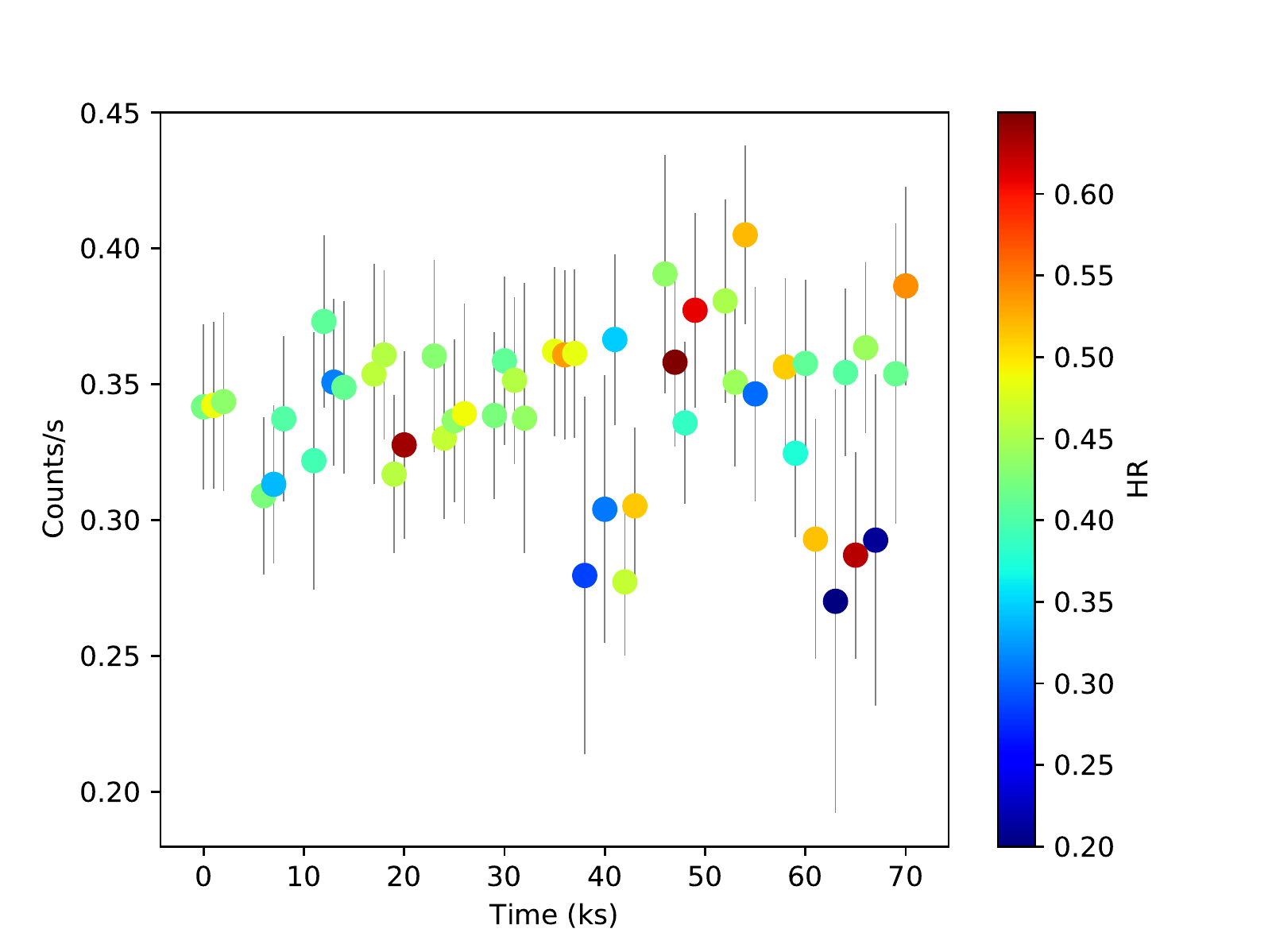}\par 
    \includegraphics[width=1.05\linewidth,angle=0]{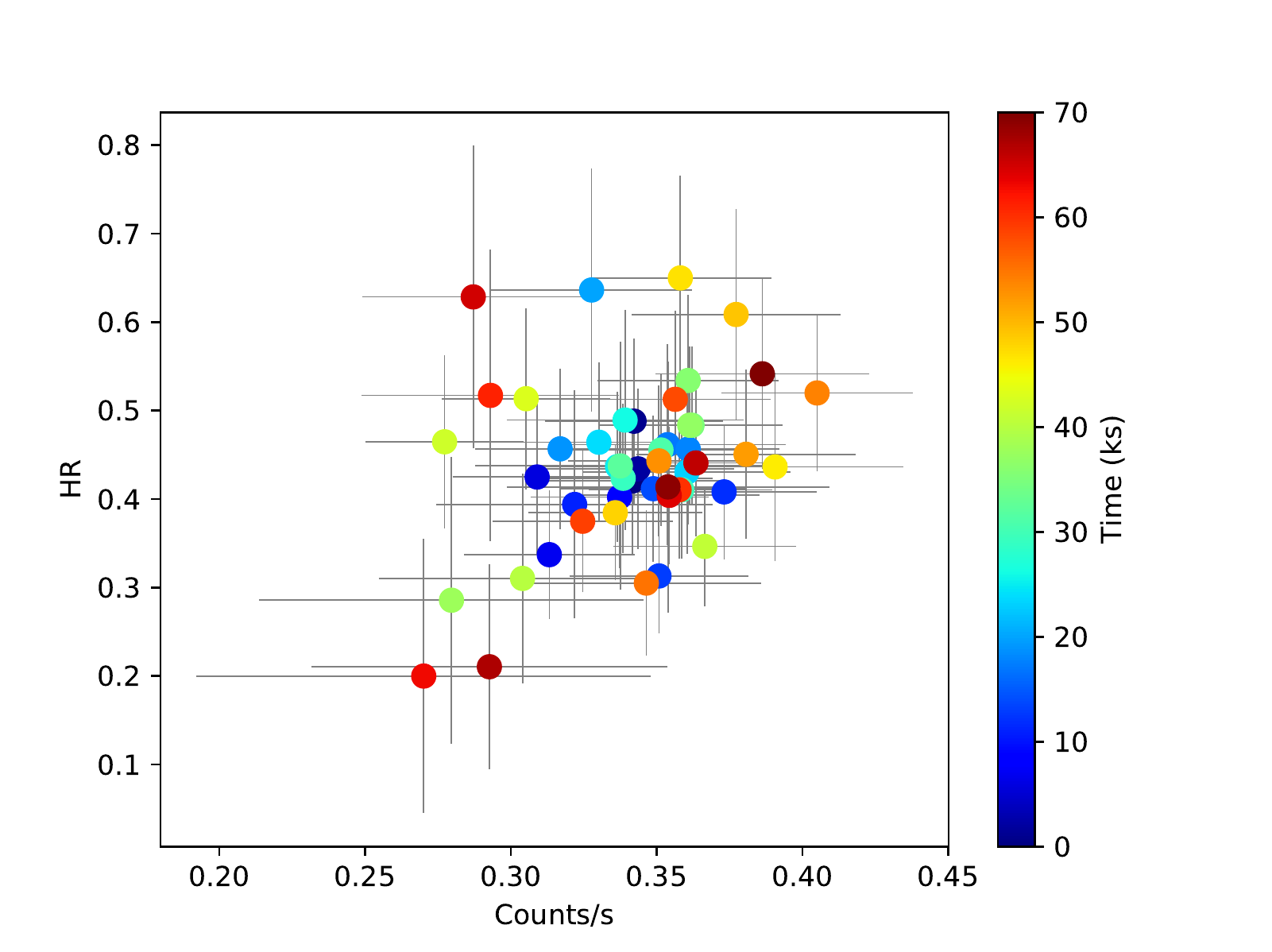}\par
    \includegraphics[width=0.72\linewidth,angle=-90]{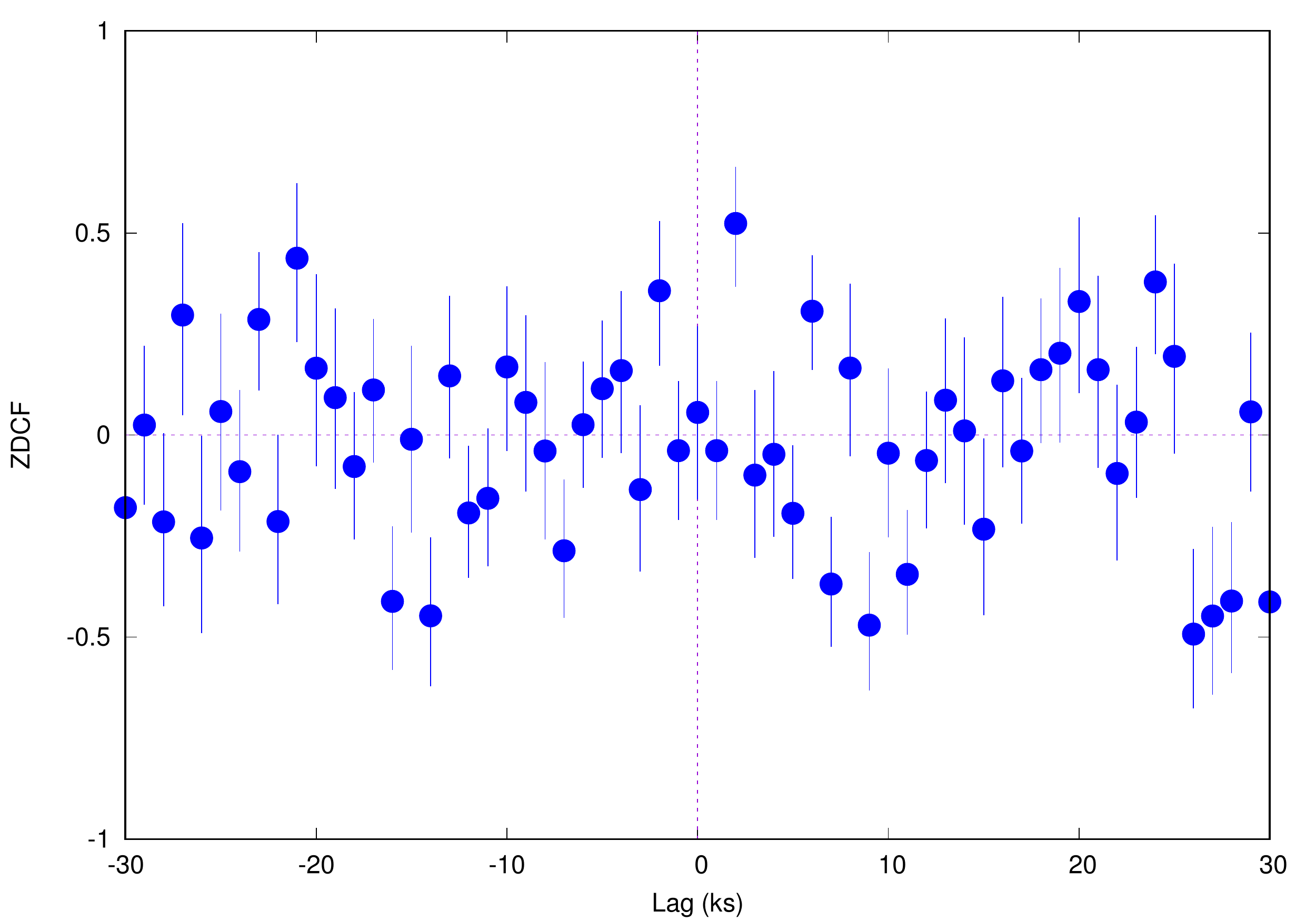}\par
    \includegraphics[width=0.73\linewidth, angle=-90]{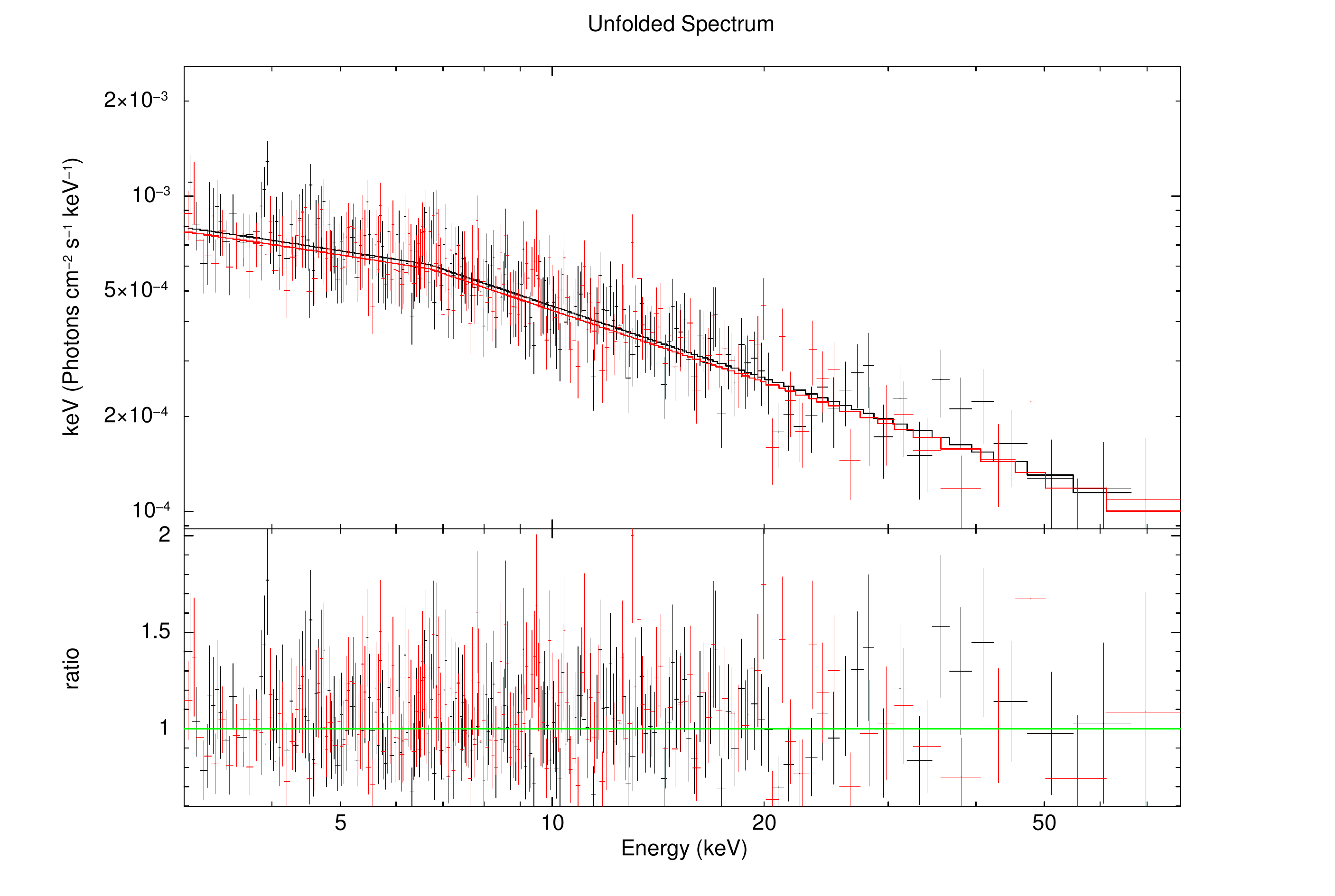}
    \end{multicols}
\center{B0222+185, 60001101004}

\begin{multicols}{4}
    \includegraphics[width=1.05\linewidth]{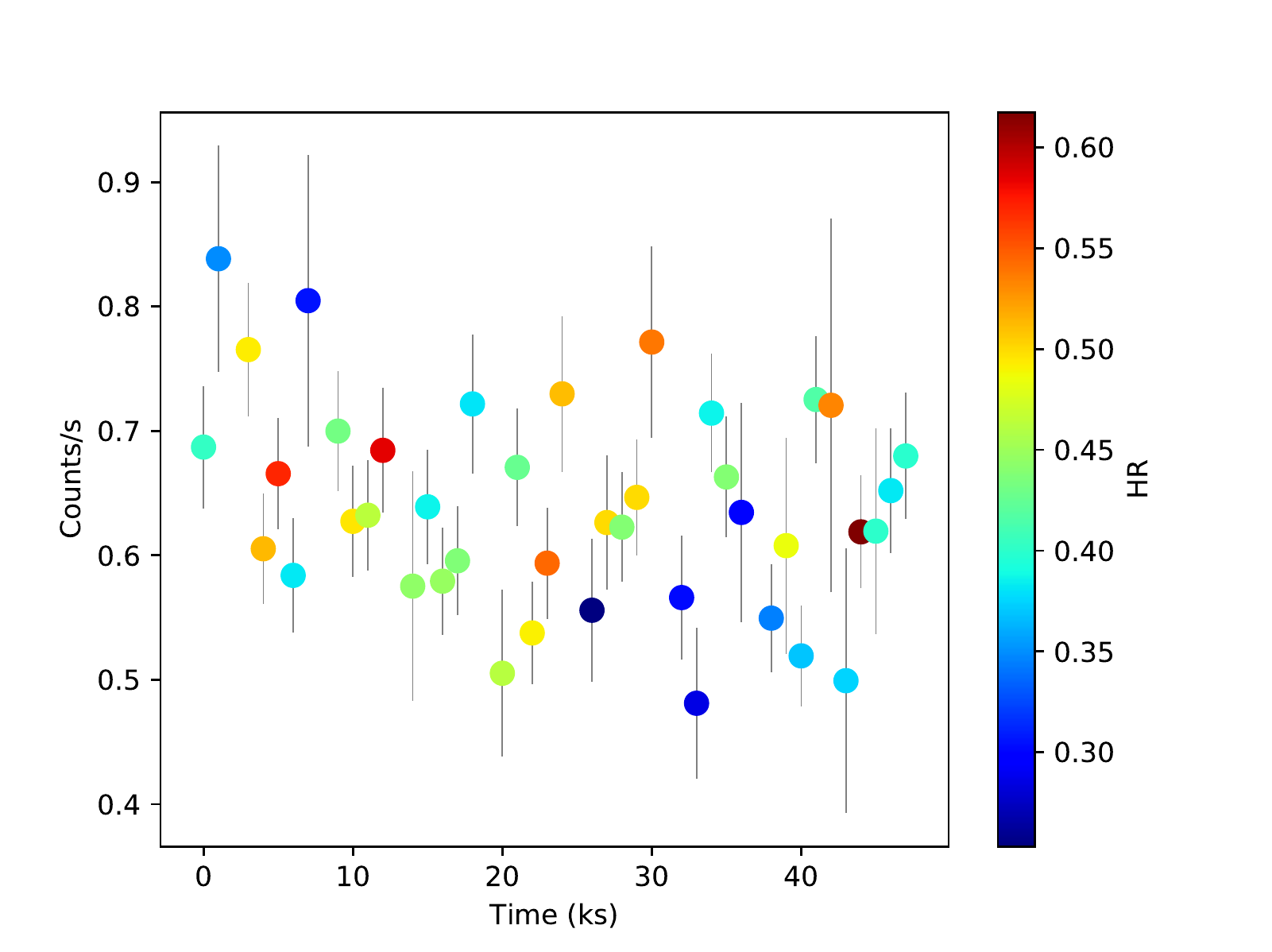}\par 
    \includegraphics[width=1.05\linewidth,angle=0]{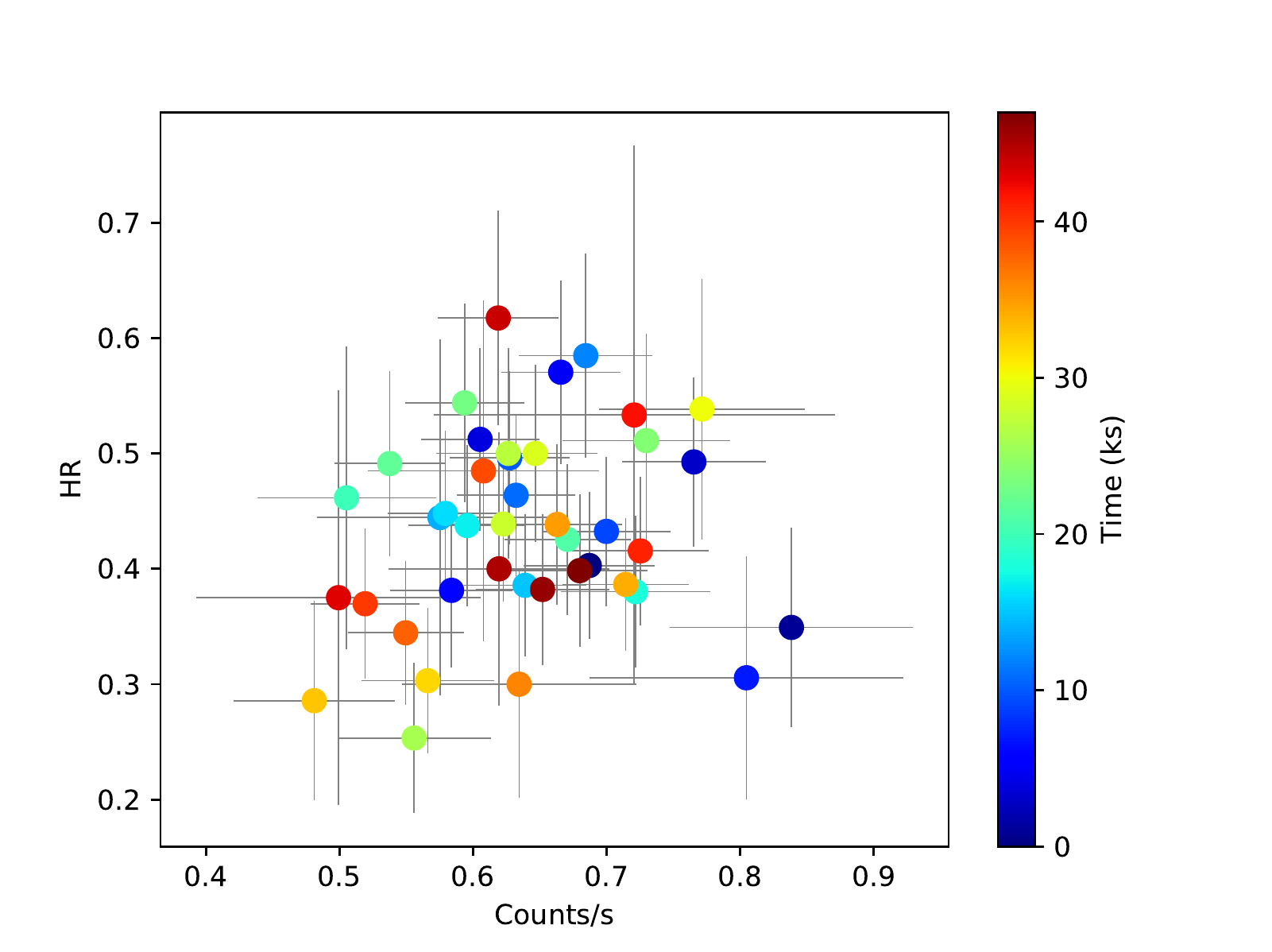}\par
    \includegraphics[width=0.72\linewidth,angle=-90]{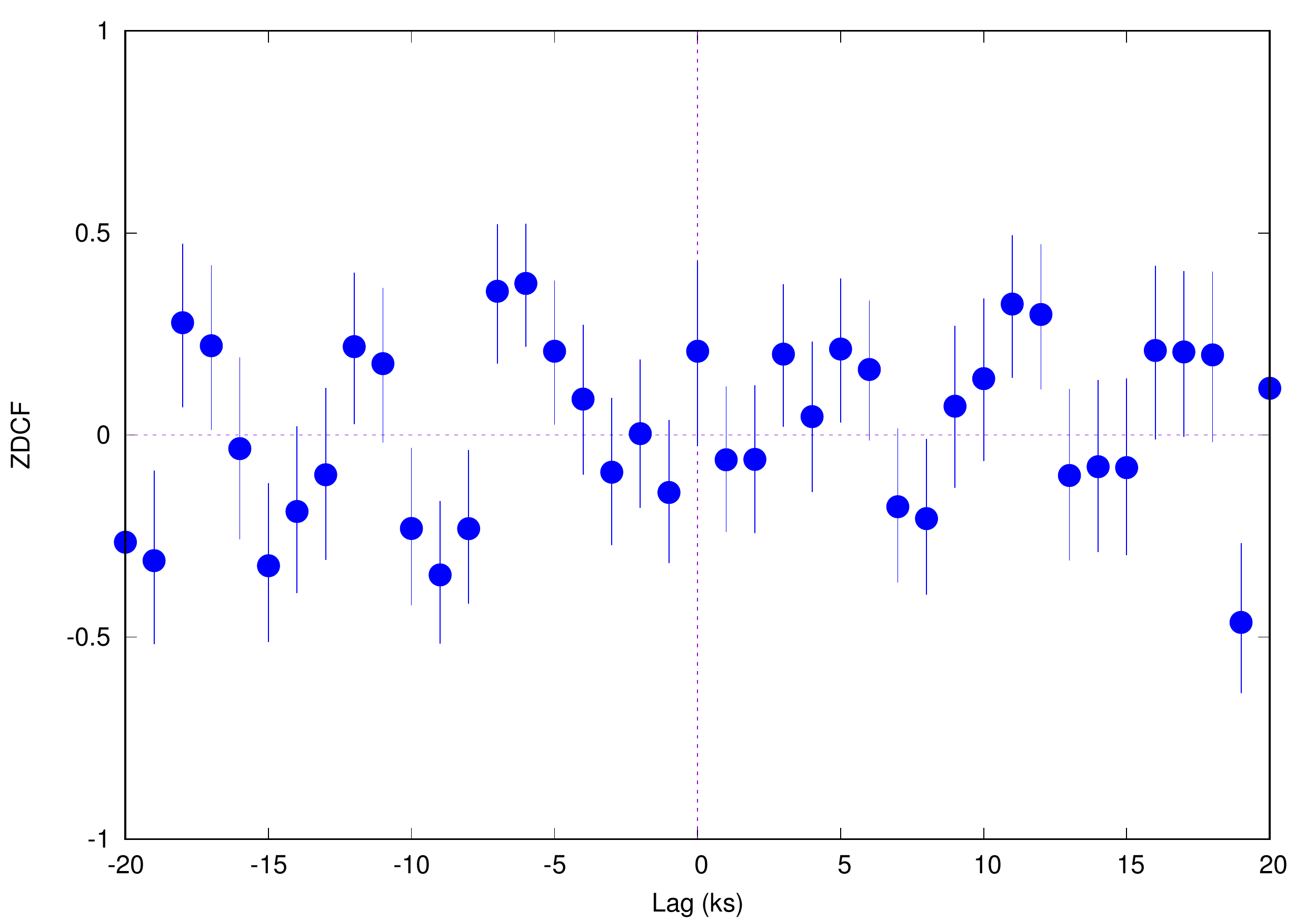}\par
    \includegraphics[width=0.73\linewidth, angle=-90]{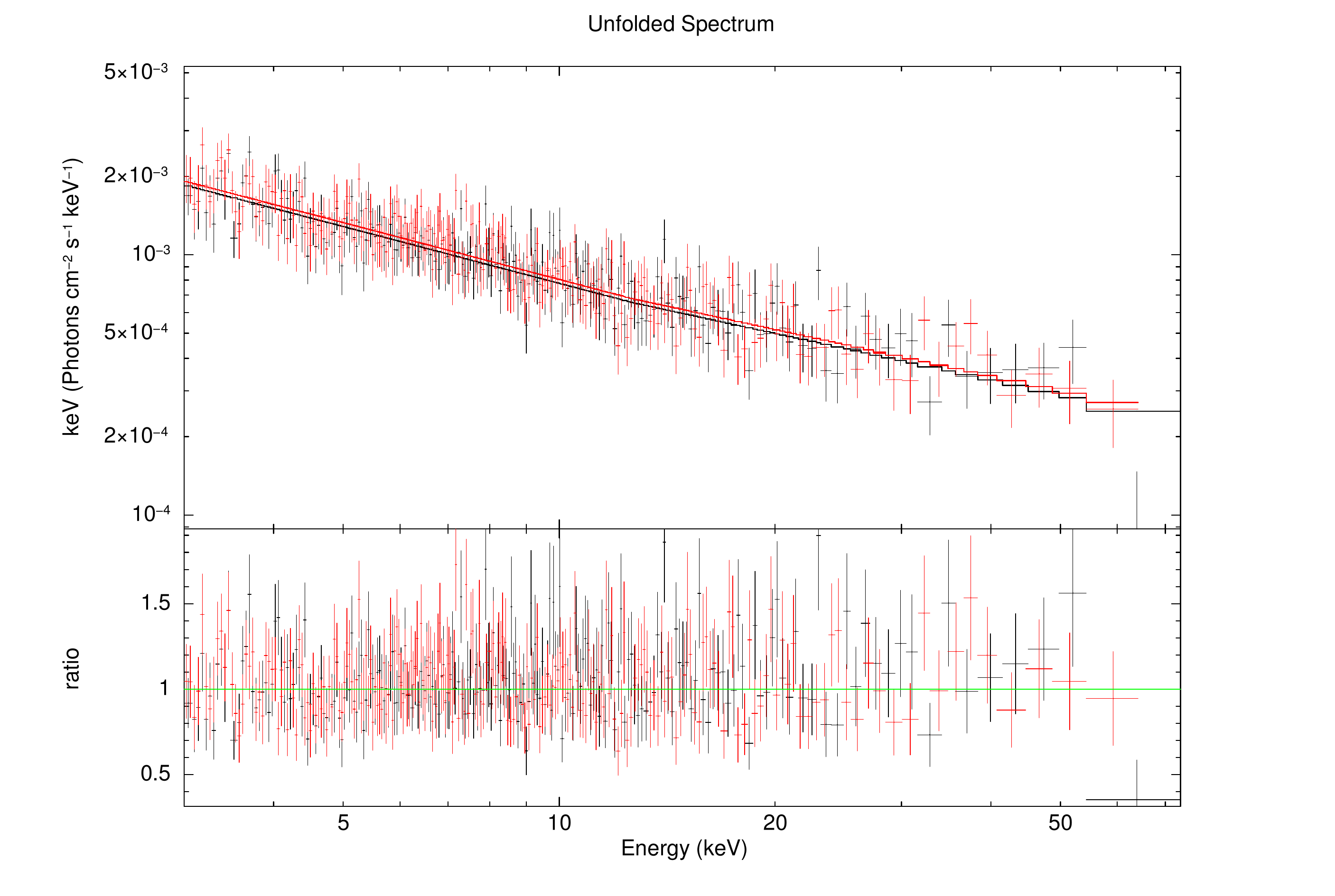}
    \end{multicols}
\center{HB 0836+710, 60002045002} \\
\caption{The plots in each row show the light curve, flux-HR relation, z-transformed discrete correlation function and the spectral fit, from the left to the right respectively, for the \textit{NuSTAR} blazar observations listed in the table 2.  The color bars in light curve and the flux-HR plots represent the HR and time, respectively.}
\label{fig:LC1}
\end{figure*}


\begin{figure*}
\begin{multicols}{4}
    \includegraphics[width=1.05\linewidth]{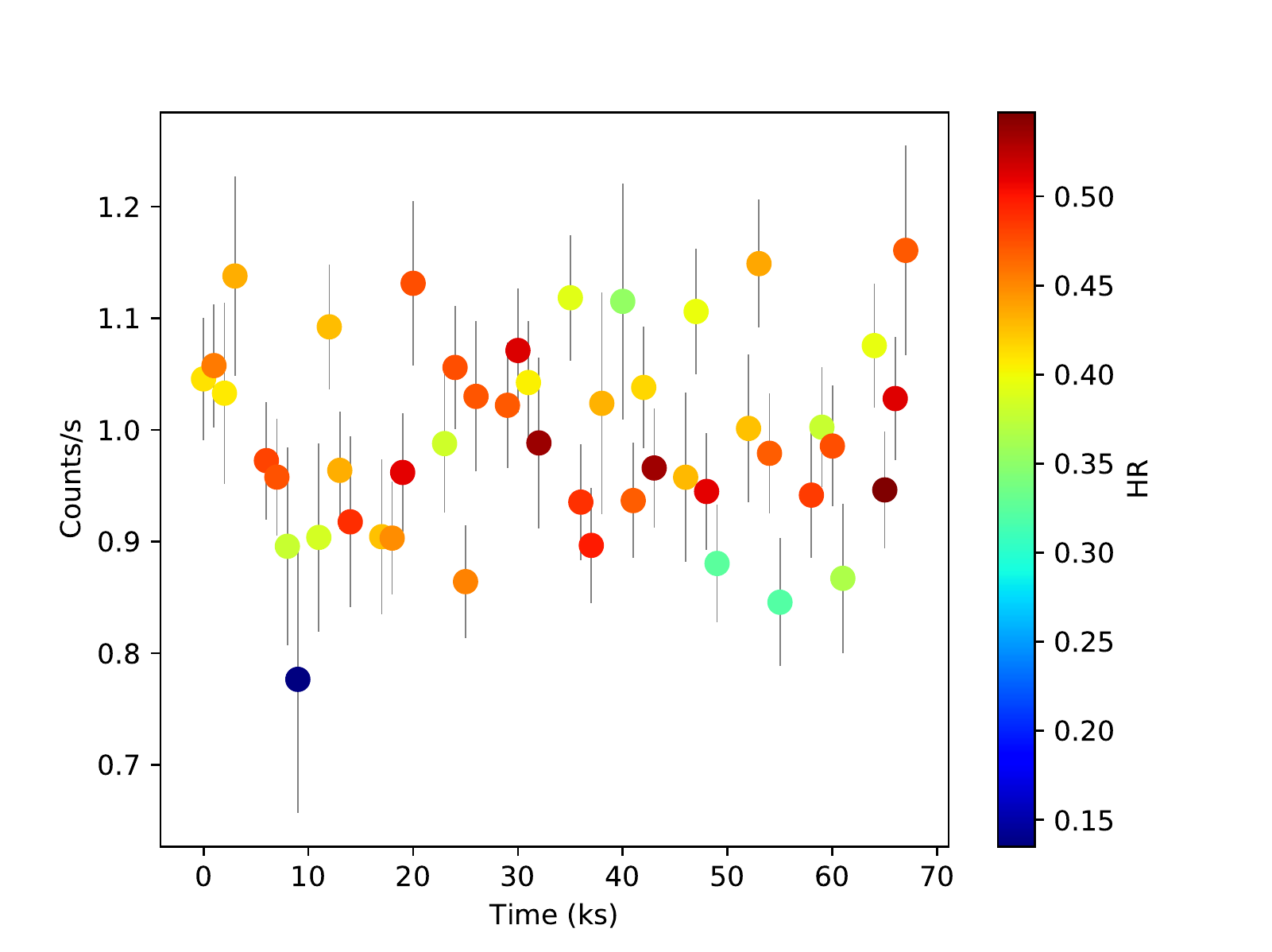}\par 
    \includegraphics[width=1.05\linewidth,angle=0]{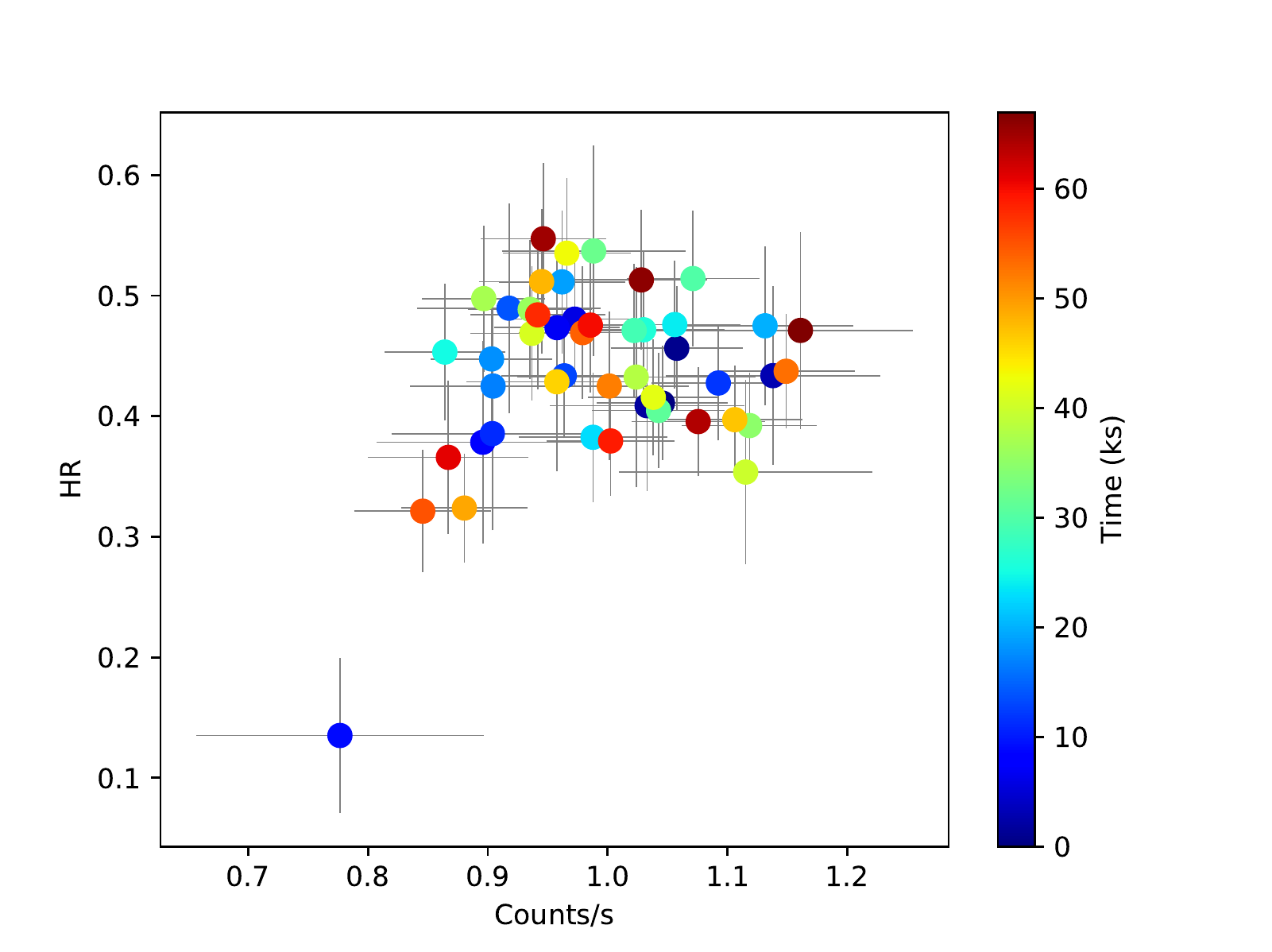}\par
    \includegraphics[width=0.72\linewidth,angle=-90]{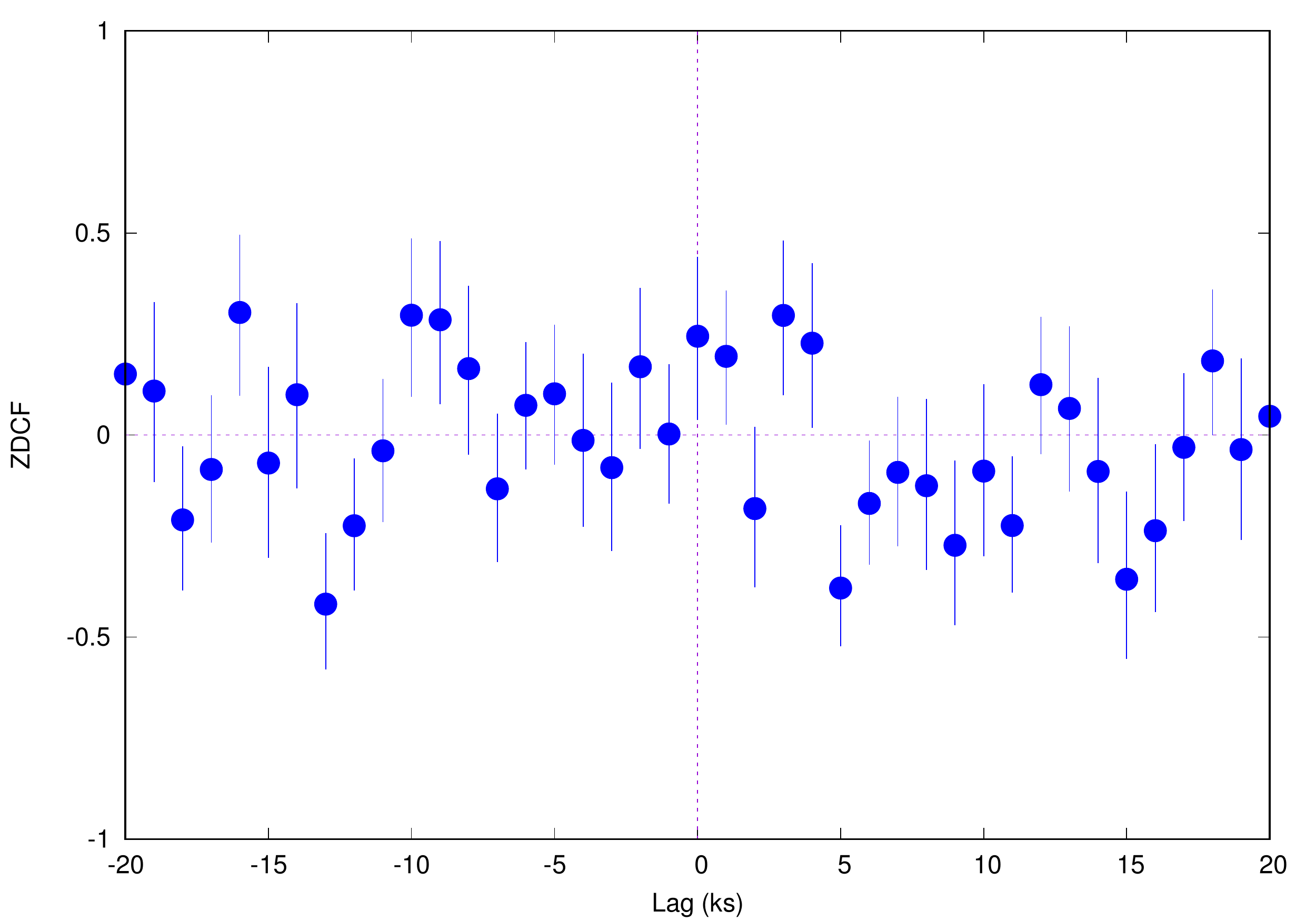}\par
    \includegraphics[width=0.73\linewidth, angle=-90]{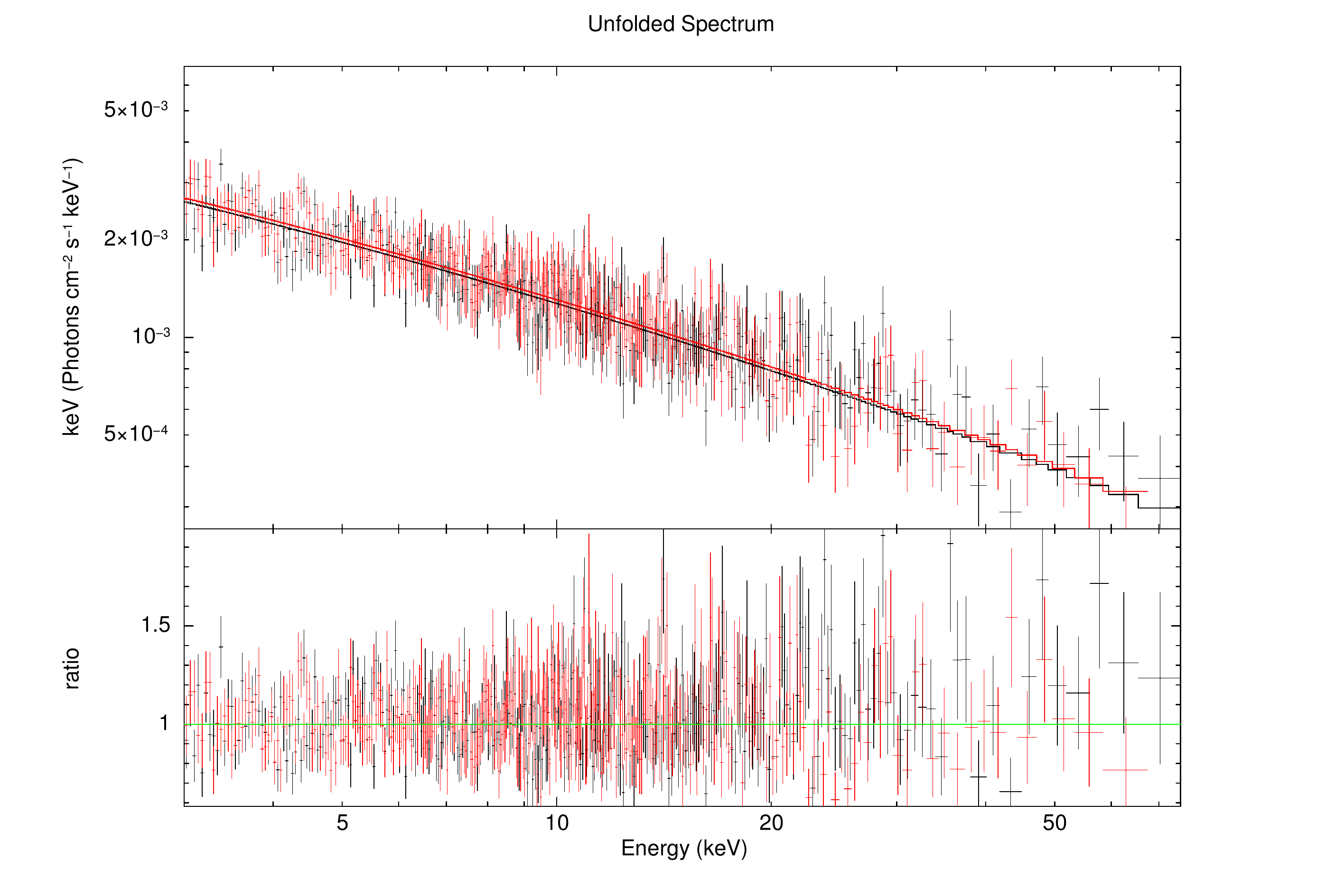}
    \end{multicols}
\center{HB 0836+710, 60002045004}

\begin{multicols}{4}
    \includegraphics[width=1.05\linewidth]{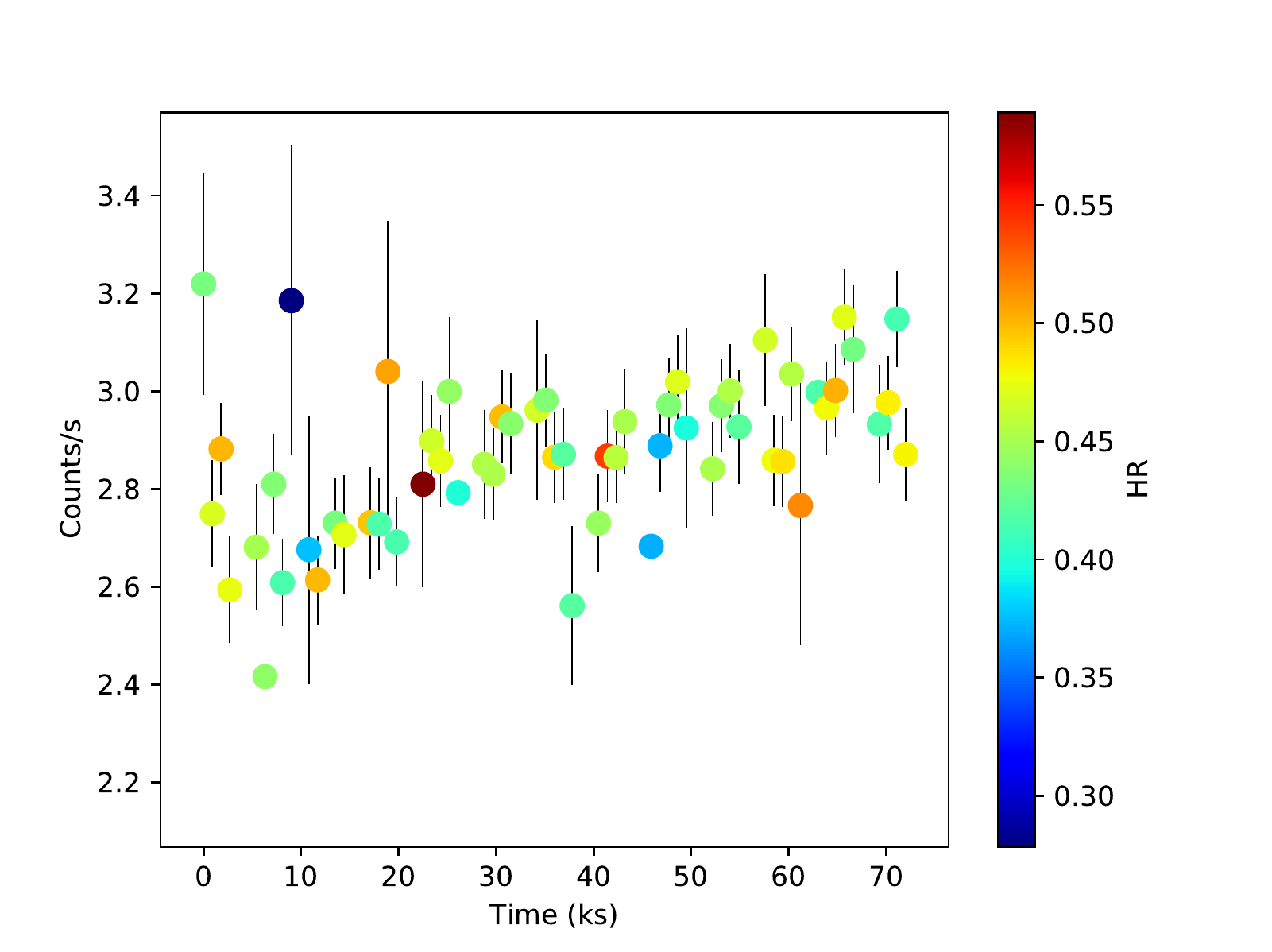}\par 
    \includegraphics[width=1.05\linewidth,angle=0]{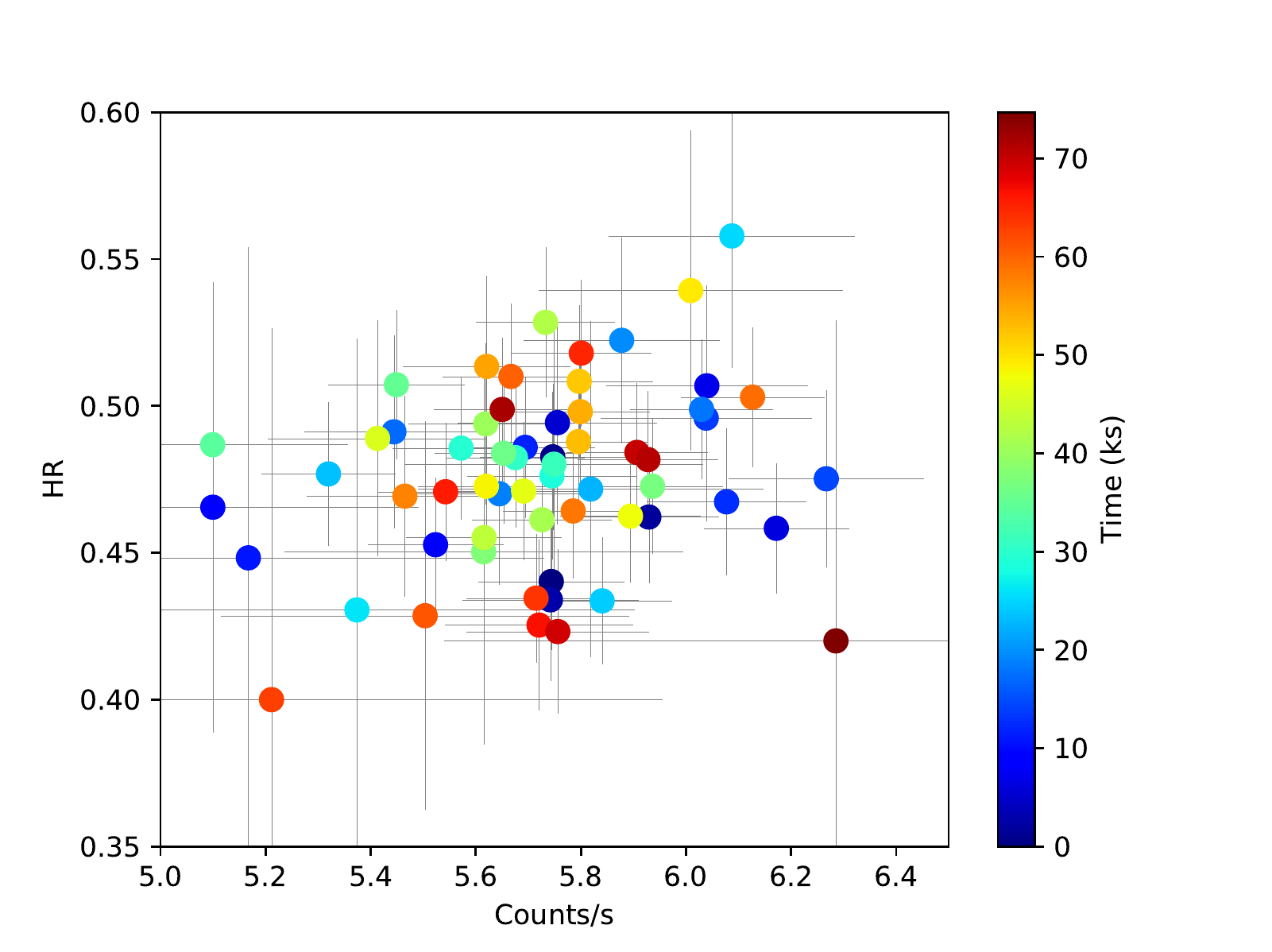}\par
    \includegraphics[width=0.72\linewidth,angle=-90]{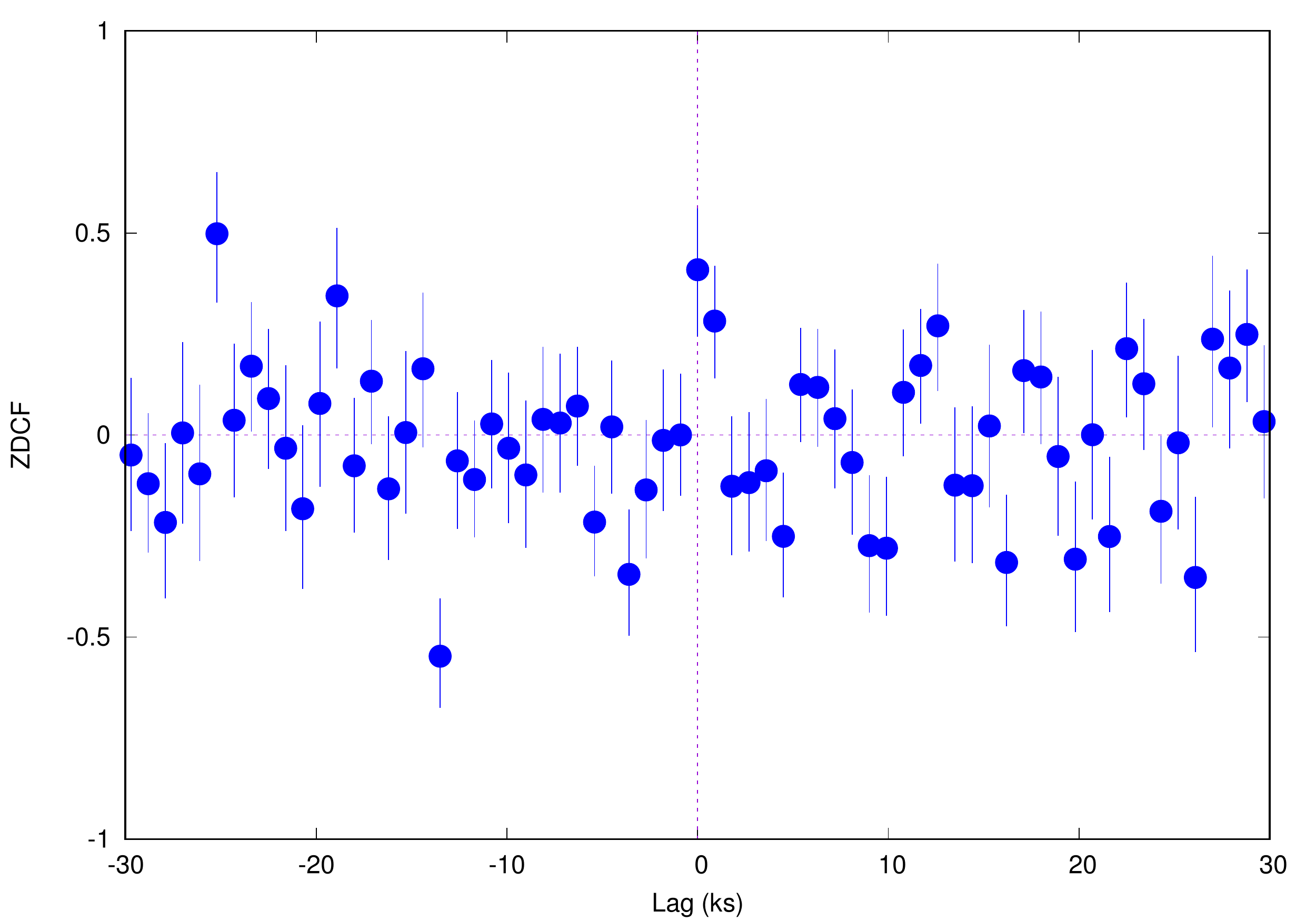}\par
    \includegraphics[width=0.73\linewidth, angle=-90]{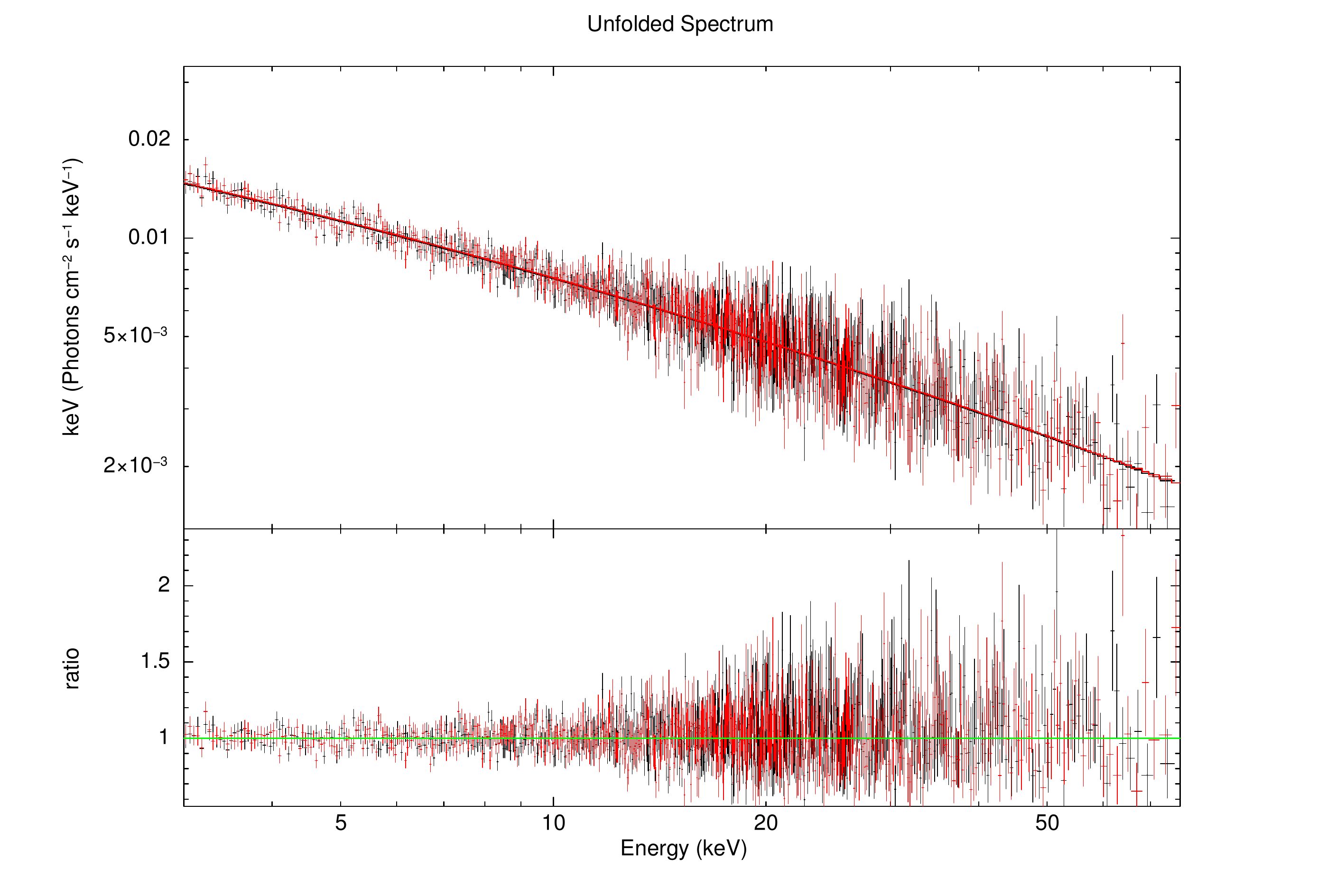}
    \end{multicols}
\center{3C 273, 10202020002}

\begin{multicols}{4}
    \includegraphics[width=1.05\linewidth]{3C3_1030_HR.pdf}\par 
    \includegraphics[width=1.05\linewidth,angle=0]{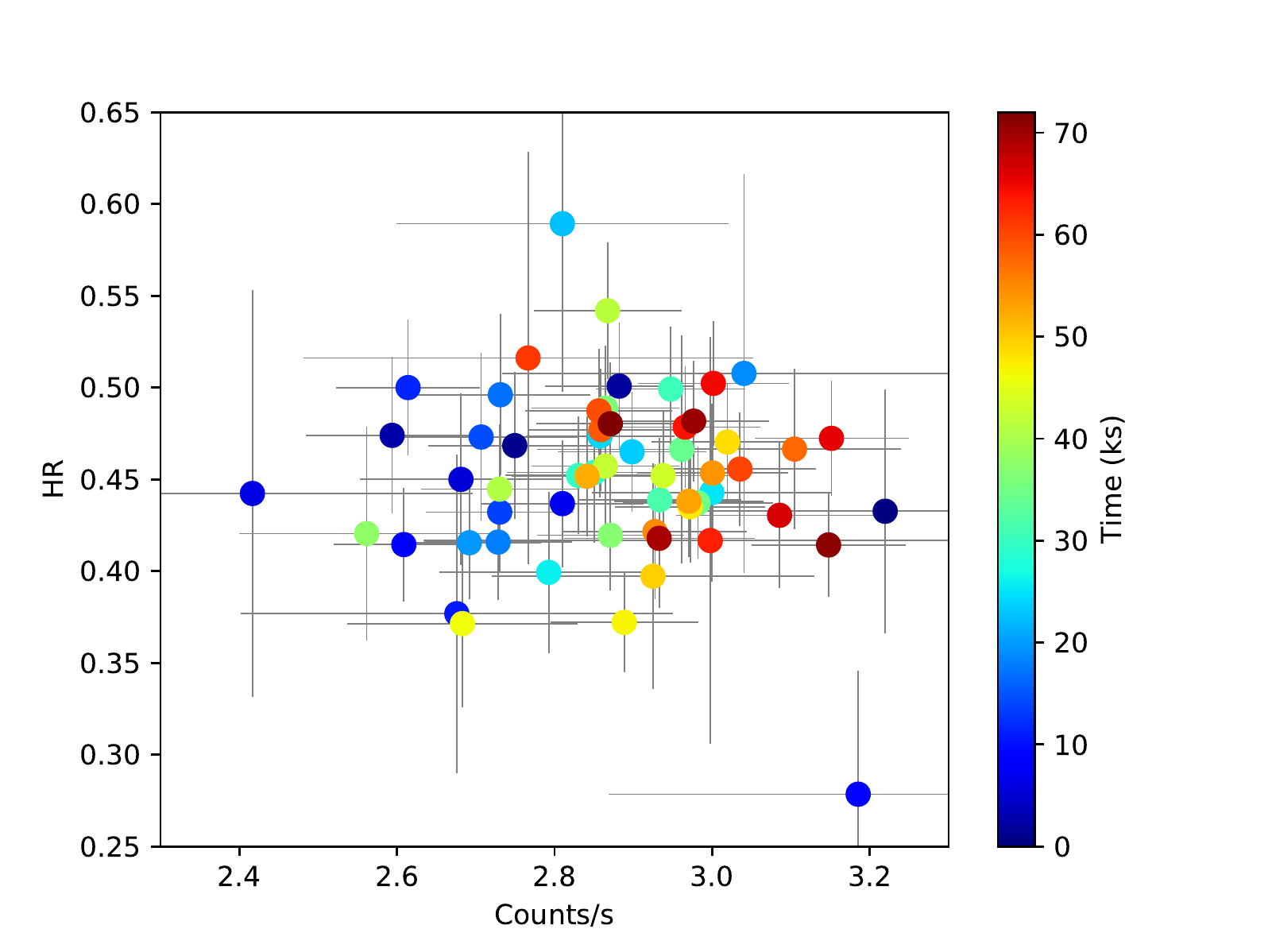}\par
    \includegraphics[width=0.72\linewidth,angle=-90]{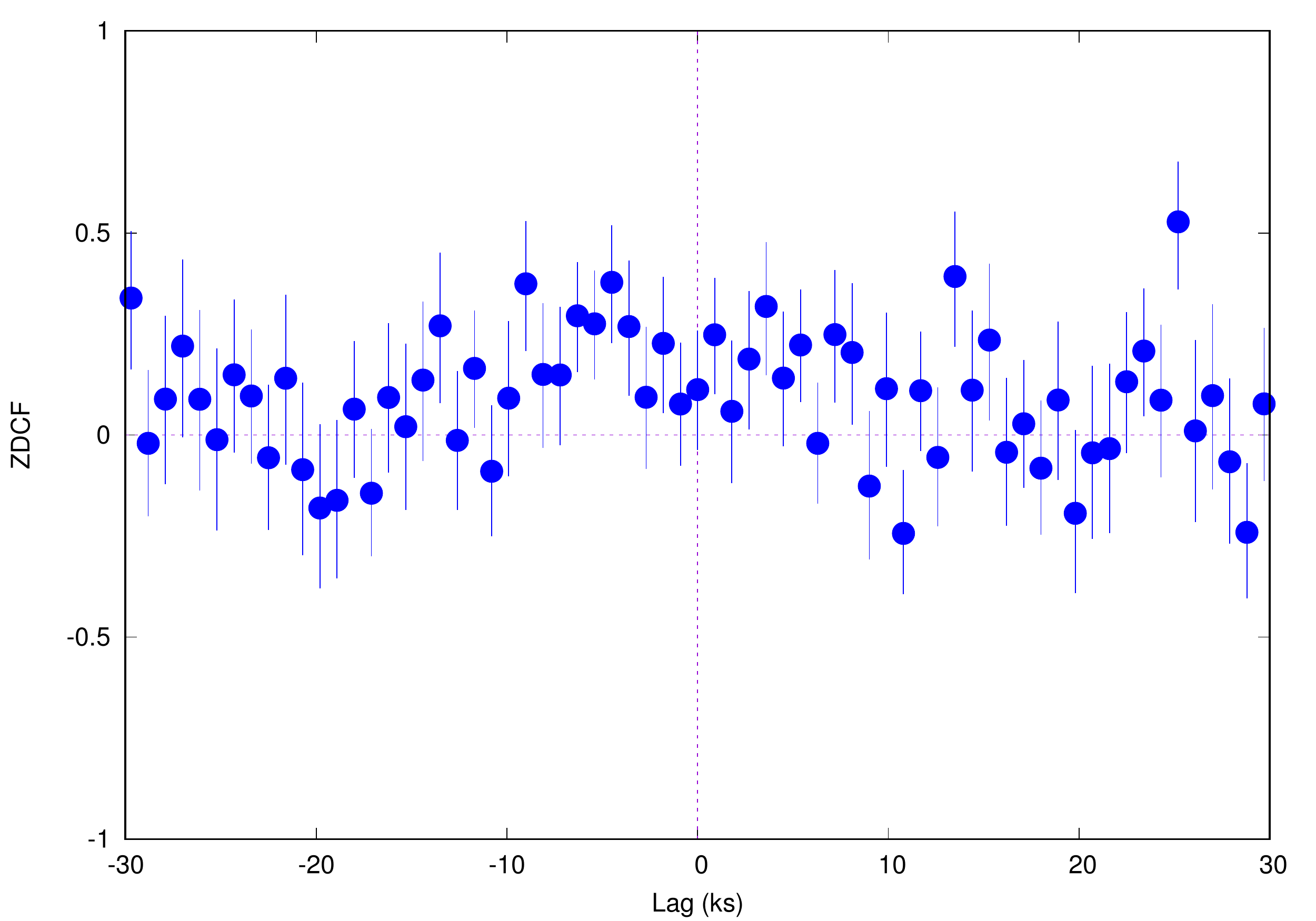}\par
    \includegraphics[width=0.73\linewidth, angle=-90]{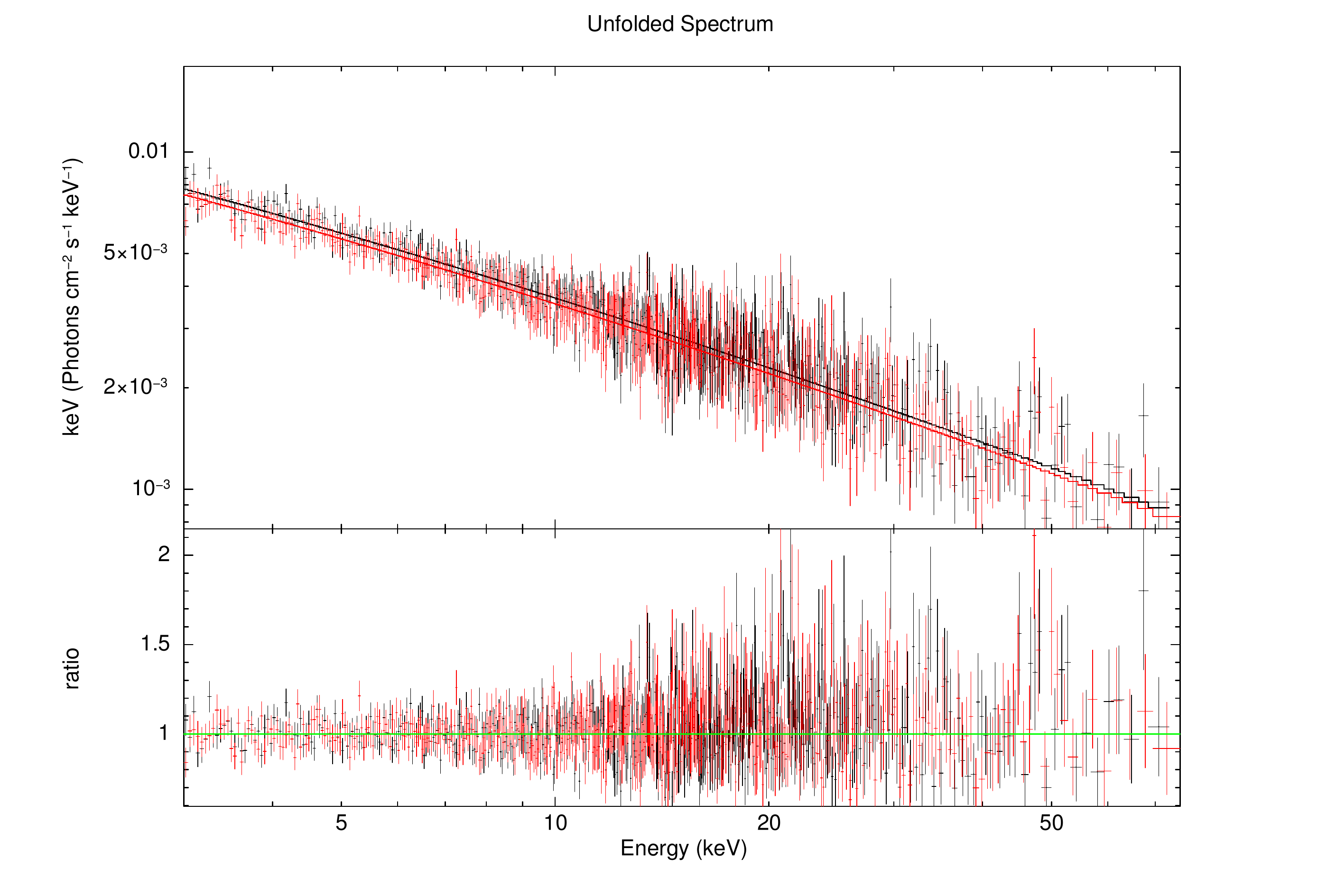}
    \end{multicols}
\center{3C 273, 10302020002}

\begin{multicols}{4}
    \includegraphics[width=1.05\linewidth]{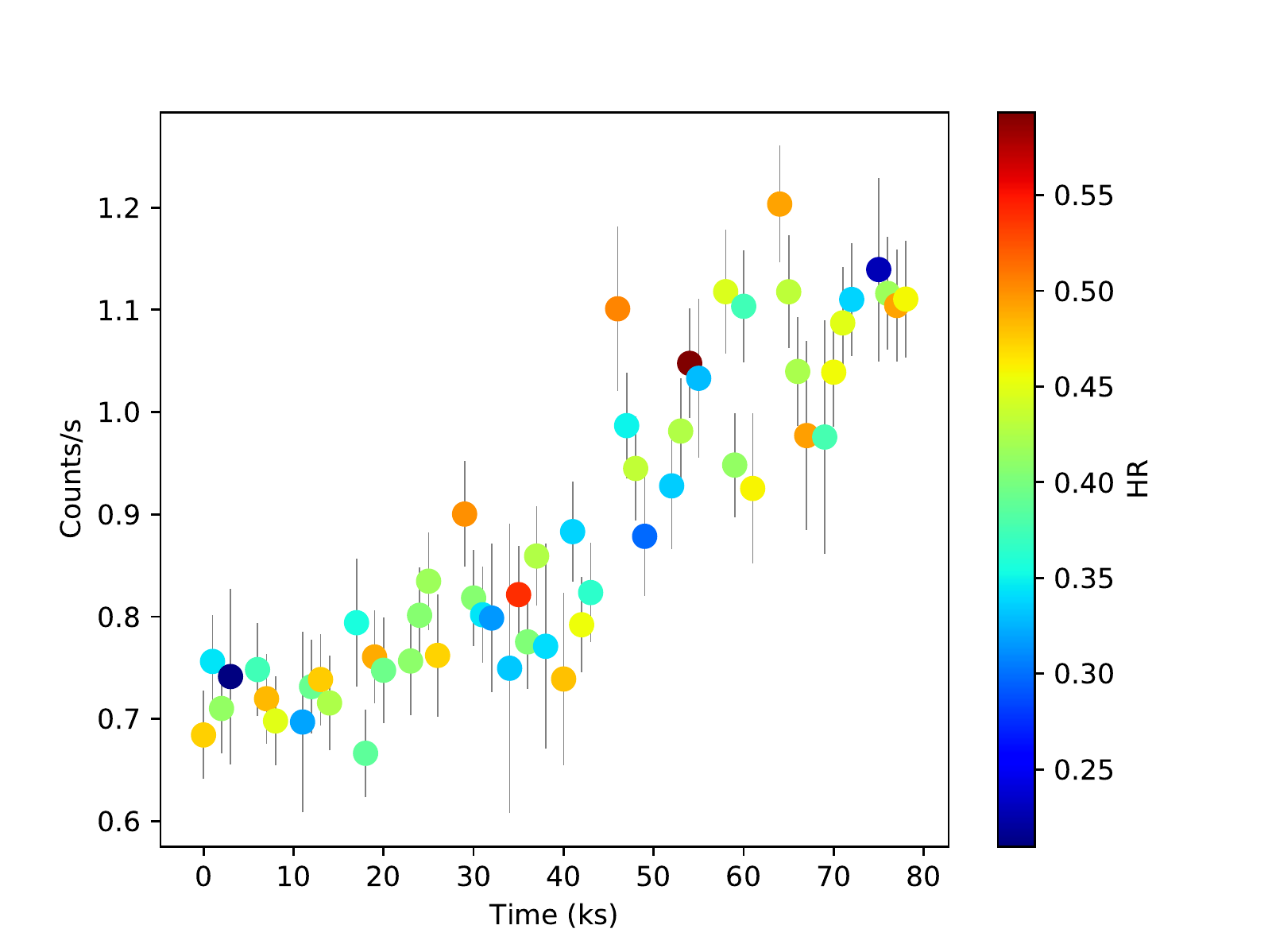}\par 
    \includegraphics[width=1.05\linewidth,angle=0]{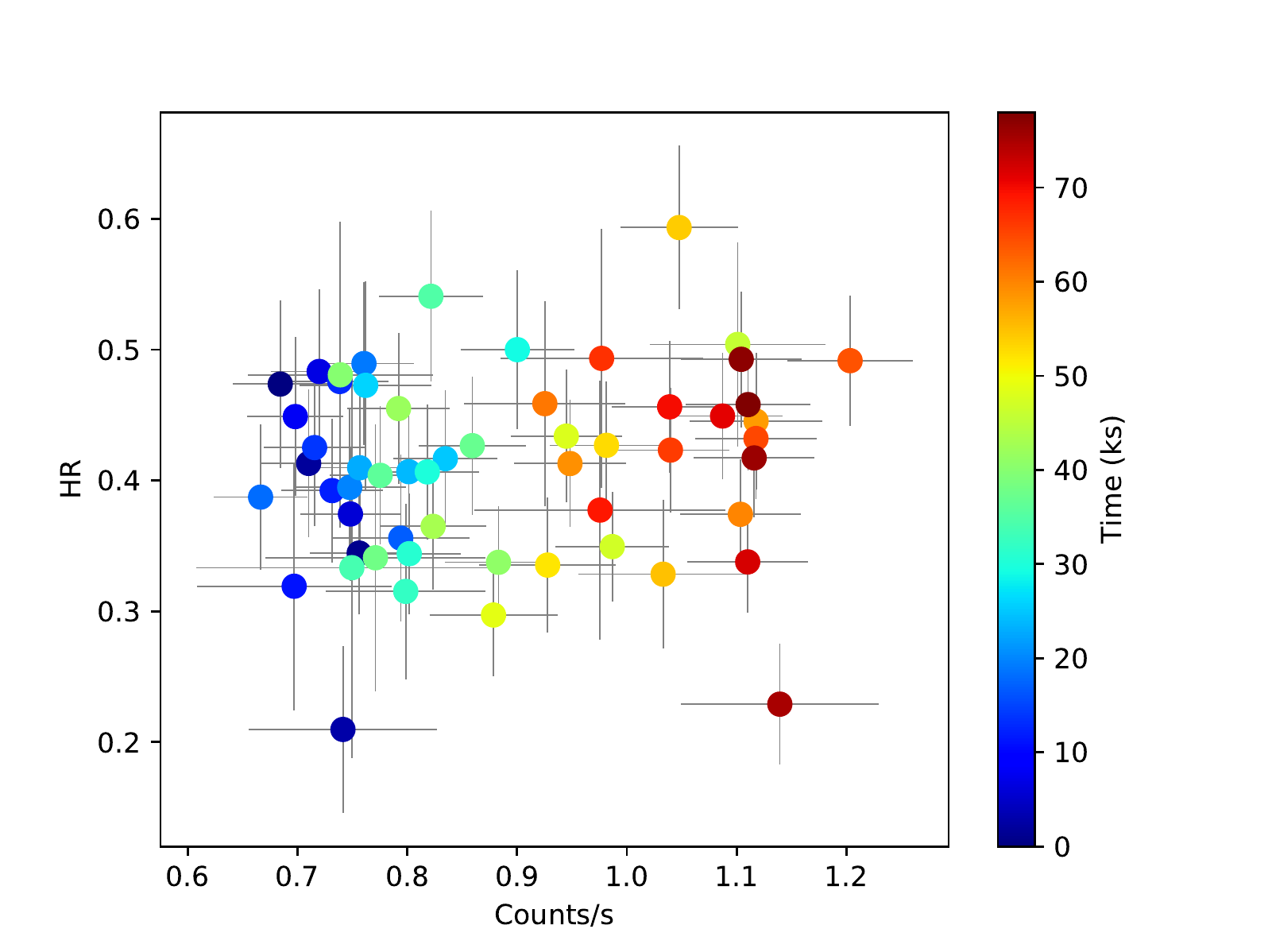}\par
    \includegraphics[width=0.72\linewidth,angle=-90]{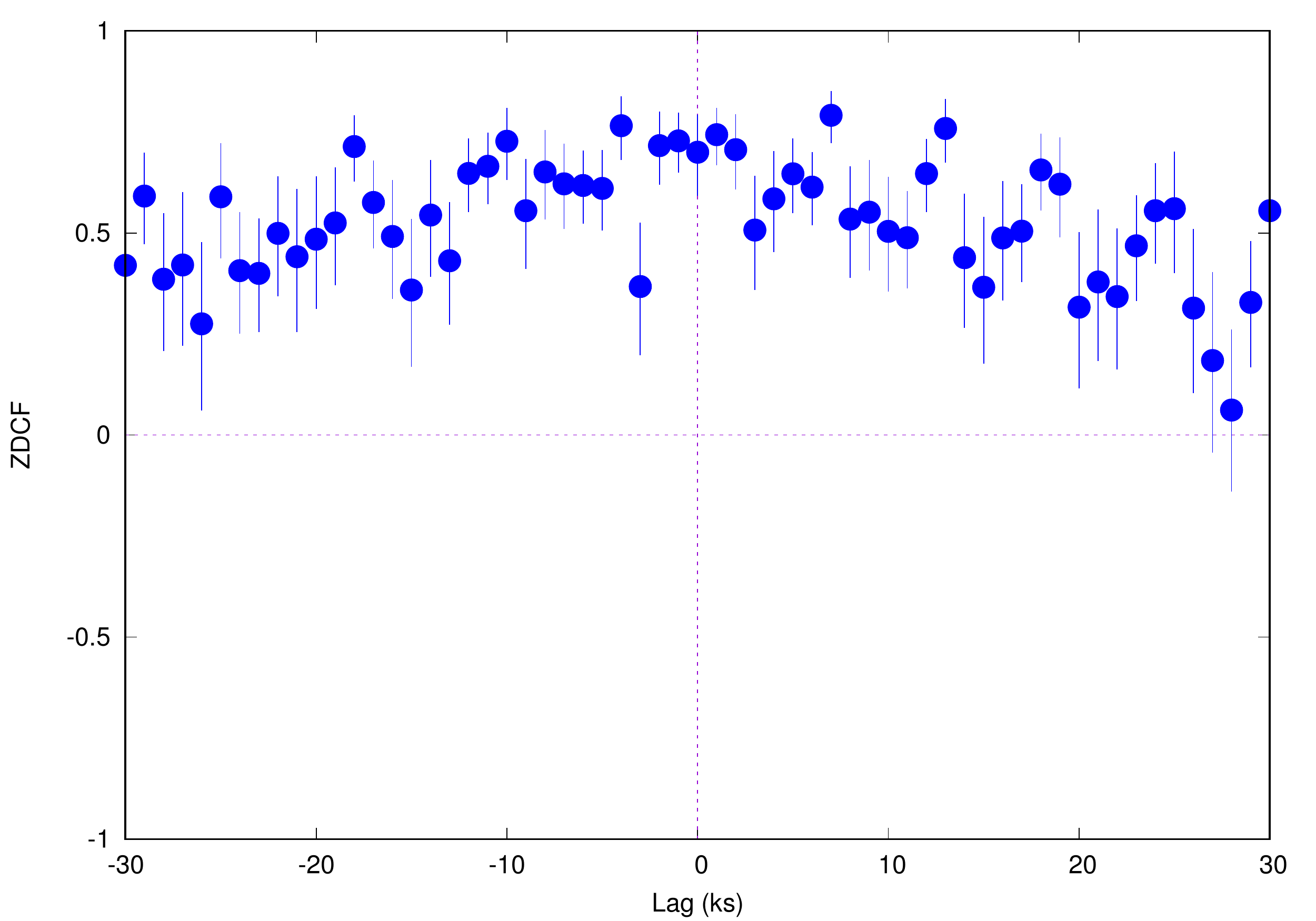}\par
    \includegraphics[width=0.73\linewidth, angle=-90]{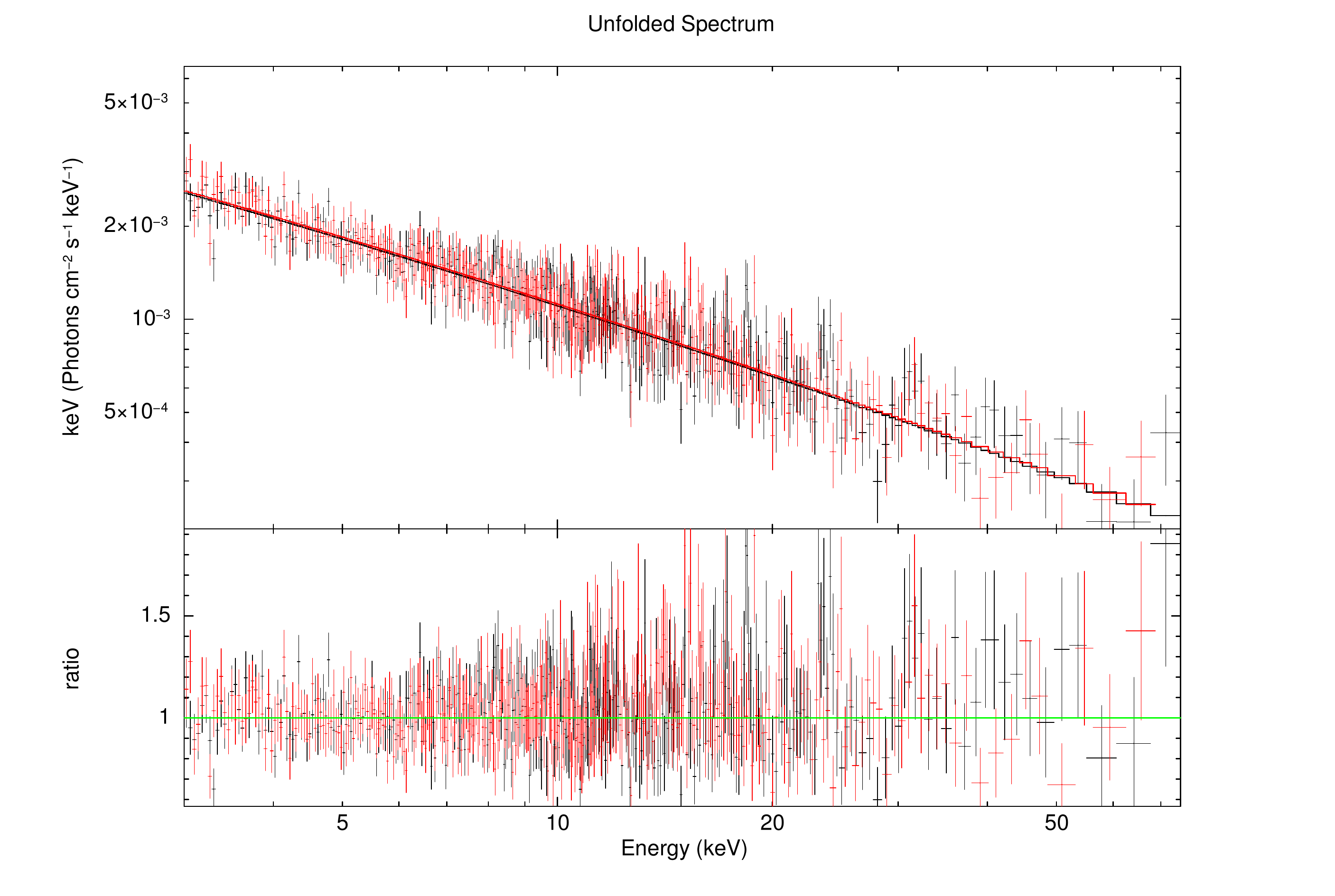}
    \end{multicols}
\center{3C 279, 60002020004}

\begin{multicols}{4}
    \includegraphics[width=1.05\linewidth]{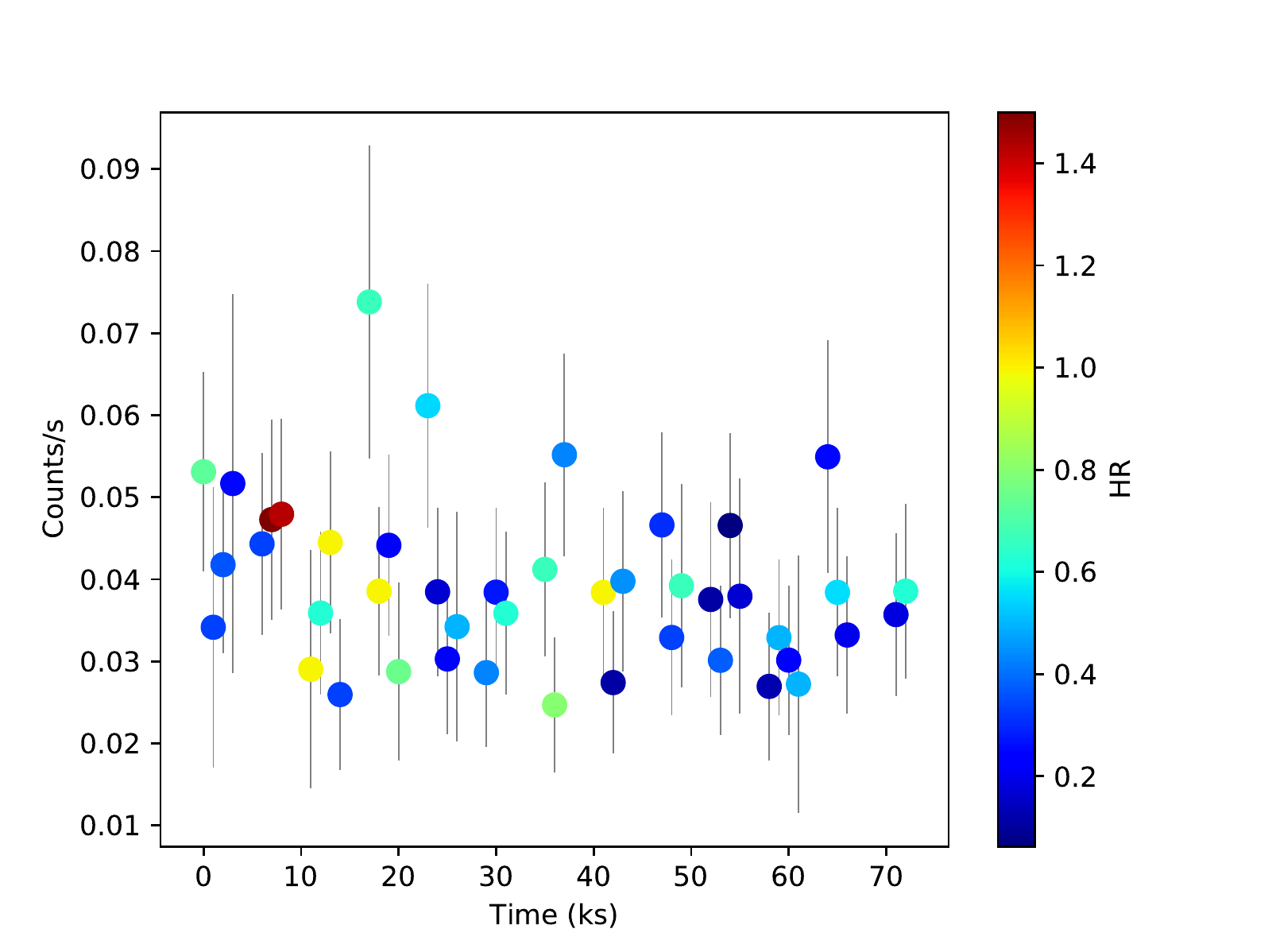}\par 
    \includegraphics[width=1.05\linewidth,angle=0]{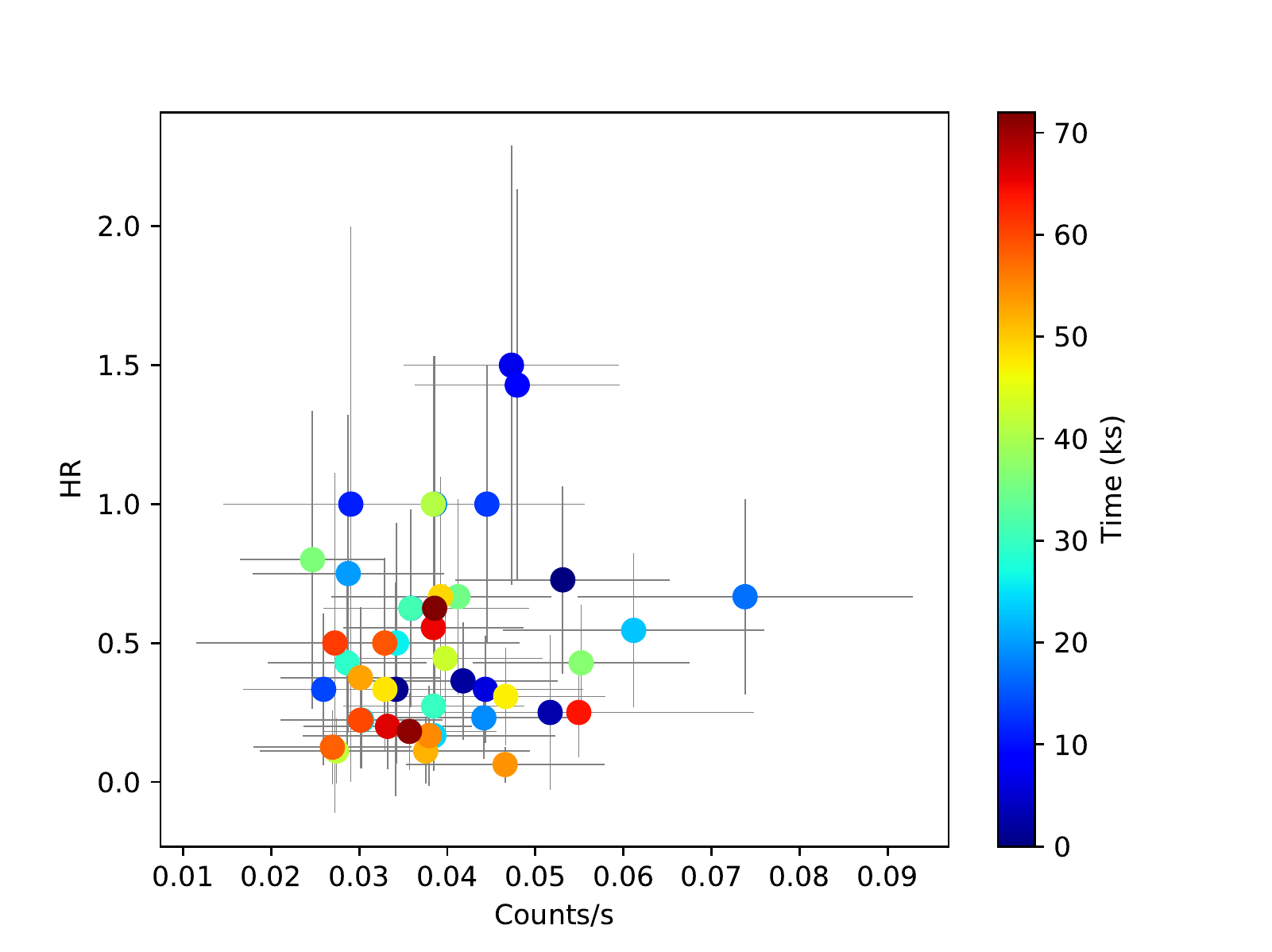}\par
    \includegraphics[width=0.72\linewidth,angle=-90]{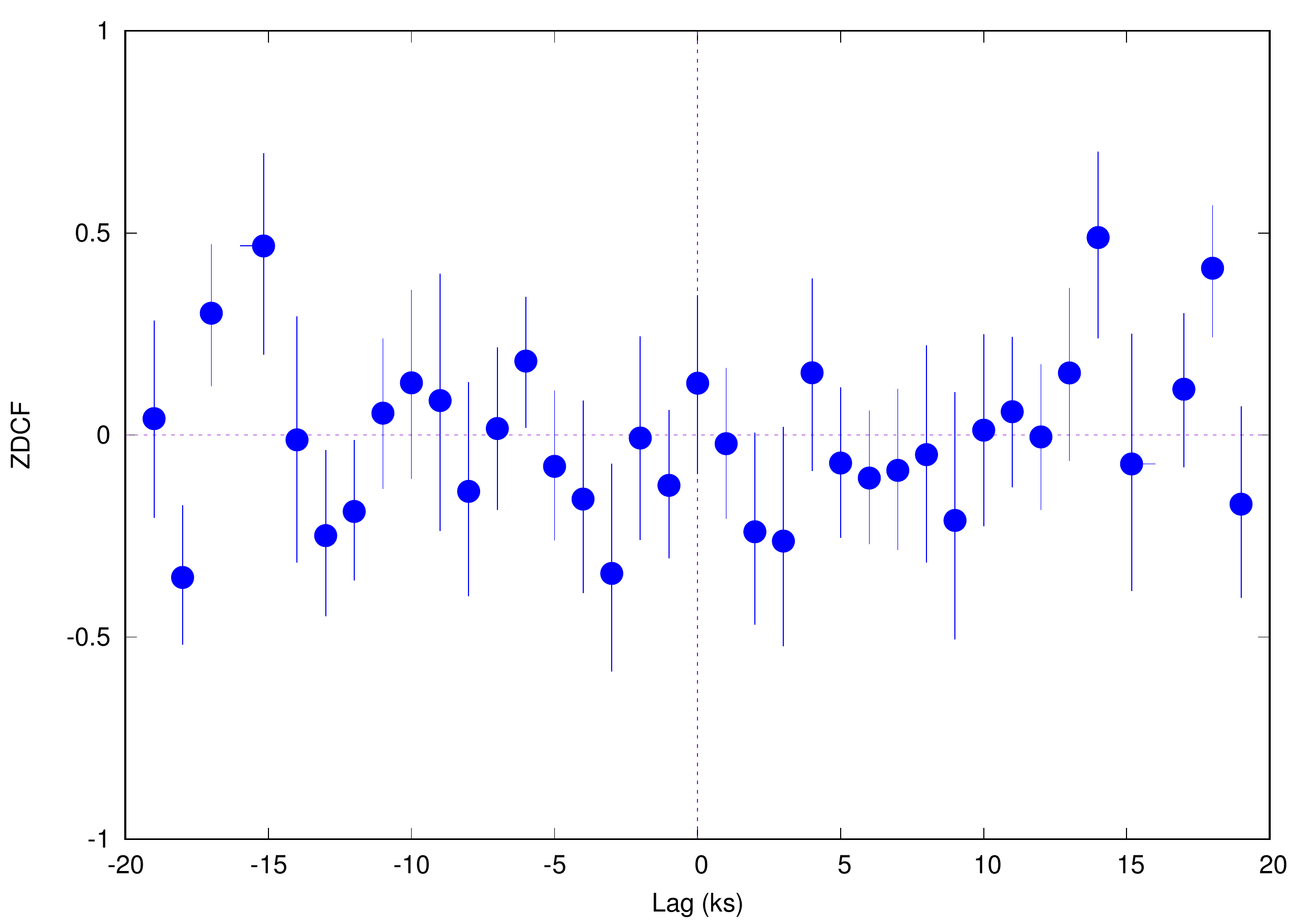}\par
    \includegraphics[width=0.73\linewidth, angle=-90]{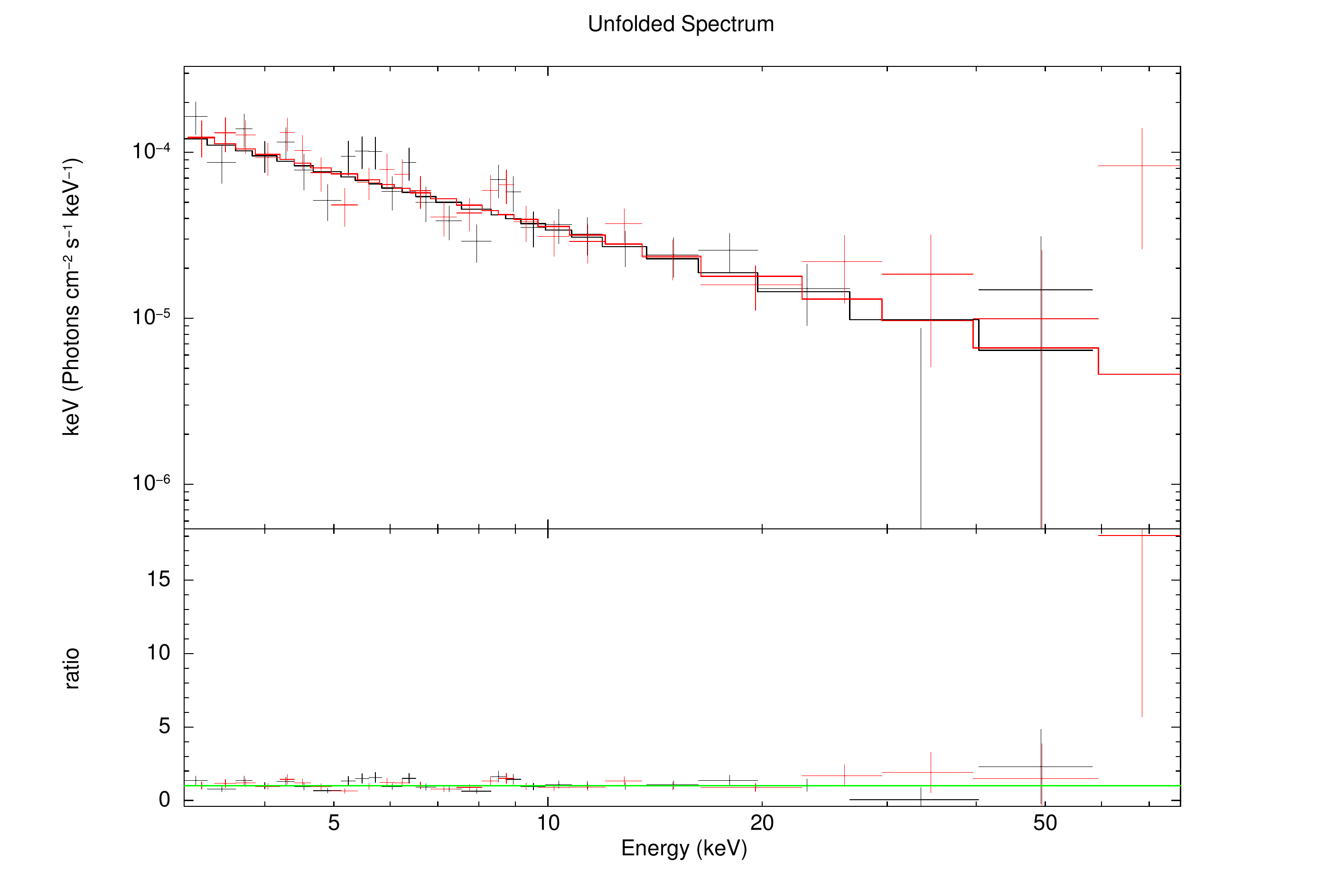}
    \end{multicols}
\center{PKS 1441+25, 90101004002} \\
\caption{Same as in Fig \ref{fig:LC1}}
\label{fig:LC10}
\end{figure*}


\begin{figure*}
\begin{multicols}{4}
    \includegraphics[width=1.05\linewidth]{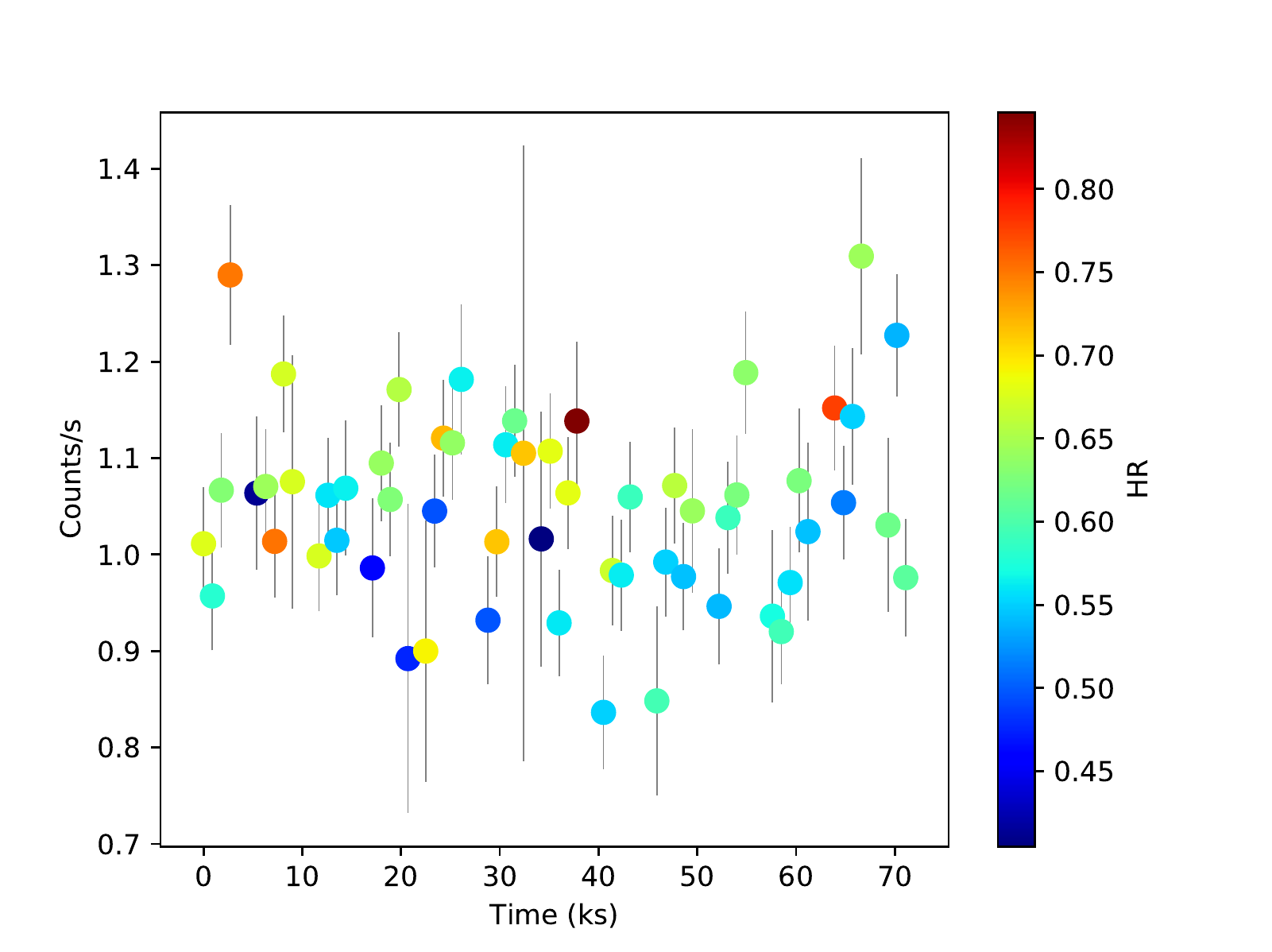}\par 
    \includegraphics[width=1.05\linewidth,angle=0]{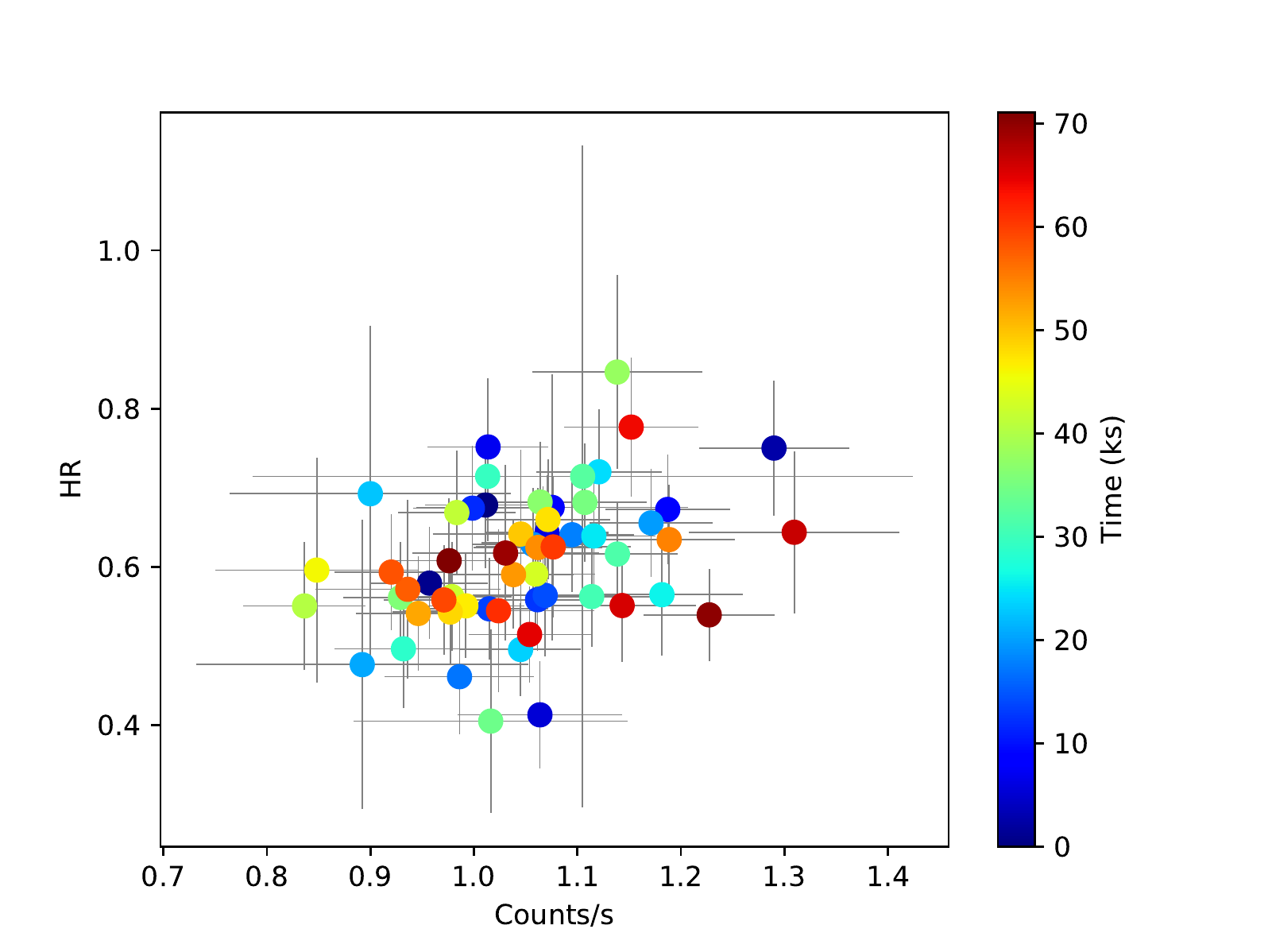}\par
    \includegraphics[width=0.72\linewidth,angle=-90]{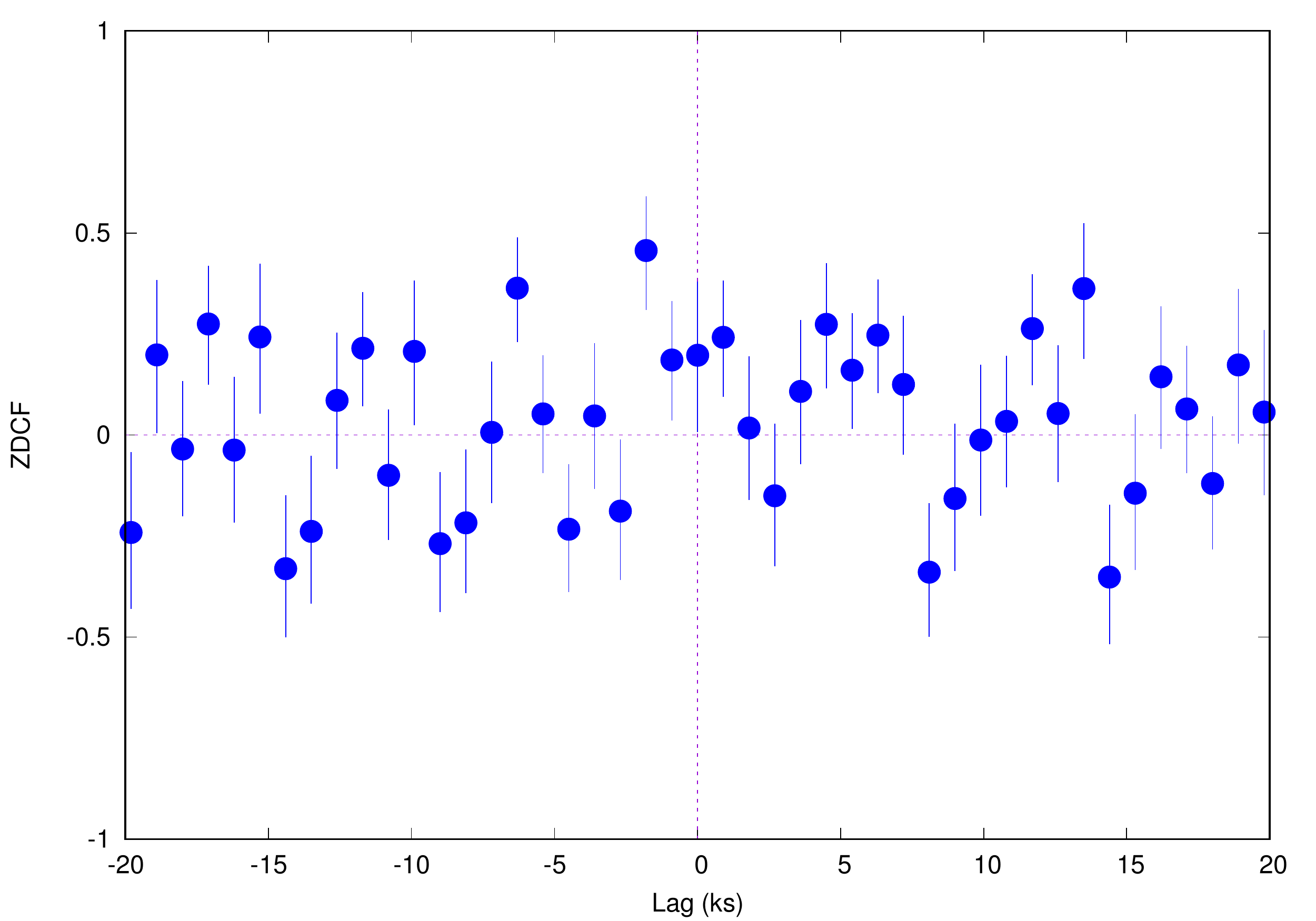}\par
    \includegraphics[width=0.73\linewidth, angle=-90]{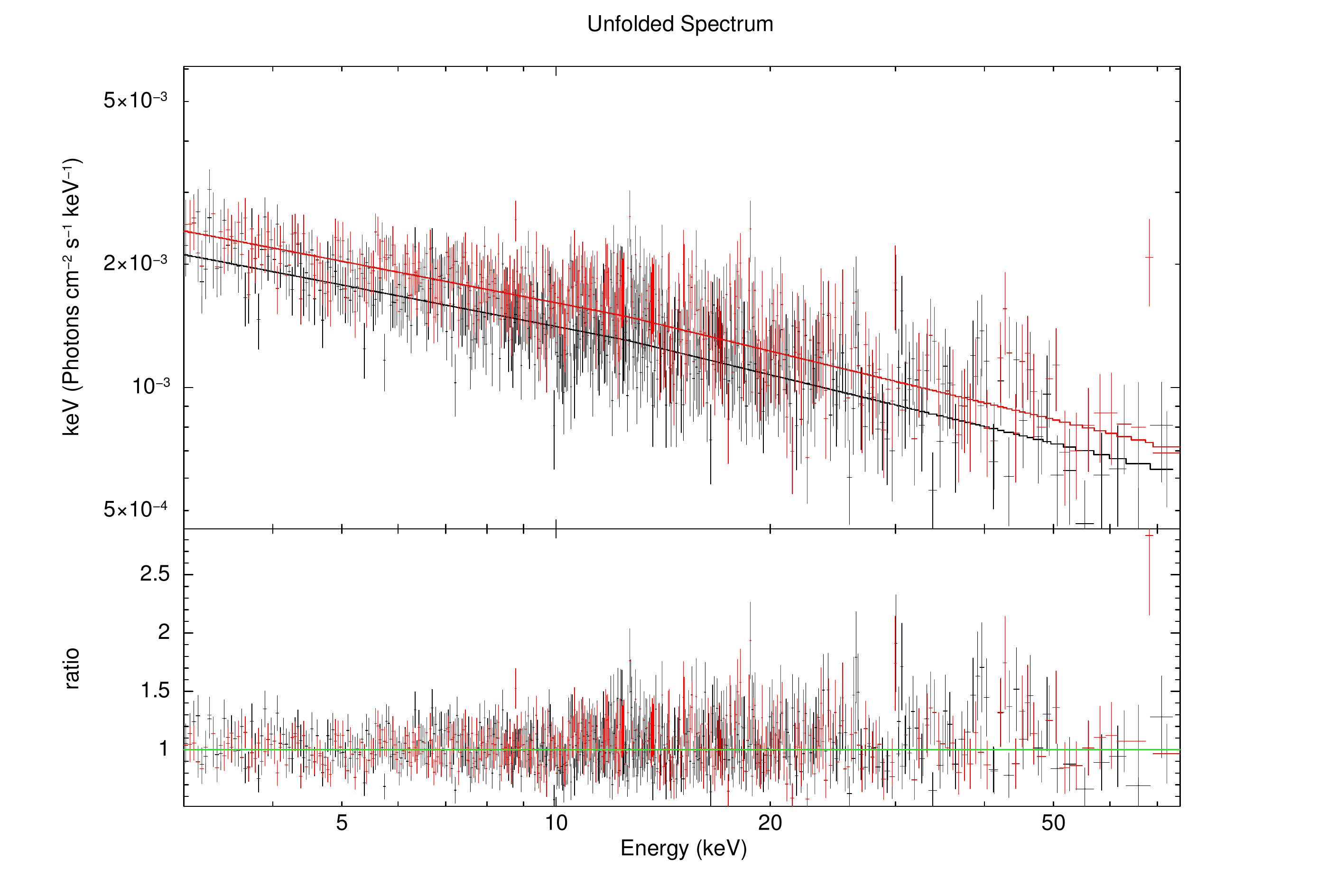}
    \end{multicols}
\center{PKS 2149--306, 60001099002}

\begin{multicols}{4}
    \includegraphics[width=1.05\linewidth]{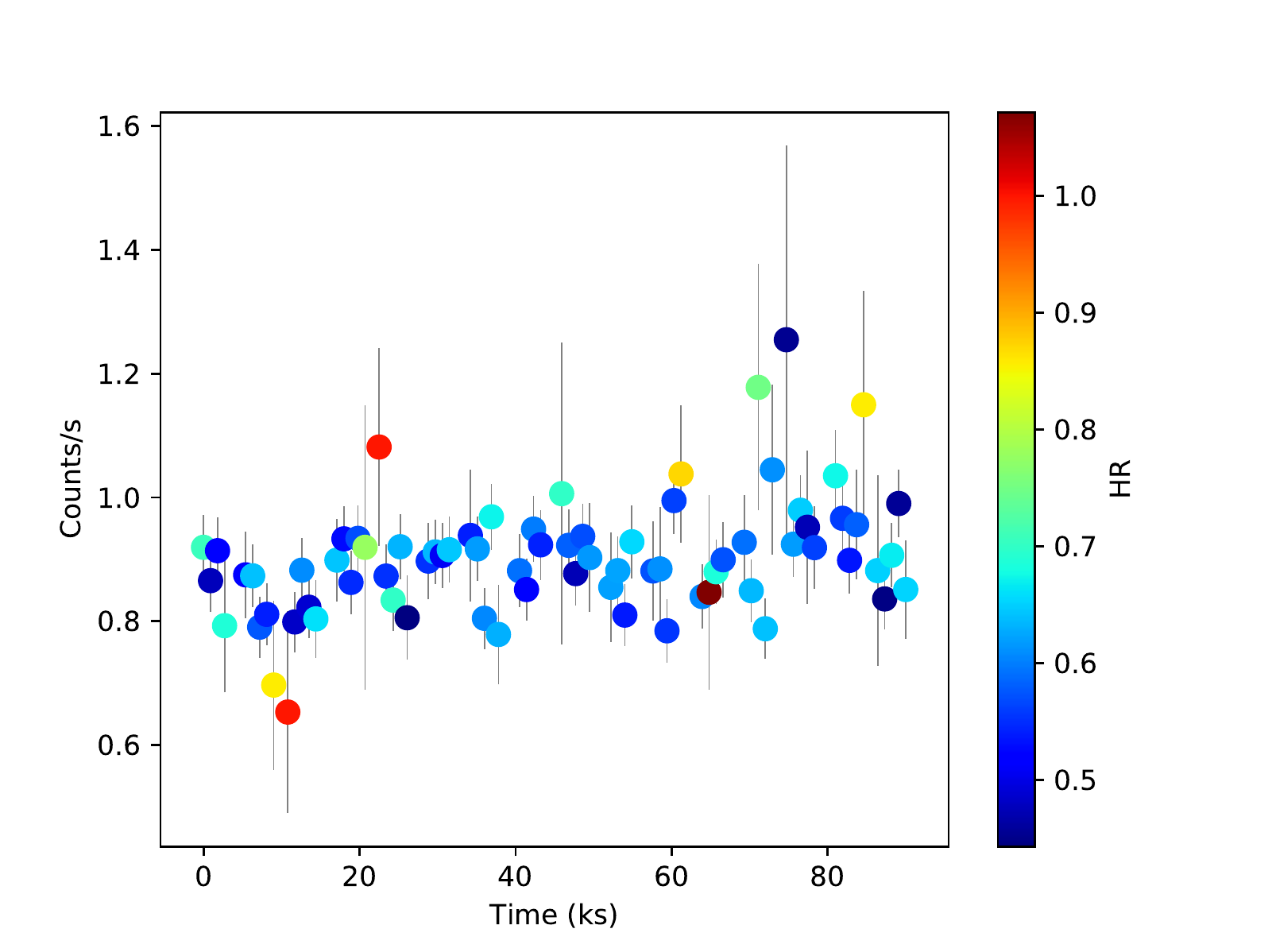}\par 
    \includegraphics[width=1.05\linewidth,angle=0]{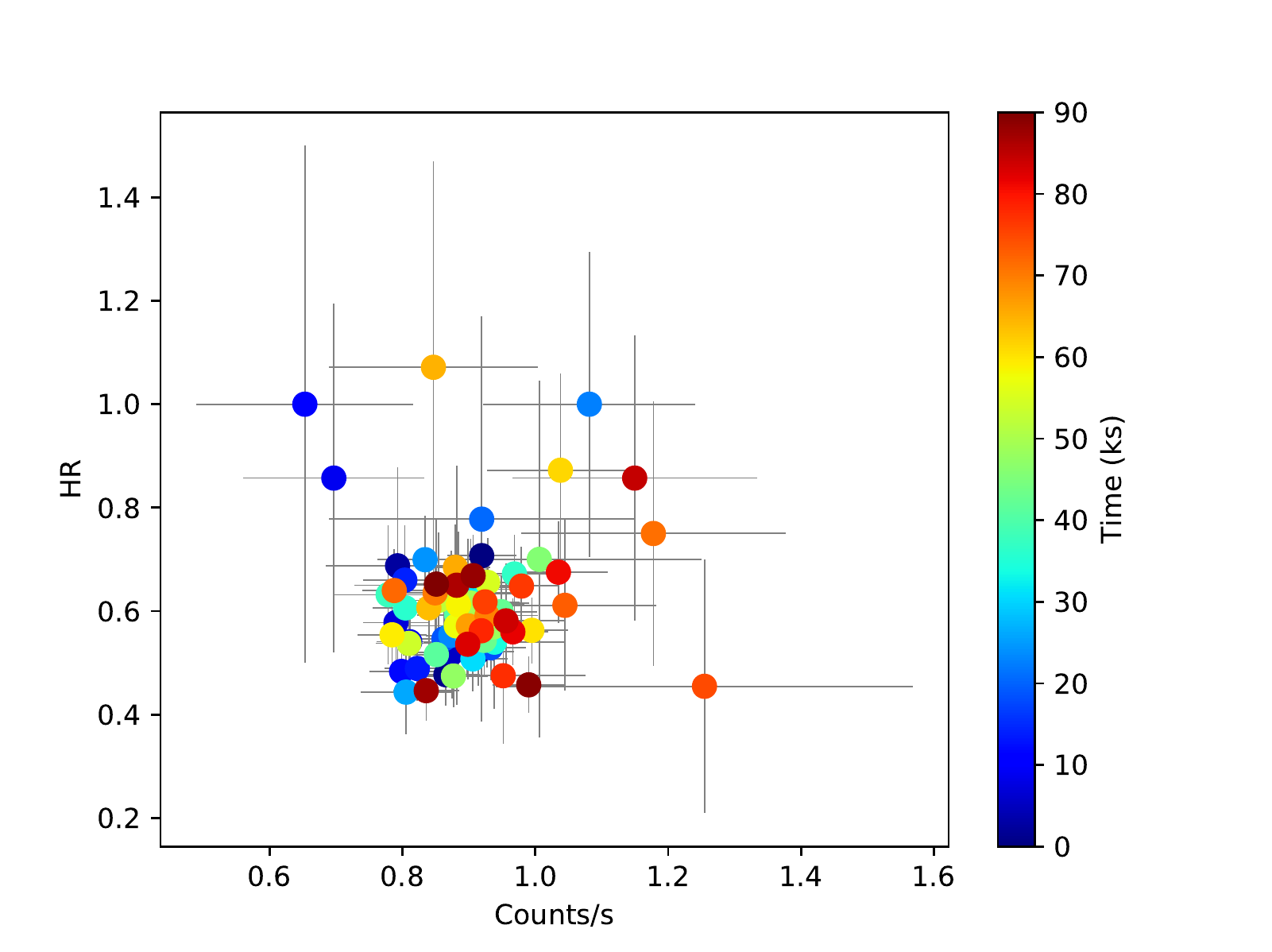}\par
    \includegraphics[width=0.72\linewidth,angle=-90]{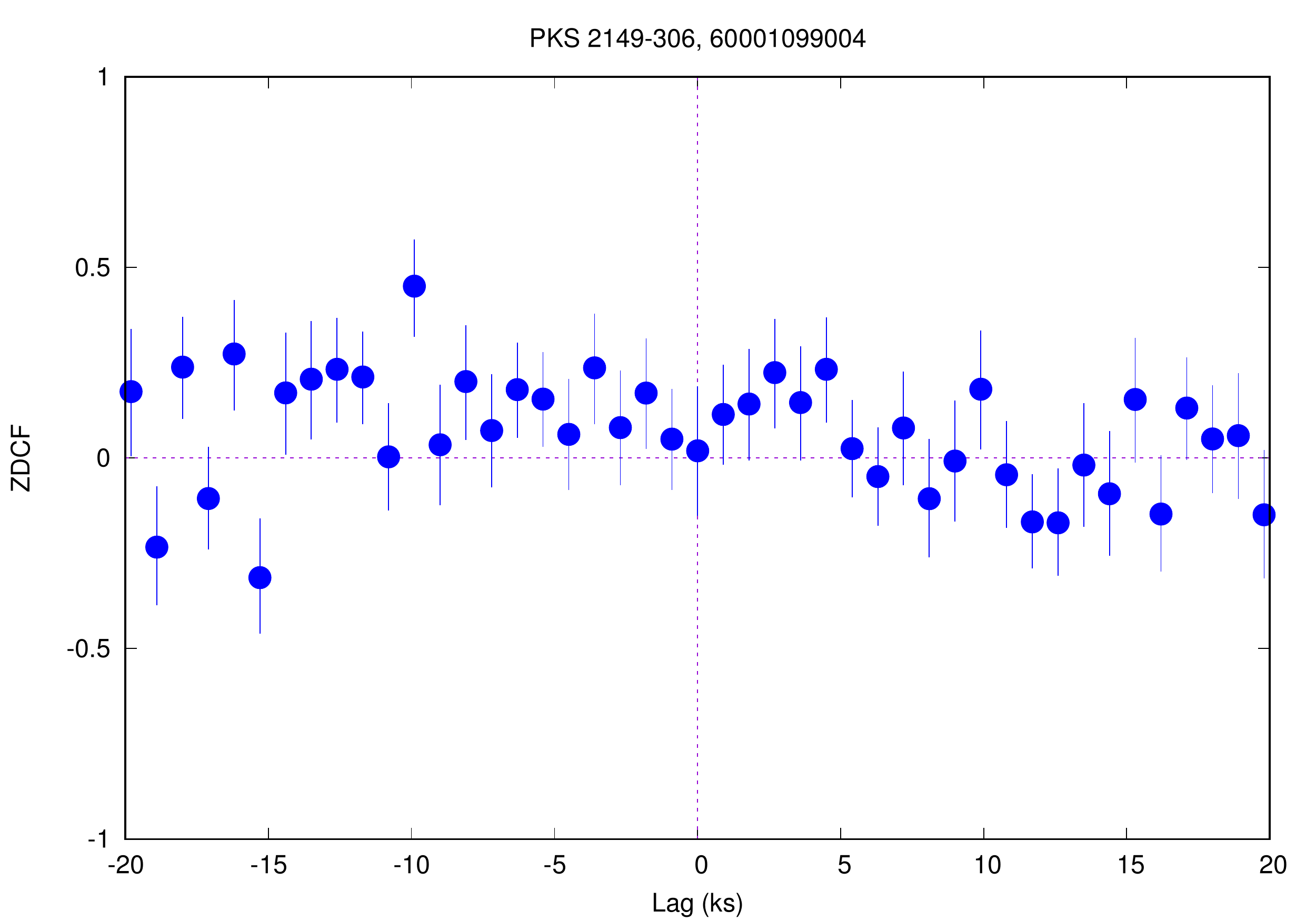}\par
    \includegraphics[width=0.73\linewidth, angle=-90]{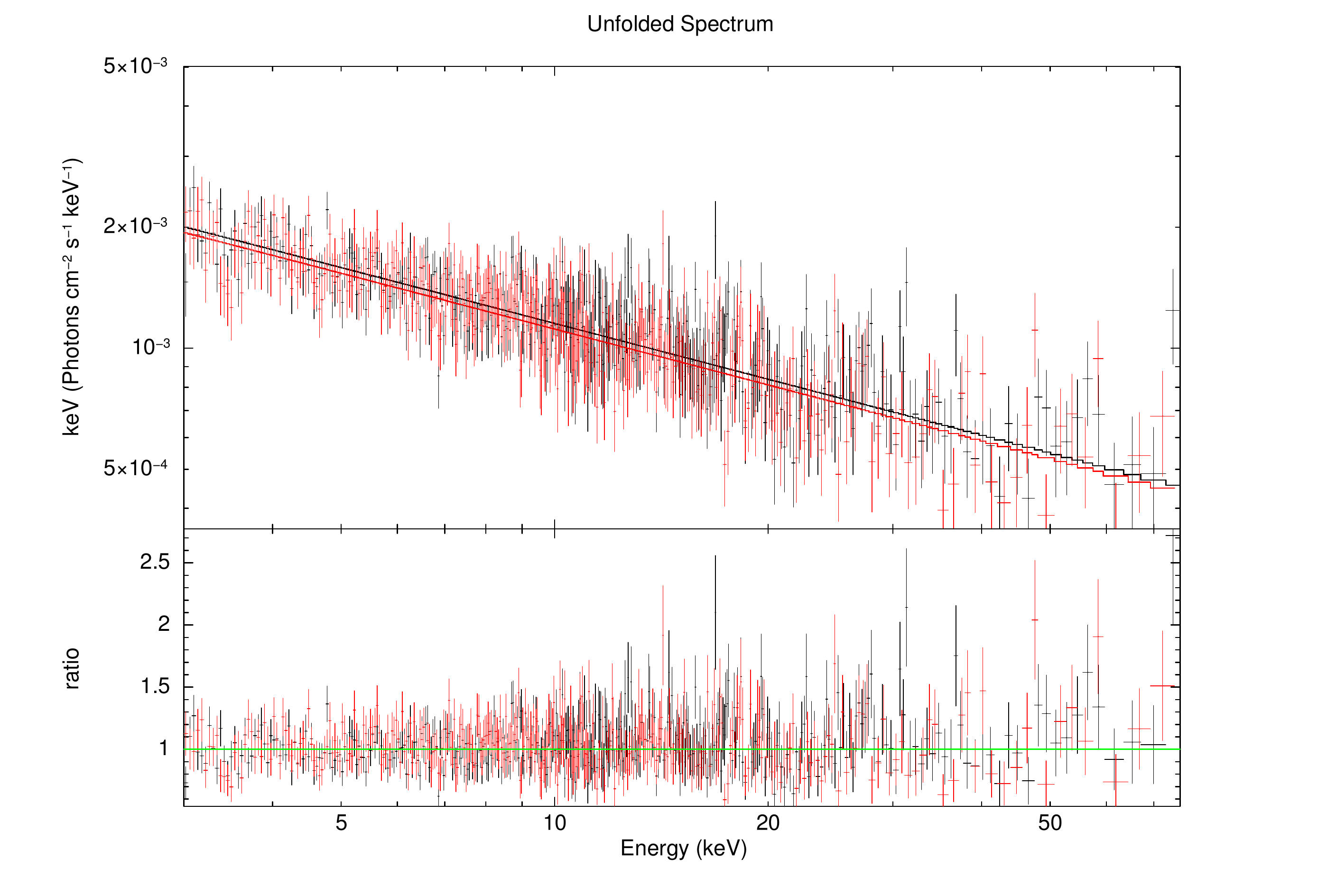}
    \end{multicols}
\center{PKS 2149--306, 60001099004}

\begin{multicols}{4}
    \includegraphics[width=1.05\linewidth]{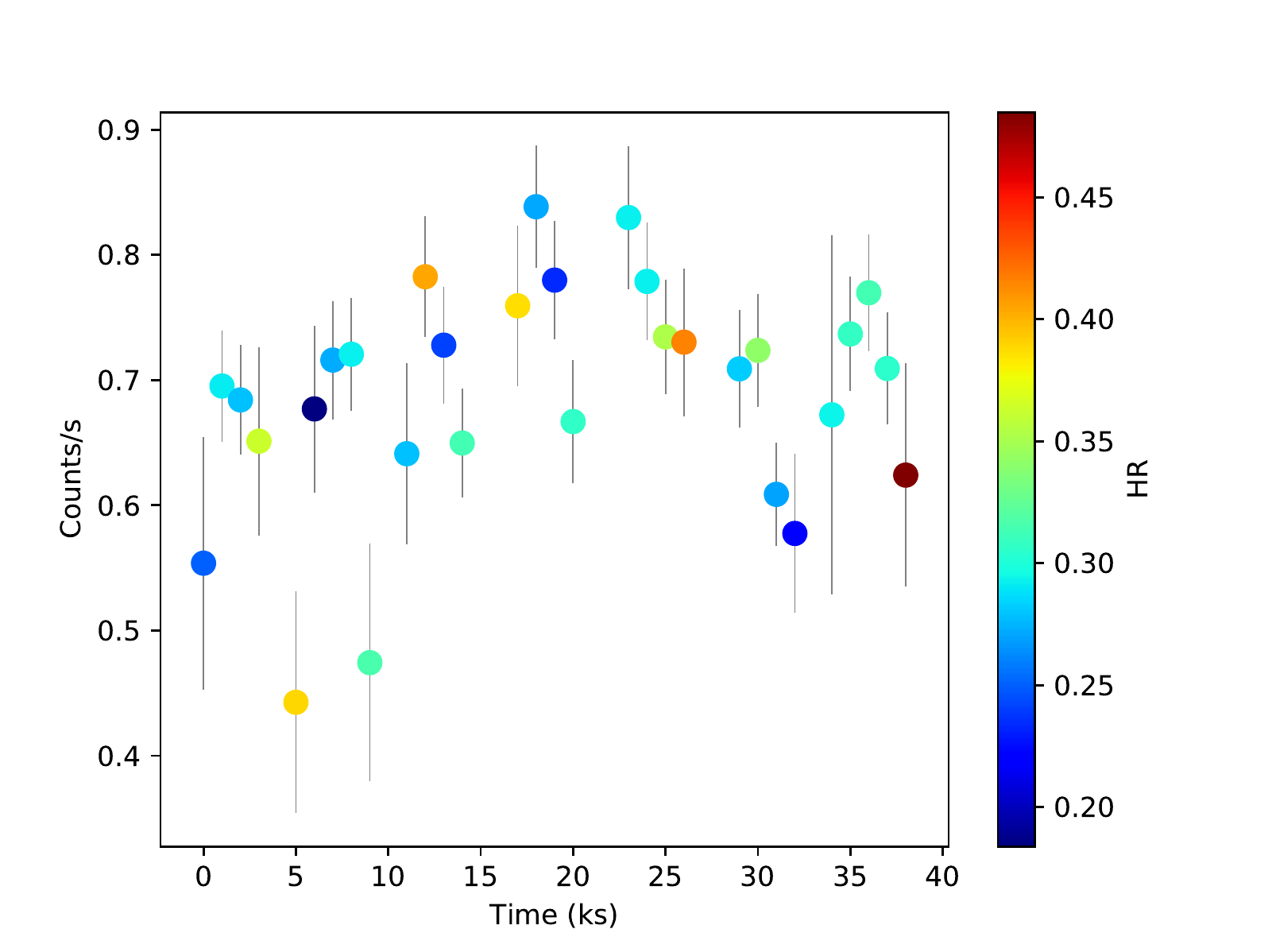}\par 
    \includegraphics[width=1.05\linewidth,angle=0]{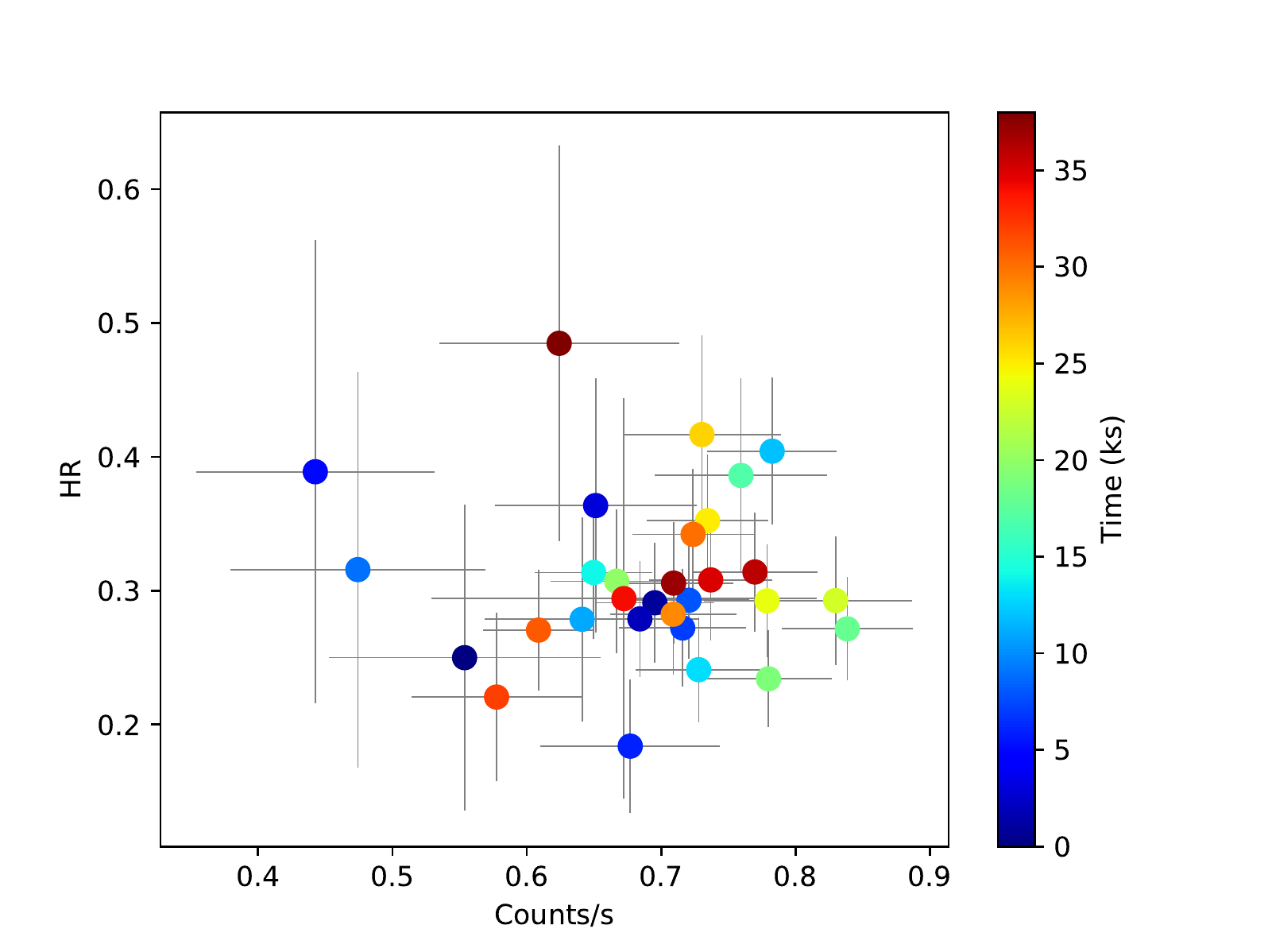}\par
    \includegraphics[width=0.72\linewidth,angle=-90]{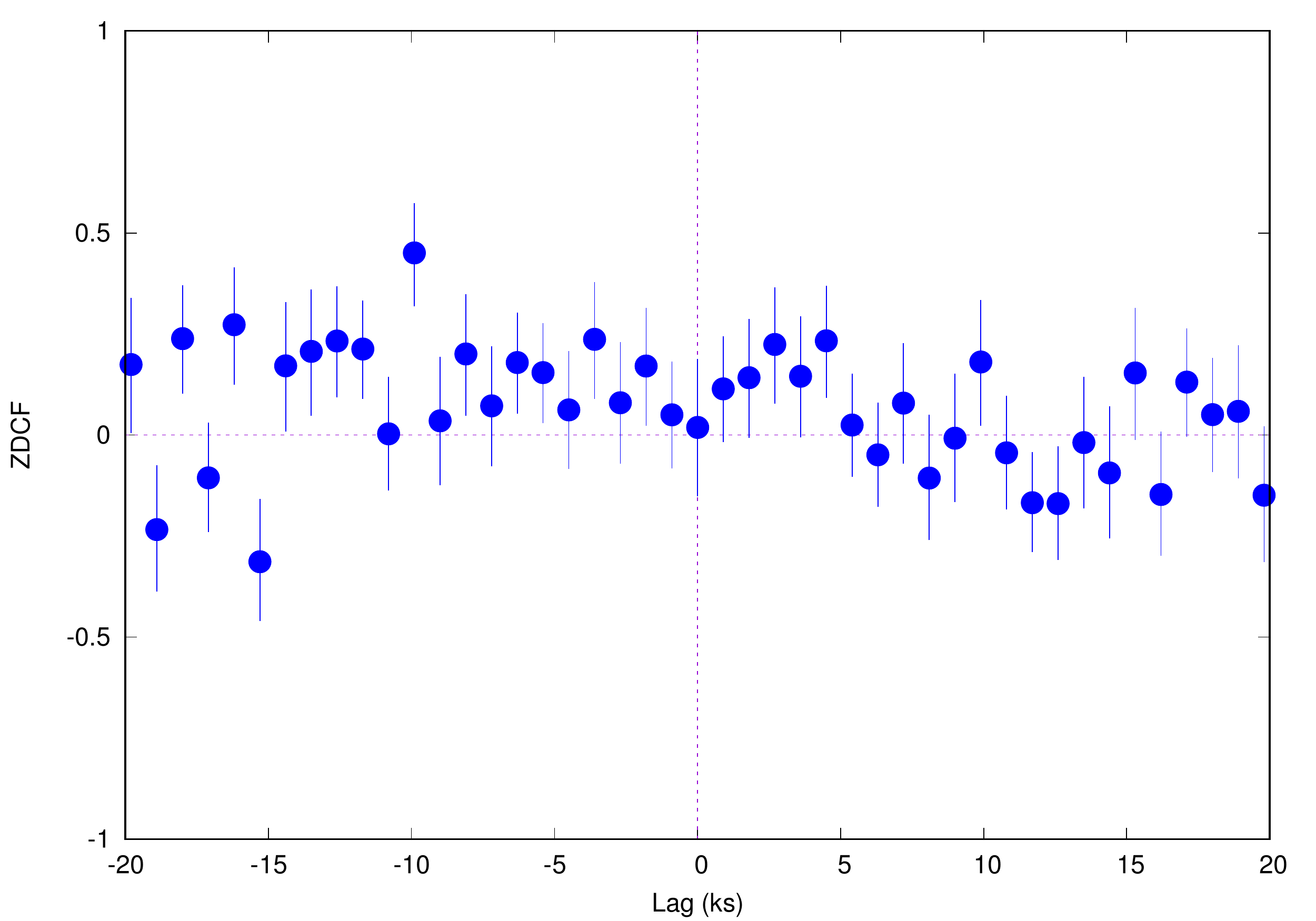}\par
    \includegraphics[width=0.73\linewidth, angle=-90]{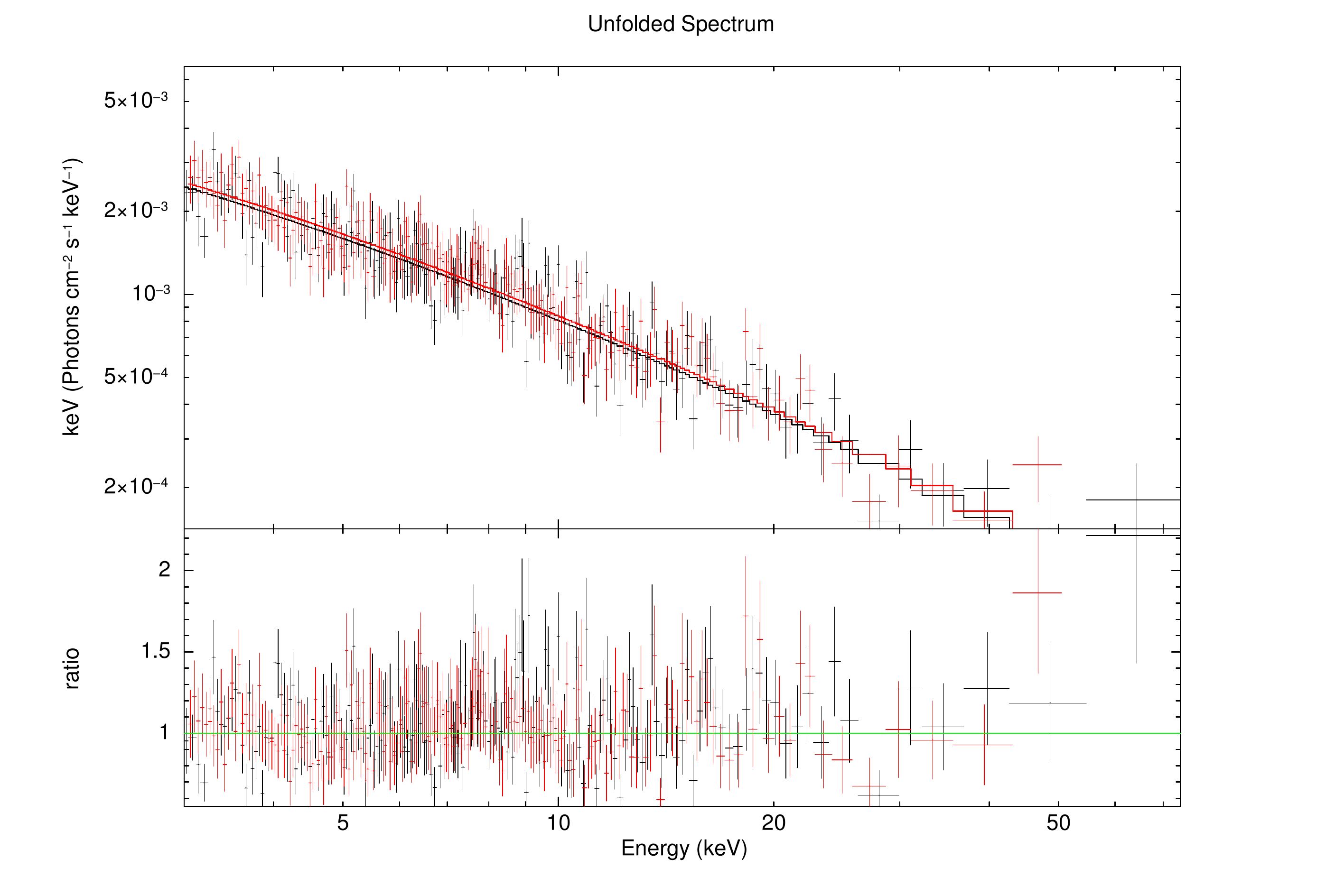}
    \end{multicols}
\center{1ES 0229+200, 60002047004}

\begin{multicols}{4}
    \includegraphics[width=1.05\linewidth]{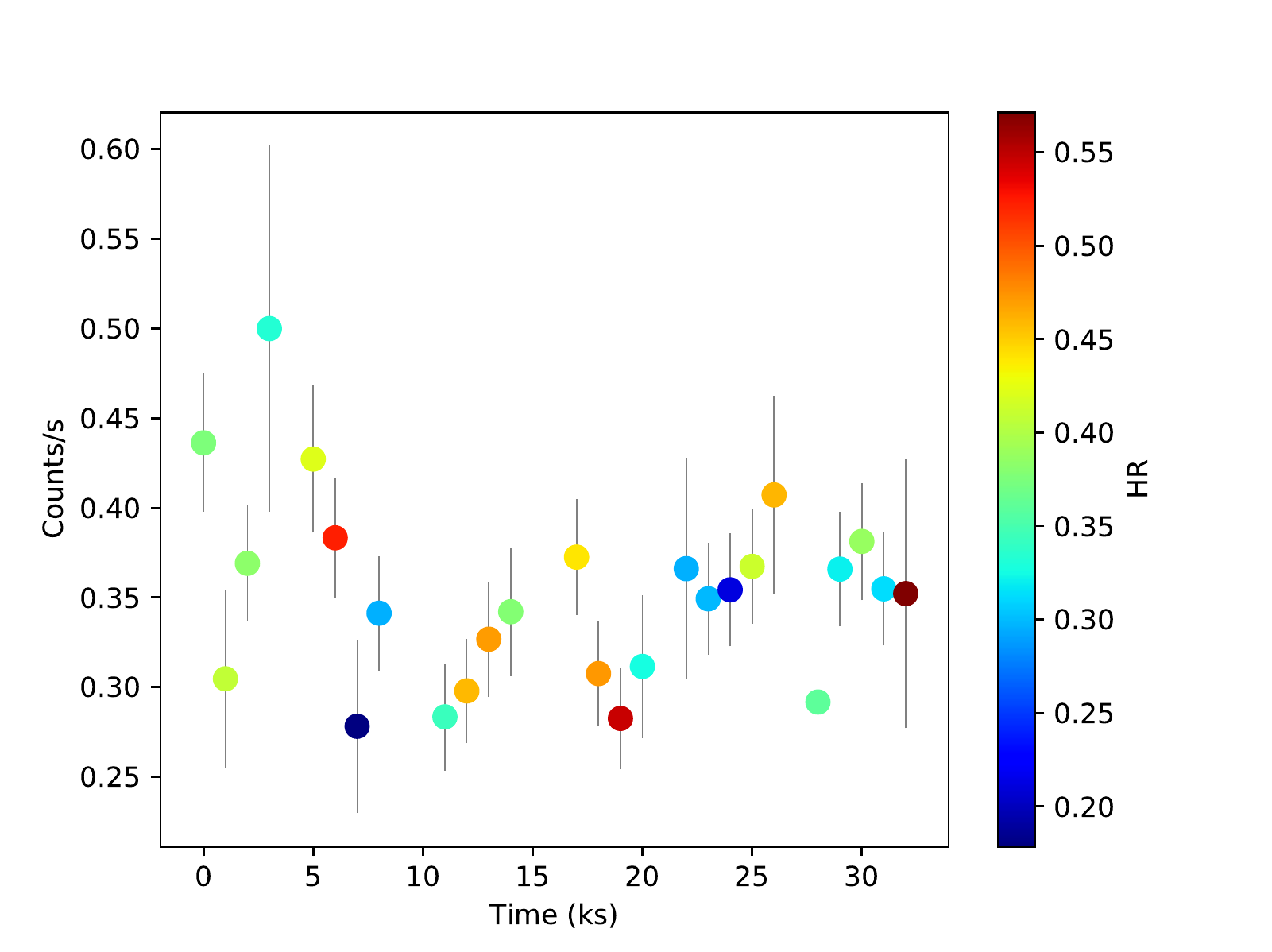}\par 
    \includegraphics[width=1.05\linewidth,angle=0]{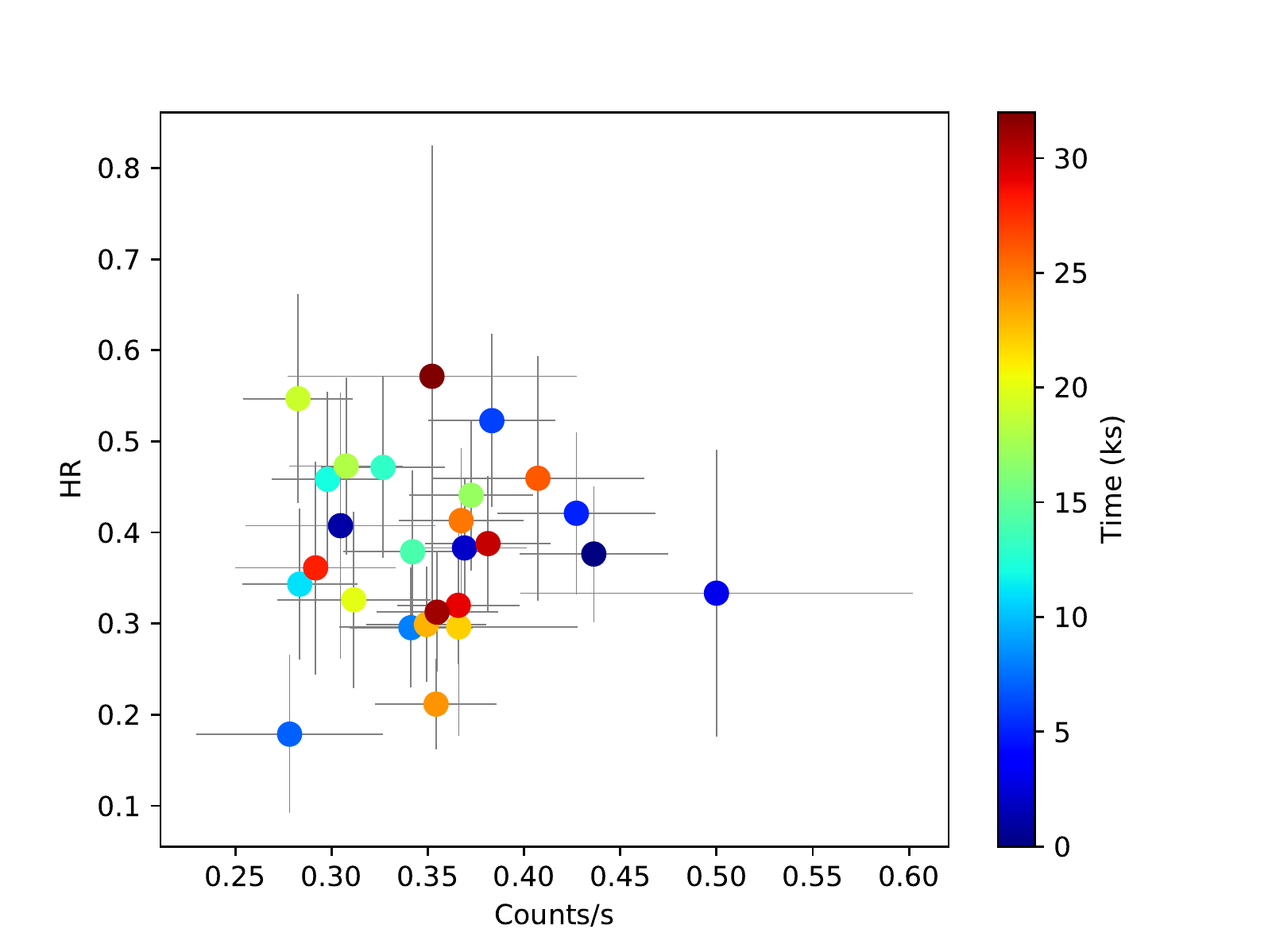}\par
    \includegraphics[width=0.72\linewidth,angle=-90]{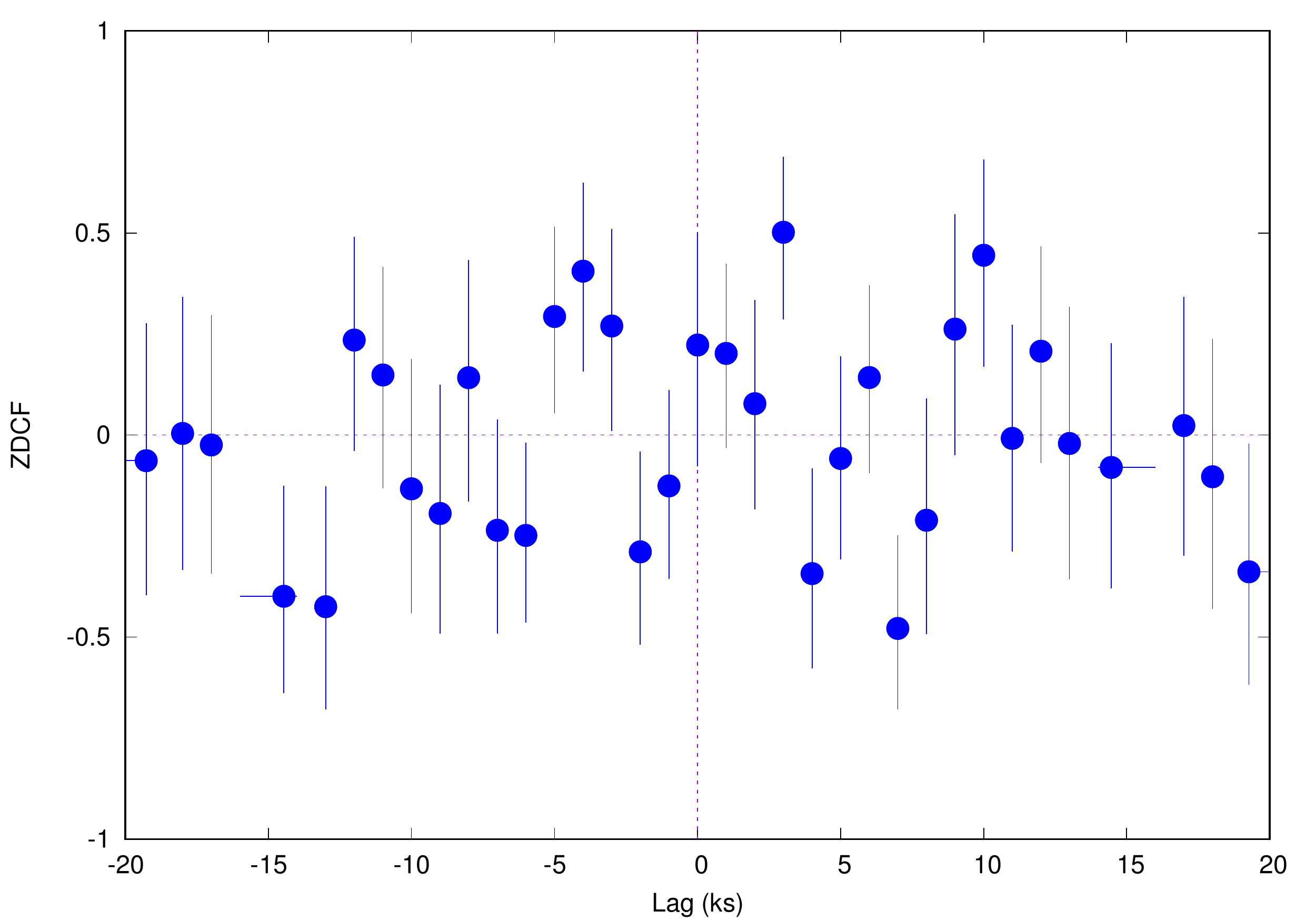}\par
    \includegraphics[width=0.73\linewidth, angle=-90]{s5_0716_02_logpar.pdf}
    \end{multicols}
\center{S5 0716+714, 90002003002}

\begin{multicols}{4}
    \includegraphics[width=1.05\linewidth]{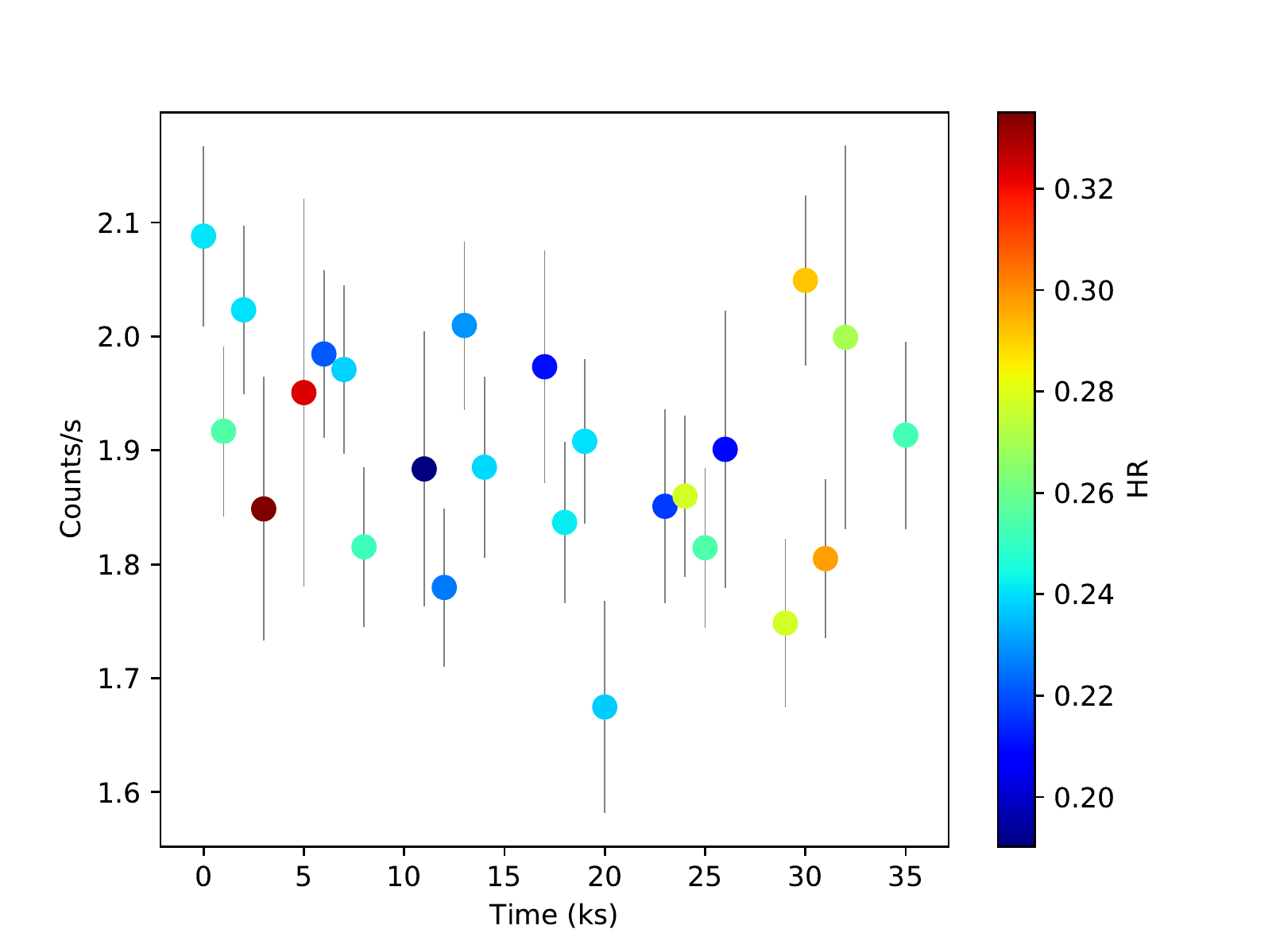}\par 
    \includegraphics[width=1.05\linewidth,angle=0]{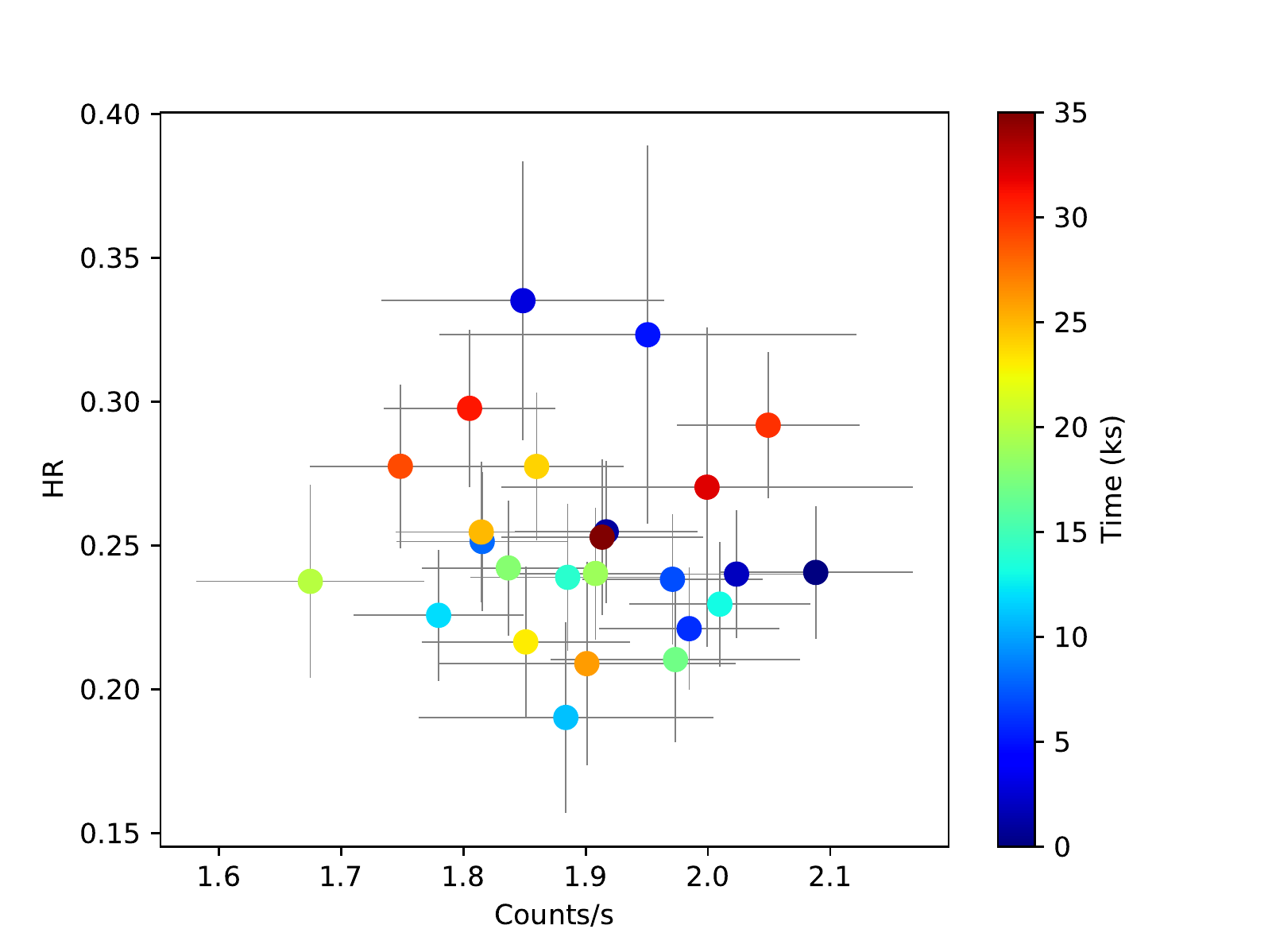}\par
    \includegraphics[width=0.72\linewidth,angle=-90]{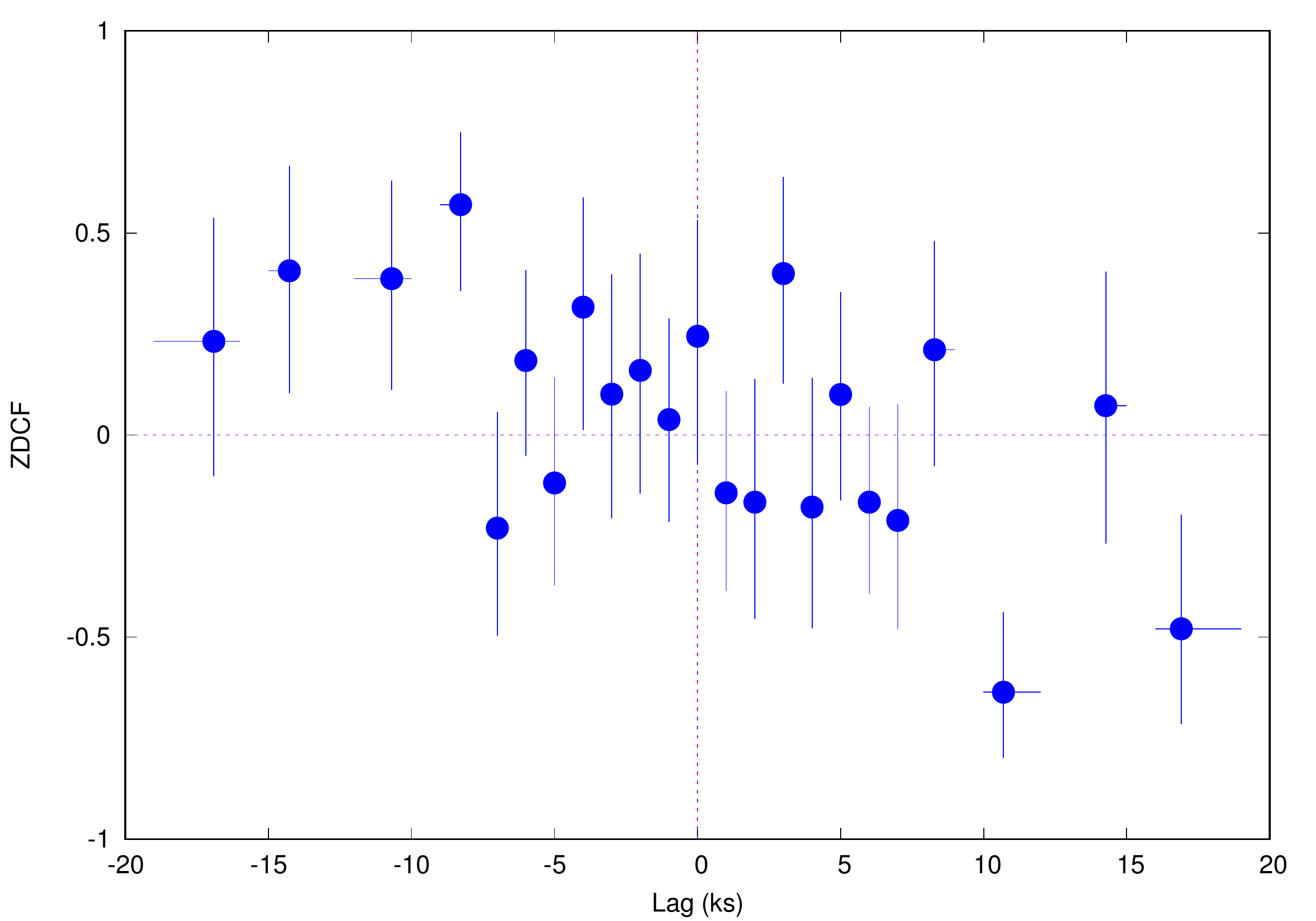}\par
    \includegraphics[width=0.73\linewidth, angle=-90]{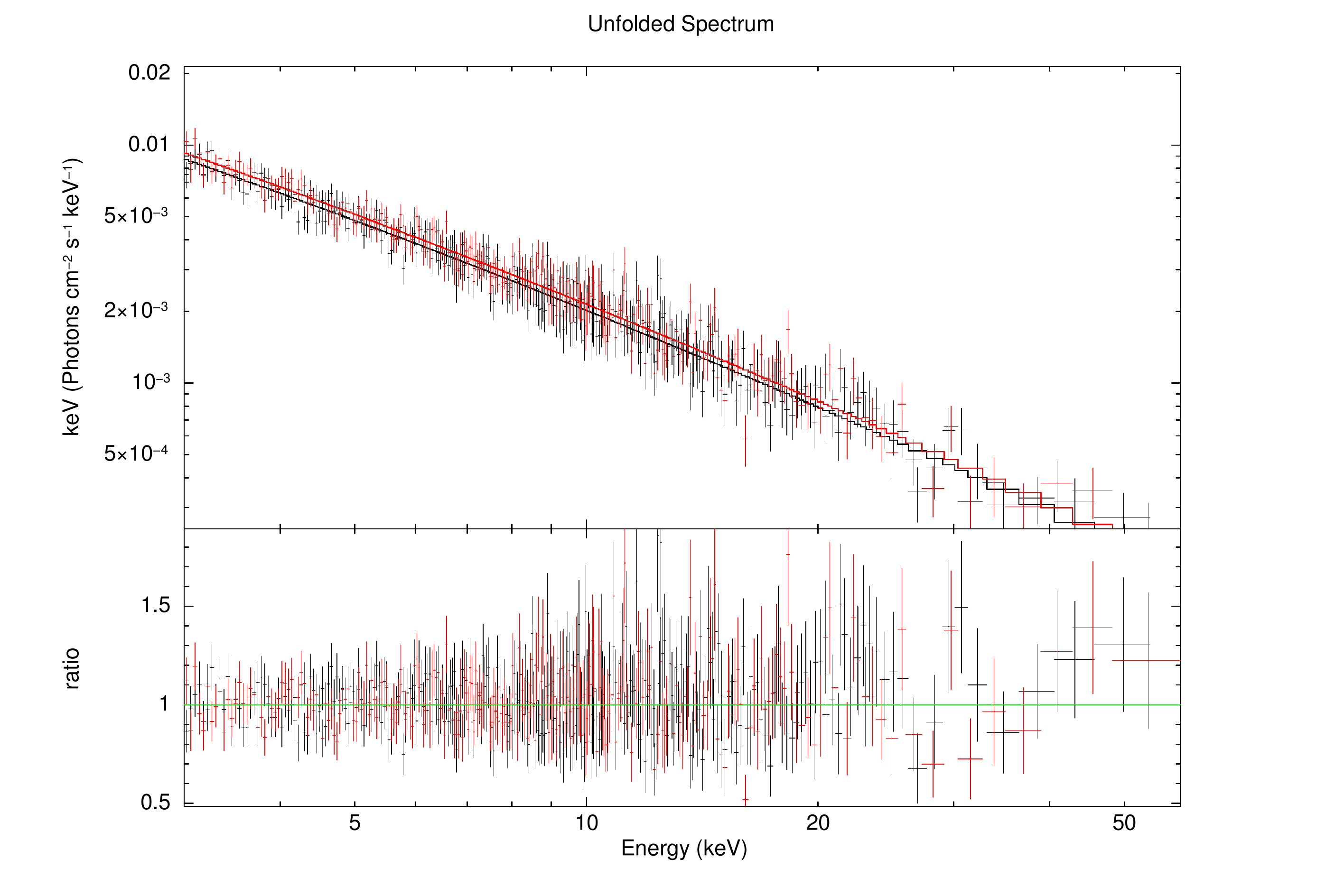}
    \end{multicols}
\center{Mrk 501, 60002024002} \\
\caption{Same as in Fig \ref{fig:LC1}}
\label{fig:LC2}
\end{figure*}


\begin{figure*}
\begin{multicols}{4}
    \includegraphics[width=1.05\linewidth]{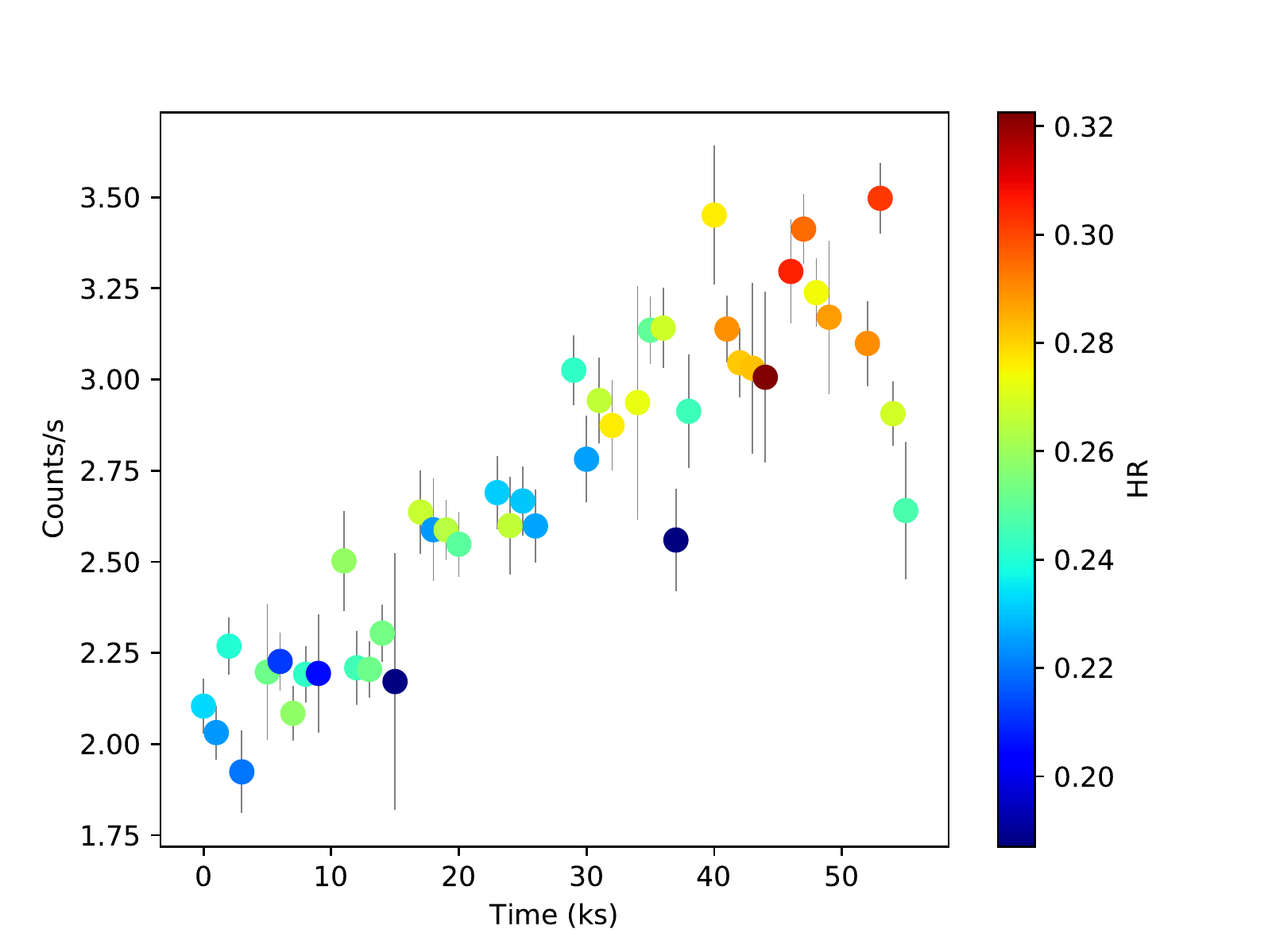}\par 
    \includegraphics[width=1.05\linewidth,angle=0]{mrk501_04_Time.pdf}\par
    \includegraphics[width=0.72\linewidth,angle=-90]{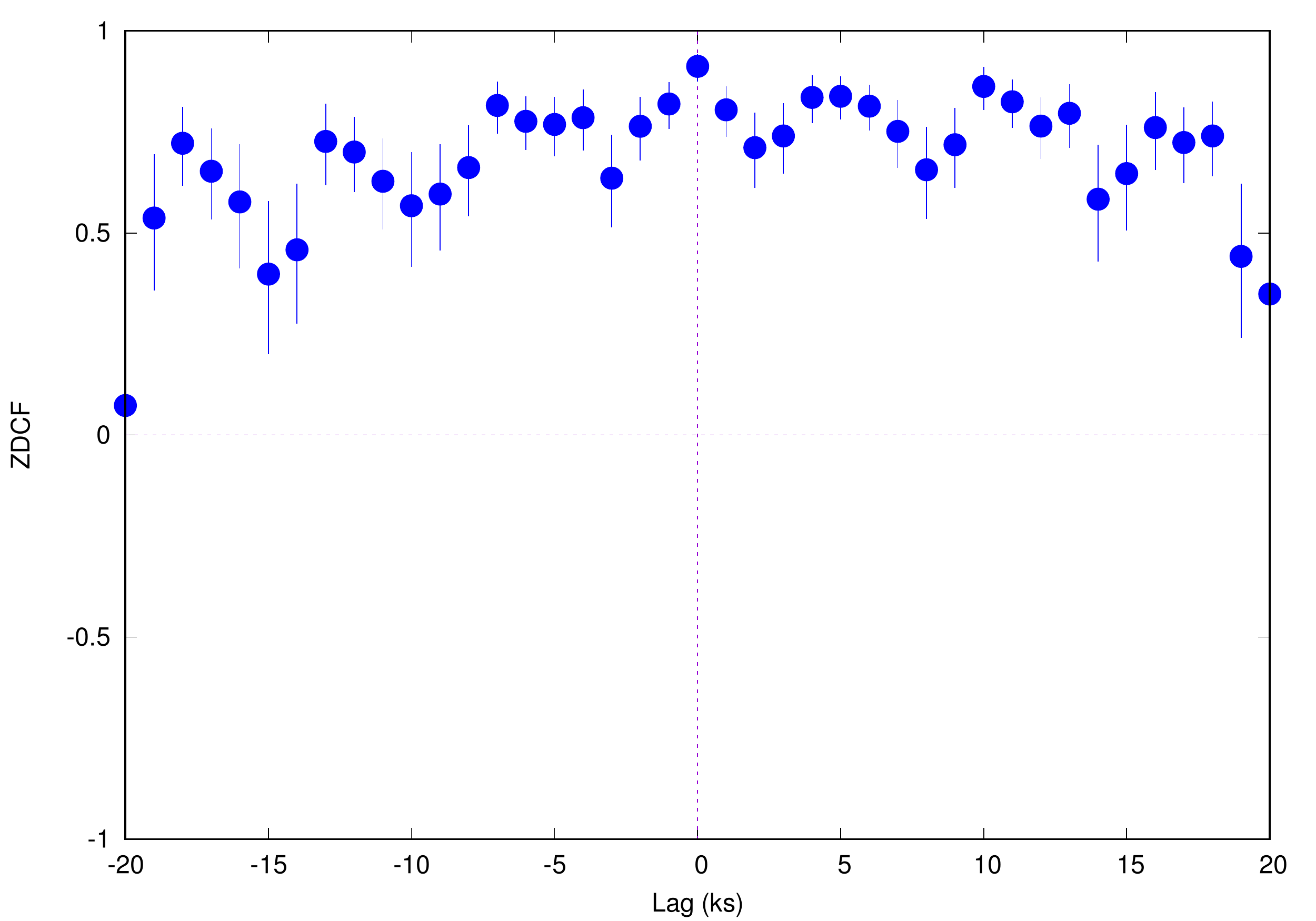}\par
    \includegraphics[width=0.73\linewidth, angle=-90]{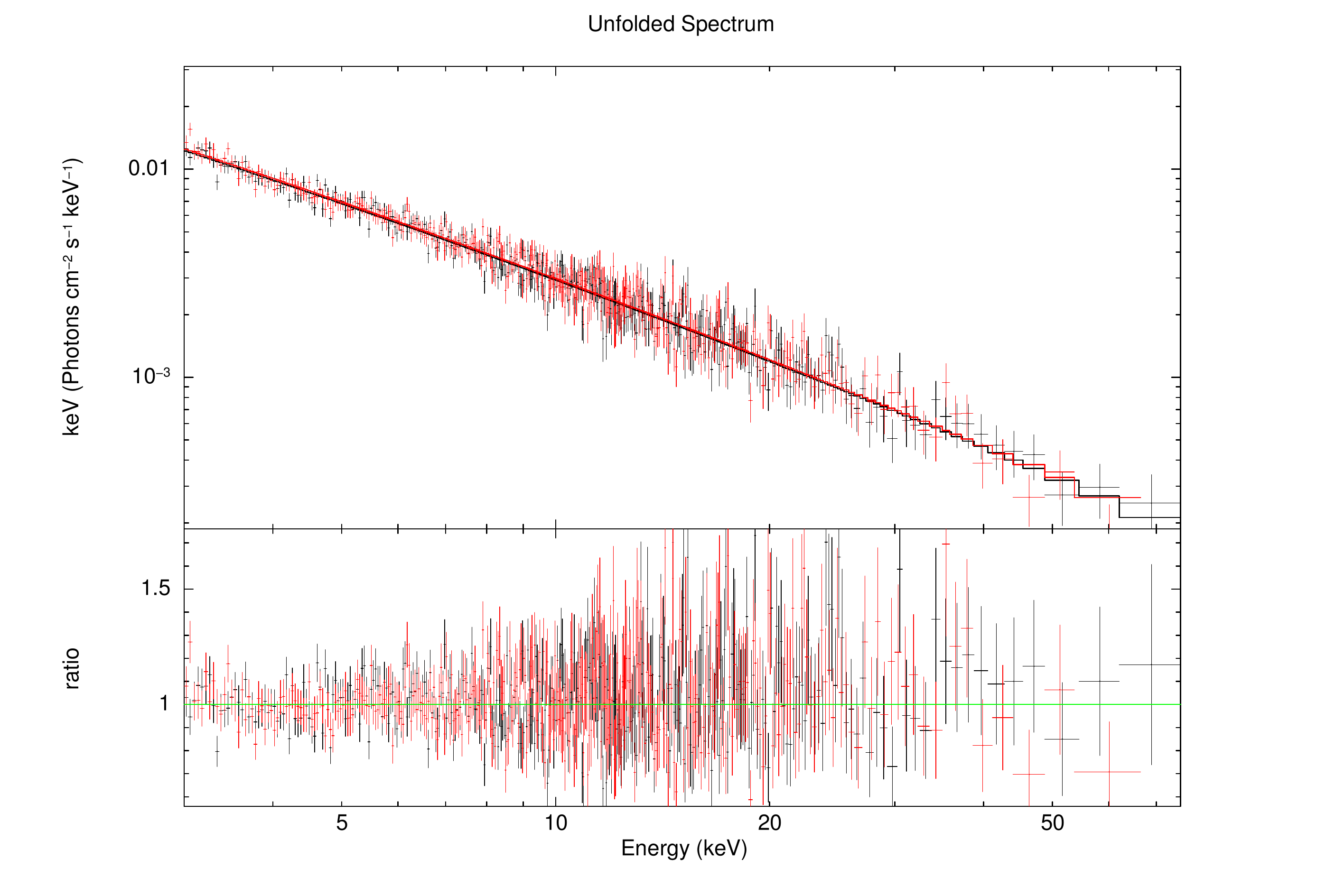}
     \end{multicols}
\center{Mrk 501, 60002024004}

\begin{multicols}{4}
    \includegraphics[width=1.05\linewidth]{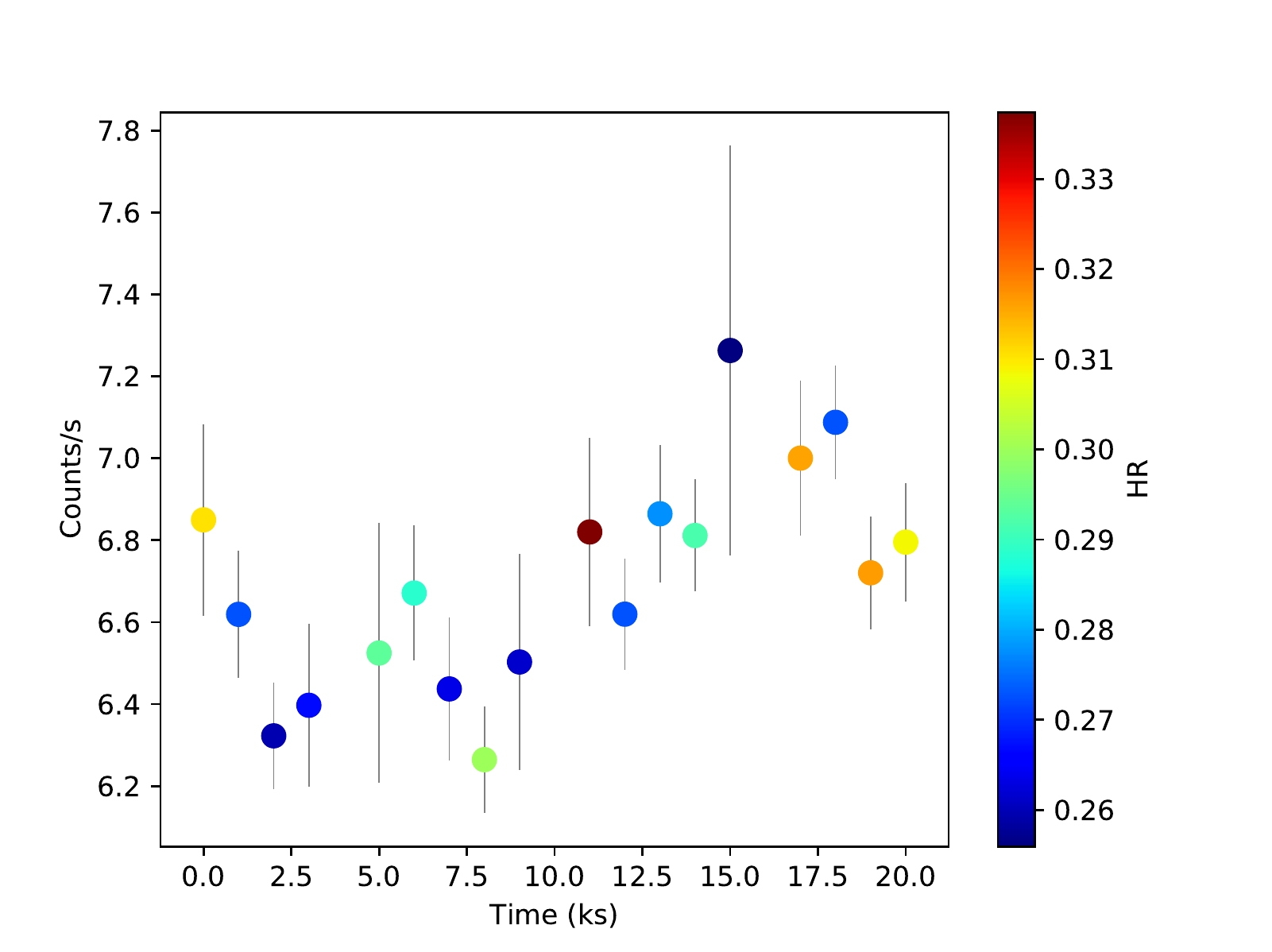}\par 
    \includegraphics[width=1.05\linewidth,angle=0]{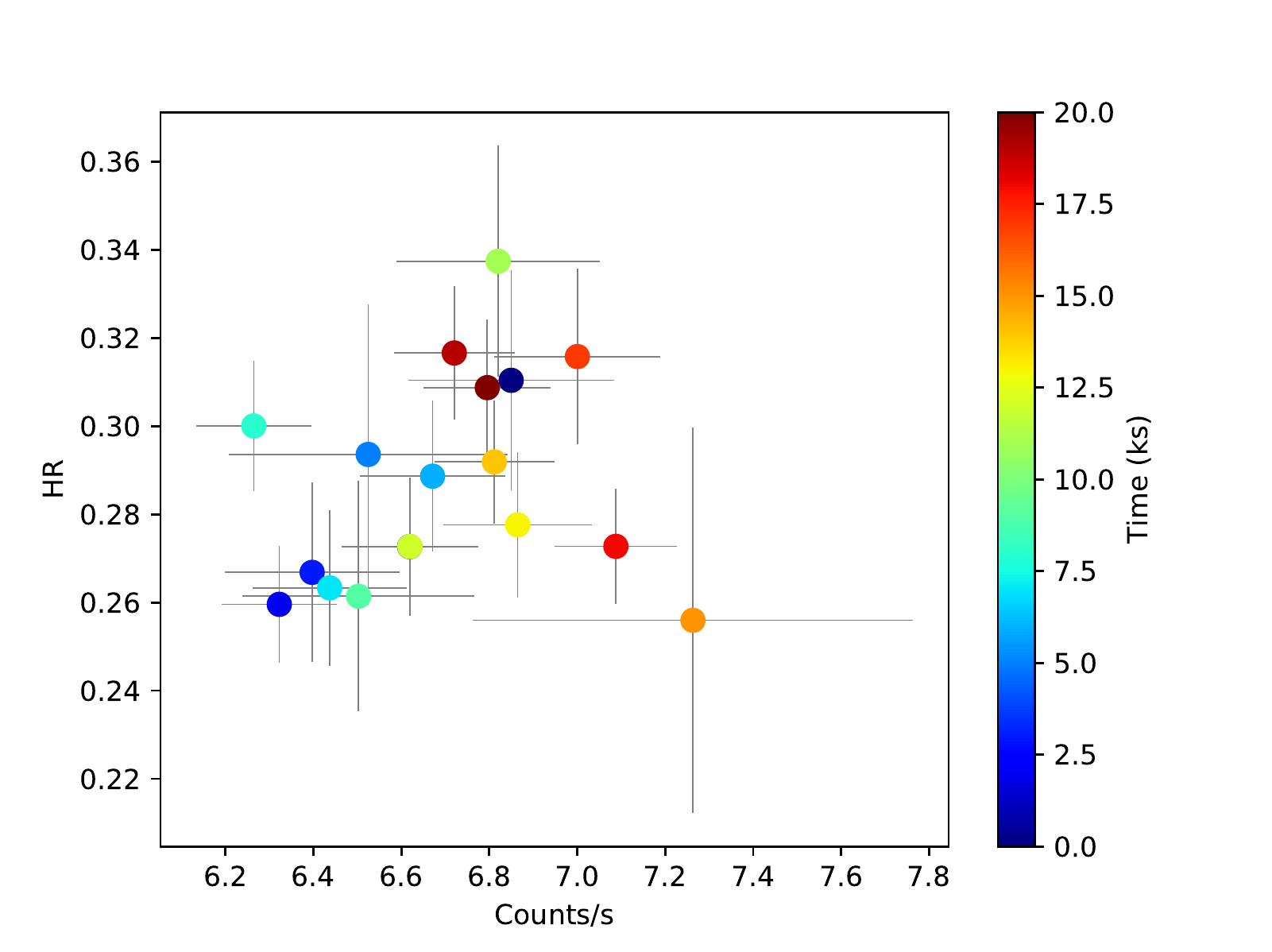}\par
    \includegraphics[width=0.72\linewidth,angle=-90]{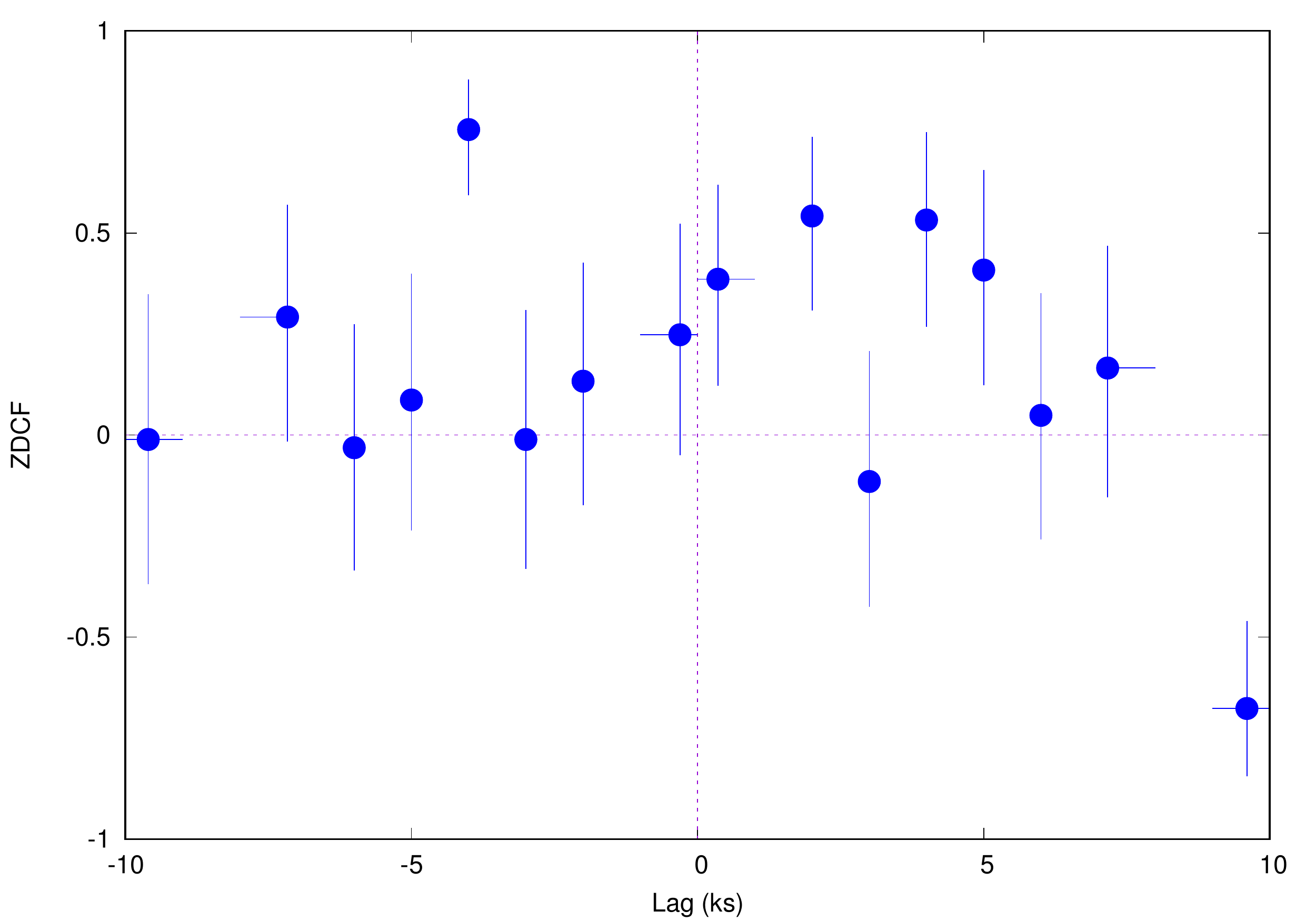}\par
    \includegraphics[width=0.73\linewidth, angle=-90]{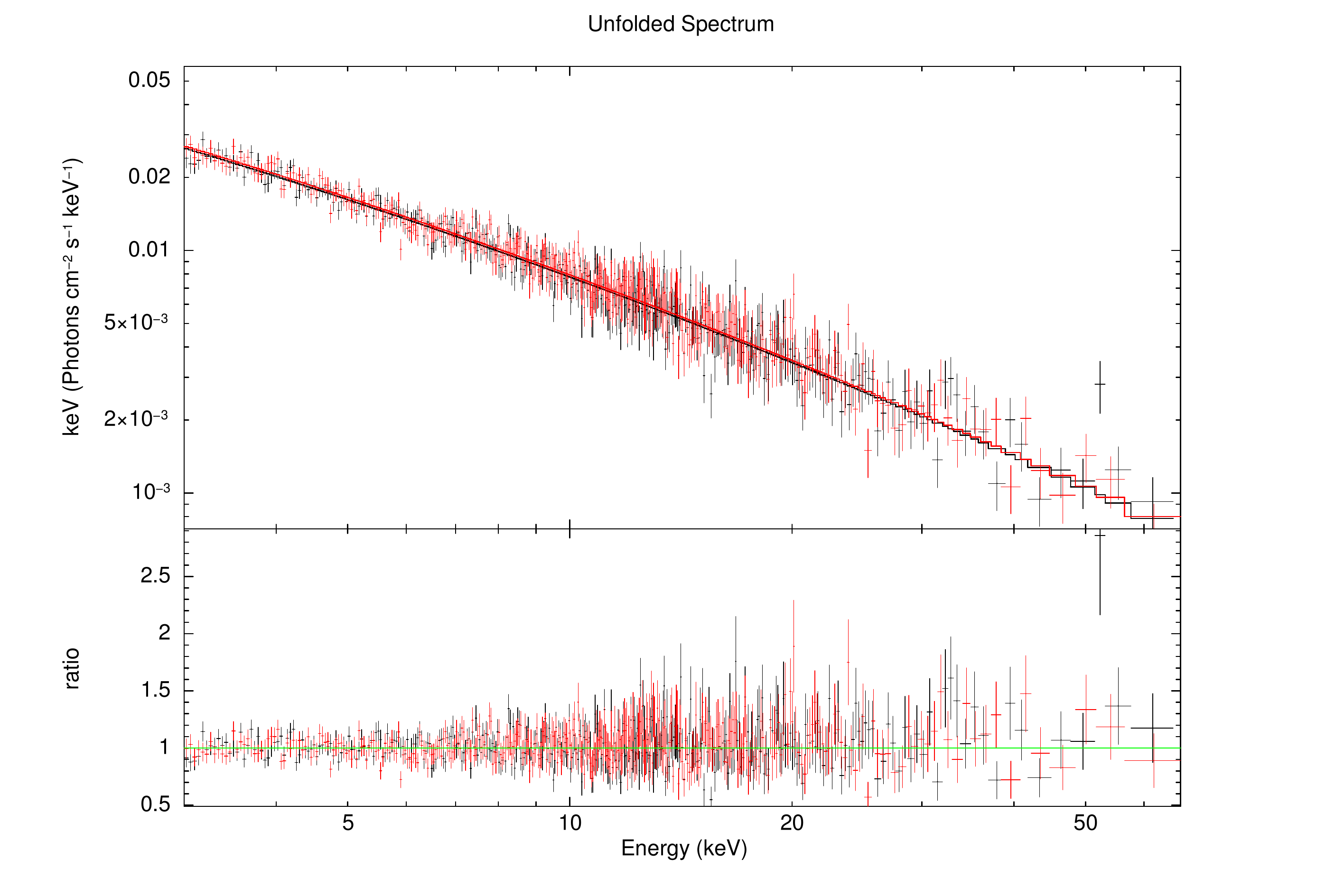}
    \end{multicols}
\center{Mrk 501, 60002024006}

\begin{multicols}{4}
     \includegraphics[width=1.05\linewidth]{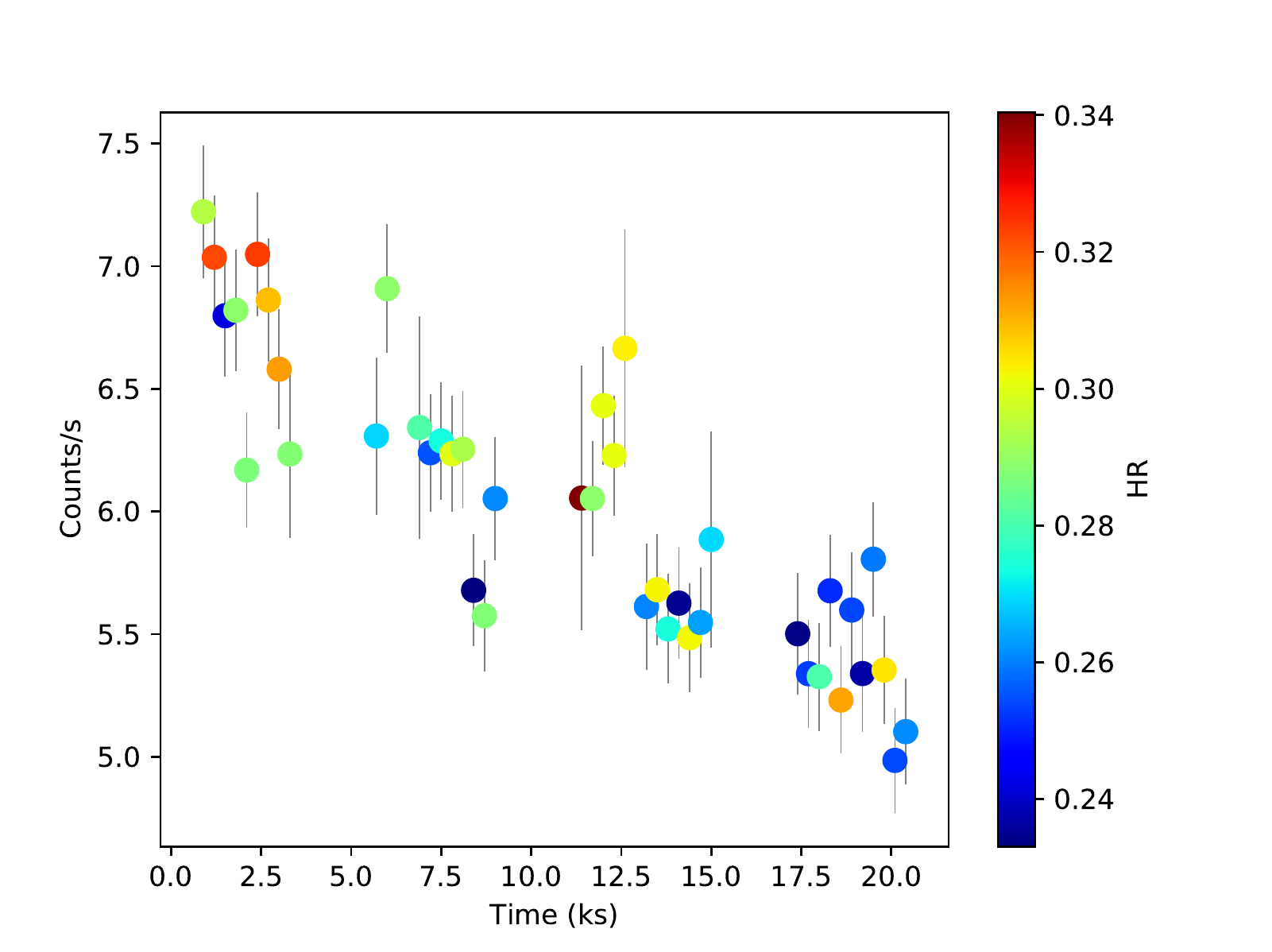}\par 
    \includegraphics[width=1.05\linewidth,angle=0]{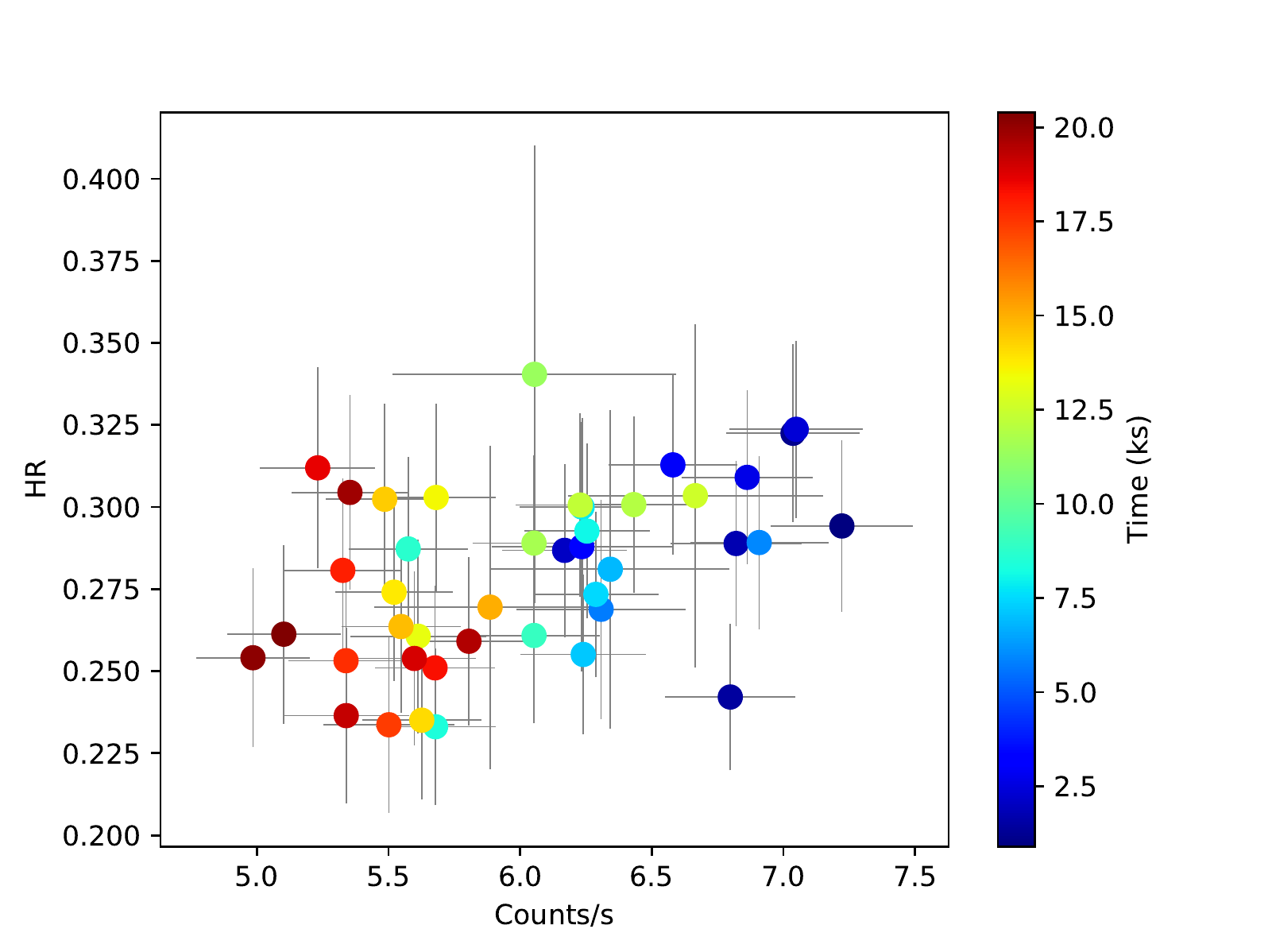}\par
    \includegraphics[width=0.72\linewidth,angle=-90]{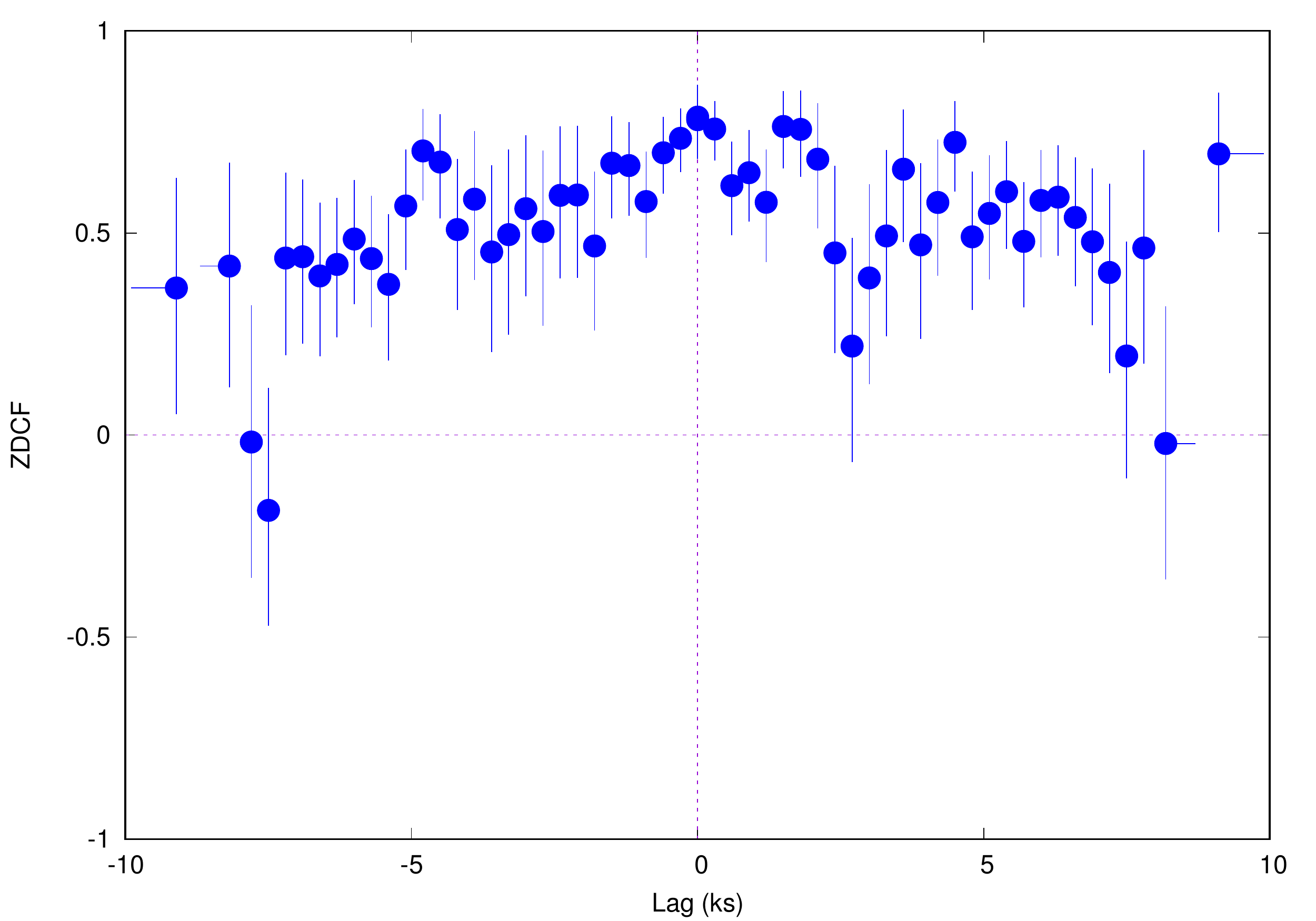}\par
    \includegraphics[width=0.73\linewidth, angle=-90]{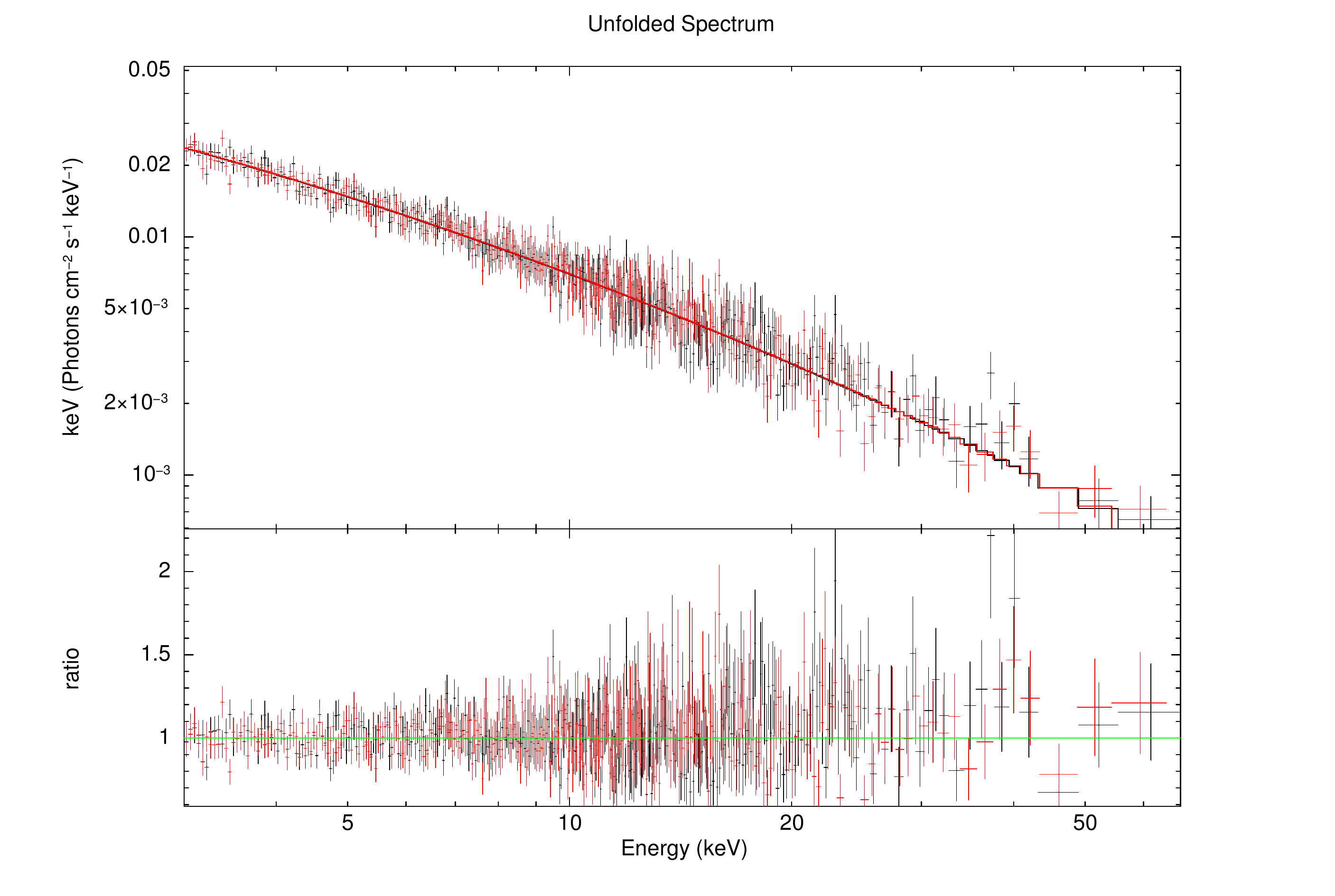}
    \end{multicols}
\center{Mrk 501, 60002024008}

\begin{multicols}{4}
    \includegraphics[width=1.05\linewidth]{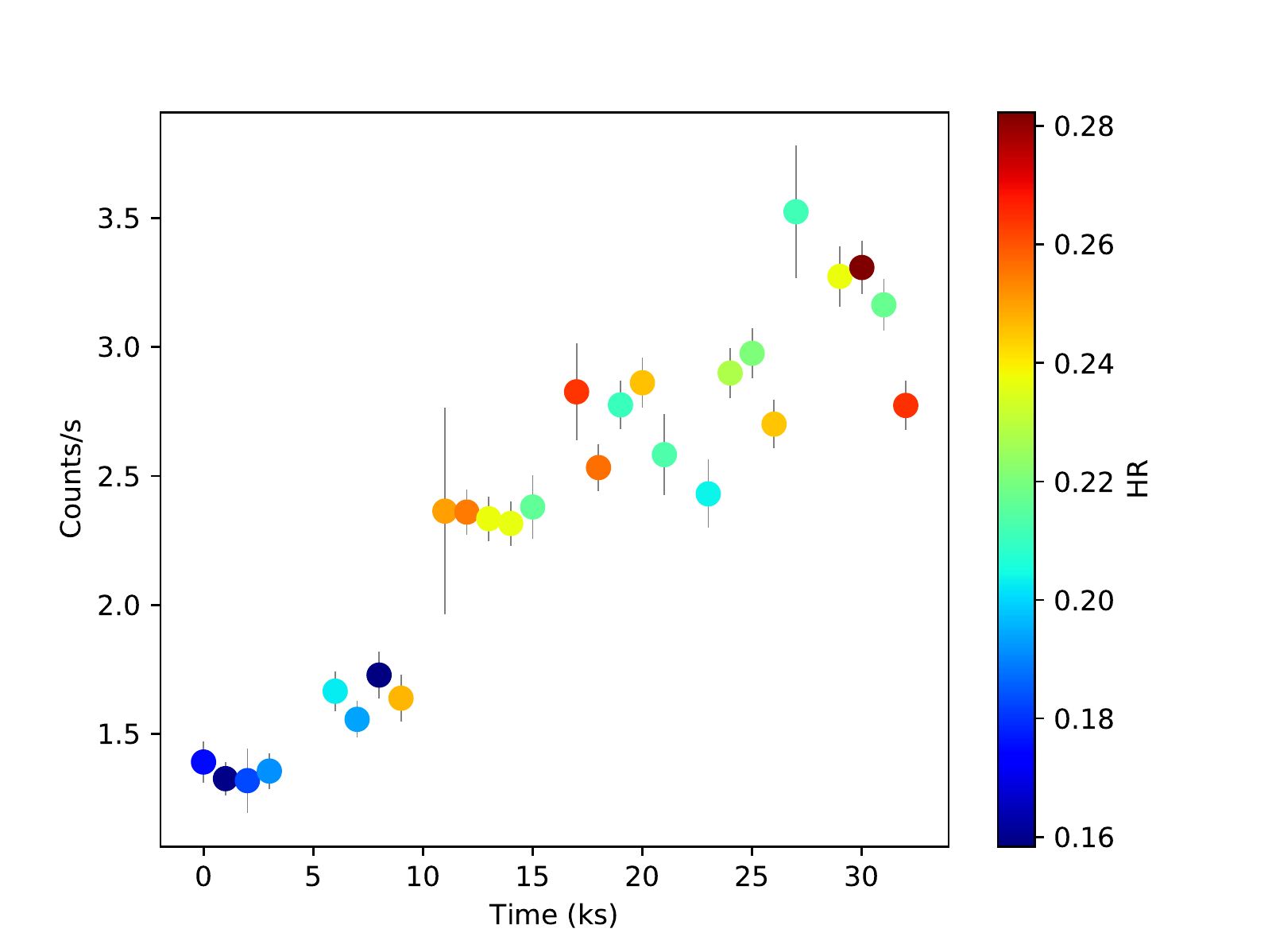}\par 
    \includegraphics[width=1.05\linewidth,angle=0]{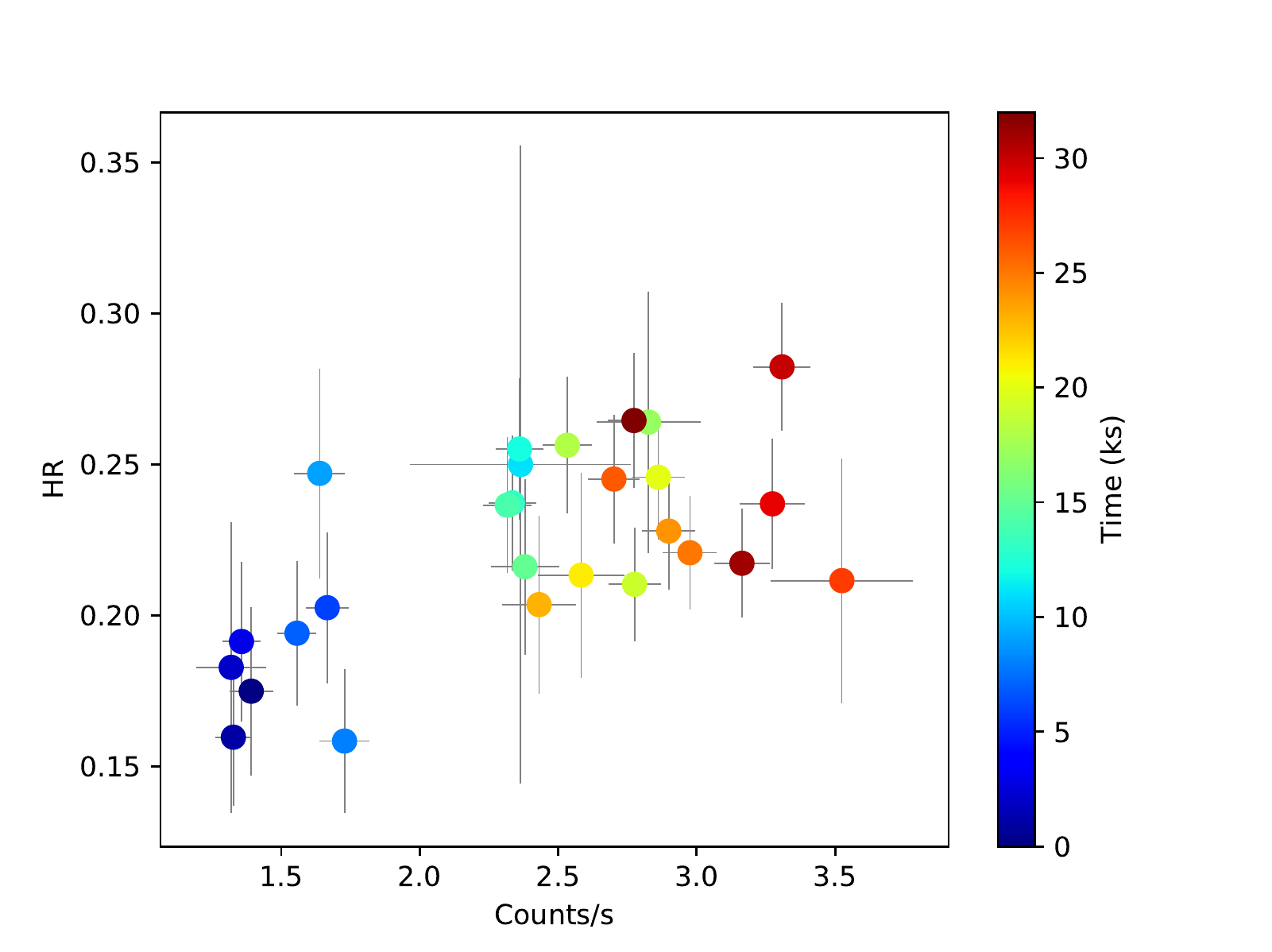}\par
    \includegraphics[width=0.72\linewidth,angle=-90]{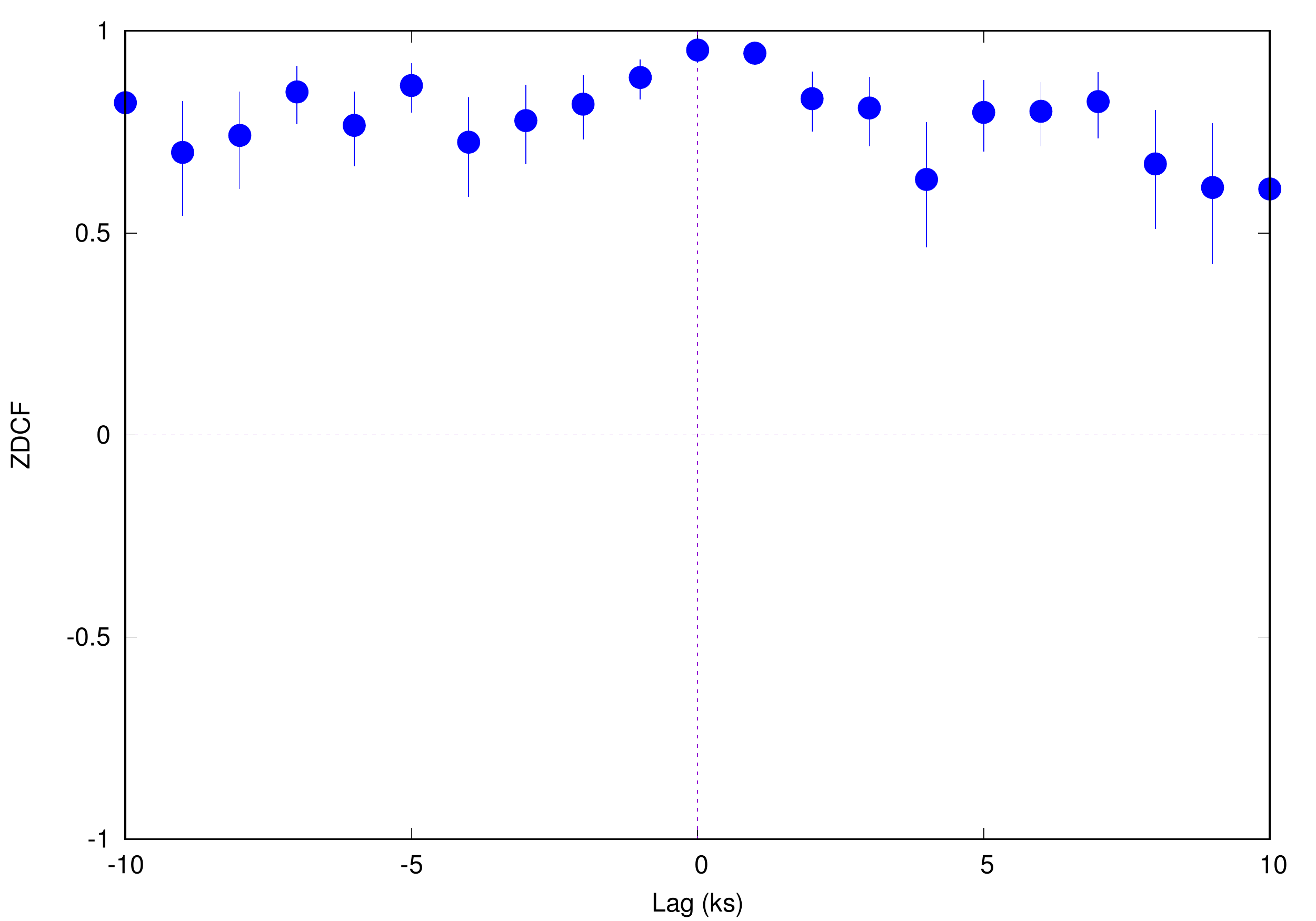}\par
    \includegraphics[width=0.73\linewidth, angle=-90]{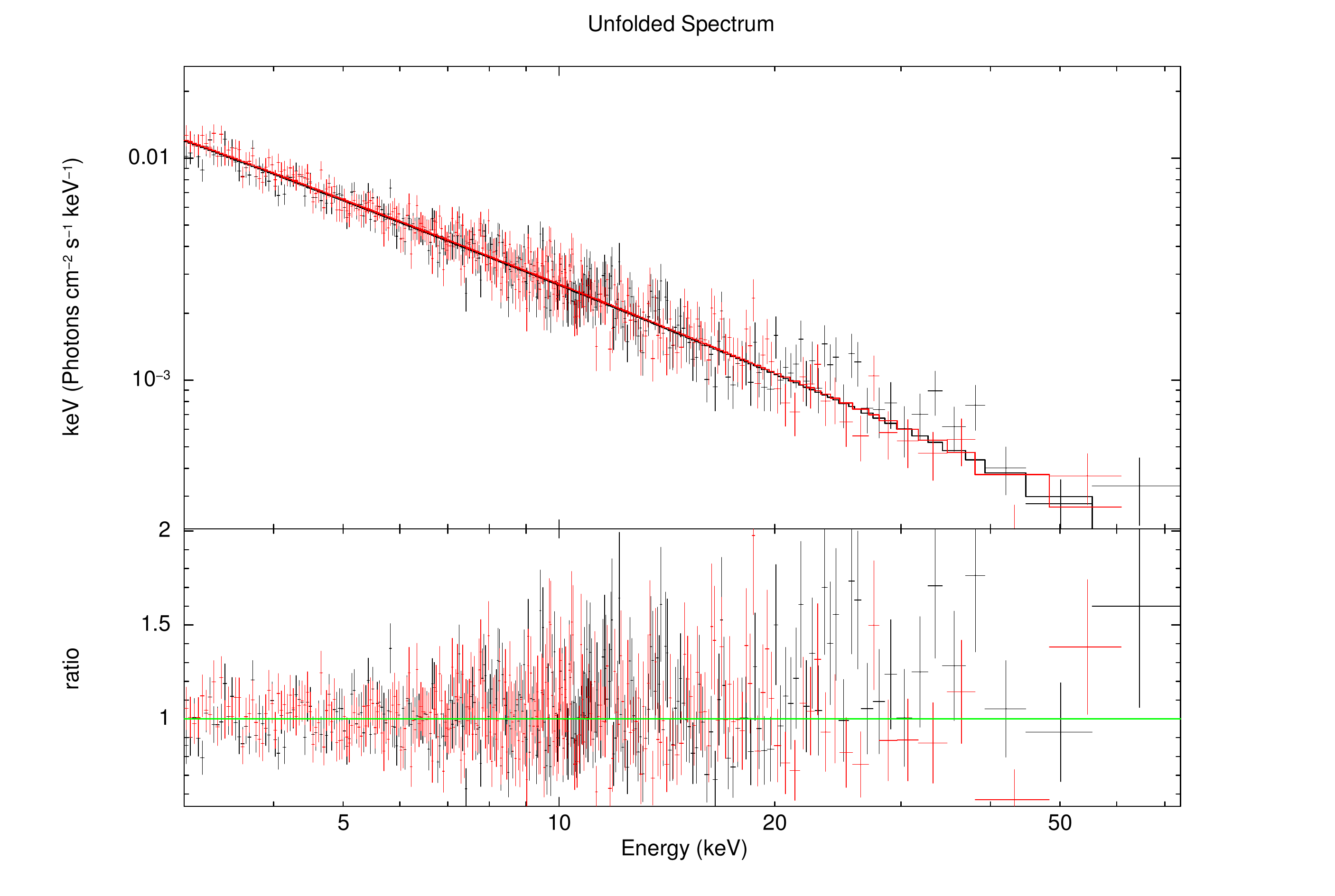}
    \end{multicols}
\center{1ES 1959+650, 60002055002}

\begin{multicols}{4}
    \includegraphics[width=1.05\linewidth]{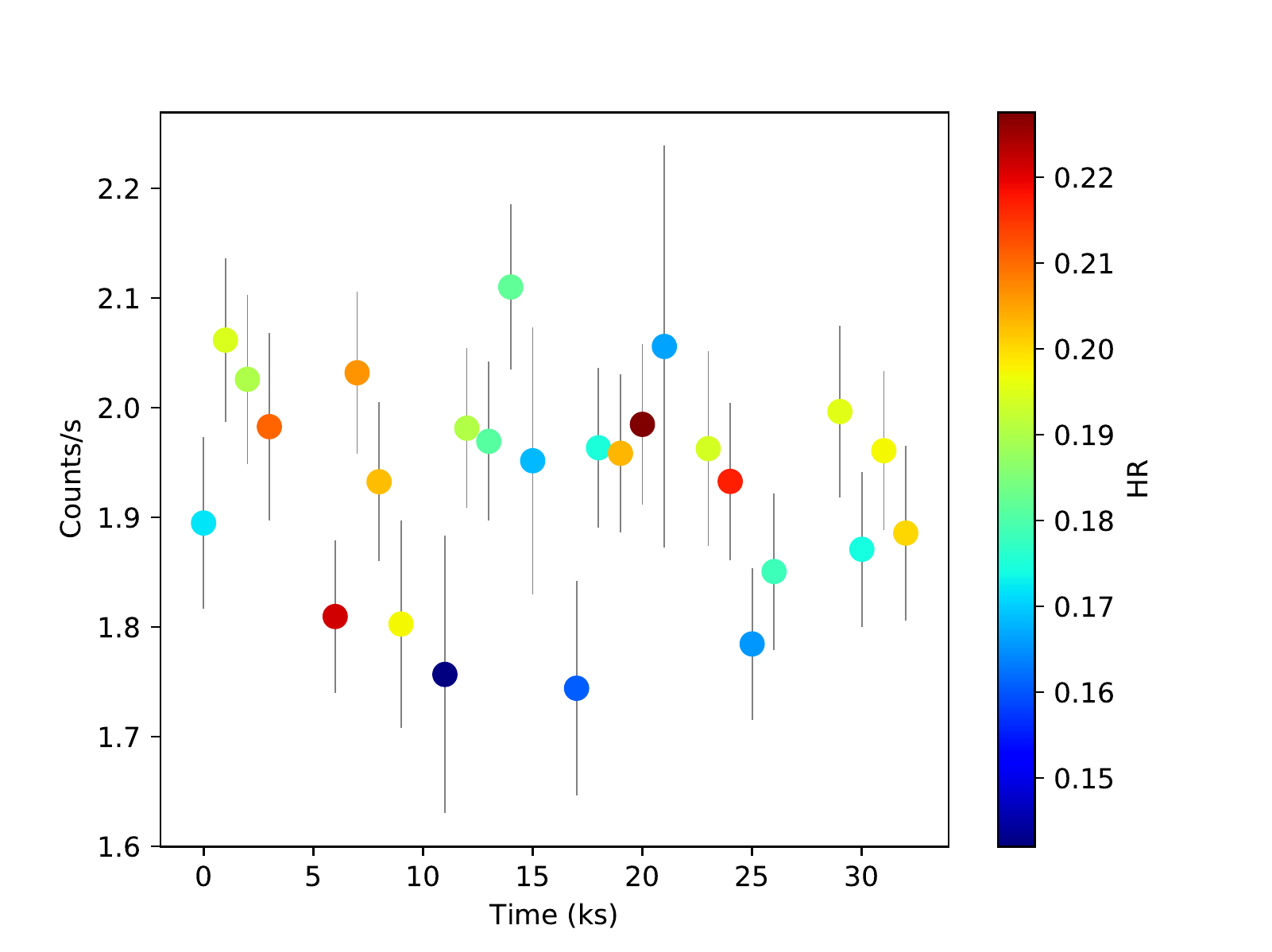}\par 
    \includegraphics[width=1.05\linewidth,angle=0]{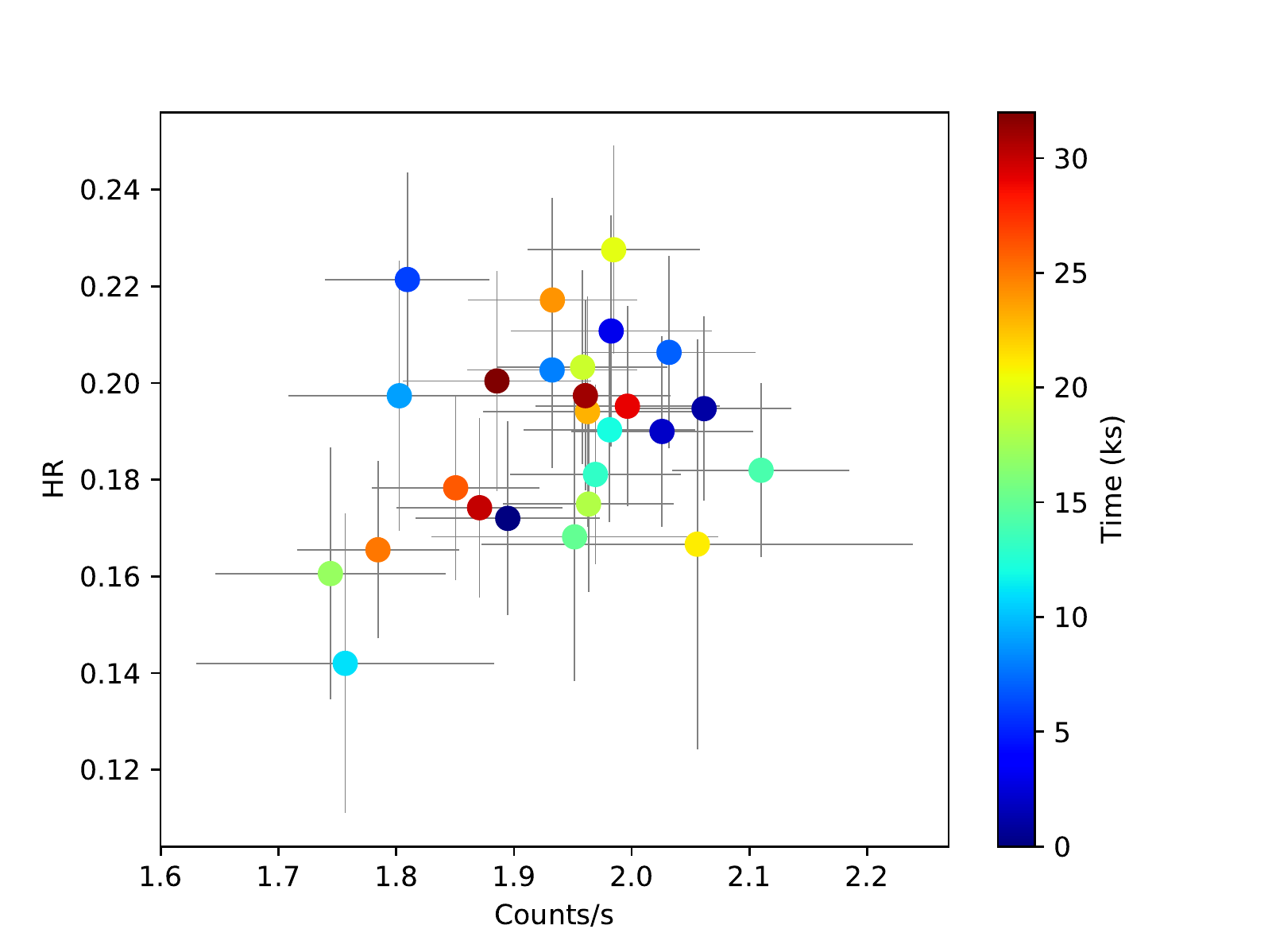}\par
    \includegraphics[width=0.72\linewidth,angle=-90]{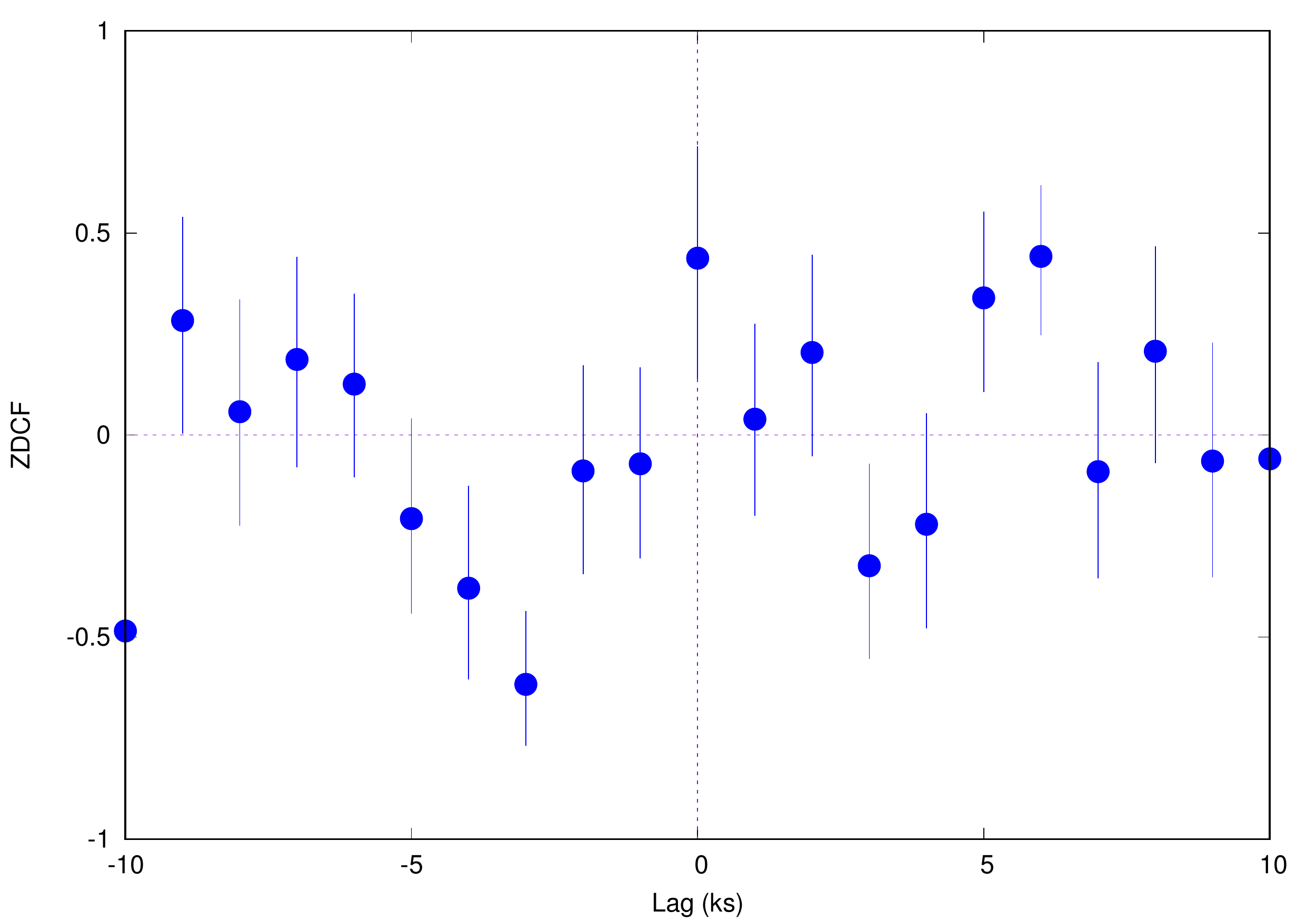}\par
    \includegraphics[width=0.73\linewidth, angle=-90]{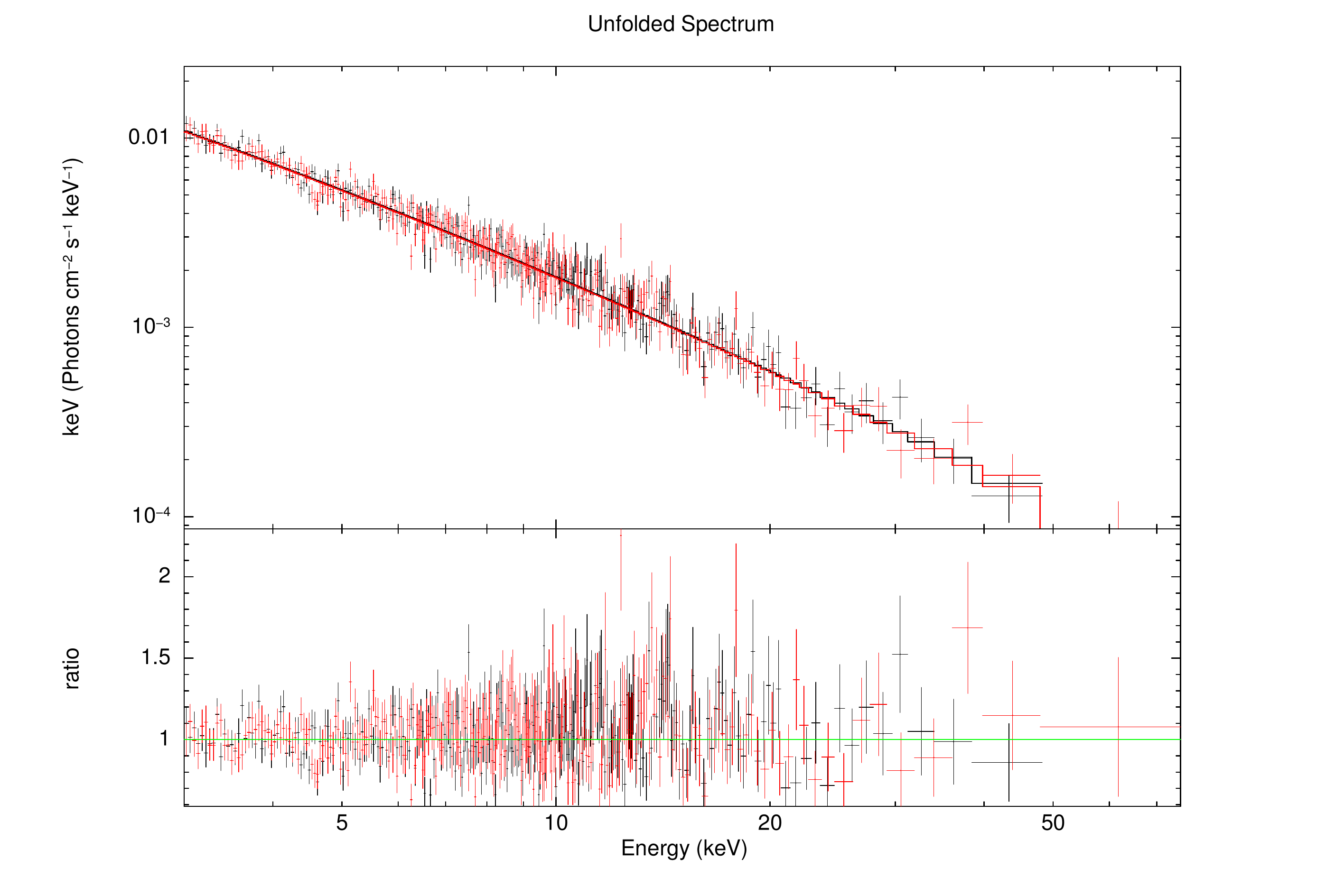}
    \end{multicols}
\center{1ES 1959+650, 60002055004} \\
\caption{Same as in Fig \ref{fig:LC1}}.
\label{fig:LC3}
\end{figure*}


\begin{figure*}
\begin{multicols}{4}
    \includegraphics[width=1.05\linewidth]{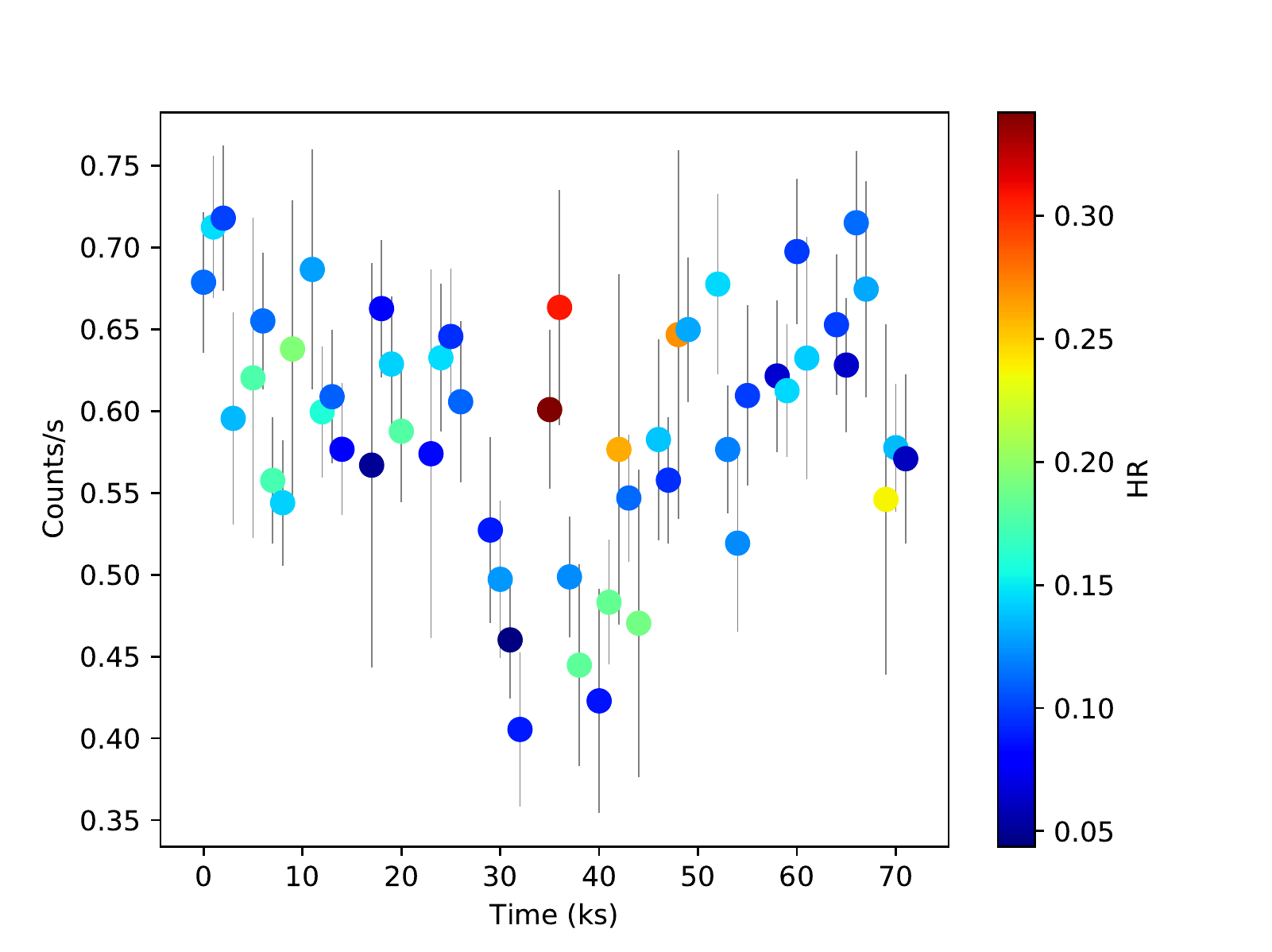}\par 
    \includegraphics[width=1.05\linewidth,angle=0]{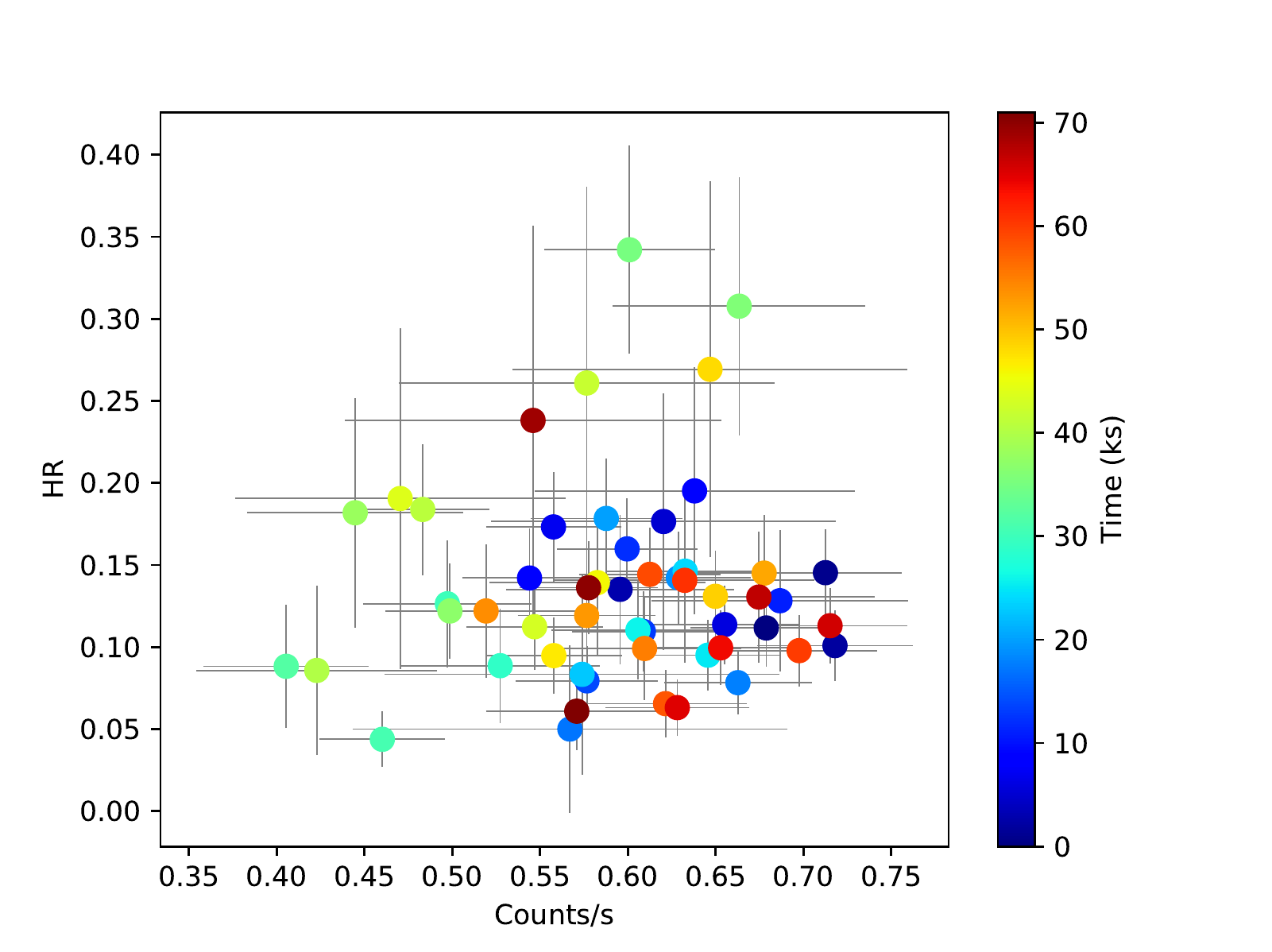}\par
    \includegraphics[width=0.72\linewidth,angle=-90]{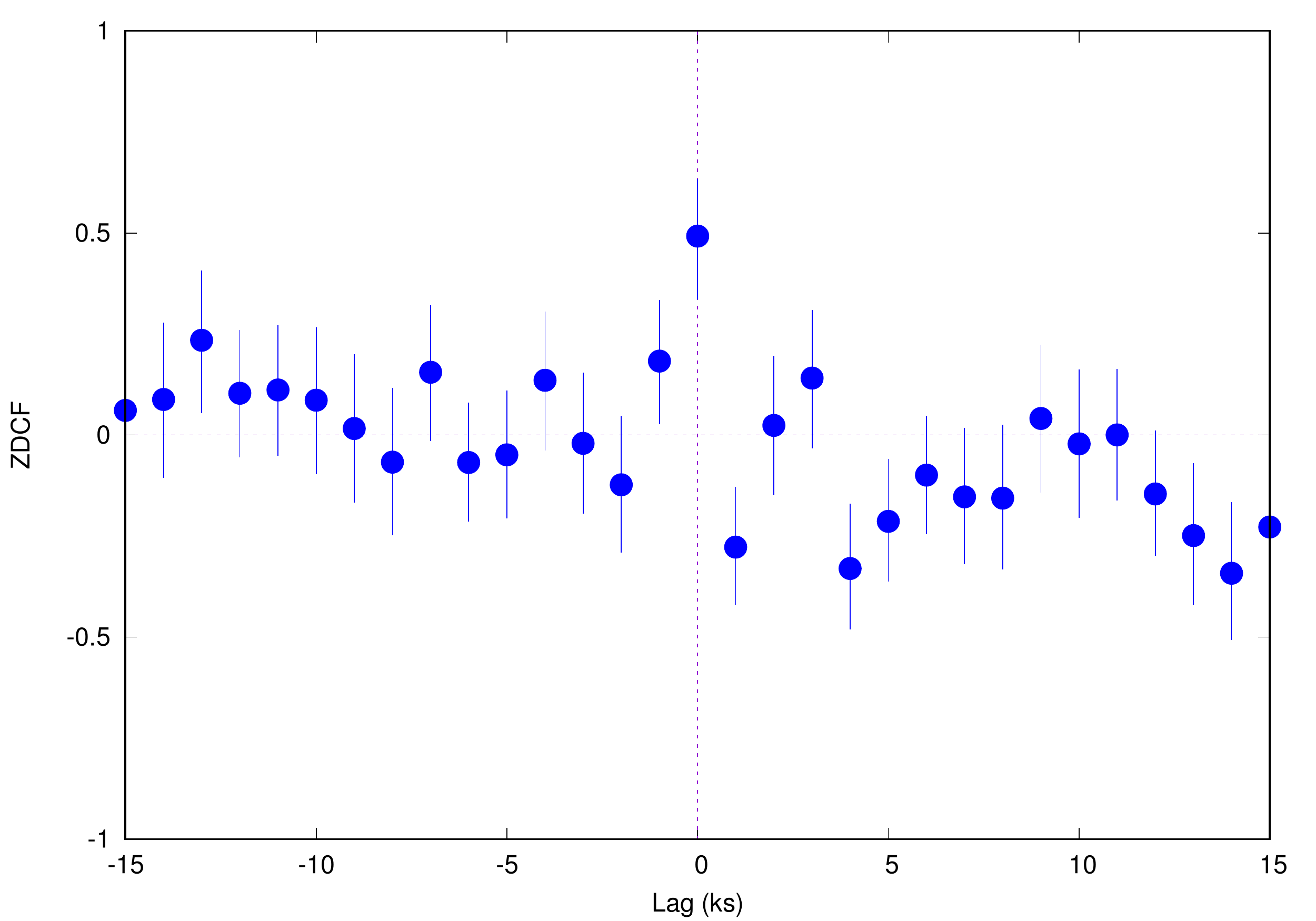}\par
    \includegraphics[width=0.73\linewidth, angle=-90]{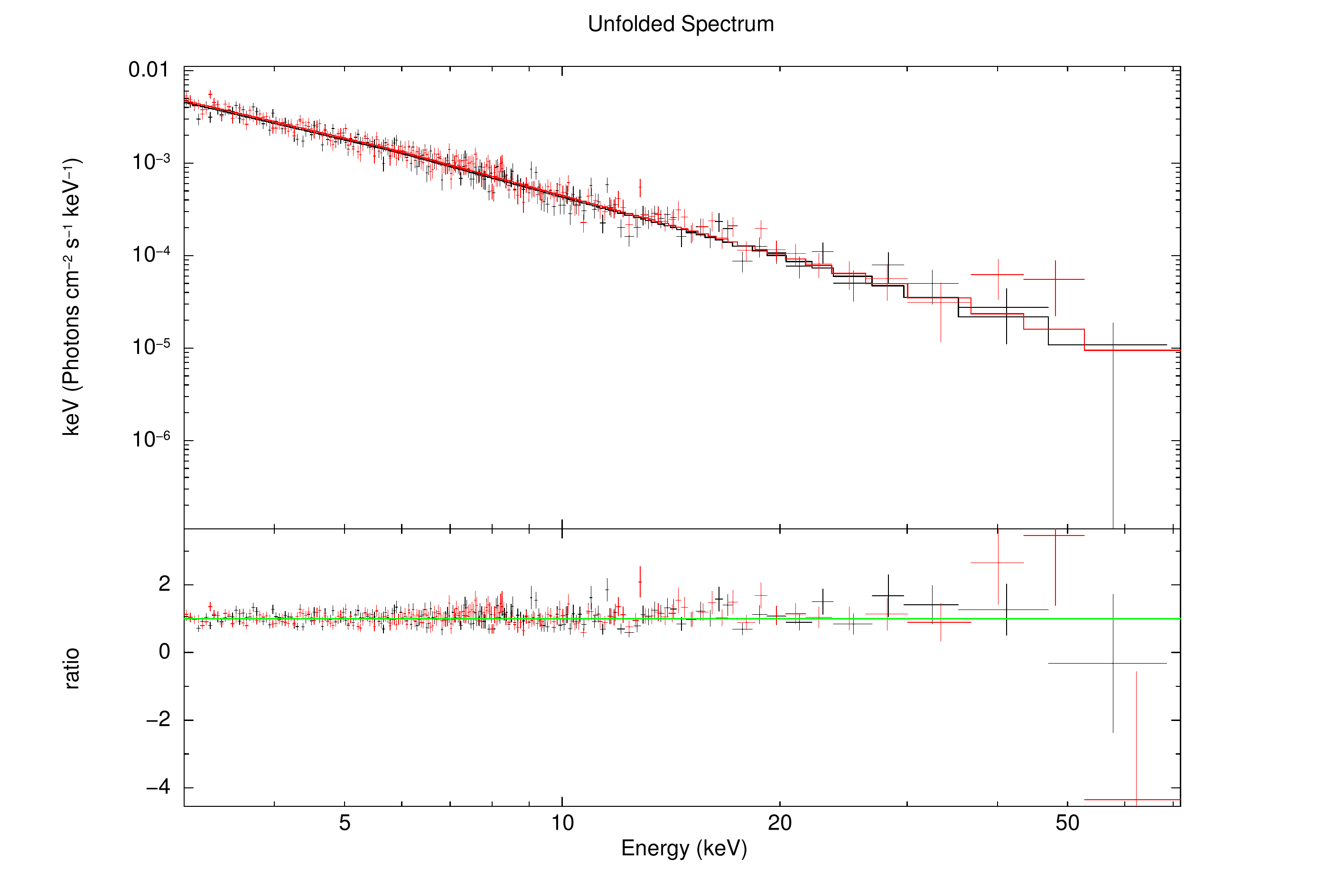}
     \end{multicols}
\center{PKS 2155--304, 10002010001 }

\begin{multicols}{4}
    \includegraphics[width=1.05\linewidth]{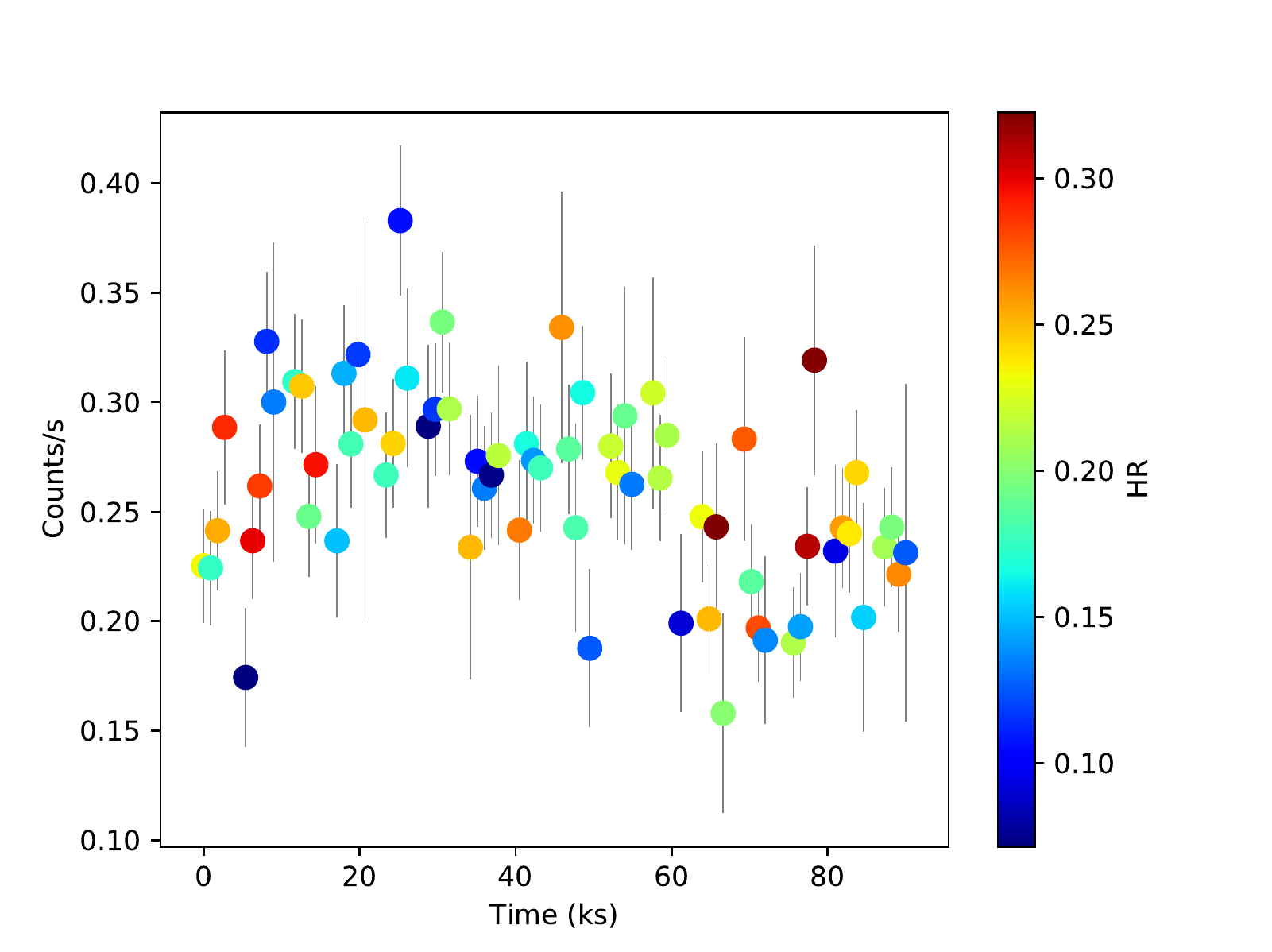}\par 
    \includegraphics[width=1.05\linewidth,angle=0]{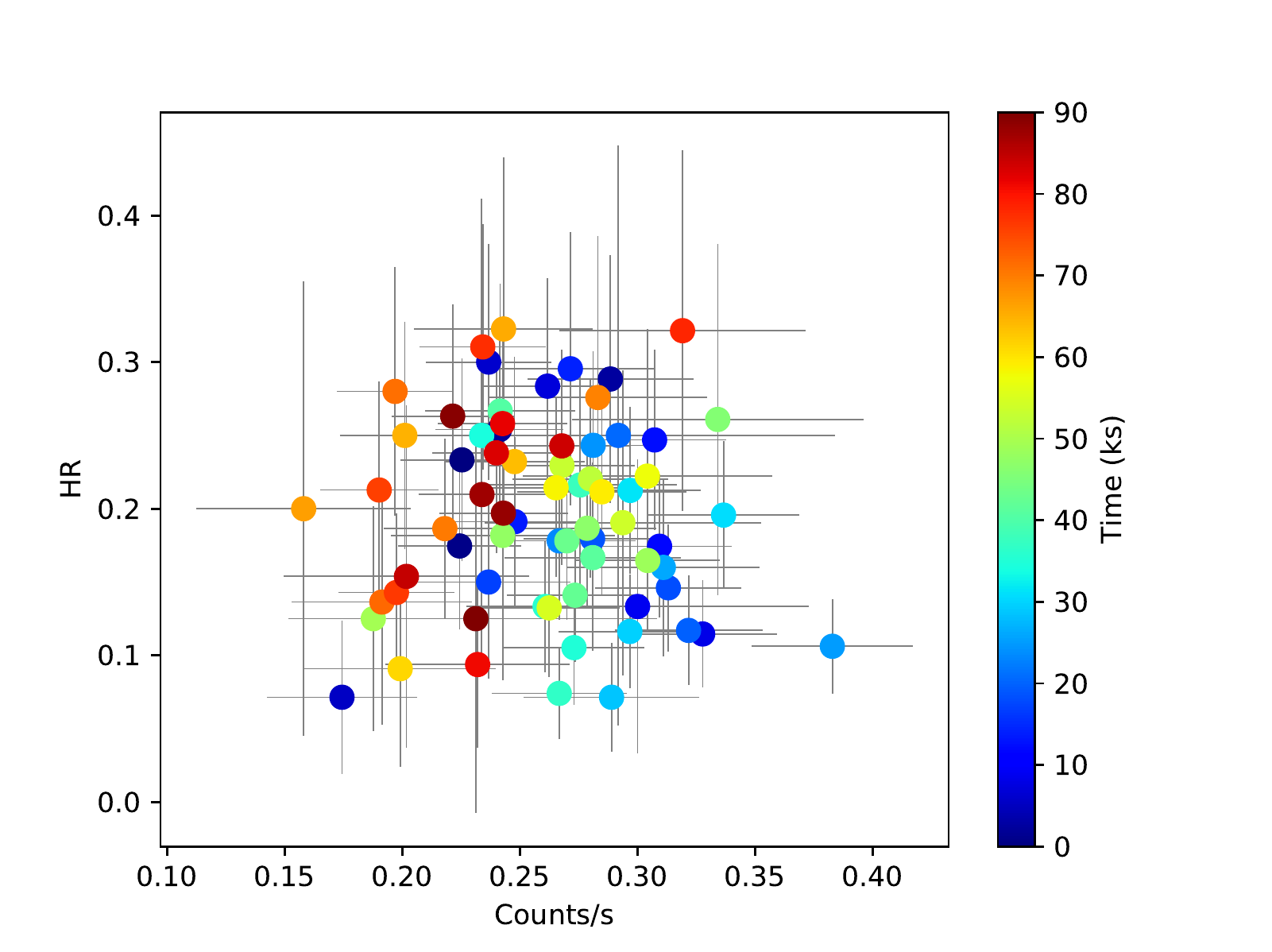}\par
    \includegraphics[width=0.72\linewidth,angle=-90]{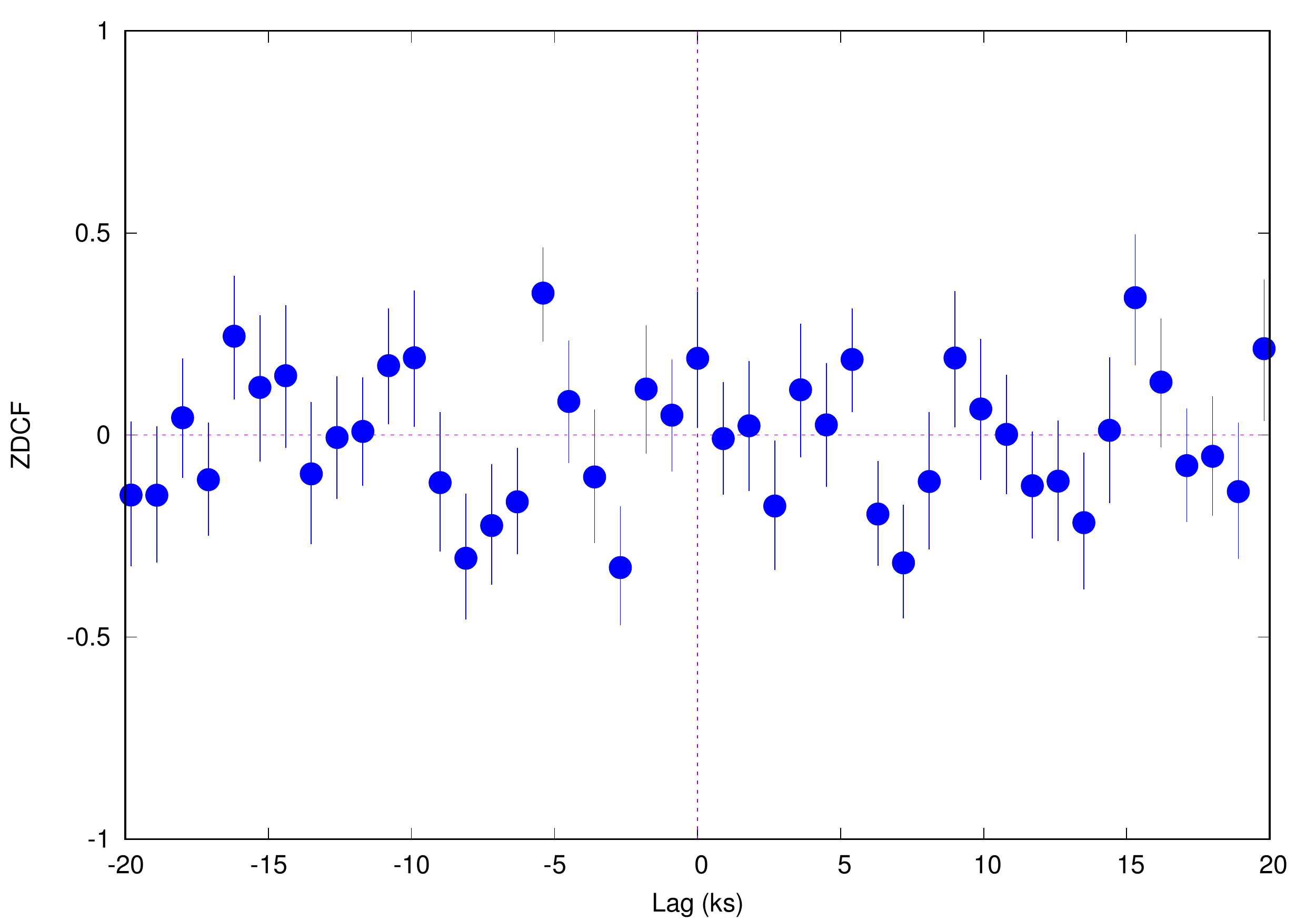}\par
    \includegraphics[width=0.73\linewidth, angle=-90]{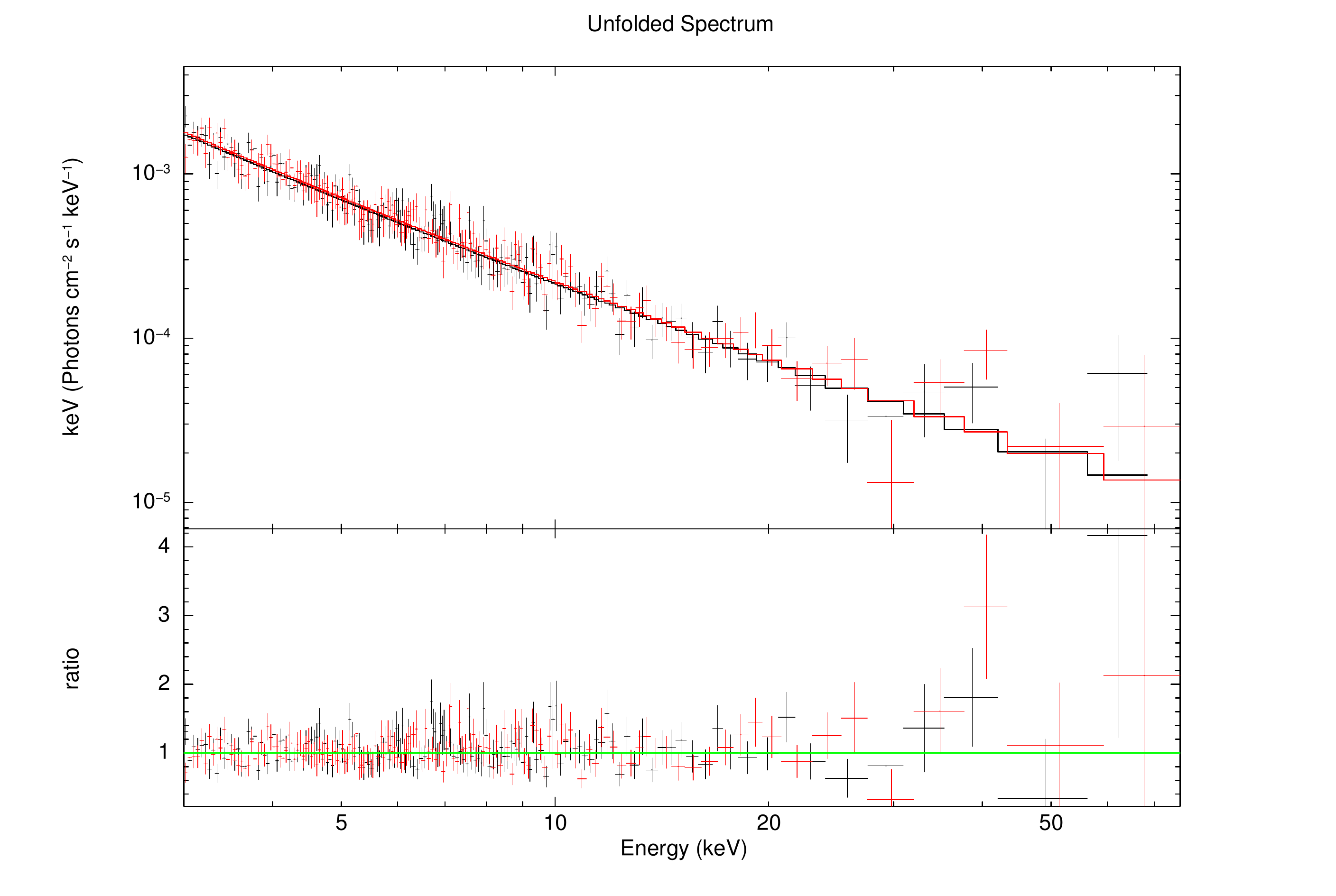}
    \end{multicols}
\center{PKS 2155--304, 60002022002 }

\begin{multicols}{4}
     \includegraphics[width=1.05\linewidth]{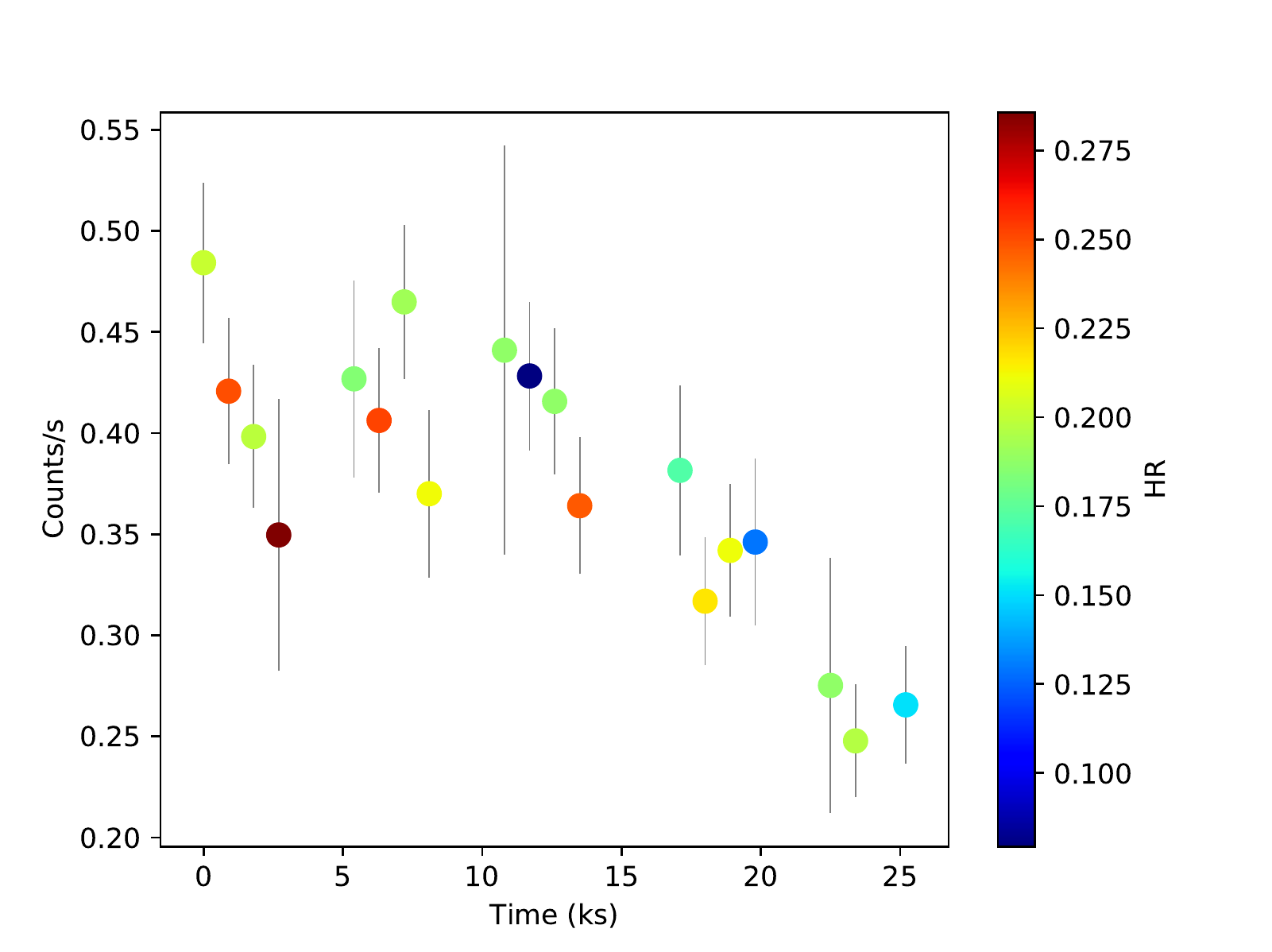}\par 
    \includegraphics[width=1.05\linewidth,angle=0]{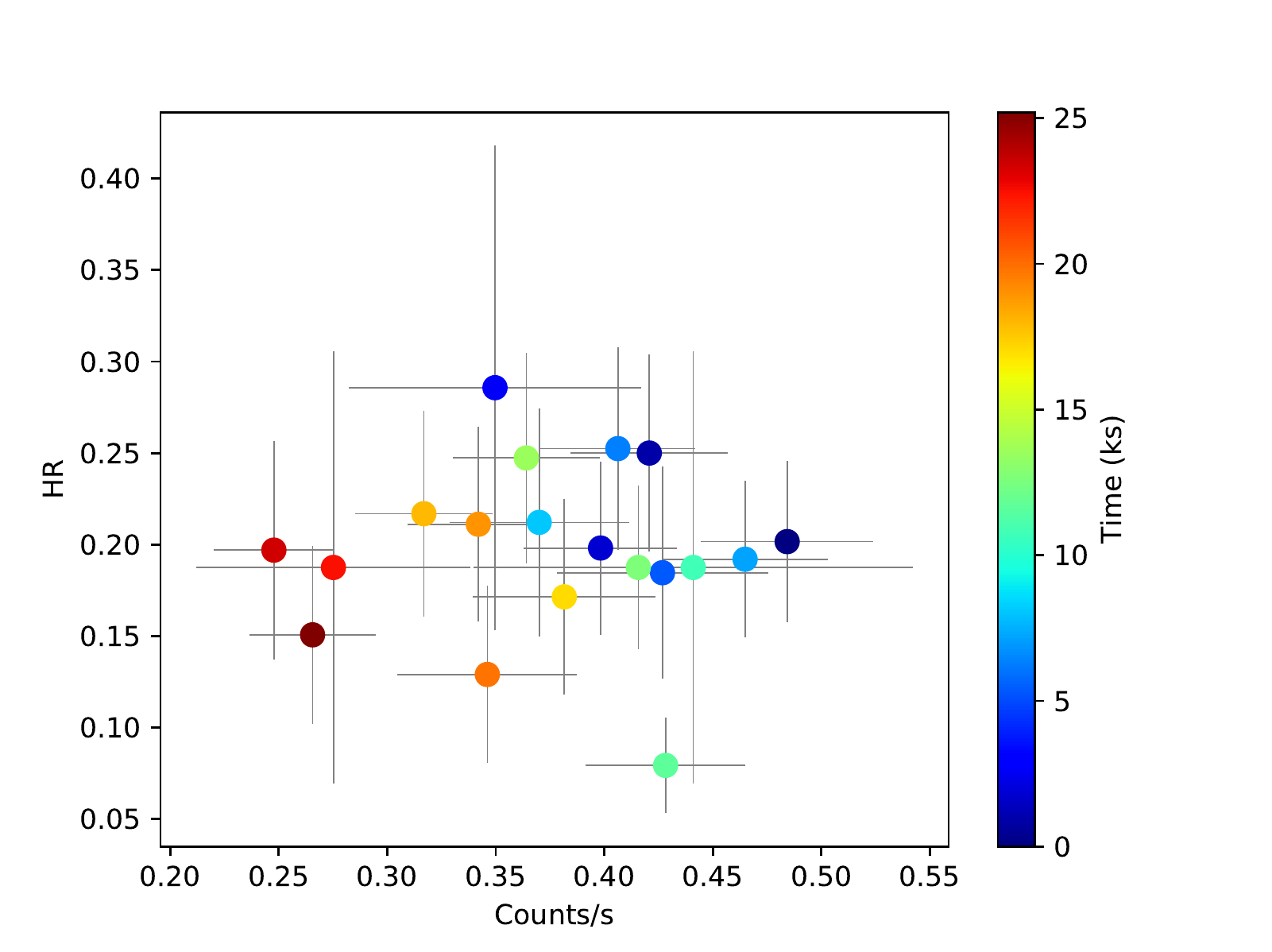}\par
    \includegraphics[width=0.72\linewidth,angle=-90]{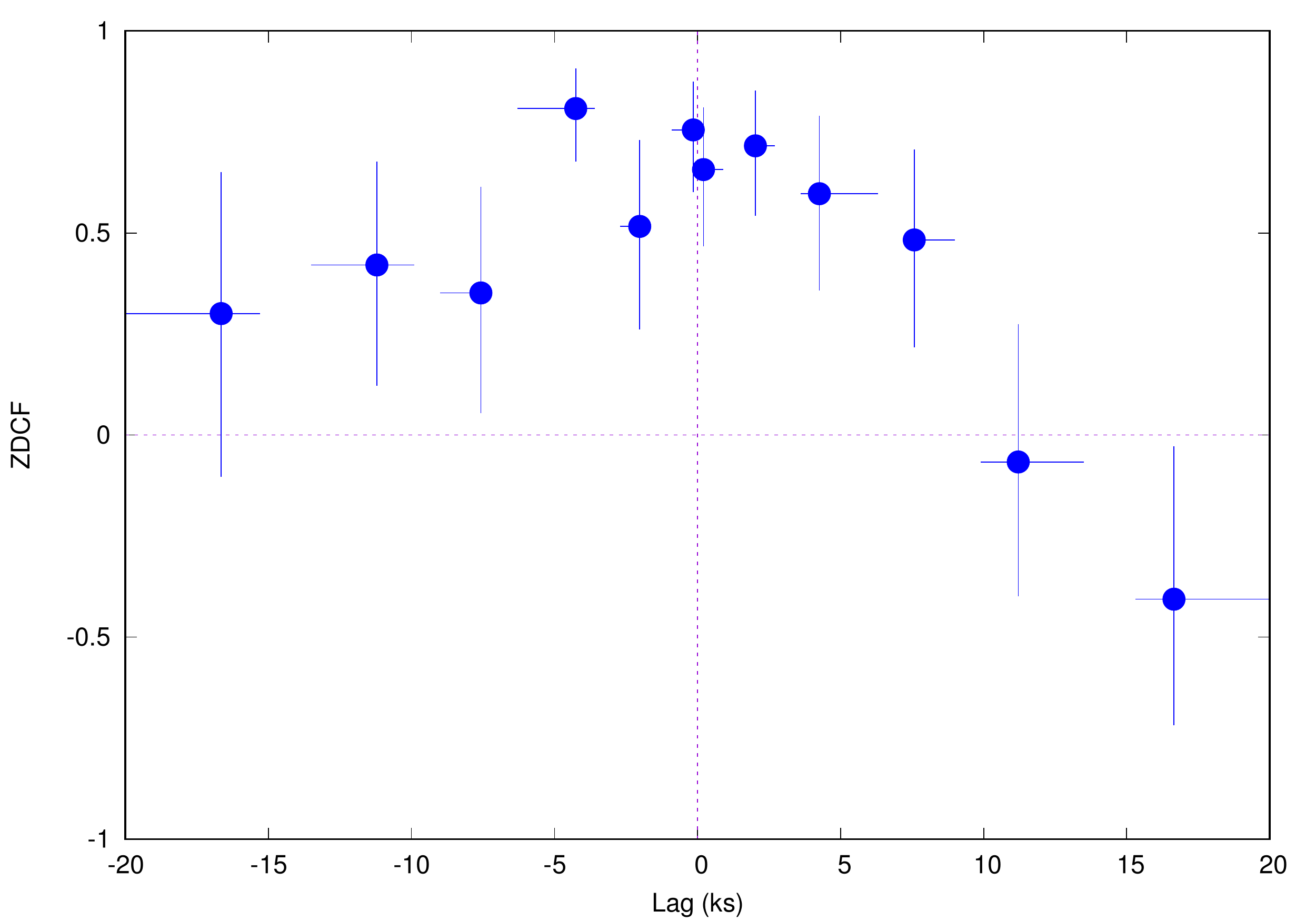}\par
    \includegraphics[width=0.73\linewidth, angle=-90]{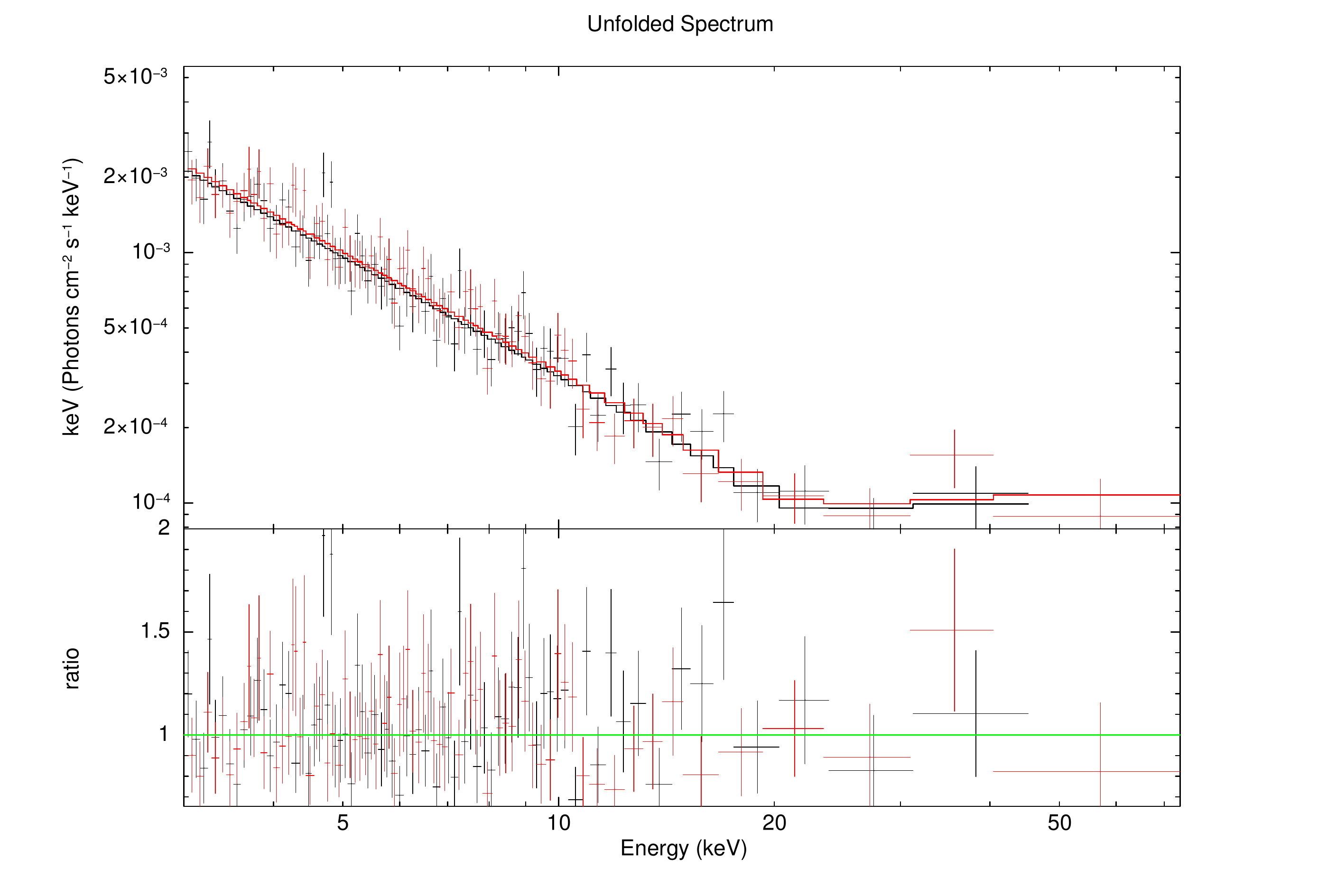}
    \end{multicols}
\center{PKS 2155--304, 60002022004 }

\begin{multicols}{4}
    \includegraphics[width=1.05\linewidth]{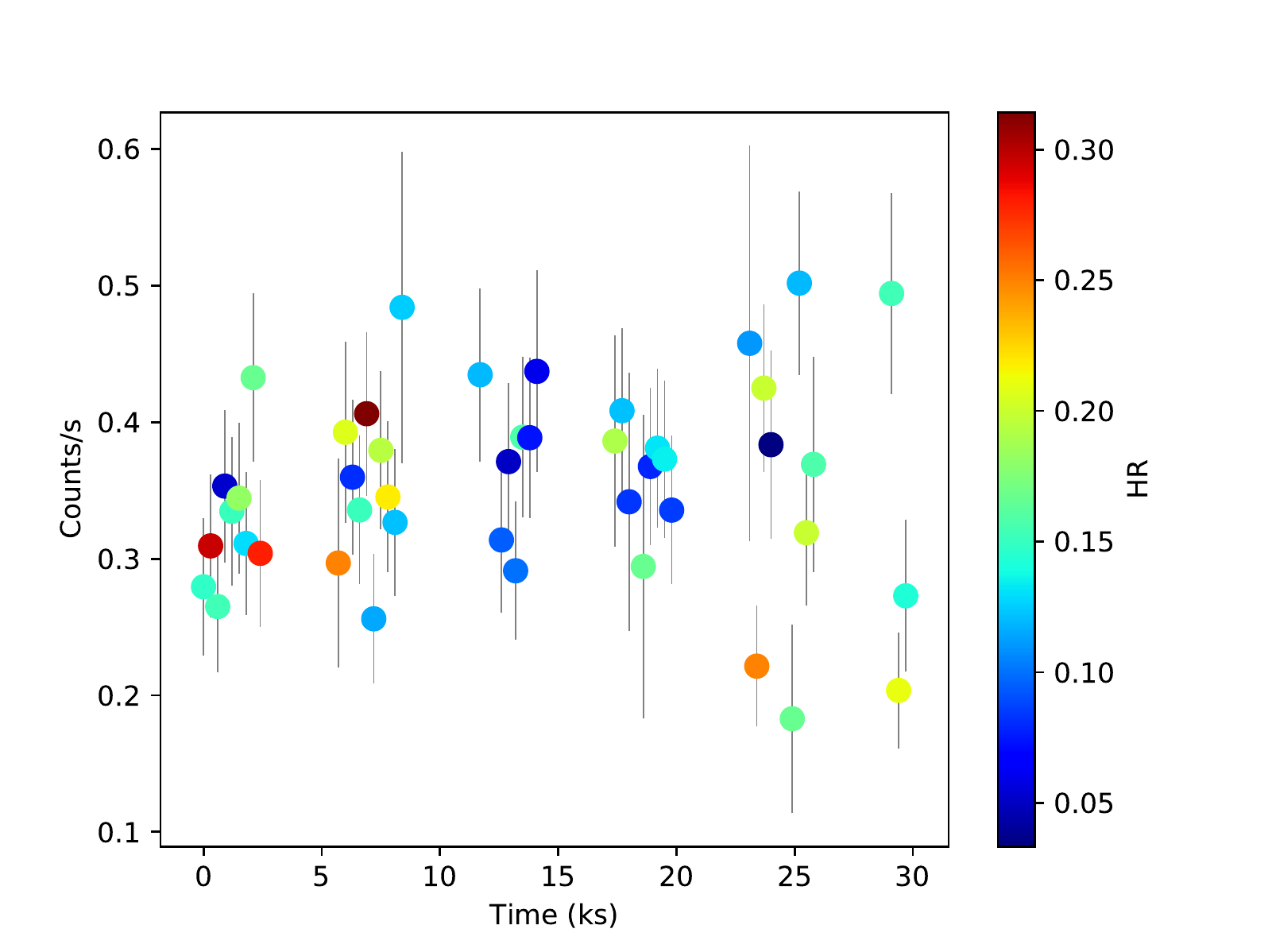}\par 
    \includegraphics[width=1.05\linewidth,angle=0]{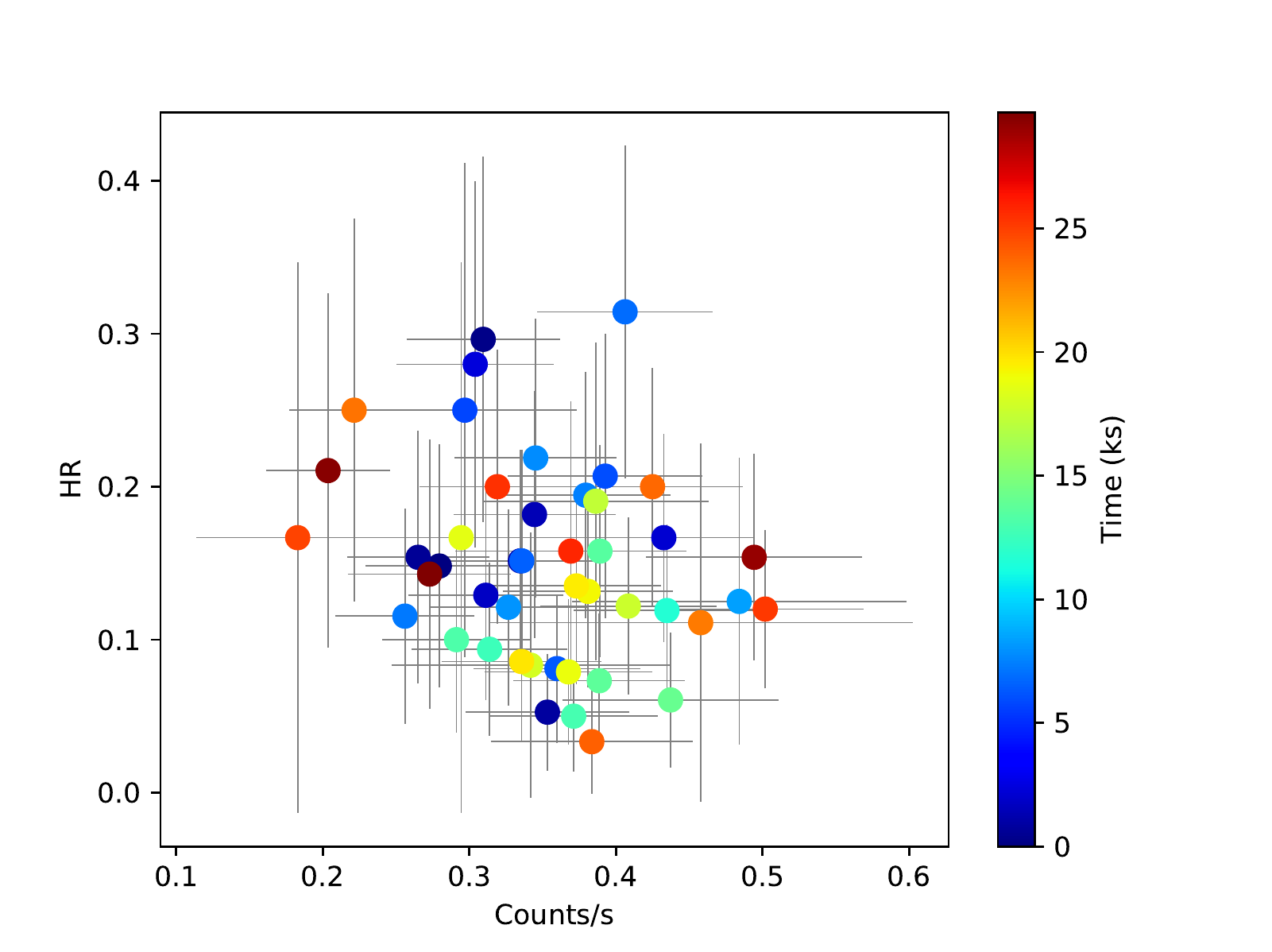}\par
    \includegraphics[width=0.72\linewidth,angle=-90]{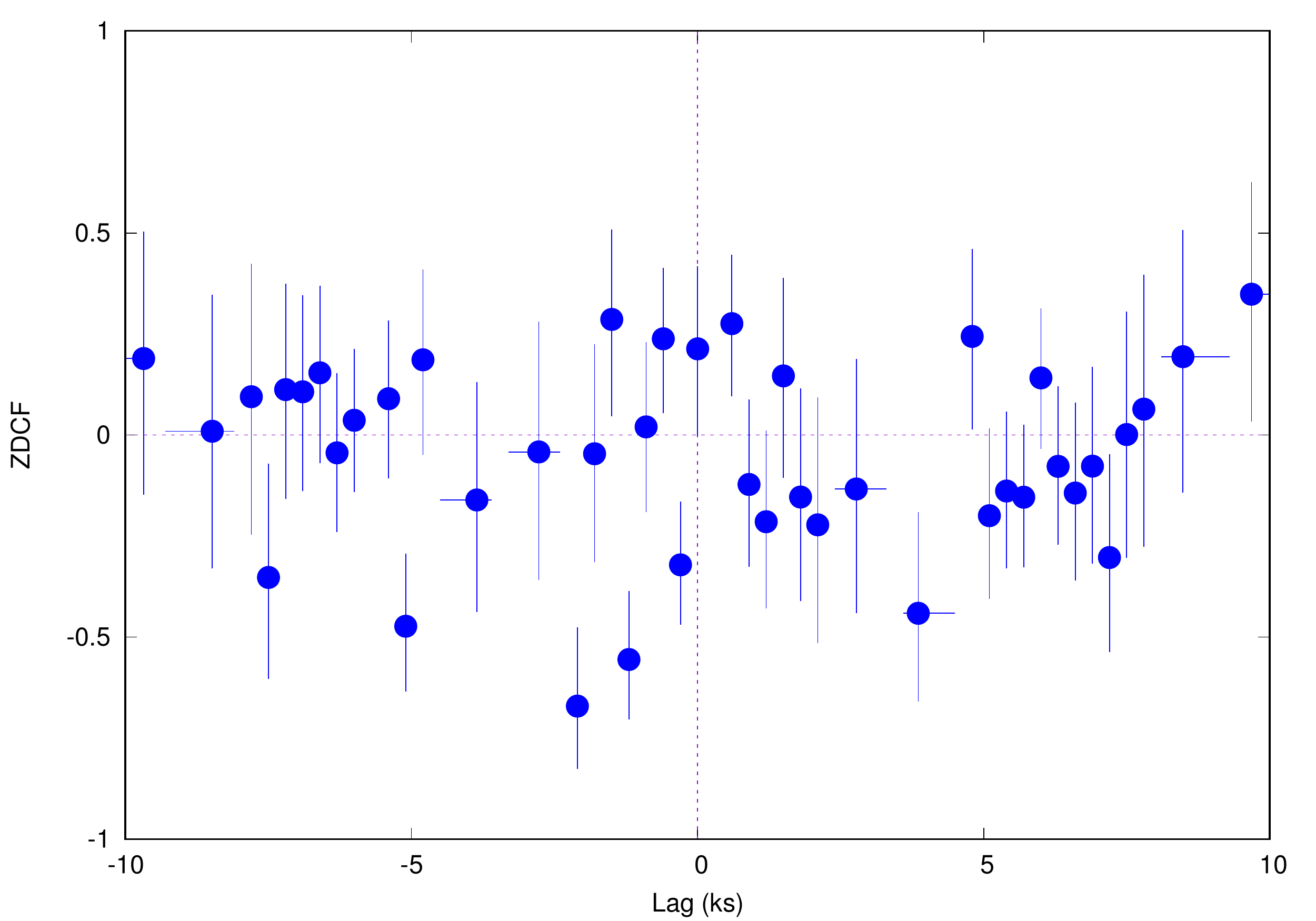}\par
    \includegraphics[width=0.73\linewidth, angle=-90]{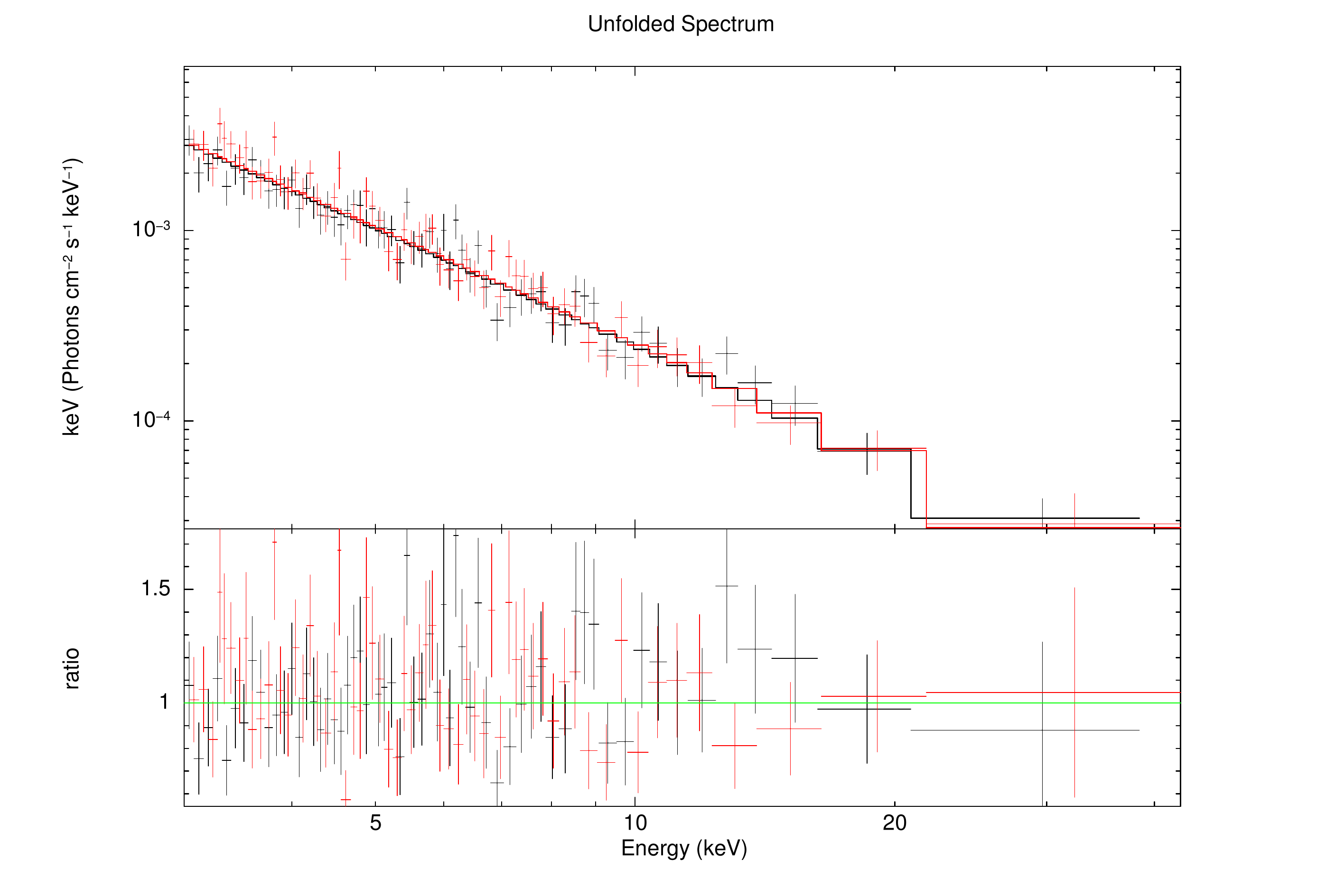}
    \end{multicols}
\center{PKS 2155--304, 60002022006 }

\begin{multicols}{4}
    \includegraphics[width=1.05\linewidth]{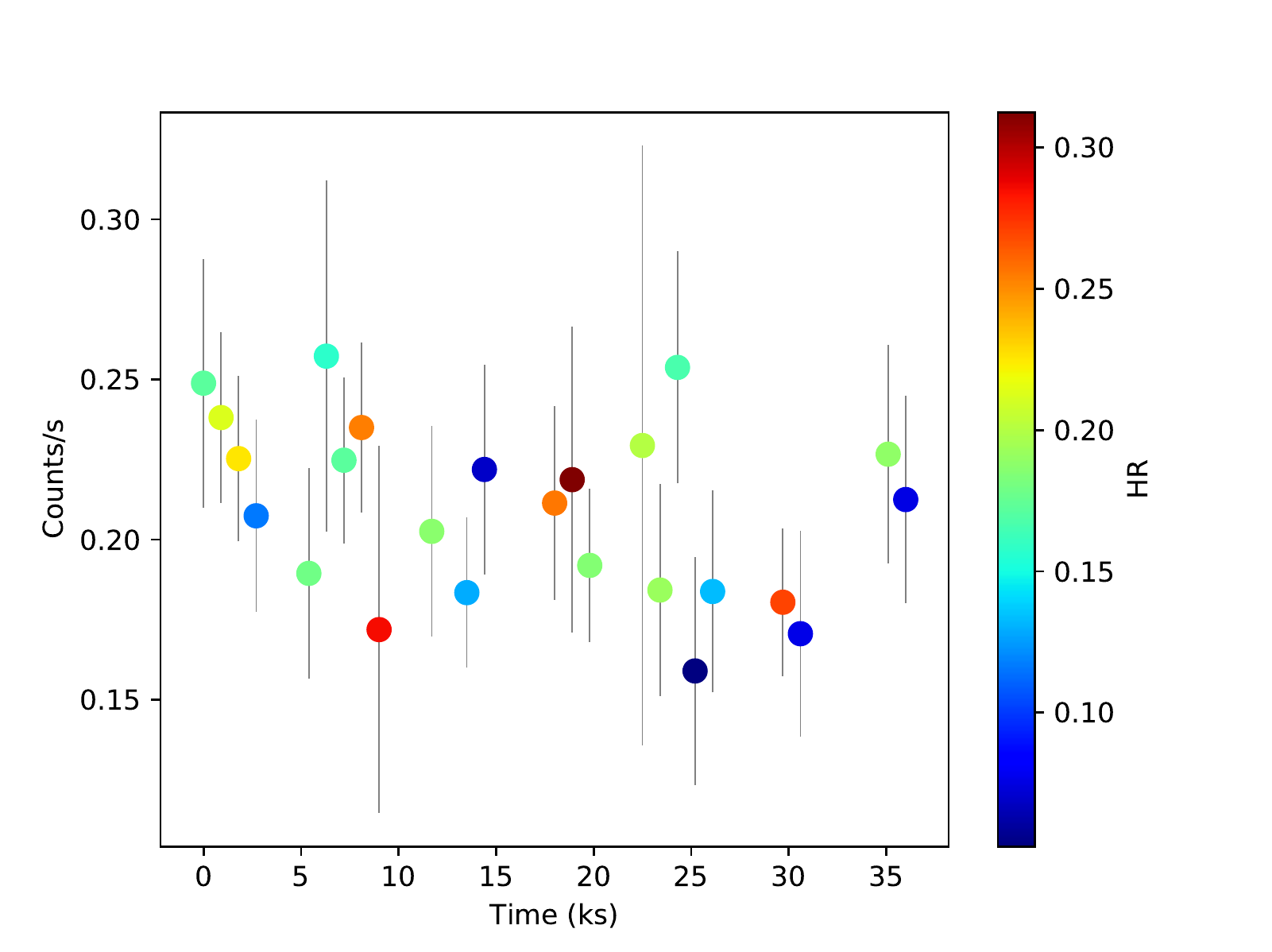}\par 
    \includegraphics[width=1.05\linewidth,angle=0]{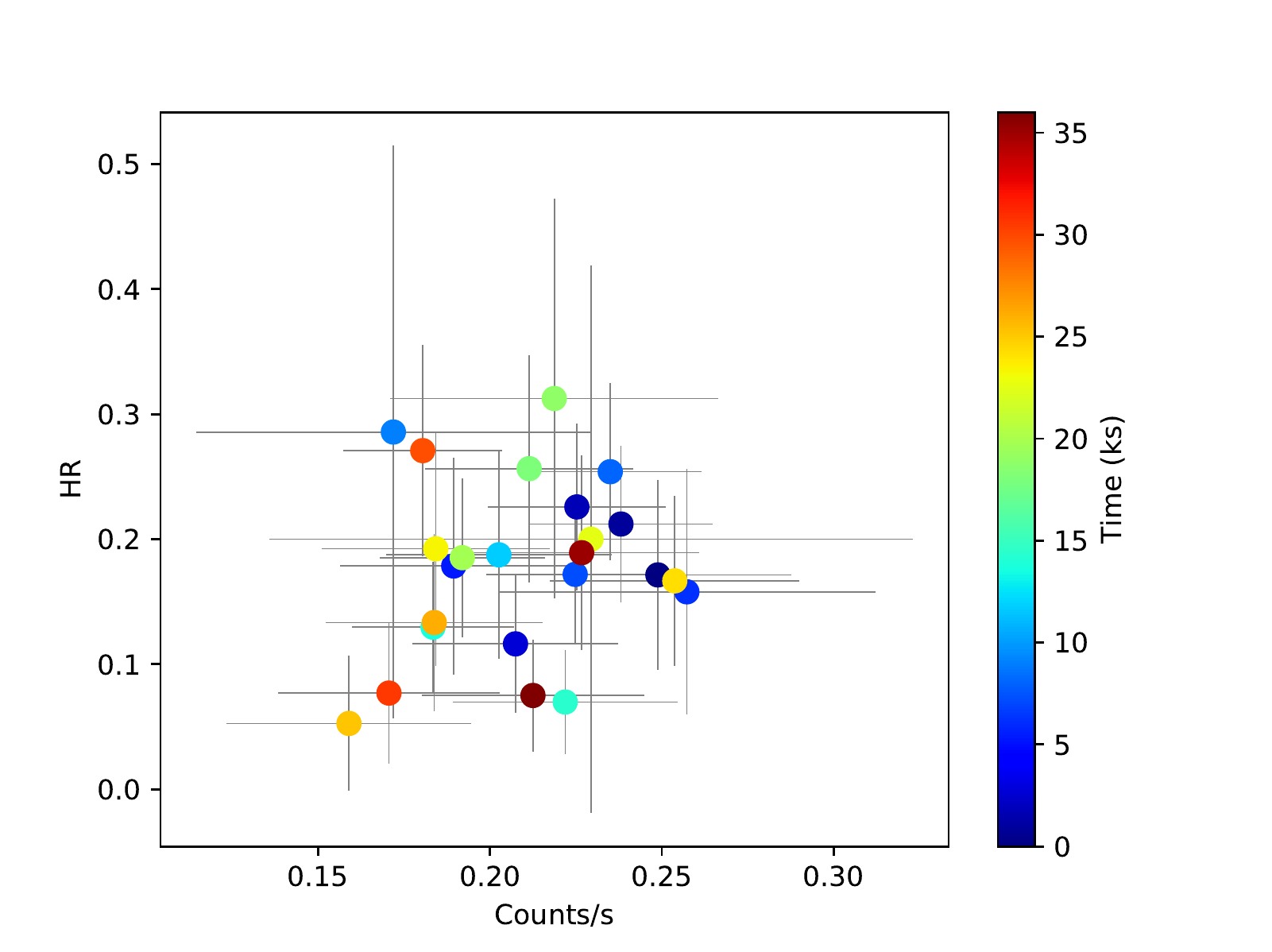}\par
    \includegraphics[width=0.72\linewidth,angle=-90]{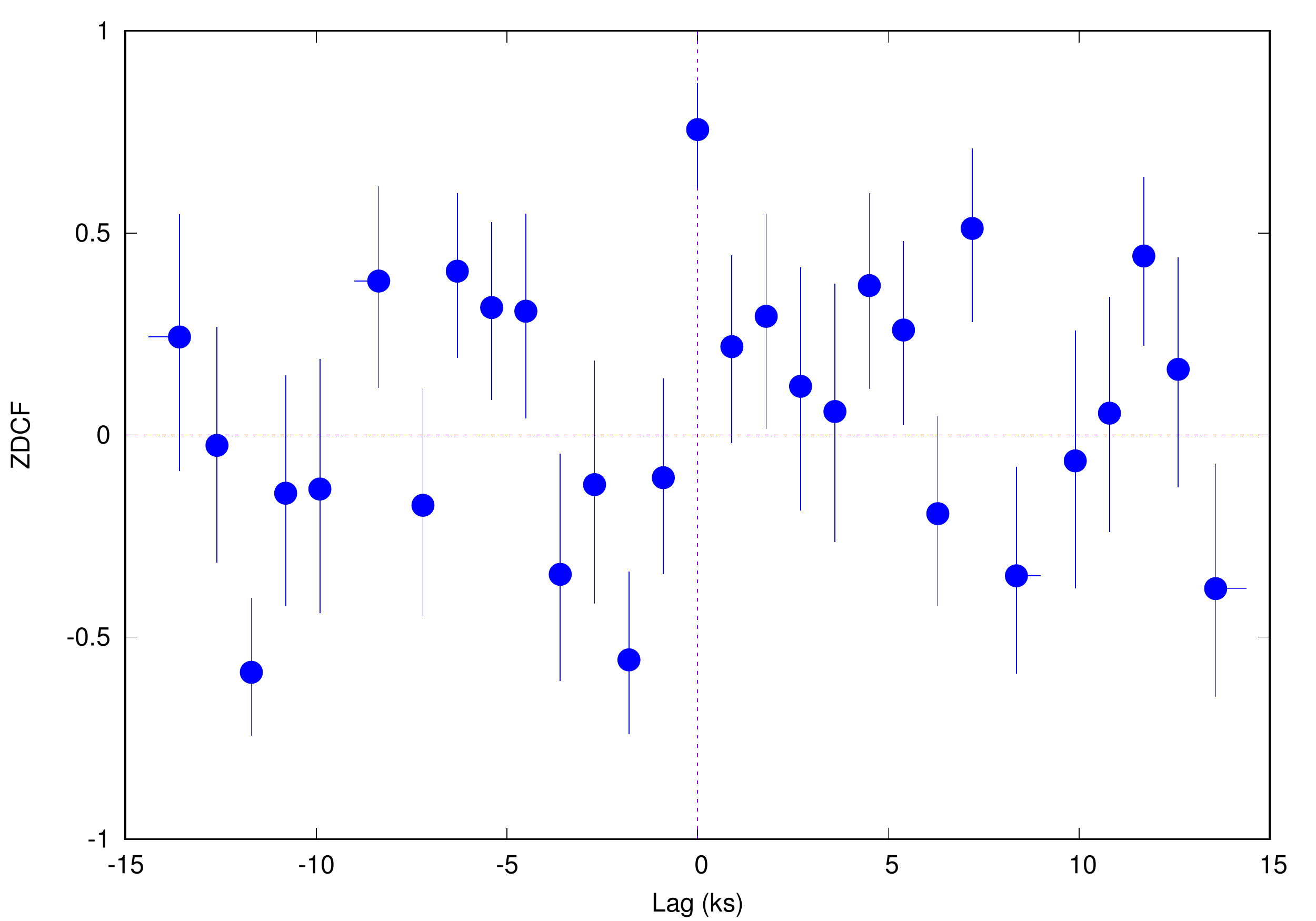}\par
    \includegraphics[width=0.73\linewidth, angle=-90]{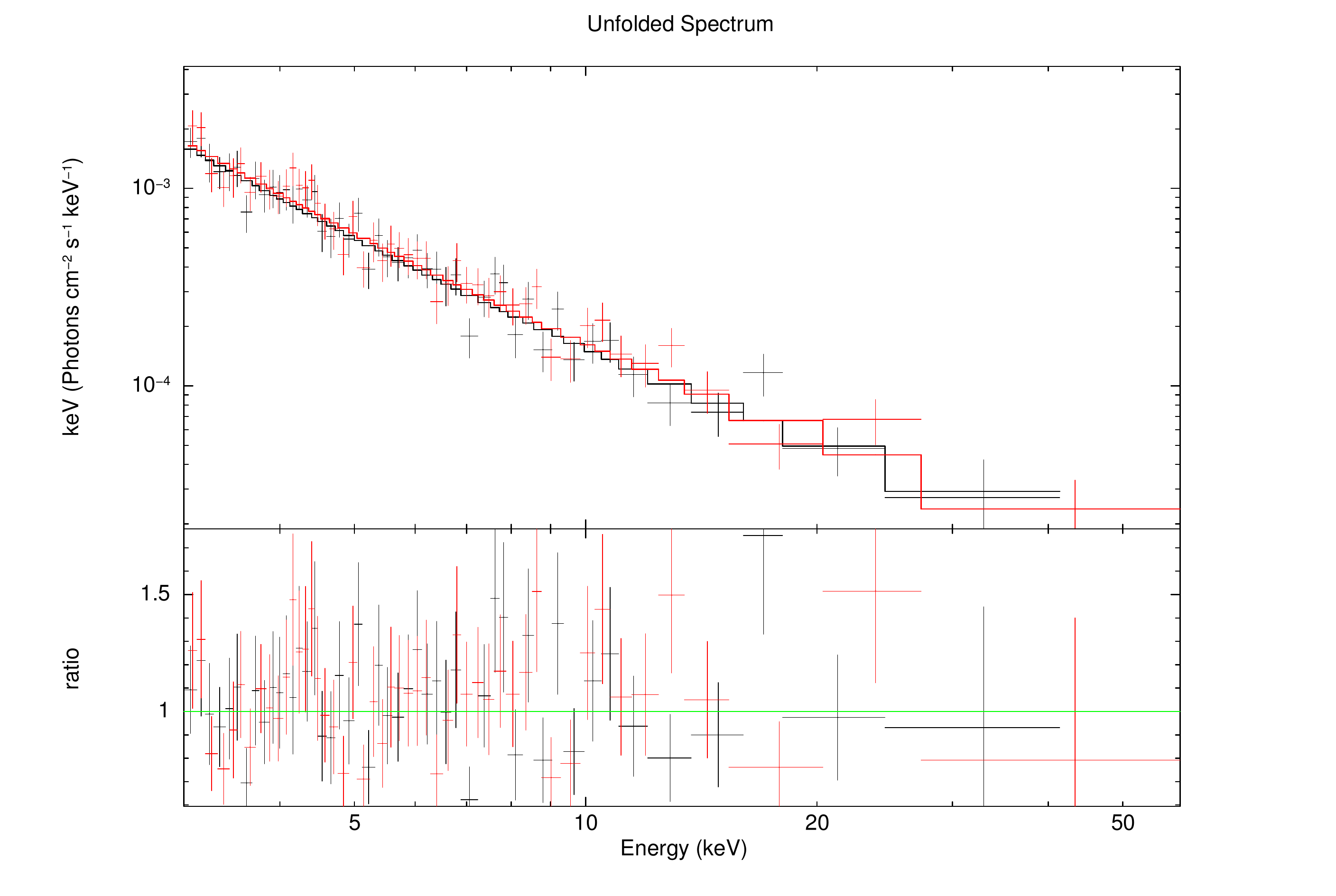}
    \end{multicols}
\center{PKS 2155--304, 60002022008 } \\
\caption{Same as in Fig \ref{fig:LC1}}
\label{fig:LC4}
\end{figure*}


\begin{figure*}
\begin{multicols}{4}
    \includegraphics[width=1.05\linewidth]{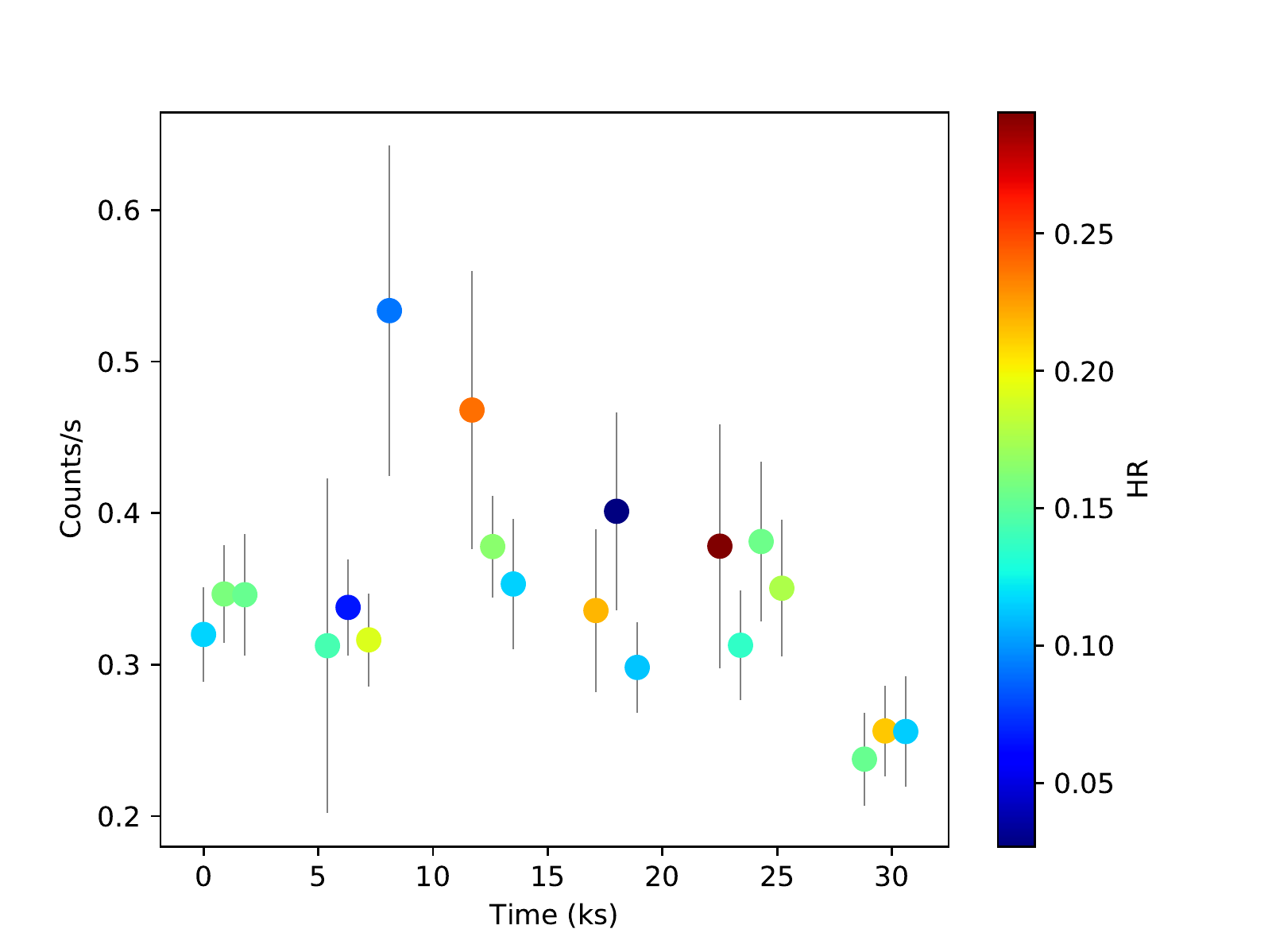}\par 
    \includegraphics[width=1.05\linewidth,angle=0]{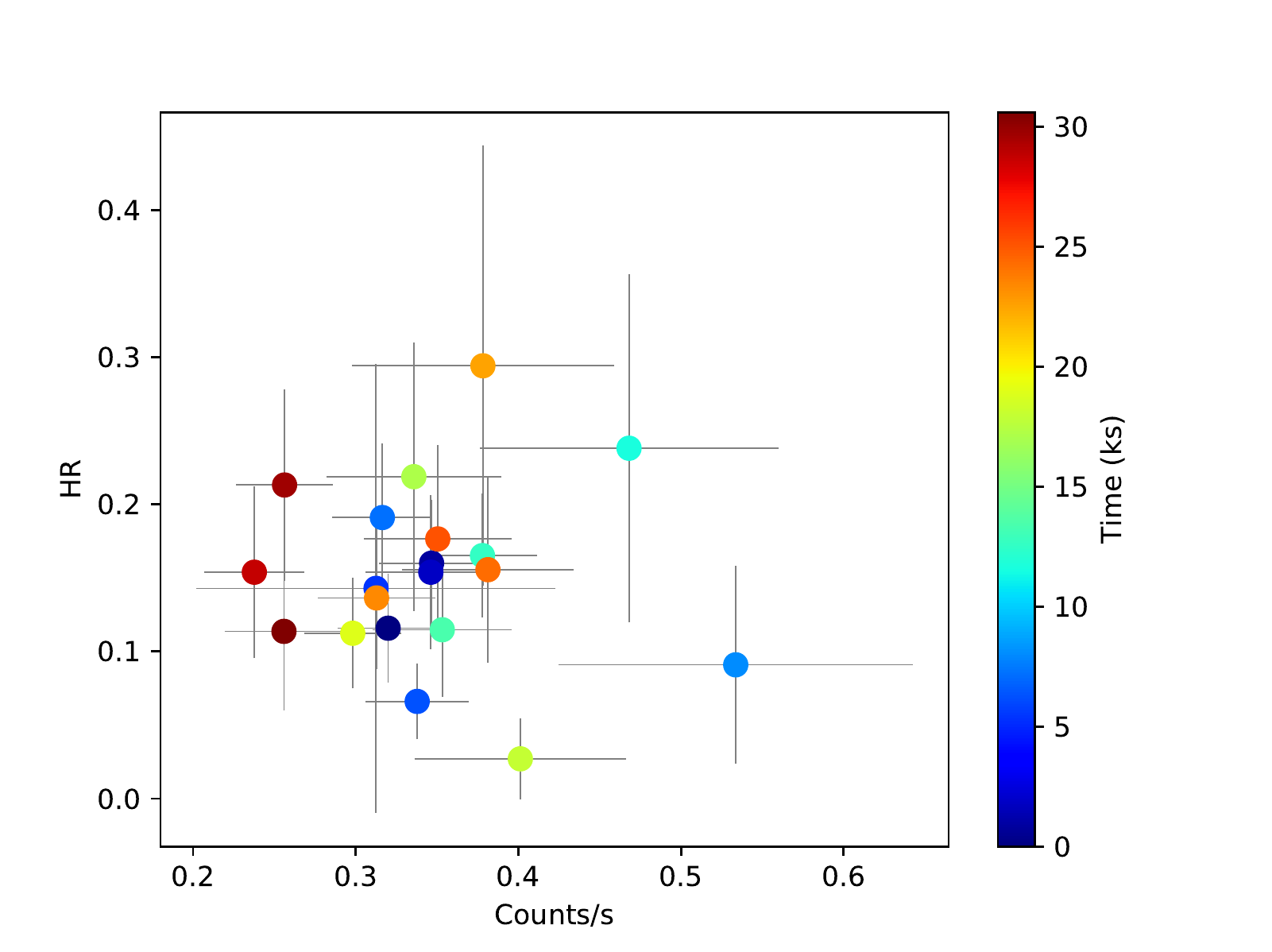}\par
    \includegraphics[width=0.72\linewidth,angle=-90]{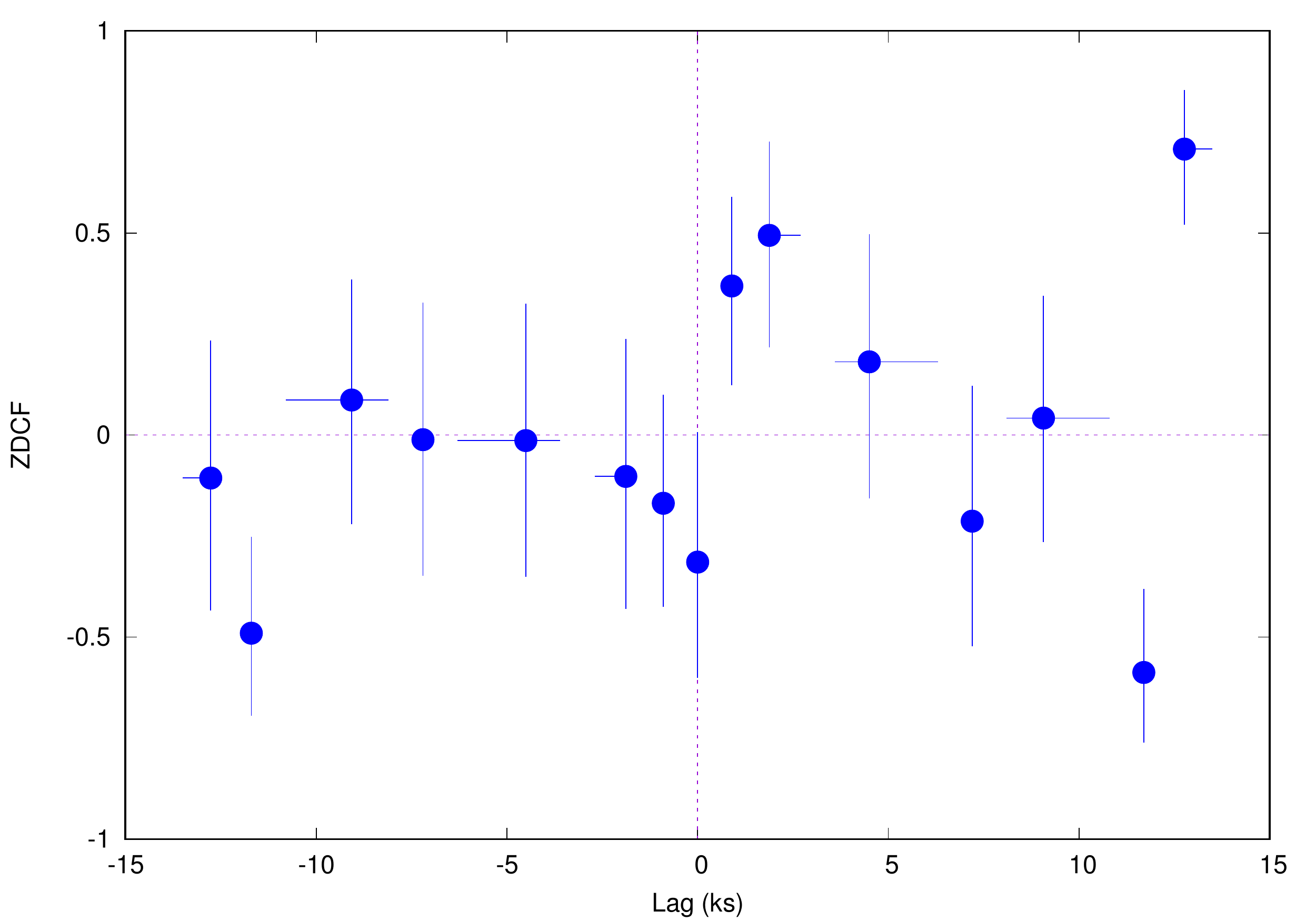}\par
    \includegraphics[width=0.73\linewidth, angle=-90]{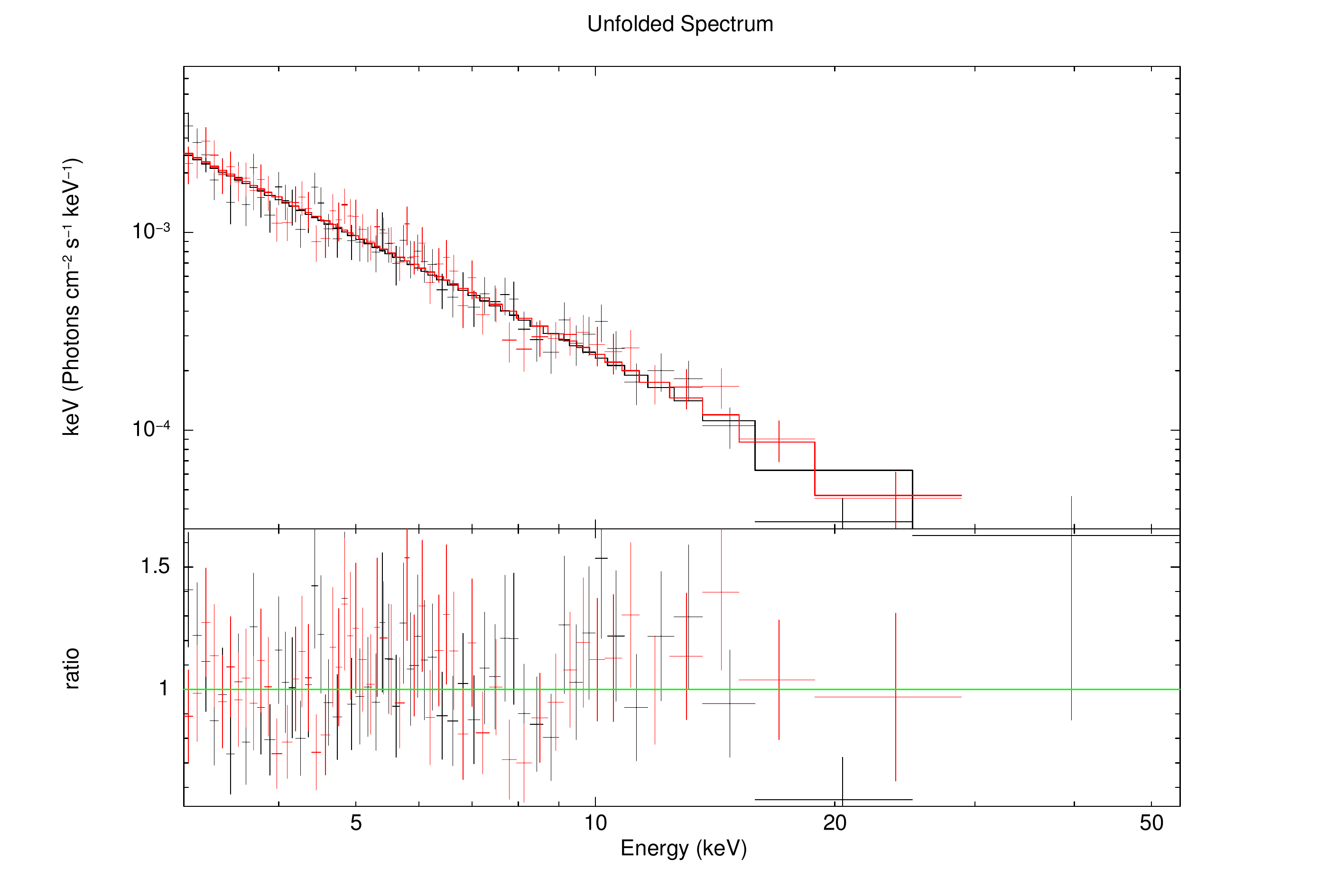}
     \end{multicols}
\center{PKS 2155--304, 60002022010 }

\begin{multicols}{4}
    \includegraphics[width=1.05\linewidth]{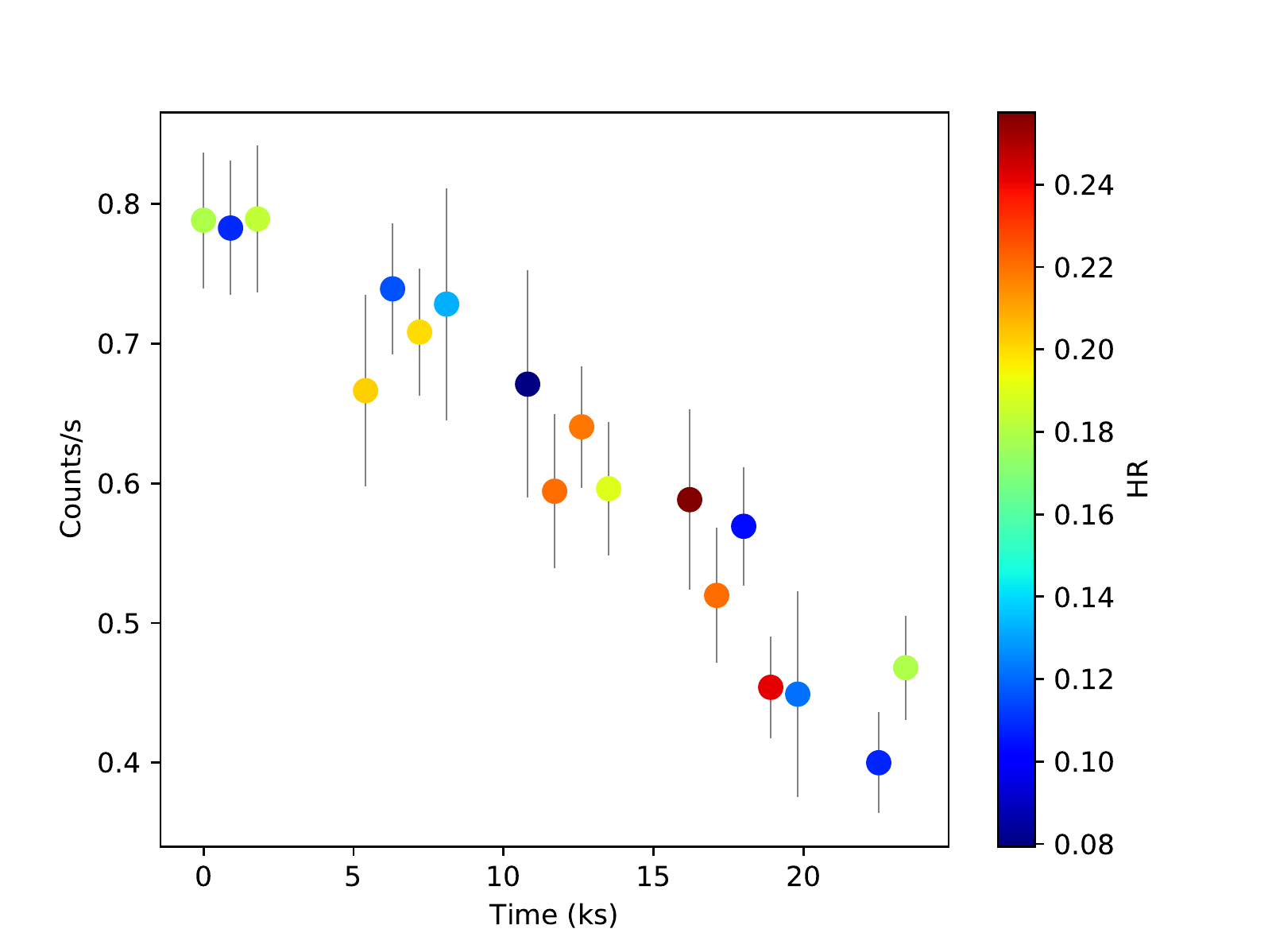}\par 
    \includegraphics[width=1.05\linewidth,angle=0]{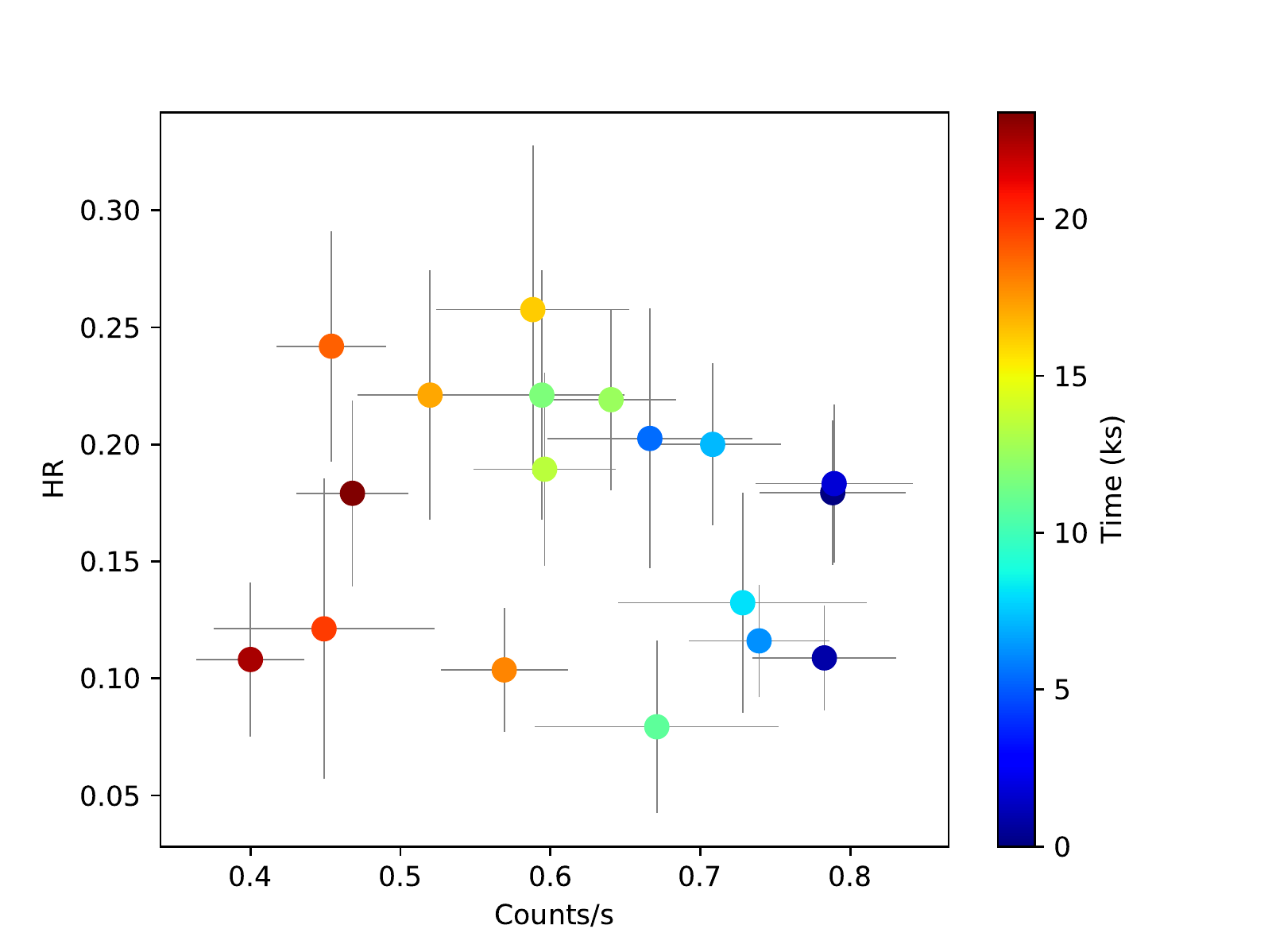}\par
    \includegraphics[width=0.72\linewidth,angle=-90]{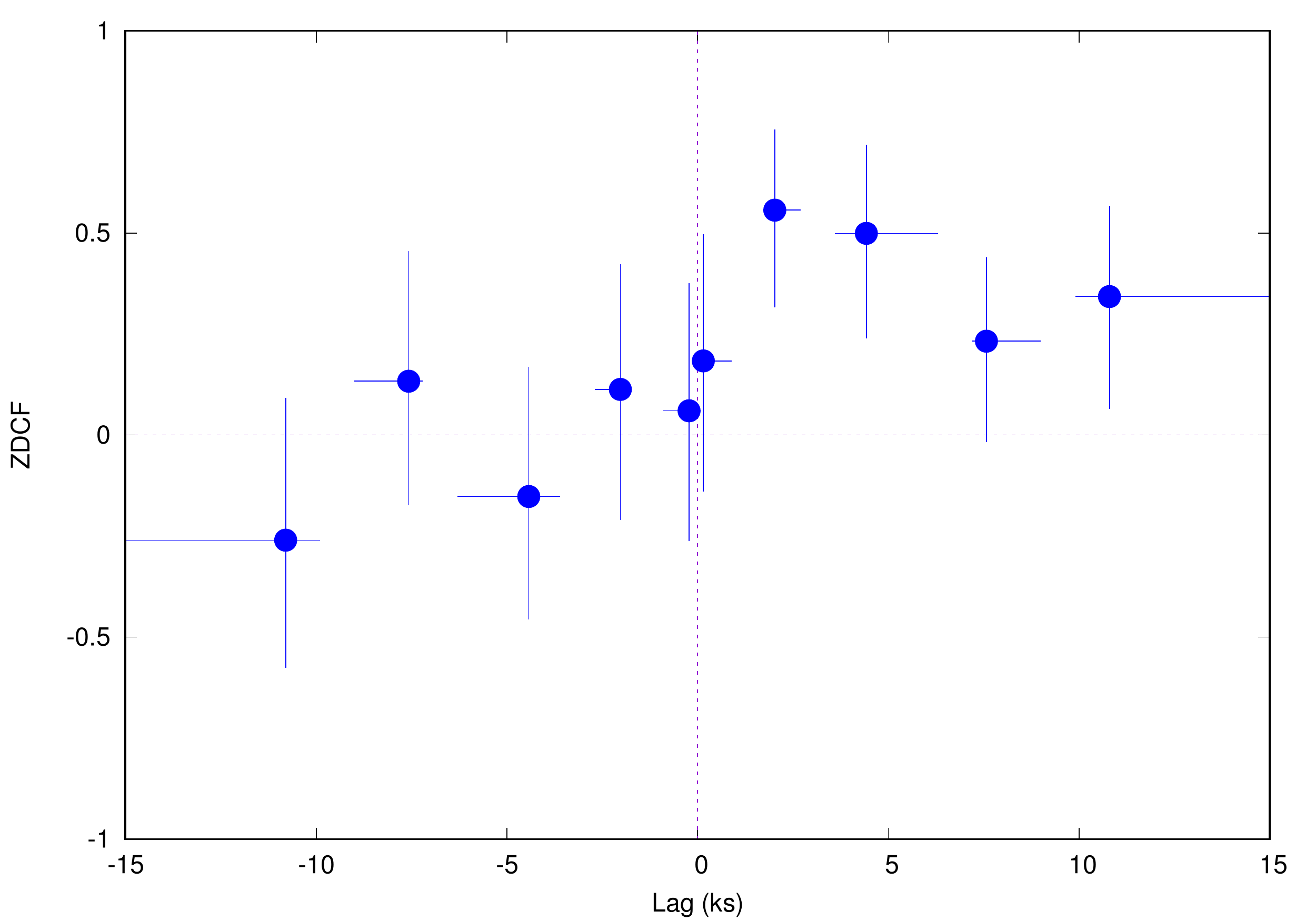}\par
    \includegraphics[width=0.73\linewidth, angle=-90]{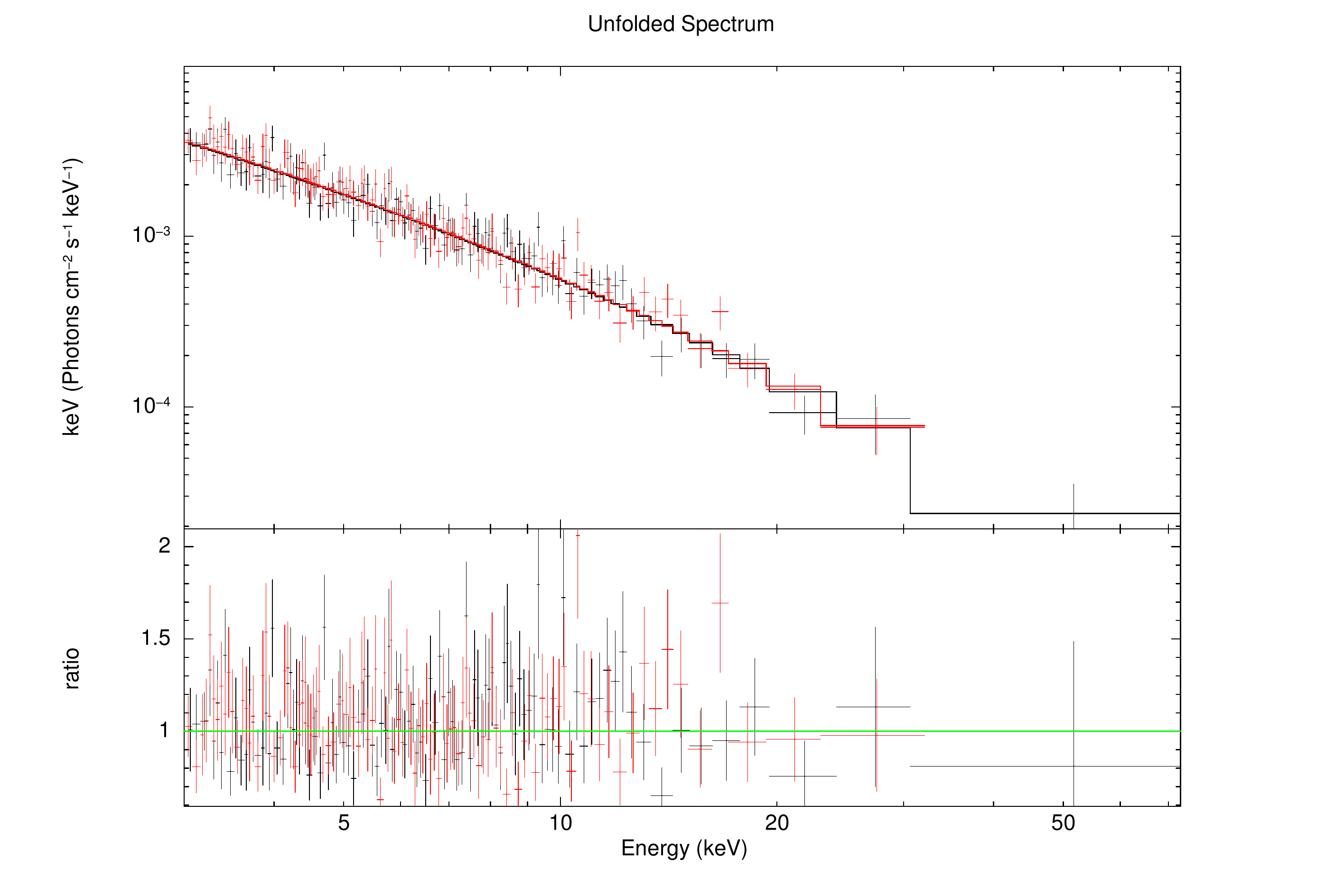}
    \end{multicols}
\center{ PKS 2155--304, 60002022012 }

\begin{multicols}{4}
     \includegraphics[width=1.05\linewidth]{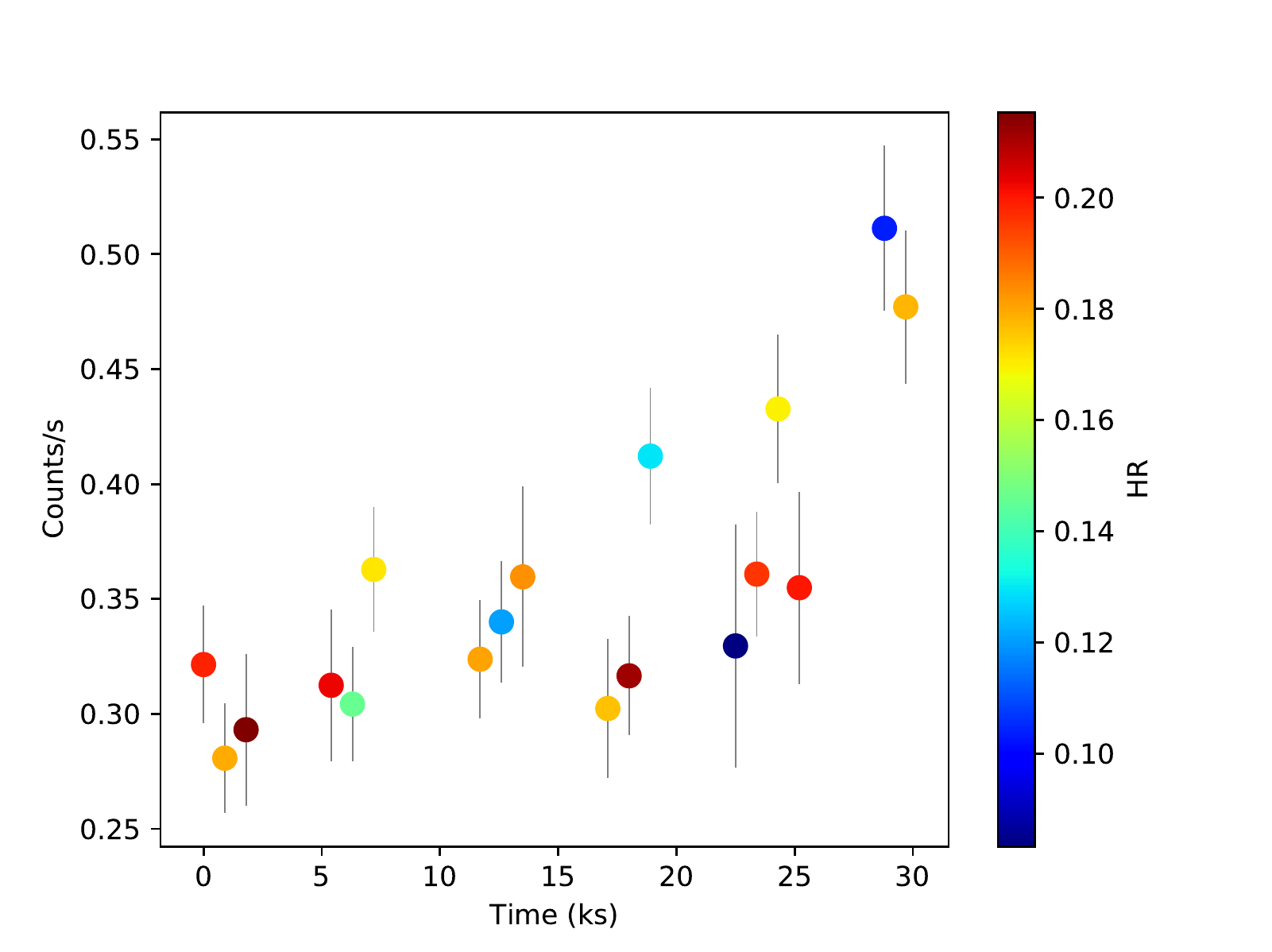}\par 
    \includegraphics[width=1.05\linewidth,angle=0]{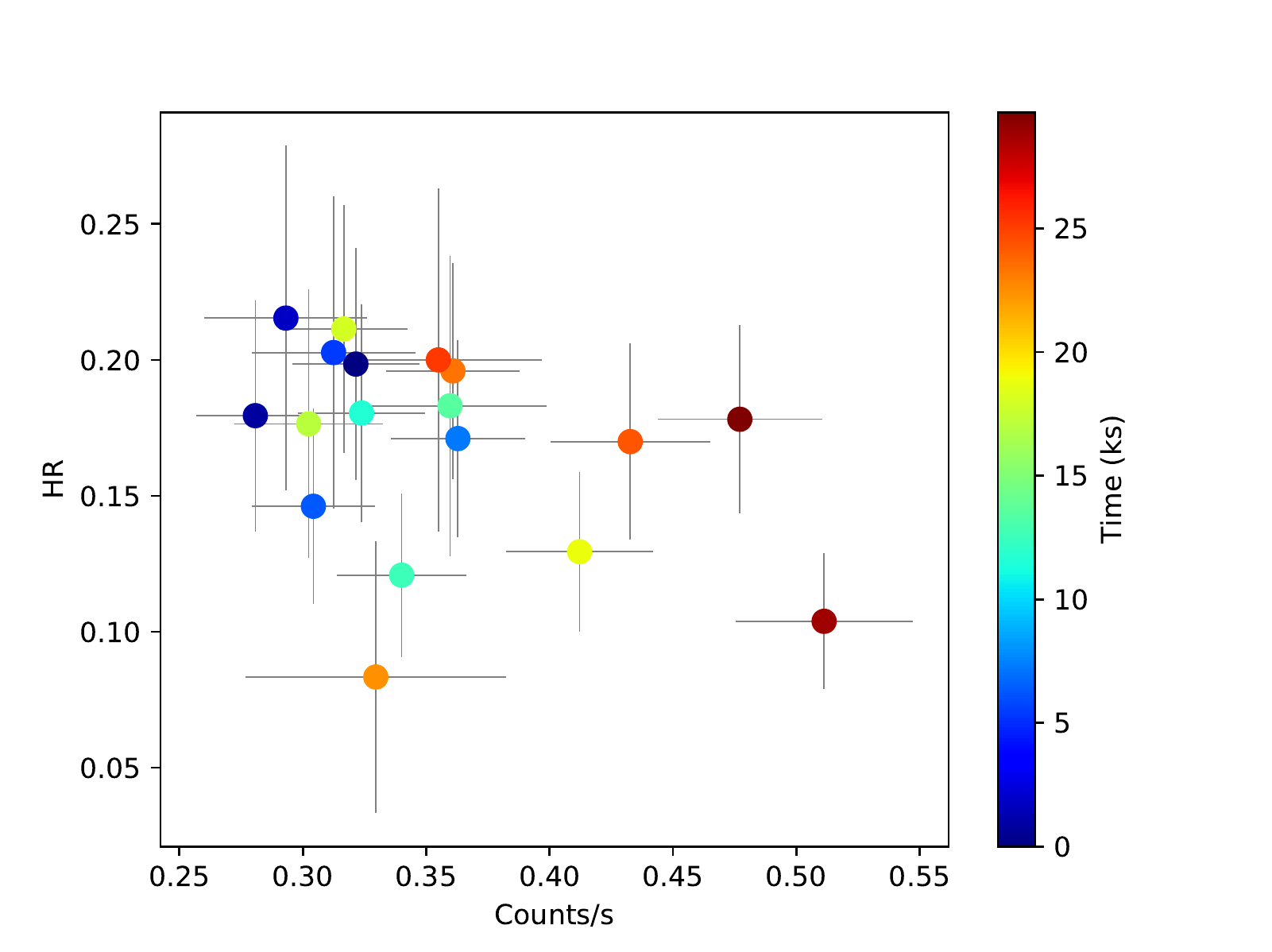}\par
    \includegraphics[width=0.72\linewidth,angle=-90]{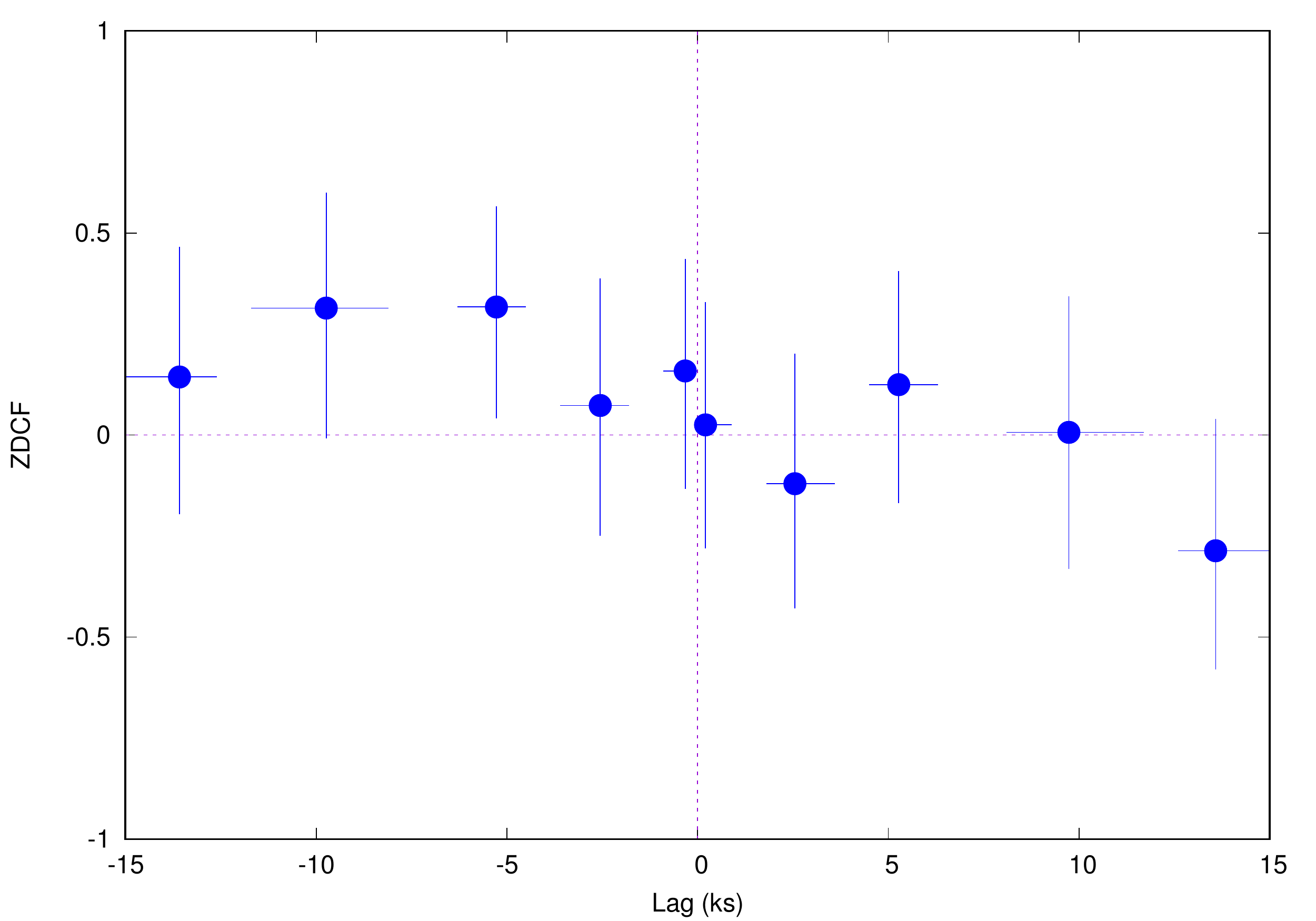}\par
    \includegraphics[width=0.73\linewidth, angle=-90]{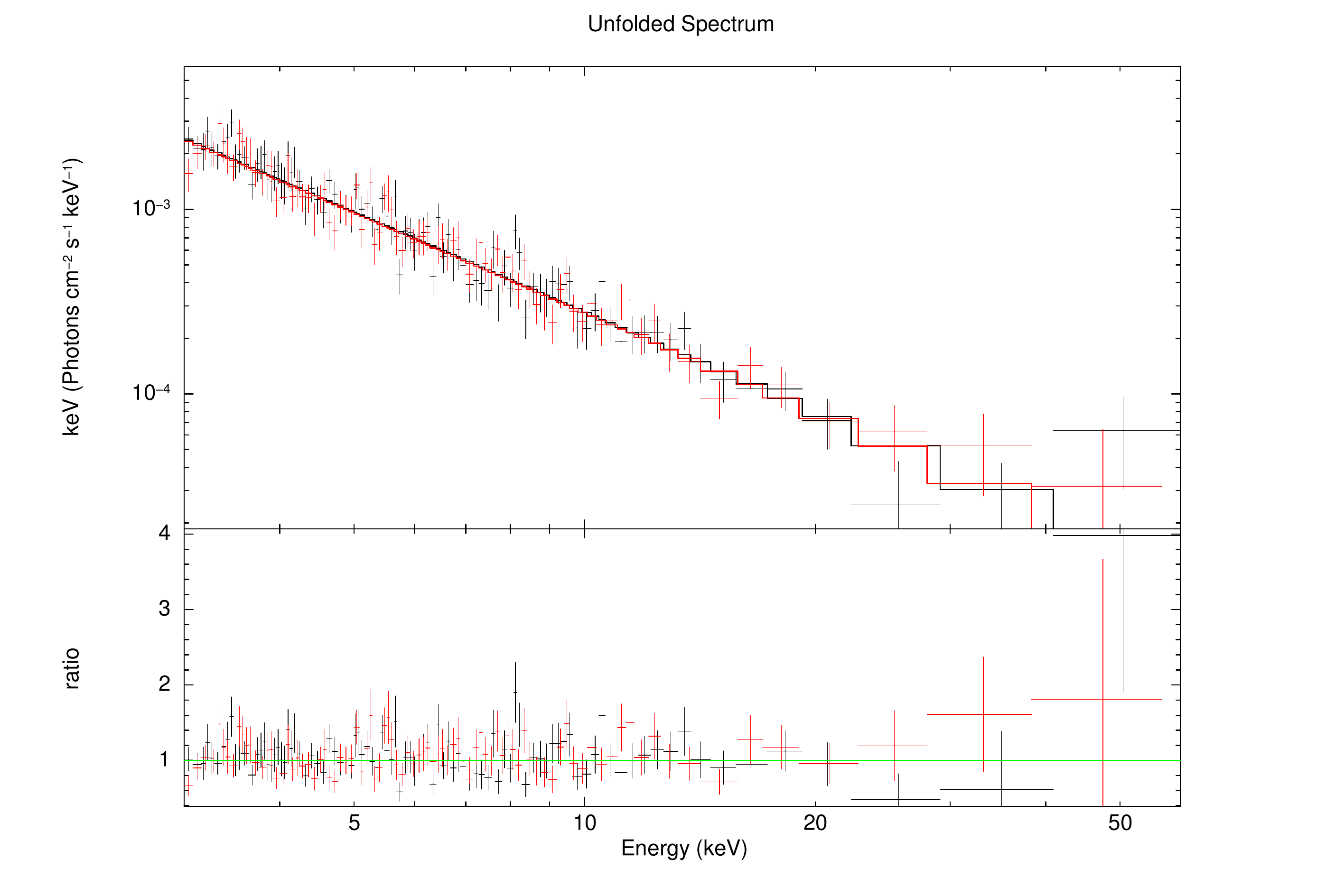}
    \end{multicols}
\center{PKS 2155--304, 60002022014}

\begin{multicols}{4}
    \includegraphics[width=1.05\linewidth]{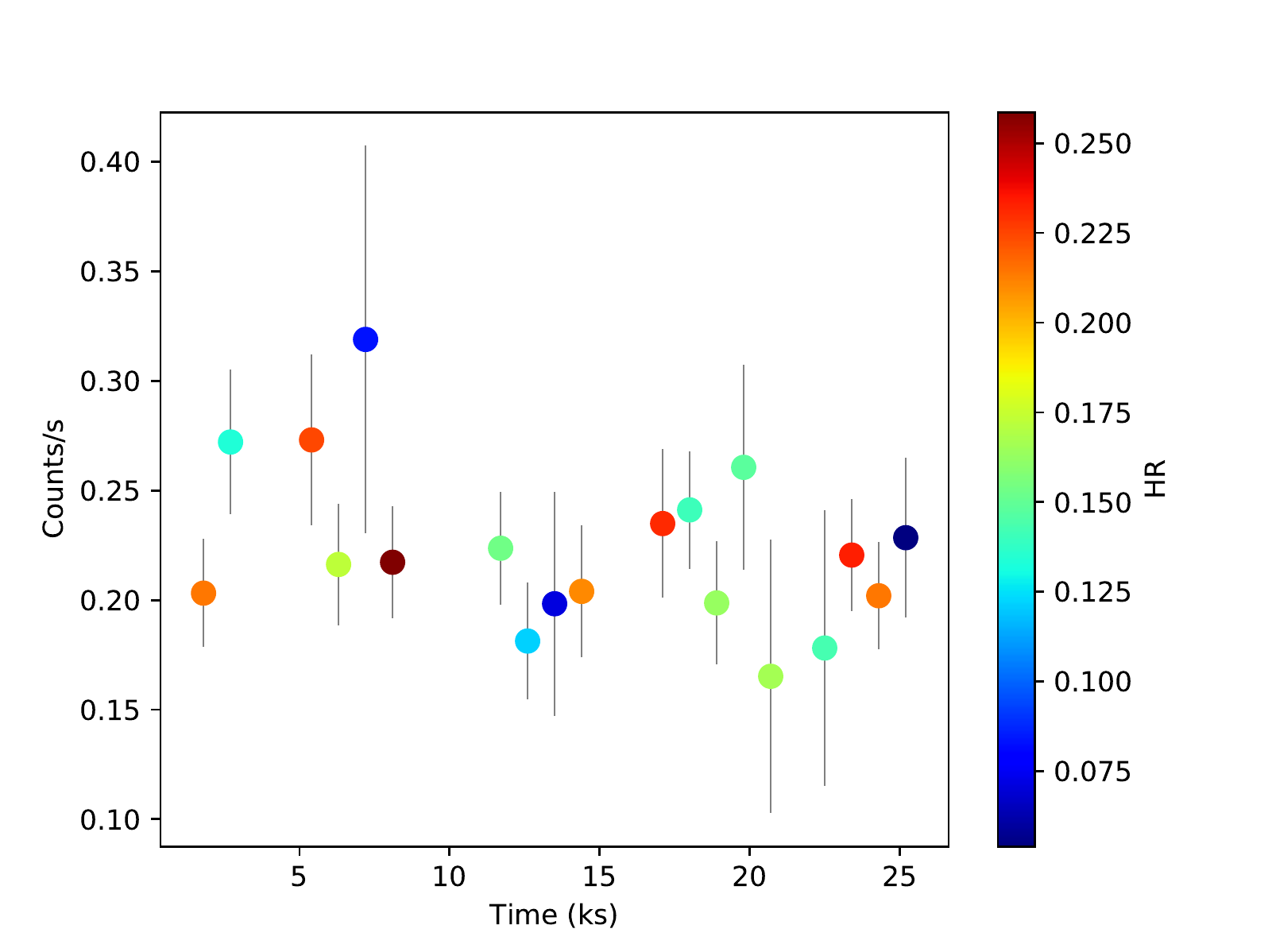}\par 
    \includegraphics[width=1.05\linewidth,angle=0]{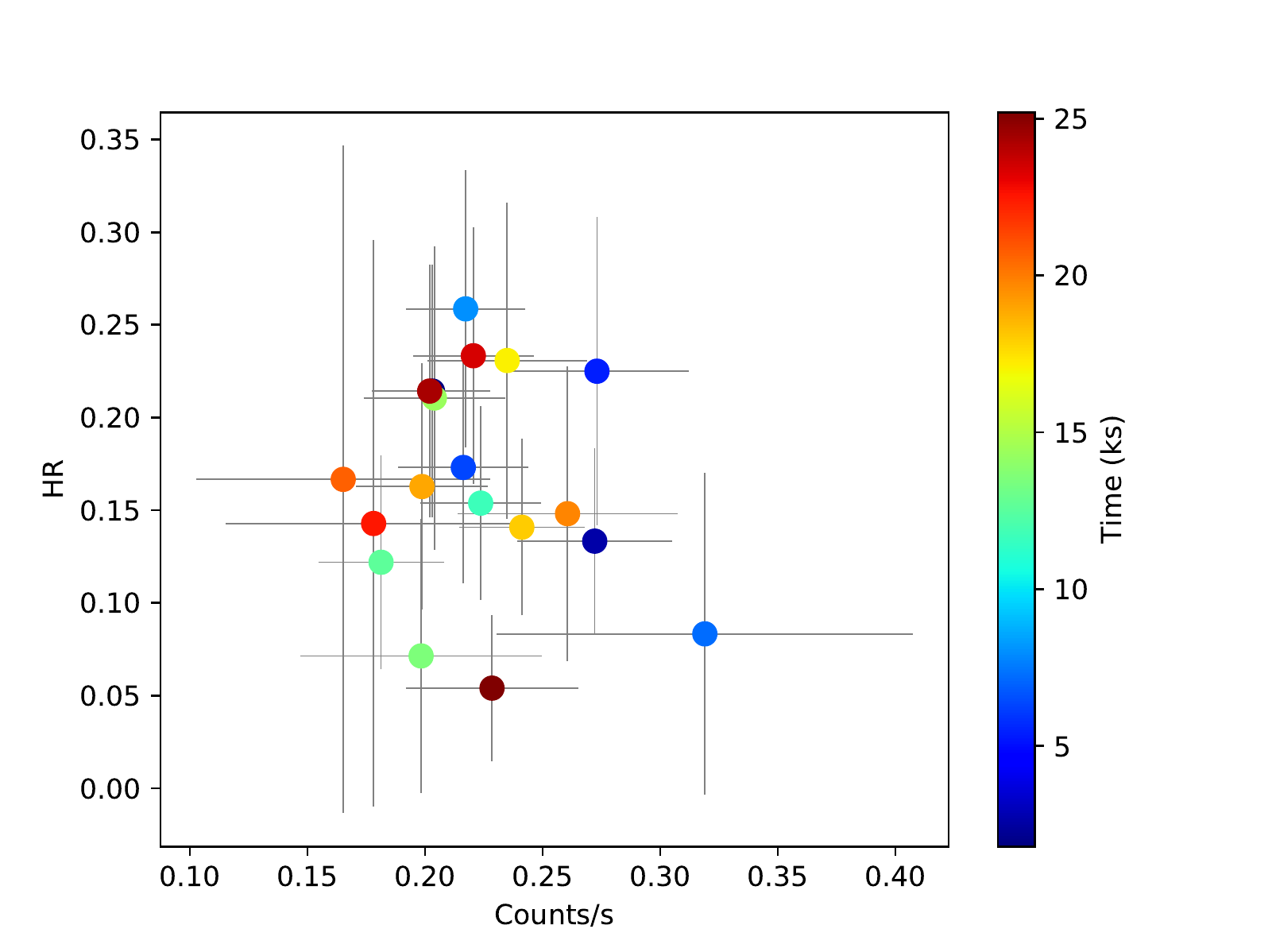}\par
    \includegraphics[width=0.72\linewidth,angle=-90]{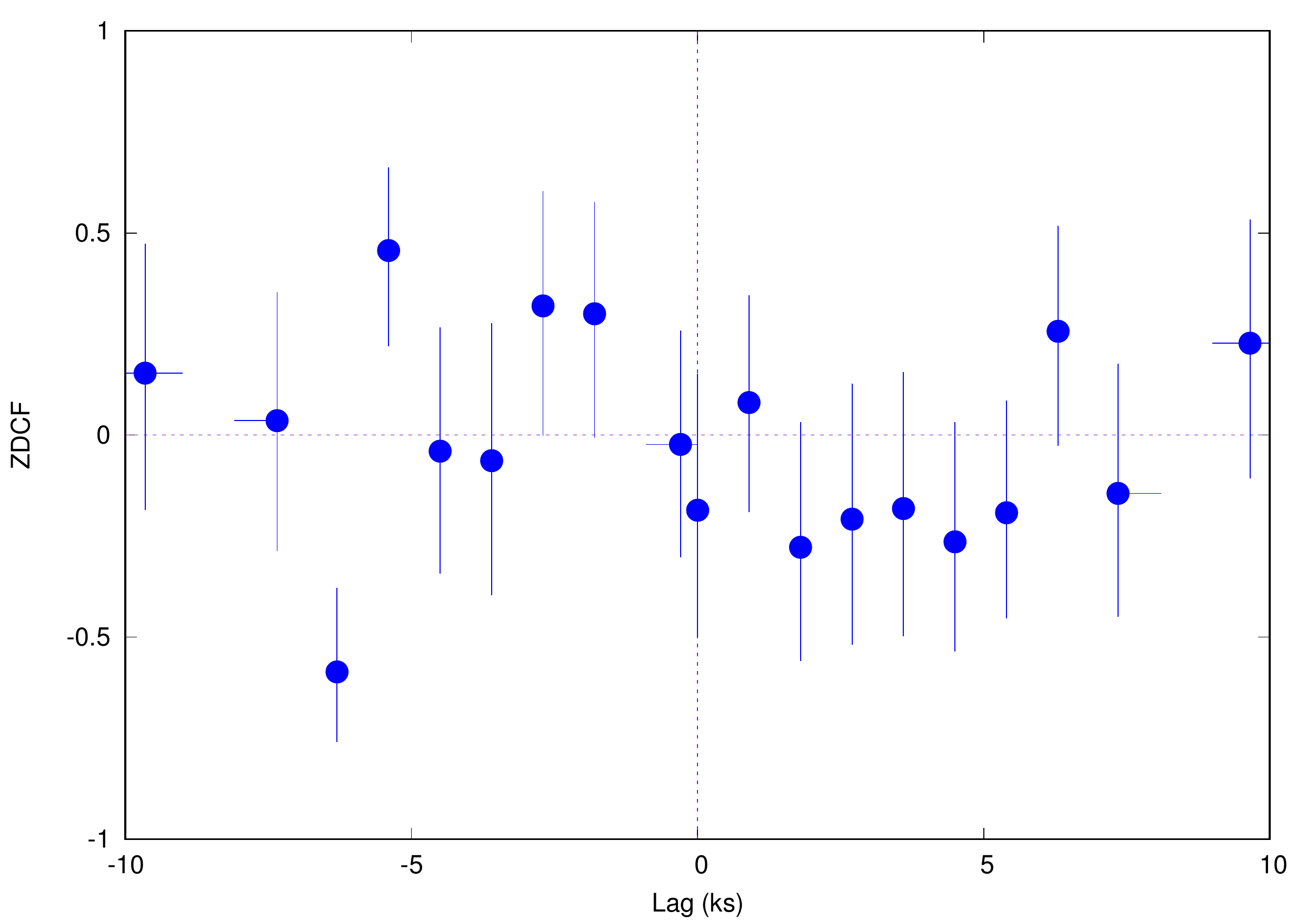}\par
    \includegraphics[width=0.73\linewidth, angle=-90]{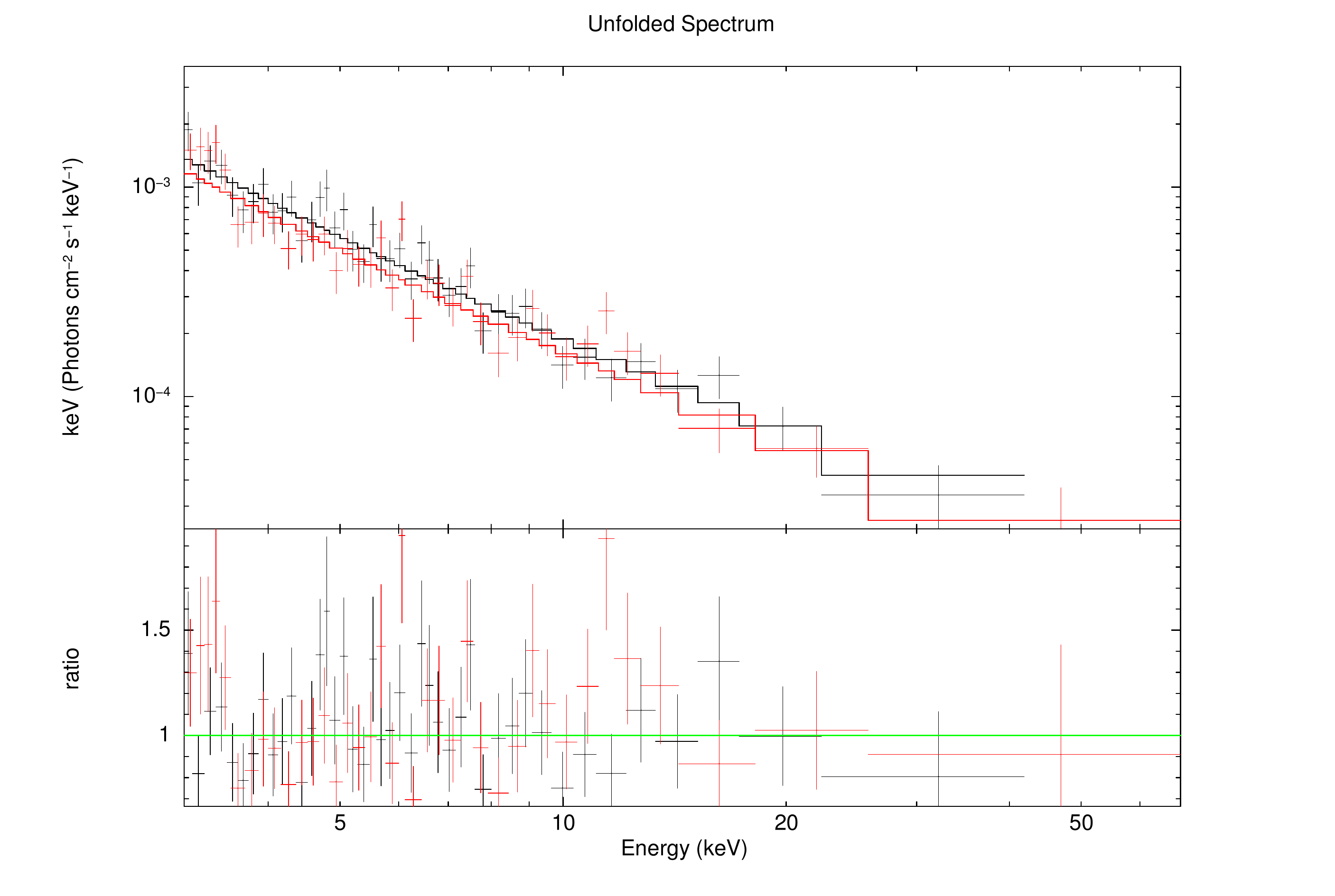}
    \end{multicols}
\center{PKS 2155--304, 60002022016 }

\begin{multicols}{4}
    \includegraphics[width=1.05\linewidth]{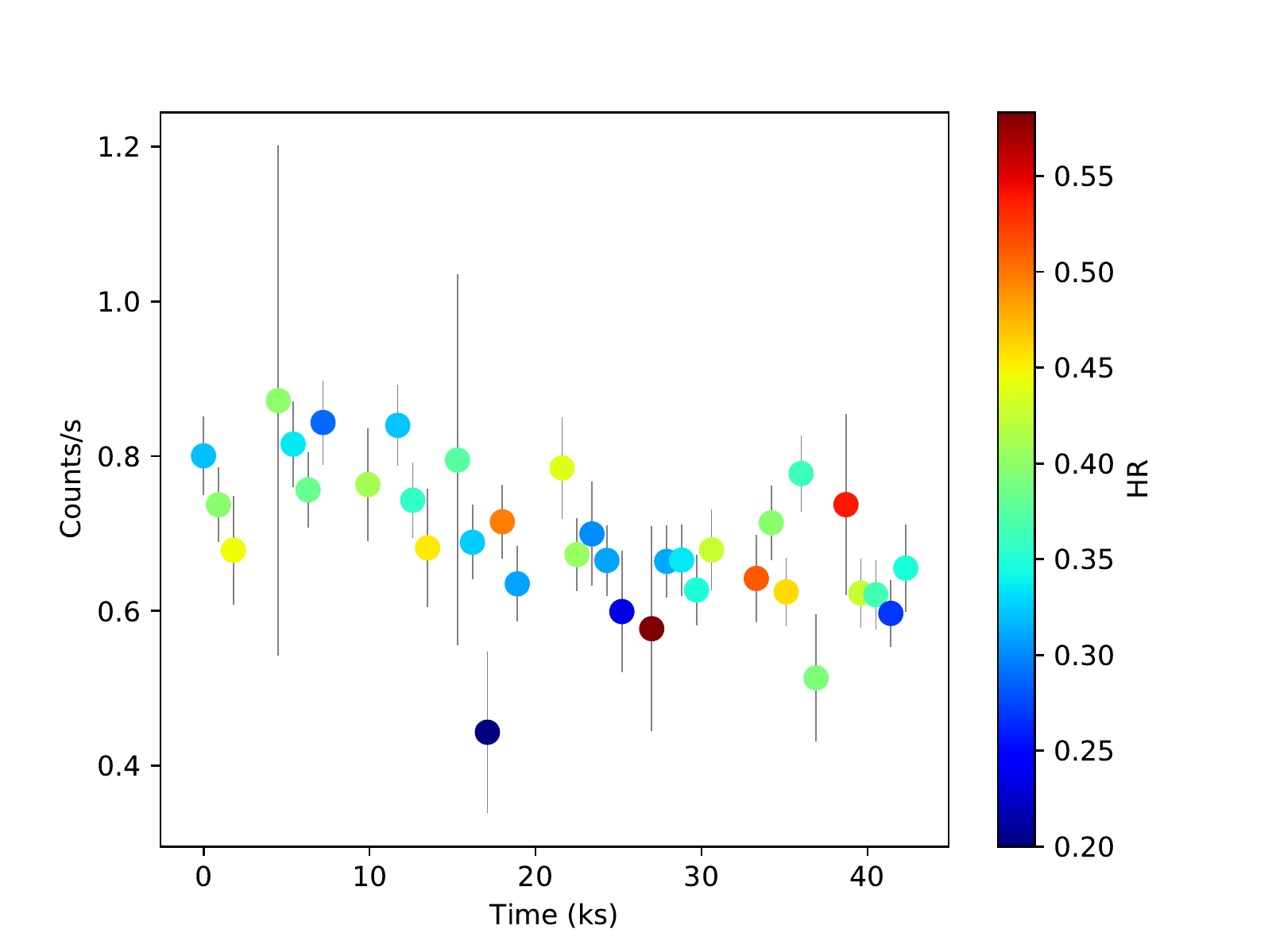}\par 
    \includegraphics[width=1.05\linewidth,angle=0]{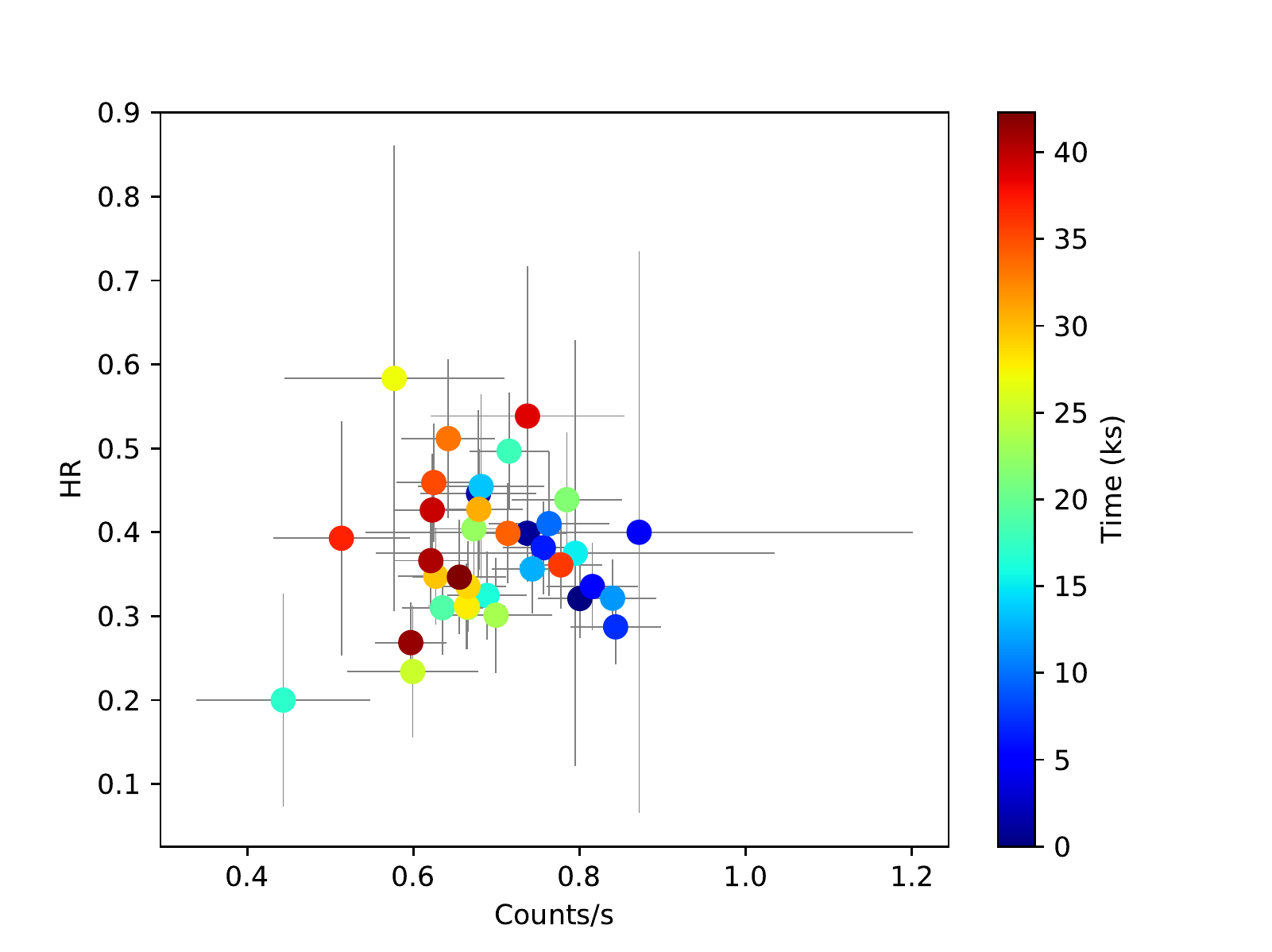}\par
    \includegraphics[width=0.72\linewidth,angle=-90]{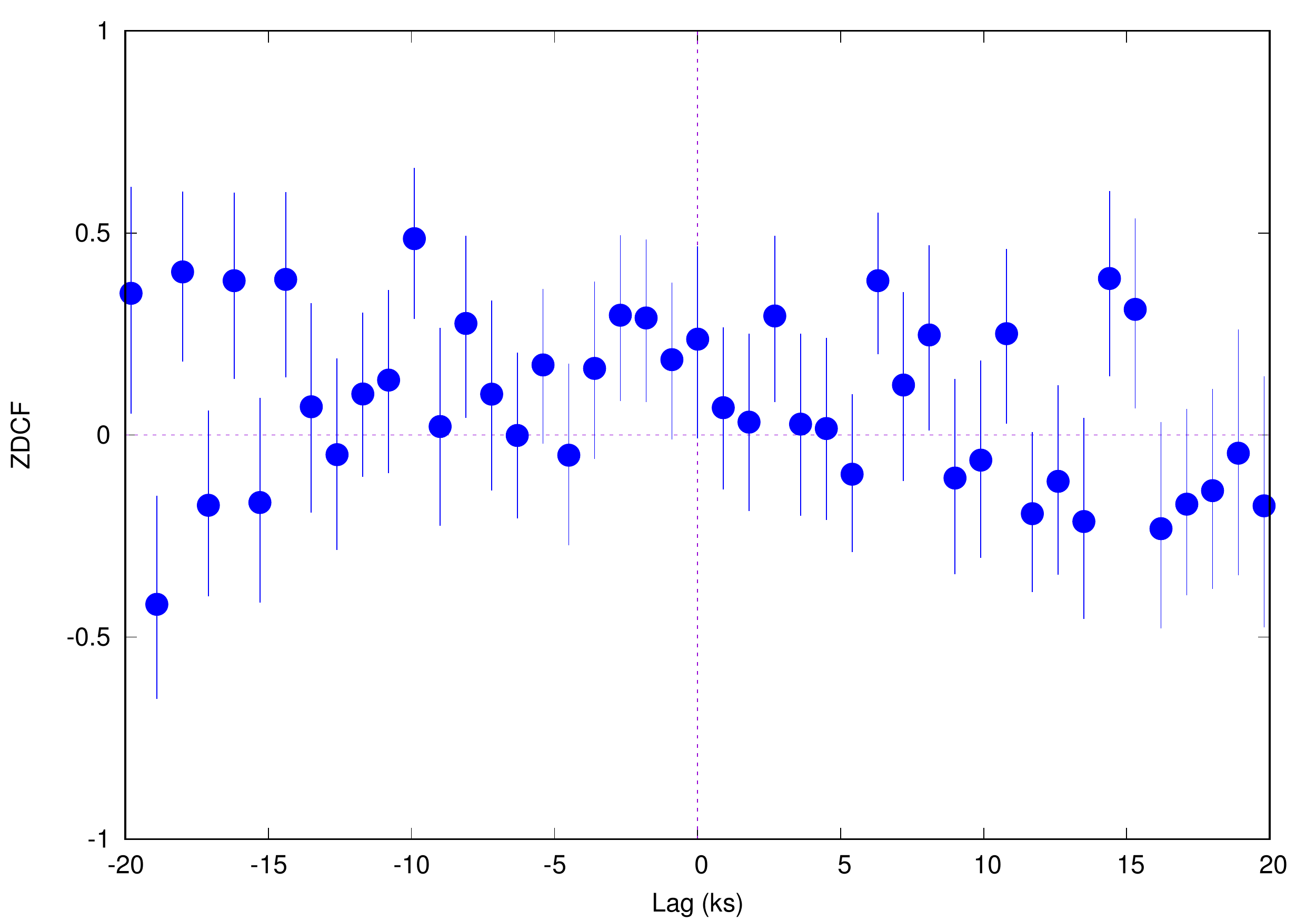}\par
    \includegraphics[width=0.73\linewidth, angle=-90]{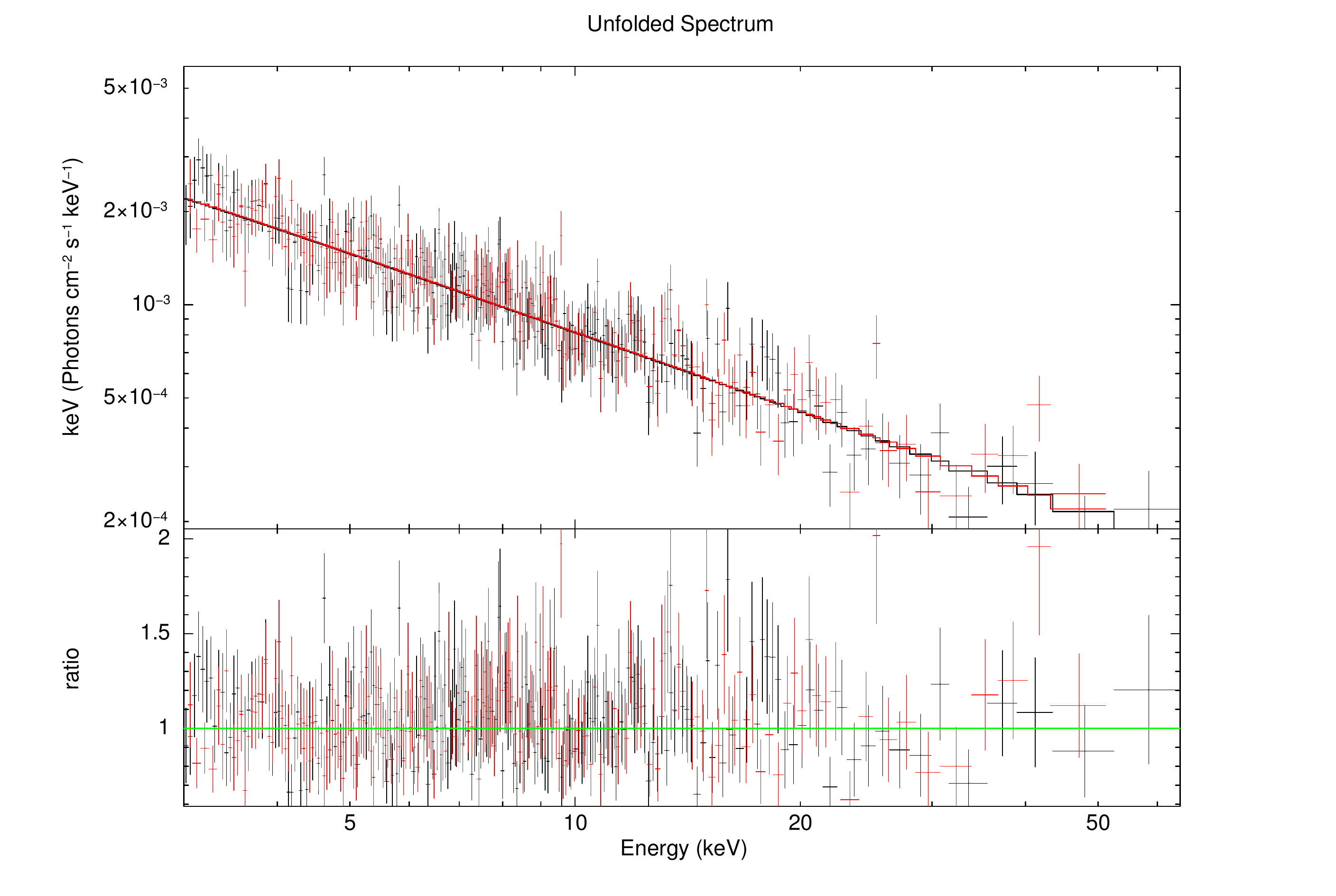}
    \end{multicols}
\center{BL Lac, 60001001002} \\
\caption{Same as in Fig \ref{fig:LC1}}
\label{fig:LC5}
\end{figure*}


\label{lastpage}
\end{document}